\documentclass[iop]{emulateapj}

\usepackage{graphicx}
\usepackage{color}
\usepackage{longtable}
\usepackage{lscape}

\def\l{\ensuremath{\lambda}}
\newcommand{\rsi}{$\mathcal{R}({\rm Si})$}
\newcommand{\rsife}{$\mathcal{R}({\rm Si,Fe})$}
\newcommand{\snia}{SN~Ia}
\newcommand{\sneia}{SN~Ia}

\let\ts=\thinspace
\newcommand{\one}{\ts {\,\sc i}}
\newcommand{\two}{\ts {\,\sc ii}}
\newcommand{\three}{\ts {\,\sc iii}}

\newcommand{\nifs}{\ensuremath{^{56}\rm{Ni}}}
\newcommand{\cofs}{\ensuremath{^{56}\rm{Co}}}
\newcommand{\fefs}{\ensuremath{^{56}\rm{Fe}}}
\newcommand{\msun}{\ensuremath{\rm{M}_{\odot}}}
\newcommand{\kms}{\ensuremath{\rm{km\,s}^{-1}}}
\newcommand{\kmsd}{\ensuremath{\rm{km\,s}^{-1}\,\rm{day}^{-1}}}

\newcommand{\dmft}{\ensuremath{\Delta m_{15}(B)}}

\def\nsntot{462}
\def\nspectot{2603}
\def\nspecfast{2447}
\def\fracspecfast{$\sim94$}

\def\fracspecfastperrymike{$\sim72$}

\def\nfastobservers{79}
\def\nspecbc{131}
\def\nspecrc{3}
\def\nspecldss{21}

\def\fracspecmultstd{50}
\def\nspecmmt9397{37}
\def\nsnnotmax{191}
\def\nsntmax{271}
\def\nsnwlcnotmax{24}

\def\nspecnotmax{397}
\def\nspectmax{2206}
\def\nspecltm10{43}
\def\nspecwithin3d{381}
\def\nsnwspecltm10{23}
\def\nsnwspecltmax{168}
\def\nspecgt150{21}
\def\nsnwspecgt150{12}
\def\nsnwtenspec{92}
\def\nsnwtwentyspec{20}
\def\nsnzgt0p015{361}
\def\nsnnot{26}
\def\nsnnobg{34}
\def\nsndmft{247}
\def\nsndmftbayesn{185}
\def\nsndmftsnpy{23}
\def\nsndmftother{39}
\def\nsnbranchclassfig{218}
\def\nspecunpub{2065}
\def\nsnunpub{406}
\def\nsnexcl{10}
\def\nspecexcl{11}
\def\nsnonespec{149}
\def\nsnwtwospec{313}
\def\meannspecpersnwtwospec{8}

\slugcomment{Accepted for publication in AJ}
\shorttitle{Diversity of SN~I\lowercase{a} Spectra}
\shortauthors{Blondin et al.}

\begin{document}

\title{The Spectroscopic Diversity of Type I\lowercase{a} Supernovae\altaffilmark{1}}

\author{
S.~Blondin,\altaffilmark{2} 
T.~Matheson,\altaffilmark{3}
R.~P.~Kirshner,\altaffilmark{4}
K.~S.~Mandel,\altaffilmark{4,5}
P.~Berlind,\altaffilmark{6}
M.~Calkins,\altaffilmark{6}
P.~Challis,\altaffilmark{4}
P.~M.~Garnavich,\altaffilmark{7}
S.~W.~Jha,\altaffilmark{8}
M.~Modjaz,\altaffilmark{9}
A.~G.~Riess,\altaffilmark{10,11}
B.~P.~Schmidt\altaffilmark{12}
}

\nopagebreak

\altaffiltext{1}{\vspace{0.00cm}
Based in part on observations obtained at the F.~L. Whipple
Observatory, which is operated by the Smithsonian Astrophysical
Observatory, the MMT Observatory, a joint facility of the
Smithsonian Institution and the University of Arizona, and with the
6.5 meter Magellan Telescopes located at Las Campanas Observatory,
Chile.}

\altaffiltext{2}{\vspace{0.00cm}
  Centre de Physique des Particules de Marseille,
  Aix-Marseille Universit\'e, CNRS/IN2P3, 163 avenue de Luminy, 13288
  Marseille Cedex 9, France; {blondin@cppm.in2p3.fr}}

\altaffiltext{3}{\vspace{0.00cm}
  National Optical Astronomy Observatory, 950 North
  Cherry Avenue, Tucson, AZ 85719, USA.}

\altaffiltext{4}{\vspace{0.00cm}
  Harvard-Smithsonian Center for Astrophysics, 60
  Garden Street, Cambridge, MA 02138, USA.}

\altaffiltext{5}{\vspace{0.00cm}
  Imperial College London, Blackett Laboratory, Prince
  Consort Road, London SW7 2AZ, UK.}

\altaffiltext{6}{\vspace{0.00cm}
  F.~L.~Whipple Observatory, 670 Mt. Hopkins Road,
  P.O. Box 97, Amado, AZ 85645, USA.}

\altaffiltext{7}{\vspace{0.00cm}
  Department of Physics, University of Notre Dame, 225 Nieuwland
  Science Hall, Notre Dame, IN 46556, USA.}

\altaffiltext{8}{\vspace{0.00cm}
  Department of Physics and Astronomy, Rutgers University,
  136 Frelinghuysen Road, Piscataway, NJ 08854, USA.}

\altaffiltext{9}{\vspace{0.00cm}
  New York University, Center for Cosmology and
  Particle Physics, 4 Washington Place, New York, NY 10003, USA.}

\altaffiltext{10}{\vspace{0.00cm}
  Department of Physics and Astronomy, Johns Hopkins University,
  Baltimore, MD 21218, USA.}

\altaffiltext{11}{\vspace{0.00cm}
  Space Telescope Science Institute, 3700 San Martin Drive, Baltimore,
  MD 21218, USA.}

\altaffiltext{12}{\vspace{0.00cm}
  The Research School of Astronomy and Astrophysics, The Australian
  National University, Mount Stromlo and Siding Spring Observatories,
  via Cotter Road, Weston Creek, PO 2611, Australia.}


\begin{abstract}
We present \nspectot\ spectra of \nsntot\ nearby Type Ia supernovae
(\sneia), including \nspecunpub\ previously-unpublished spectra,
obtained during 1993-2008 through the Center for Astrophysics
Supernova Program. There are on average \meannspecpersnwtwospec\
spectra for each of the \nsnwtwospec\ \sneia\ with at least two
spectra. Most of the spectra were obtained with the FAST spectrograph at
the FLWO 1.5\,m telescope and reduced in a consistent manner, making
this data set well suited for studies of \snia\ spectroscopic
diversity. Using additional data from the literature, we study the
spectroscopic and photometric properties of \sneia\ as a function of
spectroscopic class using the classification schemes of Branch et
al. and Wang et al. The width-luminosity relation appears to be
steeper for \sneia\ with broader lines, although the result is not
statistically significant with the present sample. Based on the
evolution of the characteristic 
Si\two\,\l6355 line, we propose improved methods for measuring
velocity gradients, revealing a larger range than previously
suspected, from $\sim0$ to $\sim400$\,\kmsd\ considering the
instantaneous velocity decline rate at maximum light. We find a
weaker and less significant correlation between Si\two\ velocity and
intrinsic $B-V$ color
at maximum light than reported by Foley et al., owing to a more
comprehensive treatment of uncertainties and host galaxy dust.
We study the extent of nuclear burning and the presence of unburnt
carbon in the outermost layers of the ejecta, and report new
detections of C\two\,\l6580 in 23 early-time \snia\ spectra.
The frequency of C\two\ detections is not higher in \sneia\ with
bluer colors or narrower light curves, in conflict with
the recent results of Thomas et al.
Based on nebular spectra of 27 \sneia, we find no relation
between the FWHM of the iron emission feature at $\sim4700$\,\AA\ and
\dmft\ after removing the two low-luminosity SN~1986G and
SN~1991bg, suggesting that the peak luminosity is not
strongly dependent on the kinetic energy of the explosion for most
\sneia. 
Finally, we confirm the correlation of velocity shifts in some
nebular lines with the intrinsic $B-V$ color of \sneia\ at maximum
light, although several outliers suggest a possible
non-monotonic behavior for the largest blueshifts.
\end{abstract}



\section{Introduction}\label{sect:intro}

Type Ia supernovae (\sneia) have received much attention due to their
use as distance indicators on cosmological scales and the subsequent
discovery of cosmic acceleration \citep{R98,P99}. Their
astrophysical nature, however, remains perplexing to this day. The
standard model of \sneia\ involves the thermonuclear disruption of a
carbon-oxygen (C/O) white dwarf (WD) star
(\citealt{Hoyle/Fowler:1960}; this has been confirmed by recent
  observations of the nearby Type Ia SN~2011fe in M101;
  \citealt{Nugent/etal:2011,Bloom/etal:2012}) as it 
approaches the Chandrasekhar mass ($M_{\rm
  Ch}\approx1.4$\,M$_{\sun}$). To do so the WD either accretes
material from a non-degenerate binary companion (the
``single-degenerate'' scenario), or merges with another WD star (the
``double-degenerate'' scenario;
\citealt{Iben/Tutukov:1984,Webbink:1984}). Both scenarios constitute
plausible progenitor channels, and can accomodate variations in
the total mass of the binary system (i.e., sub- and super-$M_{\rm Ch}$;
see \citealt{Howell:2011} for a recent review).

The favored explosion mechanism involves the transition from a
turbulent subsonic burning front propagating via thermal conductivity
(known as a deflagration; e.g., \citealt{W7}) to a supersonic
detonation propagating via a strong shock \citep{Khokhlov:1991}. The
deflagration phase synthesizes iron-peak elements, of which \nifs\ is
needed to power the light curve through the
\nifs$\rightarrow$\cofs$\rightarrow$\fefs\ decay chain. It also
pre-expands the WD so that the subsequent detonation burns the
leftover C/O fuel at lower densities and synthesizes sufficient
amounts of intermediate-mass elements (IME; e.g., Si, S, Ca) at high
expansion velocities, needed to reproduce observed spectra
\citep[e.g.,][]{Branch/etal:1982}. Recent multi-dimensional
simulations highlight the importance of hydrodynamical instabilities
during the deflagration phase
\citep[e.g.,][]{Gamezo/Khokhlov/Oran:2005,Roepke/Niemeyer:2007}, and
provide a physical basis for the deflagration-to-detonation transition
(DDT; e.g., \citealt{Woosley/etal:2009}), although this DDT is
artificially triggered based on pre-defined criteria (see, e.g.,
\citealt{KRW09}; but see \citealt{Poludnenko/etal:2011} for
3D simulations of spontaneous DDTs in unconfined media).

Since the first spectrophotometric measurements of a \snia\
(SN~1972E; \citealt{Kirshner/etal:1973}),
detailed observational studies of individual objects have revealed a
wide range of photometric and spectroscopic properties, including
luminous 1991T/1999aa-like events
\citep[e.g.,][]{Filippenko/etal:1992a,Garavini/etal:2004}, faint
1991bg-like SN
\citep[e.g.,][]{Filippenko/etal:1992b,Leibundgut/etal:1993,Garnavich/etal:2004,Taubenberger/etal:2008},
peculiar 2002cx-like events
\citep[e.g.,][]{Li/etal:2003,Phillips/etal:2007}, and high-luminosity
\sneia\ speculated to originate from super-Chandrasekhar-mass 
progenitors \citep[e.g.,][]{Howell/etal:2006,Scalzo/etal:2010}. This 
observed diversity contrasts with the apparent homogeneity of the
\snia\ class, the majority of which seems to obey an empirical
relation between the peak luminosity and the width of the light curve
(the so-called width-luminosity relation, or WLR;
\citealt{Pskovskii:1977,Phillips:1993}), needed to normalize \snia\
luminosities for precise distance measurements. Several large sets of
optical \snia\ light curves have already been published
\citep{Hamuy/etal:1996,Riess/etal:1999a,Jha/etal:2006,Hicken/etal:2009a,Contreras/etal:2010,Ganeshalingam/etal:2010,Stritzinger/etal:2011},
but until the recent publication of 432 spectra of 32 \sneia\ by
\cite{Matheson/etal:2008}, there had been no large homogeneous
spectroscopic data sets of nearby \sneia\ available.

The supernova group at the Harvard-Smithsonian Center for Astrophysics
(CfA) initiated an ambitious observational program in 1993 (PI:
R.~P.~Kirshner) using the telescopes at the Fred Lawrence Whipple
Observatory (FLWO) on Mount Hopkins, Arizona. The aims of the CfA SN
Program are to obtain well-sampled light curves and multi-epoch
spectroscopy of nearby ($z<0.015$) supernovae of all types, to study
their astrophysical nature as well as provide a homogeneous
low-redshift sample for cosmological analyses using \sneia. We follow
supernovae announced in IAU circulars\footnote{Starting in 2008
we also followed a few supernovae discovered by the SNfactory
collaboration \citep{Aldering/etal:2002} and reported on the
SNfactory Supernova Discoveries
webpage, http://snfactory.lbl.gov/snf/open\_access/snlist.php.},
discovered for the most part by the Lick Observatory Supernova
Search (LOSS) using the robotic Katzman Automatic Imaging Telescope
(KAIT; \citealt{Filippenko/etal:2001}), and by dedicated amateur
astronomers. In return we provide spectroscopic classifications for a
large fraction of supernova discoveries: during 1997-2007,
we classified over 40\%\ of all nearby SN visible from Mount Hopkins
and announced our results in 499 IAU circulars.
These prompt classifications are important to the community, and
serve as the basis for triggers of other supernova follow-up
programs (e.g., by the Swift SN group; \citealt{Milne/etal:2010}).

We were not able to follow up all SN announced in circulars,
and so had to prioritize our follow-up strategy. In a nutshell, we
assigned the highest priority to (1) \sneia\ discovered before maximum
light (we sometimes relaxed this requirement to follow up peculiar
objects) (2) bright Type Ib/c supernovae, to study both individual objects
(e.g., SN~2006aj, \citealt{Modjaz/etal:2006}; SN~2008D,
\citealt{Modjaz/etal:2009}) and the connection of broad-line SN~Ic
with gamma-ray bursts (e.g., \citealt{Modjaz/etal:2008}), and (3)
bright Type IIP supernovae discovered  within $\sim2$ weeks from
explosion for distance determinations via a variant of the Expanding
Photosphere Method \citep{Kirshner/Kwan:1974} [e.g.,
  \citealt{Dessart/etal:2008}].

We have already published several large collections of optical \snia\
light curves obtained with the FLWO 1.2\,m telescope (CfA1: 22 $BVRI$
light curves, \citealt{Riess/etal:1999a}; CfA2: 44 $UBVRI$ light
curves, \citealt{Jha/etal:2006}; CfA3: 185 $UBVRr'Ii'$ light curves,
\citealt{Hicken/etal:2009a}), as well as a sample of 21 near-infrared
(NIR; $JHK_s$) light curves obtained with the robotic PAIRITEL 1.3\,m
telescope \citep{Wood-Vasey/etal:2008}. Spectra of individual \sneia\
have been published in several papers (SN~1998aq,
\citealt{Branch/etal:2003}; SN~1998bu, \citealt{Jha/etal:1999};
SN~1999by, \citealt{Garnavich/etal:2004};
SN~2001ay, \citealt{Krisciunas/etal:2011}; SN~2001eh and SN~2001ep,
\citealt{Sauer/etal:2008}; SN~2002cx, \citealt{Li/etal:2003};
SN~2005cf, \citealt{WangX/etal:2009a}; SN~2006bt, \citealt{SN2006bt};
SN~2006gz, \citealt{SN2006gz}), including the large sample of 432
spectra of 32 \sneia\ (including SN~1998aq and SN~1998bu) published by
\cite{Matheson/etal:2008}.
A study of the use of spectroscopic
indicators in determining distances to \sneia\ in the Hubble flow was
published by \cite{Blondin/Mandel/Kirshner:2011}, based on 47 spectra
of 26 \sneia\ (including 15 spectra of 9 \sneia\ from the
\citealt{Matheson/etal:2008} sample).

Here we present the second large release of \snia\ spectra from the
CfA SN Program taken between October 1993 and August 2008. This sample
consists of \nspectot\ spectra of \nsntot\ \sneia, of which
\nspecunpub\ spectra of \nsnunpub\ \sneia\ are published here for the first
time. It includes new reductions of spectra taken during
1994-1997 of \sneia\ that were part of the CfA1 light-curve sample.
Unlike \cite{Matheson/etal:2008}, we include spectra of {\it all}
\sneia\footnote{except for \nspecexcl\ spectra of \nsnexcl\ \sneia\
with too low signal-to-noise ratio (S/N) and excessive host-galaxy
contamination, which were not used for spectroscopic classifications.} taken
during this 15-year period, regardless of sampling and whether or not
a well-calibrated light curve was available. The main reason for doing
so is that a large fraction of the \nsnonespec\ \sneia\ for which we
only have a single spectrum were classified by us, and we wish their
spectra to be publicly available. Another reason is that other groups
might have well-sampled light curves of objects for which we were
unable to determine a time of maximum light, adding value to the
``phaseless''\footnote{Throughout this paper the spectral phase
  corresponds to the number of rest-frame days from $B$-band maximum
  light.} 
spectra presented here.

The resulting sample spans a large range of \snia\ properties and is
ideal for studies of spectroscopic diversity, which is the aim of this
paper. In \S~\ref{sect:obsred} we present the spectroscopic
observations and data reduction techniques. Section~\ref{sect:sample}
comments on the general properties of the CfA sample of \nsntot\
\sneia. In \S~\ref{sect:specsub} we present the spectroscopic
classification schemes of \cite{Branch/etal:2006} and
\cite{WangX/etal:2009b}, and study the properties of \sneia\ in these
different subclasses. Section~\ref{sect:linevel} focuses on the
characteristic Si\two\,\l6355 line of \sneia, studying its velocity
evolution and relation to intrinsic color. In 
\S~\ref{sect:early} we analyze the extent of nuclear
burning and the presence of high-velocity features and unburnt carbon
in early-time spectra. Last, \S~\ref{sect:neb} examines the relation
between the FWHM of nebular iron lines and luminosity, and between
nebular line shifts and intrinsic properties of \sneia\ at maximum light.
Conclusions follow in \S~\ref{sect:ccl}.

All \nspectot\ spectra presented in this paper will be made publicly
available through the CfA Supernova
Archive\footnote{http://www.cfa.harvard.edu/supernova/SNarchive.html}.


\section{Observations and Data Reduction}\label{sect:obsred}

In this section we summarize our observational strategy and data
reduction methods. These are essentially identical to the ones
presented by \cite{Matheson/etal:2008}, but we highlight a few
differences and provide additional information when needed.


\subsection{Observations}\label{sect:obs}

As with the sample of 432 spectra published by
\cite{Matheson/etal:2008}, the majority of spectra presented here
(\nspecfast\ out of \nspectot\ spectra, or \fracspecfast\%) were obtained
with the FAST spectrograph \citep{Fabricant/etal:1998} mounted on the
Tillinghast 1.5\,m telescope at FLWO. The FAST spectrograph has been
operational since January 1994 and the first \snia\ spectrum was one
of SN~1994D taken on 1994 March 10 (all dates are given in UT). The
observations were carried out in queue-scheduled mode, for 
the most part by two professional observers (P.~Berlind and
M.~L.~Calkins, who observed \fracspecfastperrymike\% of the FAST
spectra presented here), as well as by CfA personnel. A total of
\nfastobservers\ individual observers contributed to the FAST
\snia\ sample presented here. During 1997-2008, typically 2-3 spectra
were taken each night FAST was scheduled on the FLWO 1.5\,m, namely
$\sim20$ nights per month,
excluding August which corresponds to the annual shutdown at FLWO
during the monsoon season. During 1994-1996, typically a single
spectrum was taken in any given night. Observational details of the
spectra are given in Table~\ref{tab:obs}.

The usual setup for observations with FAST consisted of a ruled
grating with 300 lines per mm and a 3\arcsec\ slit, yielding a typical
FWHM resolution of 6-7\,\AA\ over a wavelength range of $\sim3700$ to
$\sim7500$\,\AA. From September 2004 onwards (starting with
observations of SN~2004dt), we changed the standard setup,
  extending it down to $\sim$3500\,\AA\ to cover the
entire Ca\two\,H\&K absorption profile. For a few \sneia\ (SN~1996ai,
SN~1999by, SN~1999dq, SN~2001eh, SN~2001ep, and SN~2002bo) we
requested additional observations with a different grating tilt to
extend the wavelength range beyond 9000\,\AA. Other programs in the
FAST queue would sometimes request different instrument setups, either
with narrower slits (1.5\arcsec\ and 2\arcsec) or with a
higher-resolution grating (cf. spectra of SN~1995ac, SN~1998aq,
SN~1998bu, SN~1999by, and SN~1999cl). For most of the observations
obtained during 1994-1998, the slit was oriented at a position angle
of 90\degr. From late 1998 onwards, the slit was generally positioned
at the parallactic angle (unless the object was at airmass
$\lesssim1.1$) so as to minimize the effects of atmospheric dispersion
\citep{Filippenko:1982}. Table~\ref{tab:obs} gives the slit position
angle for each spectrum as well as the absolute difference,
$|\Delta\Phi|$, with the actual parallactic angle.

Additional spectra were obtained during classically-scheduled nights
at the MMT Observatory 6.5\,m telescope with the Blue Channel
(\nspecbc\ spectra) and Red Channel (\nspecrc\ spectra) spectrographs
\citep{Schmidt/etal:1989}. A number of different spectrograph settings
were used, 
yielding FWHM resolutions ranging between $\sim3$ and $\sim13$\,\AA,
with a wavelength range typically extending below 3500\,\AA\ and
beyond 8000\,\AA. A few spectra were taken with two settings of the
Blue Channel spectrograph with non-overlapping wavelength ranges. This
only concerns the five MMT spectra taken on 1994 June 12 (SN~1994D,
SN~1994M, SN~1994Q, SN~1994S, and SN~1994T), all of which have
$\sim400$\,\AA-wide gaps ($\sim$6150-6550\,\AA) between the two
portions of the spectra. These spectra have been clearly marked in
Table~\ref{tab:obs}.

Last, a few spectra were obtained with the Magellan 6.5\,m Clay
(+LDSS-2/LDSS-3; \nspecldss\ spectra) and Baade (+IMACS; one spectrum
of SN~2003kf) telescopes. The FWHM resolution varies between $\sim9$
and $\sim18$\,\AA\ for LDSS-2, 9-12\,\AA\ for LDSS-3, and
5-6\,\AA\ for IMACS. The spectra taken with the Magellan 6.5\,m
telescopes typically do not reach bluer wavelengths than the FAST
spectra, but they generally extend beyond 9000\,\AA.


\subsection{Data Reduction}\label{sect:red}

The data reduction methods are the same as those presented by
\cite{Matheson/etal:2008}, and we refer the reader to that paper for
complementary information. The FAST data were all reduced in the same
consistent manner. This also applies to FAST data from 1994-1997 that
were entirely re-reduced for the purposes of this paper (this enabled
the recovery of two spectra of SN~1997do and SN~1997dt accidently
omitted from the data set published by \citealt{Matheson/etal:2008}). The
resulting FAST sample of \nspecfast\ spectra makes this by far the
largest homogeneous \snia\ spectroscopic dataset to date.

We used standard routines in IRAF\footnote{IRAF is distributed by the
National Optical Astronomy Observatory, which is operated by the
Association of Universities for Research in Astronomy (AURA) under
cooperative agreement with the National Science Foundation.} to
correct the CCD frames for overscan. The bias frames are sufficiently
uniform so that we do not subtract them to avoid introducing additional
noise. We did not correct for dark current as it is typically negligible
with FAST. As noted by \cite{Matheson/etal:2008}, however, a few
spectra are affected by dark-current problems after UV flashing,
resulting in a small emission feature at $\sim7100$\,\AA\ (observed)
and increased noise in the red portion of the spectrum. For a handful
of spectra taken in 1998 (SN~1998cs, SN~1998de, SN~1998ec, and
SN~1998eg), the S/N degradation is large enough that we have simply
trimmed off affected portions of the spectra (these have been clearly
marked in Table~\ref{tab:obs}). The CCD frames are then flat-fielded
using a combined normalized flat-field image. One-dimensional spectra
are then optimally extracted using the algorithm of \cite{Horne:1986}
as implemented in the IRAF {\it apall} package. These spectra are then
wavelength calibrated using HeNeAr lamps taken immediately after
(occasionally before) each exposure.
The same procedure was applied to spectra taken with the MMT and
Magellan 6.5\,m telescopes, although we subtracted a combined bias
frame before the flat-fielding stage. Subsequent reduction steps
described below are common to all telescope/instrument combinations.

We use our own set of routines in IDL to flux calibrate the 
extracted one-dimensional wavelength-calibrated spectra. Aside from
the actual flux calibration, these routines also apply small
adjustments to the wavelength calibration based on night-sky lines in
the SN frames, and apply a heliocentric correction. The
spectrophotometric standard stars observed on the same night (see
Table~\ref{tab:obs}) are also used to remove telluric 
absorption features from the spectra \citep[see,
  e.g.,][]{Wade/Horne:1988,Matheson/etal:2000}. We did not re-reduce
the \nspecmmt9397\ MMT spectra from 1993-1997 which were flux
calibrated using standard procedures in IRAF.

Approximately \fracspecmultstd\% of the spectra listed in
Table~\ref{tab:obs} were cross-calibrated with two spectrophotometric
standard stars of different colors to overcome the impact of
second-order light contamination. Prior to the FAST refurbishment in
August 2003, the spectrograph blocked blue light such that the FAST
spectra until SN~2003gq do not suffer from this effect (FAST spectra
taken between September 1997 and July 2003 were usually calibrated
with a single standard star). Starting in September 2003, two standard
stars were systematically used for the flux calibration. Our new
reductions of the 1994-1997 data also uses two standard stars for
consistency with the 2003-2008 data set, although the impact on the
resulting flux calibration is minor.

\begin{figure}
\plotone{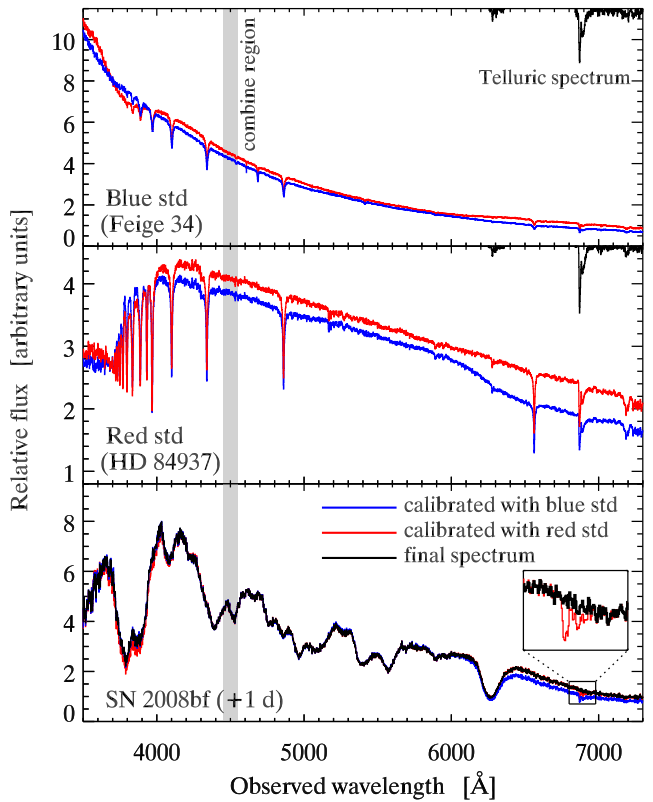}
\caption{\label{fig:secondorder}
Illustration of our flux-calibration procedure with one ``blue''
(Feige~34; {\it top panel}) and one ``red'' (HD~84937; {\it middle
panel}) spectroscopic standard star, applied to the FAST spectrum of
SN~2008bf at one day past $B$-band maximum ({\it bottom panel}). In
all three panels, the blue (red) spectrum corresponds to a flux
calibration using the blue (red) standard star. The gray shaded area
highlights the wavelength interval used to combine the blue- and
red-calibrated spectra of SN~2008bf. Blueward (redward) of this
interval, the blue-(red-) calibrated spectrum is used to generate the
final spectrum ({\it black line}). Both standard stars are also used
to generate normalized telluric absorption spectra ({\it black line in
top and middle panels}) which serve to remove these features (here the
atmospheric B-band) in the final SN spectrum ({\it see inset in bottom
  panel; black line}).
}
\end{figure}

Figure~\ref{fig:secondorder} illustrates our flux calibration
technique with two standard stars on the FAST spectrum of SN~2008bf at
1\,d past $B$-band maximum. On the same night as a given SN
observation, we observed a relatively blue standard star (here
Feige~34, sdO spectral type; {\it top panel}) and a relatively red
standard star (here HD~84937, sdF spectral type; {\it middle
  panel}). The blue standard yields a more accurate flux calibration
in the blue due to a greater number of counts and the absence of a
Balmer jump which complicates the calibration around 4000\,\AA, while
the red standard yields a better calibration in the red where it does
not suffer as much from second-order contamination. One clearly sees that the
blue standard Feige~34 is poorly calibrated by the red standard
blueward of $\sim4000$\,\AA\ ({\it top panel; red line}), while
applying the blue standard to calibrate the red standard HD~84937
results in a flux deficit redward of $\sim6000$\,\AA\ ({\it middle
  panel; blue line}). The SN spectrum is calibrated using both
standard stars ({\it bottom panel}), and the two resulting spectra are
combined in a $\sim100$\,\AA-wide region around
$\sim4500$\,\AA\ (i.e., slightly redward of the Balmer jump). Blueward
(redward) of this region, the spectrum calibrated using the blue (red)
standard is used. The impact on the final spectrum ({\it bottom panel;
black line}) may not seem spectacular, but we clearly recover flux
in the Ca\two\,H\&K absorption region and redward of the
Si\two\,\l6355 line. 

\epsscale{1.15}
\epsscale{1.0}

We also illustrate the removal of atmospheric absorption features,
here the O$_2$ B-band around 6880\,\AA. Each standard star is
used to derive a normalized telluric spectrum ({\it top and middle
panels; black line}) which is then divided out from the SN spectrum
after appropriate scaling (ratio of airmasses raised to some power
$\alpha\approx0.6$; see \citealt{Wade/Horne:1988}). 
The red standard star is more effective in removing these
features, since the blue standard suffers from second-order
contamination in these atmospheric bands.
In this particular
example the B-band leaves a small imprint on the SN spectrum, but the
inset in the lower panel of Fig.~\ref{fig:secondorder} shows that this
technique is successful in removing the unwanted absorption.

\begin{figure}
\plotone{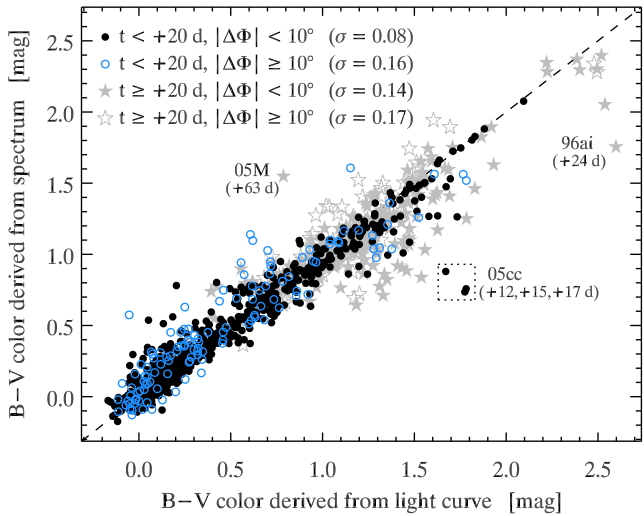}
\caption{\label{fig:bmvcomp}
Comparison of $B-V$ color derived from spectra with that derived from
photometric measurements \citep[see][their
  Fig.~4]{Matheson/etal:2008}. We show measurements for spectra with
phases less than 20\,d past $B$-band maximum ({\it circles}), for
which the difference between the position angle of the spectroscopic
slit and the parallactic angle, noted $|\Delta\Phi|$, is less than
({\it black}) or greater than ({\it blue}) 10$^\circ$. Spectra taken
at low airmass ($\sec{z}<1.1$) are included in the
$|\Delta\Phi|<10^\circ$ measurements. The stars correspond to
spectra at phases greater than 20\,d past $B$-band maximum. In each
case we indicate the magnitude scatter about the 1:1 relation ({\it
  dashed line}).
The +63\,d spectrum of SN~2005M has low S/N, SN~1996ai is the most 
highly-reddened \snia\ in our sample ($E(B-V)\approx1.8$\,mag), and
the spectra of SN~2005cc are contaminated by host-galaxy light.
}
\end{figure}

We can check the accuracy of the relative flux calibration of our
spectra (no attempt was made to put the spectra on an absolute flux
scale) by comparing the $B-V$ color derived from photometry with that
derived directly from the spectra (see \citealt{Matheson/etal:2008},
their Fig.~4). We interpolated the corresponding $B$- and $V$-band
light curves at the time each spectrum was taken, unless the difference was
larger than 3\,d (the interpolated measurements were visually
cross-checked). FAST spectra taken before September 2004 do not
extend to the blue edge of the $B$ filter (3600\,\AA). We have run
simulations based on the \snia\ spectral template of
\cite{Hsiao/etal:2007} that show that even for spectra extending
only down to 3750\,\AA\ the error on the inferred $B-V$ color is less than
0.005\,mag, so we include all the 1994-2008 FAST spectra in the
comparison. The resulting 1128 $B-V$ measurements are displayed in
Fig.~\ref{fig:bmvcomp}. For SN spectra at phases less than 20\,d past
$B$-band maximum light taken within 10\degr\ of the parallactic angle
(or at low airmass: $\sec{z}<1.1$; {\it filled circles}), the scatter
around zero difference is $\sim0.08$\,mag, slightly larger than the
$\sim0.06$\,mag scatter found by \cite{Matheson/etal:2008}. The reason
for this larger scatter is the inclusion of low-S/N spectra in our
sample (${\rm S/N}<10$ per pixel). Three
significant outliers ($\sim1$\,mag difference in $B-V$ color)
correspond to spectra of the 2002cx-like SN~2005cc that are
contaminated by host-galaxy light. 

The scatter is $\sim2$ times larger
for spectra not observed at the parallactic angle
($|\Delta\Phi\ge10\degr|$), regardless of phase ({\it open circles and
  open stars}). The scatter is also significantly larger for SN
spectra at phases greater than 20\,d past maximum observed at the
parallactic angle ({\it filled stars}). Possible reasons for this were
already noted by \cite{Matheson/etal:2008}: the spectra become
dominated by prominent emission features, giving rise to systematic
errors when multiplied by a filter that is not precisely matched to
the photometry. The spectra are also fainter, increasing the impact of
host-galaxy contamination. 

\epsscale{1.15}
\epsscale{1.0}

\begin{figure*}
\resizebox{\textwidth}{!}{
\includegraphics{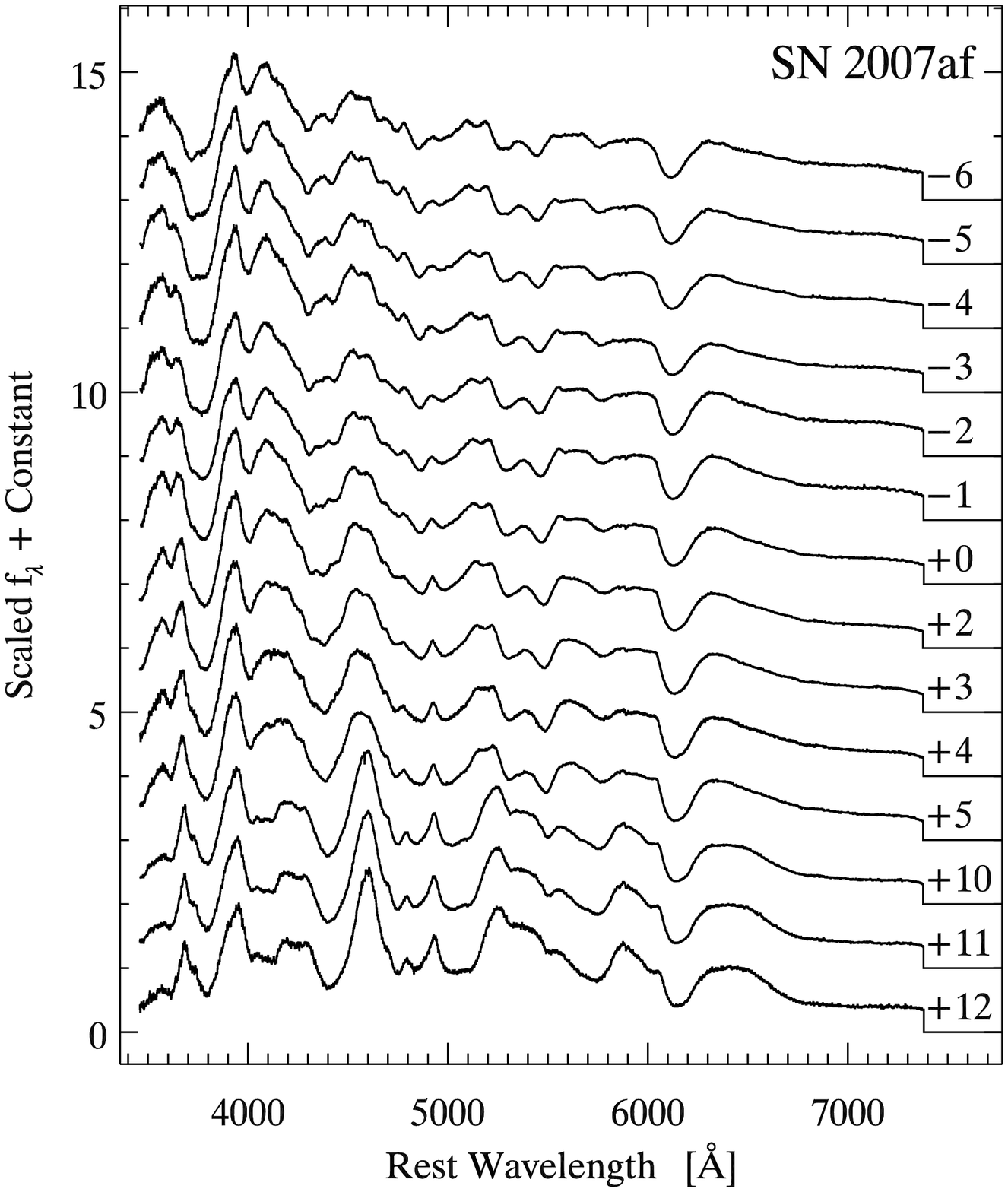}\hspace{.5cm}
\includegraphics{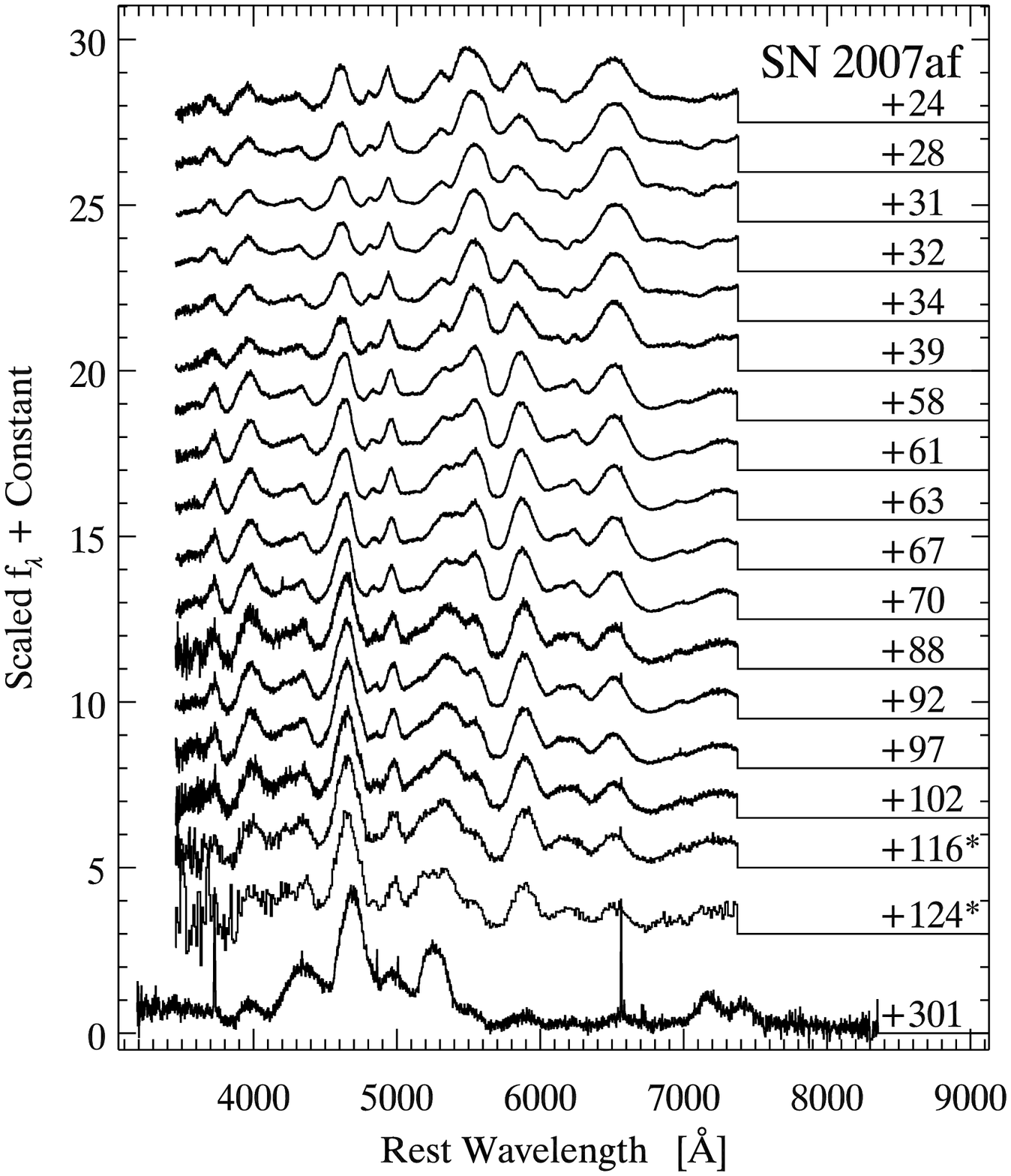}\hspace{.5cm}
\includegraphics{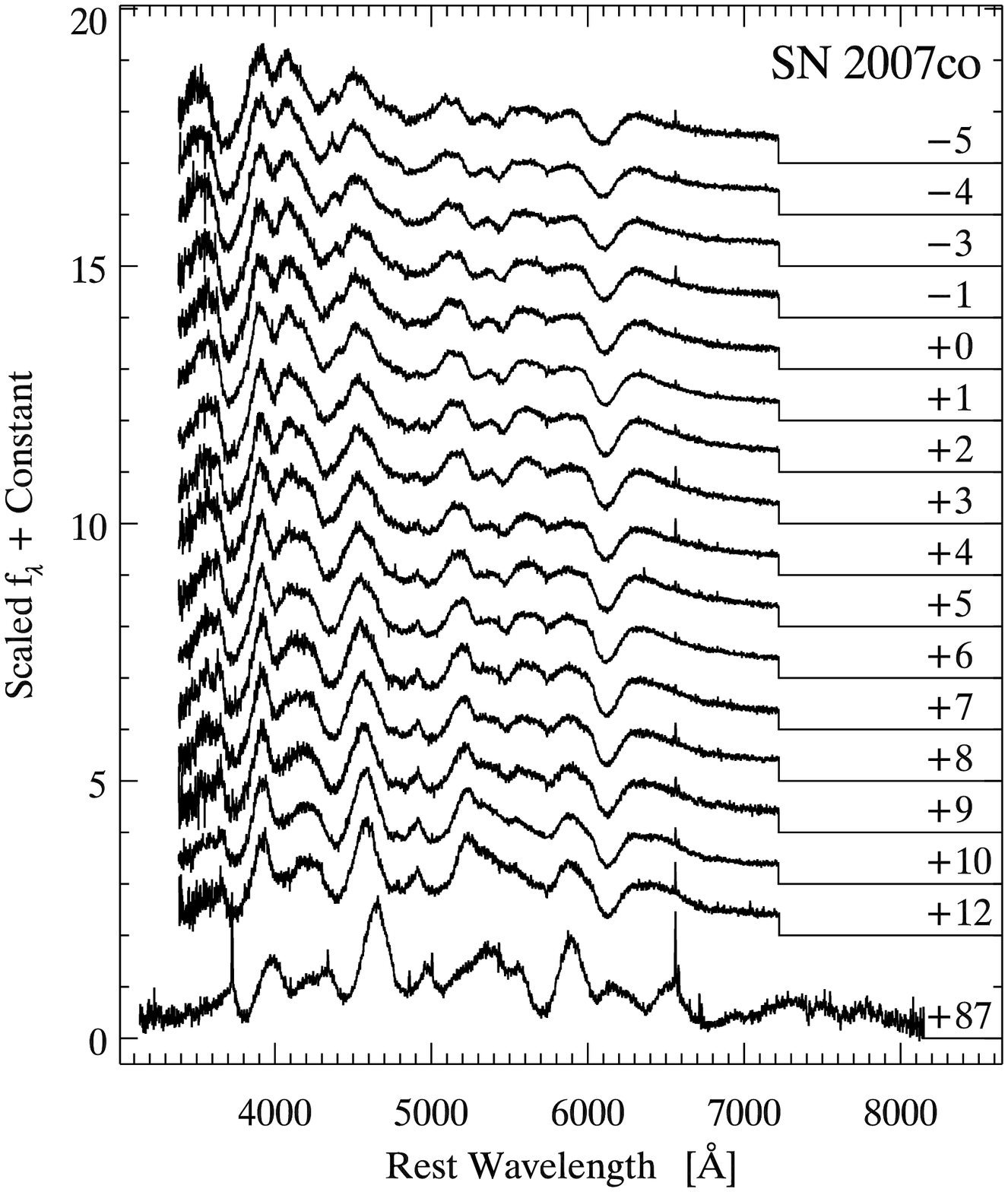}
}

\vspace{1cm}
\resizebox{\textwidth}{!}{
\includegraphics{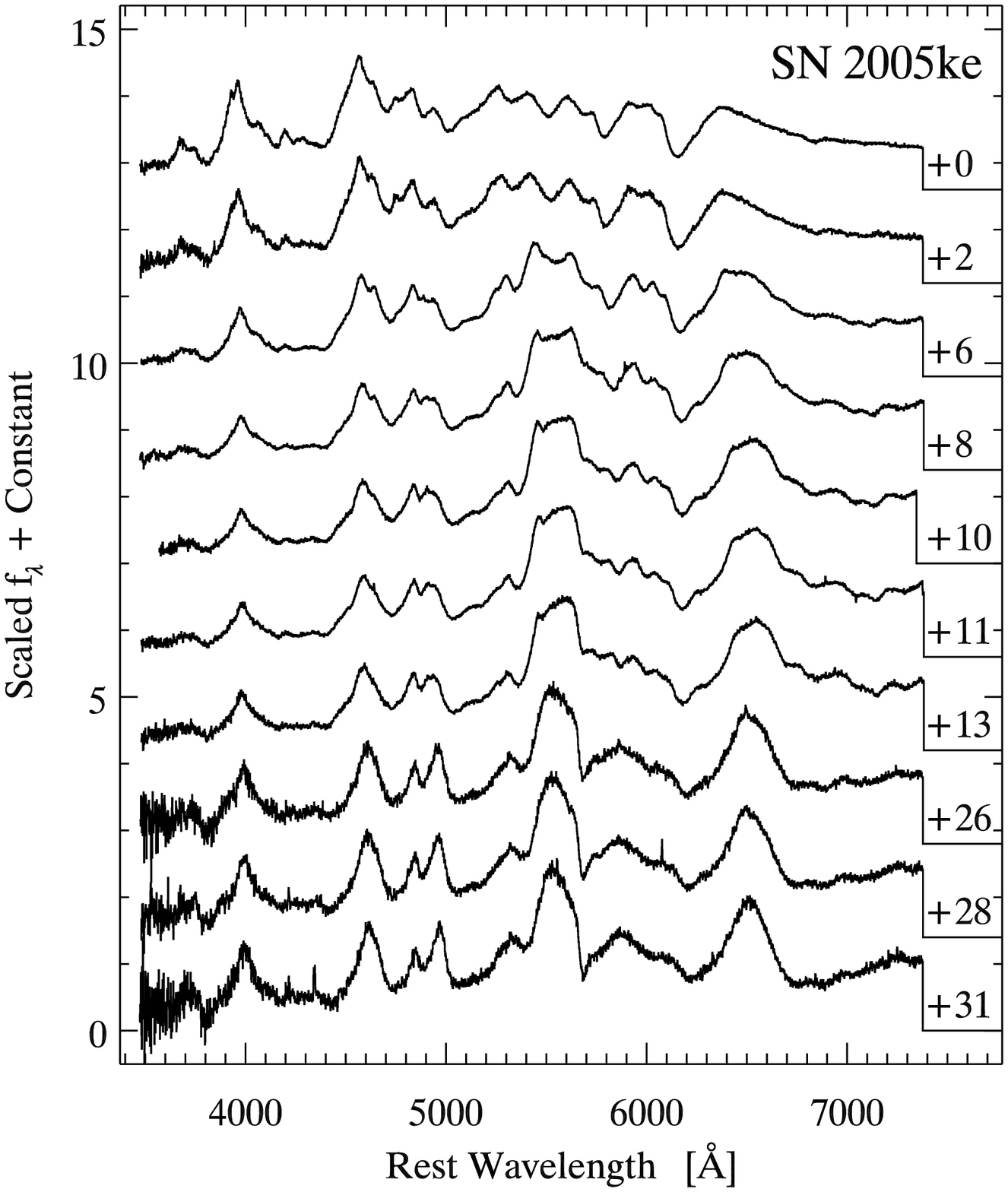}\hspace{.5cm}
\includegraphics{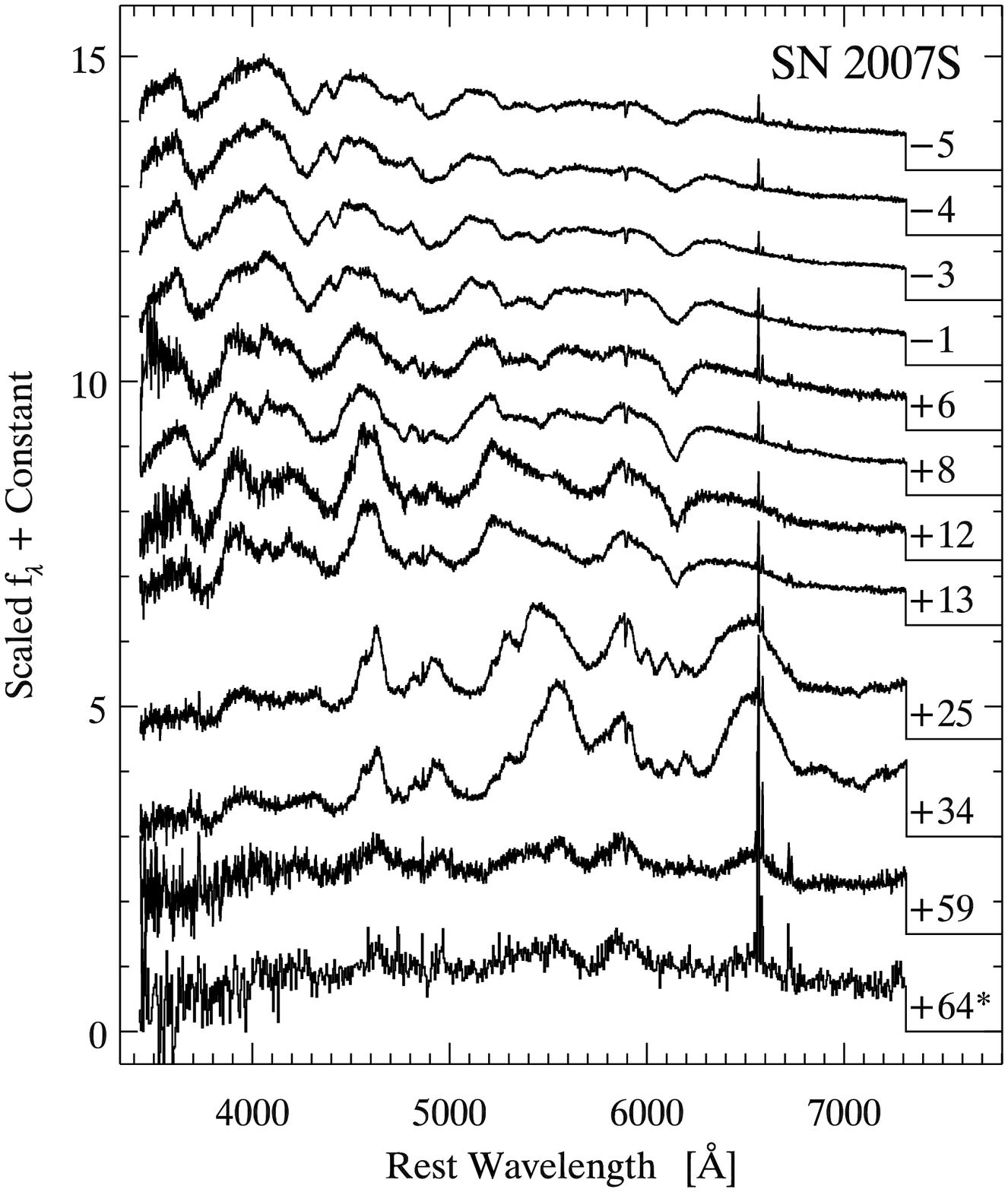}\hspace{.5cm}
\includegraphics{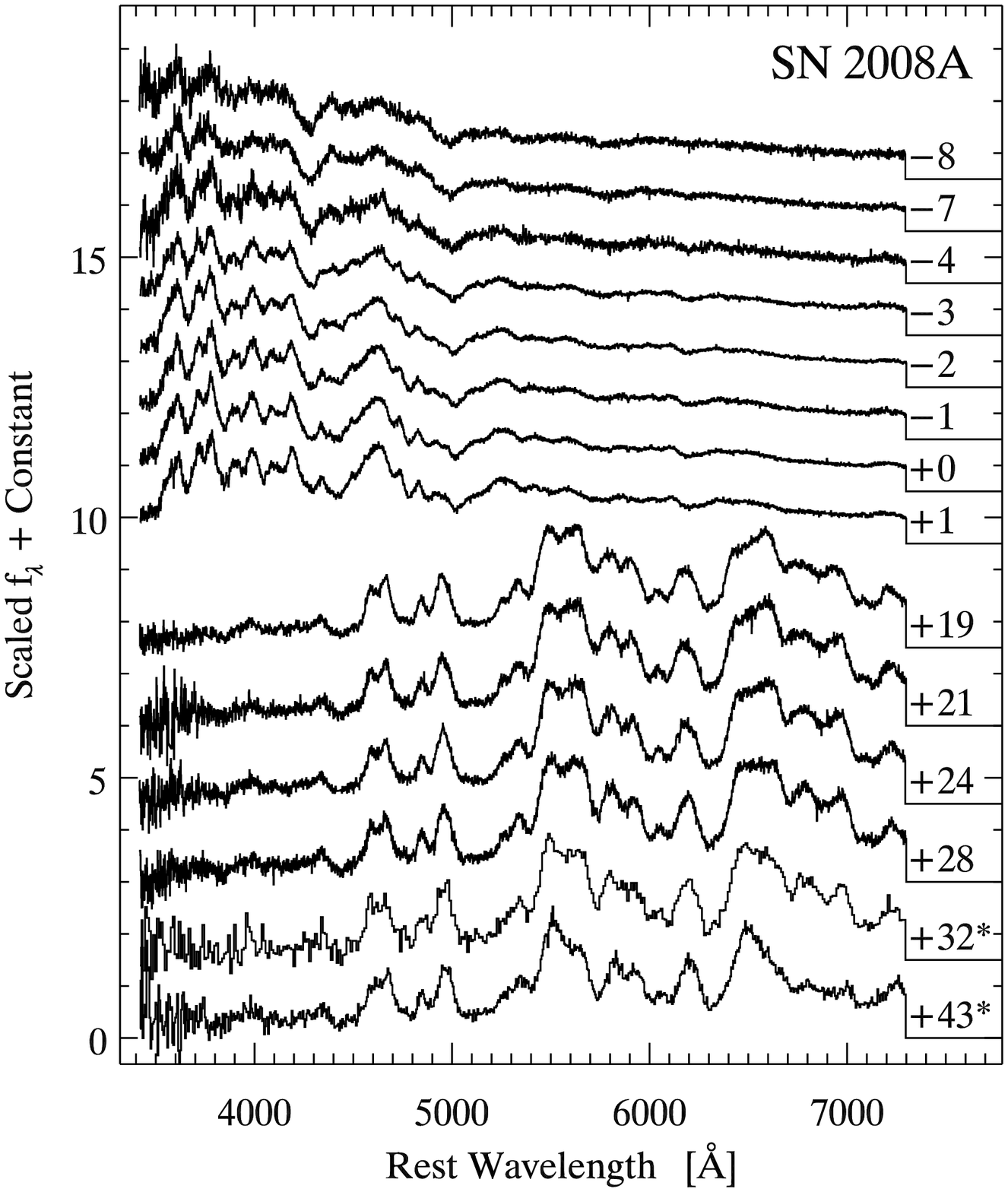}
}
\caption{\label{fig:cfaspec}
Example \snia\ spectral series from the CfA SN Program. 
The flux units are $f_{\lambda}$ (erg s$^{-1}$ cm$^{-2}$ \AA$^{-1}$)
that have been normalized and then additive offsets applied for
clarity. The zero-flux level for each spectrum is marked with an
extension on the red edge. The wavelength axis is corrected for the
recession velocity of the host galaxy. The number associated with each
spectrum indicates the phase in (rest-frame) days from $B$-band
maximum. Spectra with low S/N have been binned; they are indicated
with an asterisk appended to the phase label. Shown are
spectra of two ``normal'' \sneia\ in the top row (SN~2007af and
SN~2007co); one 1991bg-like (SN~2005ke), one 1991T-like (SN~2007S; the
+59\,d and +64\,d spectra are strongly contaminated by the
host-galaxy), and one 2002cx-like (SN~2008A) \snia\ in the bottom
row.
}
\end{figure*}

We show example spectral series in Fig.~\ref{fig:cfaspec}. Plots of
all \sneia\ from the CfA SN Program will be made available alongside
the actual data on the CfA SN Archive website.

\epsscale{1.15}
\epsscale{1.0}


\section{Sample properties}\label{sect:sample}

We briefly summarize the properties of the \snia\ sample presented in
this paper in terms of phase, redshift, and \dmft\ distributions, as 
well as the typical number of spectra per SN. We cross-checked the
classifications of all supernovae in our sample with those reported in
circulars using the SNID code of \cite{SNID}. Three \sneia\ have no
previously reported classifications: SN~2001es, SN~2003de, and
SN~2004cu. One \snia\ (SN~1999bv) was mis-classified as a ``probable
Type Ib/c'' supernova by \cite{Jha_iauc7150}, although its spectrum
shows it to be a Type Ia SN at $\sim3$\,months past maximum light (as
noted by \citealt{Matheson/etal:2001}). All other SN in our sample are
classified as Type Ia in circulars.

We were able to determine a reliable time of $B$-band maximum ($t_{\rm
max}(B)$; see Table~\ref{tab:snparam}) for \nsntmax\ out of \nsntot\
\sneia, based for the most part on photometry from the CfA
\citep{Riess/etal:1999a,Jha/etal:2006,Hicken/etal:2009a}, and from the
Lick Observatory Supernova Search (LOSS; \citealt{Ganeshalingam/etal:2010}). 
We rely on a number of
light-curve fitters for this purpose: MLCS2k2 \citep{MLCS2k2}, SALT2
\citep{SALT2}, SNooPy \citep{SNooPy}, and the BayeSN statistical
model of \cite{Mandel/etal:2009,Mandel/etal:2011}. When more than one
estimate for $t_{\rm max}(B)$ is available, we simply take the mean
value. For a few SN we used values from the literature, in particular
results from SNooPy fits to light curves from the Carnegie Supernova
Program (CSP; \citealt{Contreras/etal:2010,Stritzinger/etal:2011}) by
\cite{Burns/etal:2011} and \cite{Stritzinger/etal:2011}.
Estimates for six \sneia\ (SN~2001gc, SN~2002es, SN~2007fb, SN~2007fs,
SN~2007kk, and SN~2008ae) were provided by the Berkeley SN group based
on LOSS light curves (Ganeshalingam, Li, \& Silverman 2011;
priv. comm.), while $t_{\rm max}(B)$ values for 4 of the 5 SNF \sneia\
in our sample were provided by the SNfactory collaboration (Bailey
2011; priv. comm.).
We thus have
\nsnnotmax\ \sneia\ with no reliable time of maximum (\nsnwlcnotmax\
of which have 
$B$-band light curves). While this represents a sizeable fraction of
the total number of individual \sneia\ in our sample, it is only a
modest percentage of the total number of spectra (\nspecnotmax\ in
\nspectot\ spectra, or $\sim16$\%). 

\begin{deluxetable*}{lcllcrrc}
\tabletypesize{\scriptsize}
\tablewidth{0pt}
\tablecaption{\label{tab:snparam}Basic SN parameters}
\tablehead{\colhead{SN} & \colhead{$z$\tablenotemark{a}} & \colhead{$t_{\rm max}(B)$\tablenotemark{b}} & \colhead{\dmft\tablenotemark{c}} & \colhead{$M_B$\tablenotemark{d}} & \colhead{$B-V$\tablenotemark{e}} & \colhead{$B^{\rm max}-V^{\rm max}$\tablenotemark{f}} & \colhead{Phot. Ref.\tablenotemark{g}}}
\startdata
1993ac               & 0.0503 & 49269.2 (1.2) [Jha07]          & 1.19 (0.10) [Phi99]            & \nodata                        & \nodata                        & \nodata                        & CfA1                           \\
1993ae               & 0.0190 & \nodata\tablenotemark{h}       & \nodata                        & \nodata                        & \nodata                        & \nodata                        & CfA1                           \\
1994D                & 0.0029 & 49432.5 (0.1) [B,M,SA,SN]      & 1.37 (0.03) [B]                & \nodata                        & \nodata                        & \nodata                        & Jha07                          \\
1994M                & 0.0232 & 49474.5 (0.9) [B,M,SA,SN]      & 1.26 (0.06) [B]                & $-$19.21 (0.08)                & $-$0.11 (0.06)                 & $-$0.10 (0.06)                 & CfA1                           \\
1994Q                & 0.0295 & 49496.2 (1.1) [Jha07]          & 1.03 (0.10) [Phi99]            & \nodata                        & \nodata                        & \nodata                        & CfA1                           \\
1994S                & 0.0151 & 49518.3 (0.3) [B,M,SA,SN]      & 0.94 (0.06) [B]                & $-$19.55 (0.09)                & $-$0.12 (0.05)                 & $-$0.11 (0.05)                 & CfA1                           \\
1994T                & 0.0346 & 49514.4 (0.4) [M,SN]           & 1.51 (0.08) [SN]               & \nodata                        & \nodata                        & \nodata                        & CfA1                           \\
1994ae               & 0.0043 & 49685.5 (0.8) [B,M,SA,SN]      & 0.96 (0.04) [B]                & \nodata                        & \nodata                        & \nodata                        & CfA1,Jha07                     \\
1995D                & 0.0067 & 49768.7 (0.1) [B,M,SA,SN]      & 1.05 (0.03) [B]                & \nodata                        & \nodata                        & \nodata                        & CfA1,Jha07                     \\
1995E                & 0.0120 & 49775.2 (0.5) [B,M,SA,SN]      & 1.16 (0.05) [B]                & \nodata                        & \nodata                        & \nodata                        & CfA1                           \\[-.25cm]
\enddata
\tablecomments{Table~\ref{tab:snparam} is published in its entirety in the electronic edition of The Astronomical Journal. A portion is shown here for guidance regarding its form and content.}
\tablenotetext{a}{Heliocentric redshift as reported in NED (except for the redshifts given in Table~\ref{tab:z}).}
\tablenotetext{b}{MJD at $B$-band maximum light. The $1\sigma$ error is given in between parentheses.
The fitters used to determine $t_{\rm max}(B)$ are given in between square brackets:
B=BayeSN \citep{Mandel/etal:2009}; M=MLCS2k2 \citep{MLCS2k2}; SA=SALT2 \citep{SALT2}; SN=SNooPy \citep{SNooPy}.
References to estimates taken from the literature are given below.
In the case of multiple $t_{\rm max}(B)$ estimates the error is given as the standard deviation of the individual values.
For cases where no error information was available we assume a typical 0.5\,d uncertainty.
}
\tablenotetext{c}{\dmft\ decline rate in magnitudes. The $1\sigma$ error is given in between parentheses.
The fitter used to determine \dmft\ is given in the square brackets (see above).
We assign a minimum 0.05\,mag uncertainty to \dmft\ estimates with SNooPy (which typically yield a formal $\sim0.01$\,mag uncertainty).
References to estimates taken from the literature are given below.
}
\tablenotetext{d}{Intrinsic absolute peak $B$-band magnitude from BayeSN (assuming $H_0=72$\,\kms\,Mpc$^{-1}$) for \sneia\ with $z_{\rm CMB}\ge0.01$ and $A_V<1$\,mag.}
\tablenotetext{e}{Intrinsic $B-V$ color at $B$-band maximum light from BayeSN for \sneia\ with $z_{\rm CMB}\ge0.01$ and $A_V<1$\,mag.}
\tablenotetext{f}{Intrinsic $B^{\rm max}-V^{\rm max}$ pseudo-color from BayeSN for \sneia\ with $z_{\rm CMB}\ge0.01$ and $A_V<1$\,mag.}
\tablenotetext{g}{References for published photometric data used to determine $t_{\rm max}(B)$ and \dmft\ (see below).
References to ``Jha07" are meant to point the reader to references given in Table~1 of \cite{MLCS2k2}.}
\tablenotetext{h}{\cite{MLCS2k2} give $t_{\rm max}(B)=49288.08\pm1.19$, but the fit is uncertain (first photometric point $>+10$\,d past $B$-band maximum).}
\tablenotetext{i}{\cite{MLCS2k2} give $t_{\rm max}(B)=50630.25\pm0.98$ and $\dmft=1.02\pm0.10$, but the fit is uncertain (first photometric point $>+10$\,d past $B$-band maximum).}
\tablenotetext{j}{\cite{MLCS2k2} give $t_{\rm max}(B)=50840.57\pm2.02$, but the fit is qualified by them as ``poor".}
\tablerefs{
(BSNIP) Berkeley \snia\ Program (2011, priv. comm.); 
(Buf05) \cite{Bufano/etal:2005};
(Bur11) \cite{Burns/etal:2011};
(CfA1) \cite{Riess/etal:1999a}; (CfA2) \cite{Jha/etal:2006}; (CfA3) \cite{Hicken/etal:2009a}; 
(Con10) \cite{Contreras/etal:2010}; 
(Gan10) \cite{Ganeshalingam/etal:2010};
(Gan12) \cite{Ganeshalingam/etal:2012};
(Hic07) \cite{SN2006gz};
(Jha07) \cite{MLCS2k2} (and references therein); 
(Kas08) \cite{Kasliwal/etal:2008};
(Kri10) \cite{SN2001ay};
(Phi99) \cite{Phillips/etal:1999};
(Phi07) \cite{Phillips/etal:2007};
(Pri09) J.~L.~Prieto (2009, priv. comm.; based on the method of \cite{Prieto/Rest/Suntzeff:2006}; 
(Sca10) \cite{Scalzo/etal:2010};
(SNF) SNfactory collaboration (2011, priv. comm.); 
(Str11) \cite{Stritzinger/etal:2011};
(Tau08) \cite{Taubenberger/etal:2008}.
}
\end{deluxetable*}

\begin{figure*}
\resizebox{\textwidth}{!}{
\includegraphics{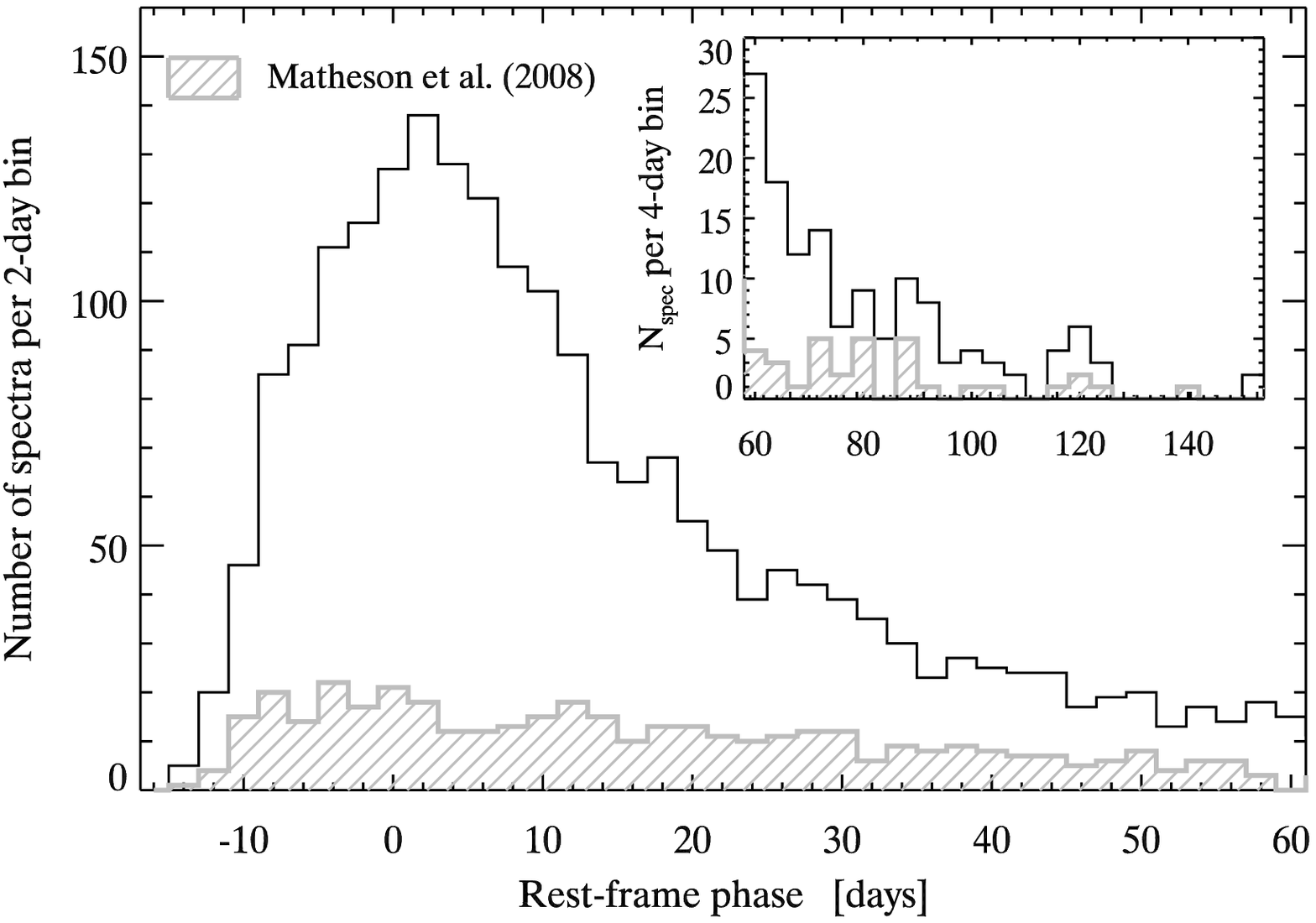}\hspace{1cm}
\includegraphics{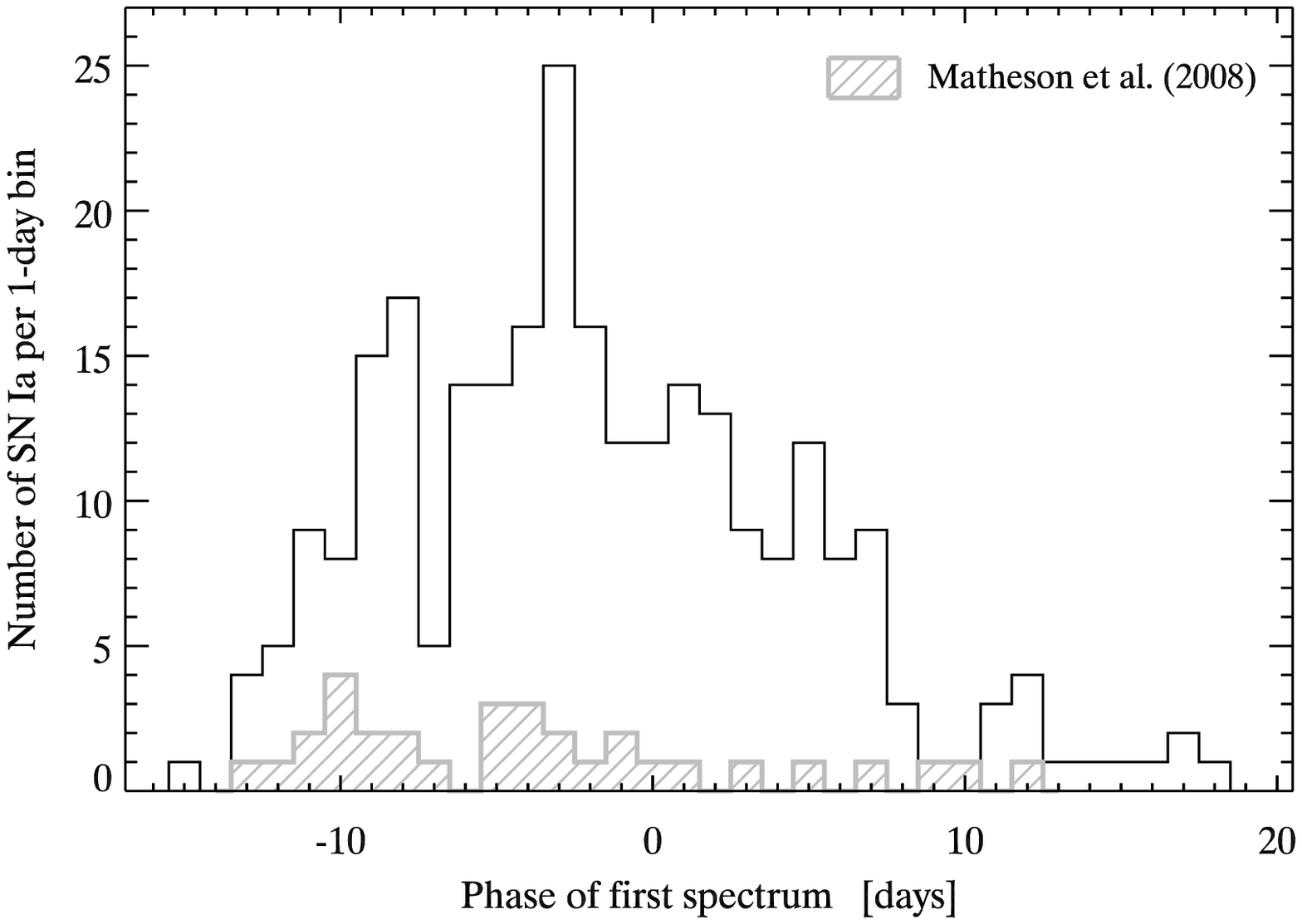}
}
\caption{\label{fig:nspecage}
Distributions of \sneia\ for which we have a reliable estimate for the
time of maximum light.
{\it Left:} Distribution of the number of spectra as a function of
phase out to 60\,d past maximum light. The hatched histogram
corresponds to the \snia\ sample published by
\cite{Matheson/etal:2008}. The inset shows spectra at phases in the
range [+60,+150]\,d.
{\it Right:} Number of \sneia\ {\it vs.} the phase of the first
spectrum.
}
\end{figure*}

The resulting phase distribution of the \nspectmax\ spectra
for which we have a reliable $t_{\rm max}(B)$ estimate is shown in the
left panel of Fig.~\ref{fig:nspecage} out to 150\,d past maximum. Also
shown is the corresponding distribution for the sample of spectra
published by \cite{Matheson/etal:2008}, including 9 spectra
that were accidently omitted from their sample (see
Table~\ref{tab:obs}). There are \nspecltm10\ spectra at phases 
less than $-10$\,d, and \nspecwithin3d\ spectra within 3\,d from maximum
light. Many of the \sneia\ in this sample have observations starting
before maximum light. The right panel of Fig.~\ref{fig:nspecage} shows
the distribution of SN phases of the first spectrum. There are \nsnwspecltm10\
\sneia\ for which the spectroscopic follow-up was initiated earlier
than $-10$\,d, and \nsnwspecltmax\ \sneia\ with spectra starting prior
to maximum light. We also have \nspecgt150\ spectra of \nsnwspecgt150\
\sneia\ with spectra taken later than 150\,d past maximum (see
Table~\ref{tab:nebspec}).

\begin{deluxetable}{ll}
\tablewidth{0pt}
\tablecaption{\label{tab:nebspec}CfA \snia\ spectra at 150\,d past
  maximum or later (see Table~\ref{tab:obs} for more observational details).}
\tablehead{\colhead{SN} & \colhead{Phase (d)}}
\startdata
1994D  & +611 \\
1994ae & +152,+366 \\
1995D  & +275,+283 \\
1998aq & +209,+229,+239 \\
1998bu & +178,+189,+207,+216,+242 \\
2000cx & +154,+181 \\
2002bo & +310 \\
2002cx & +308 (low S/N) \\
2003kf & +397 \\
2006X  & +150 \\
2007af & +301 \\
2007sr & +190
\enddata
\end{deluxetable}

\begin{figure}
\plotone{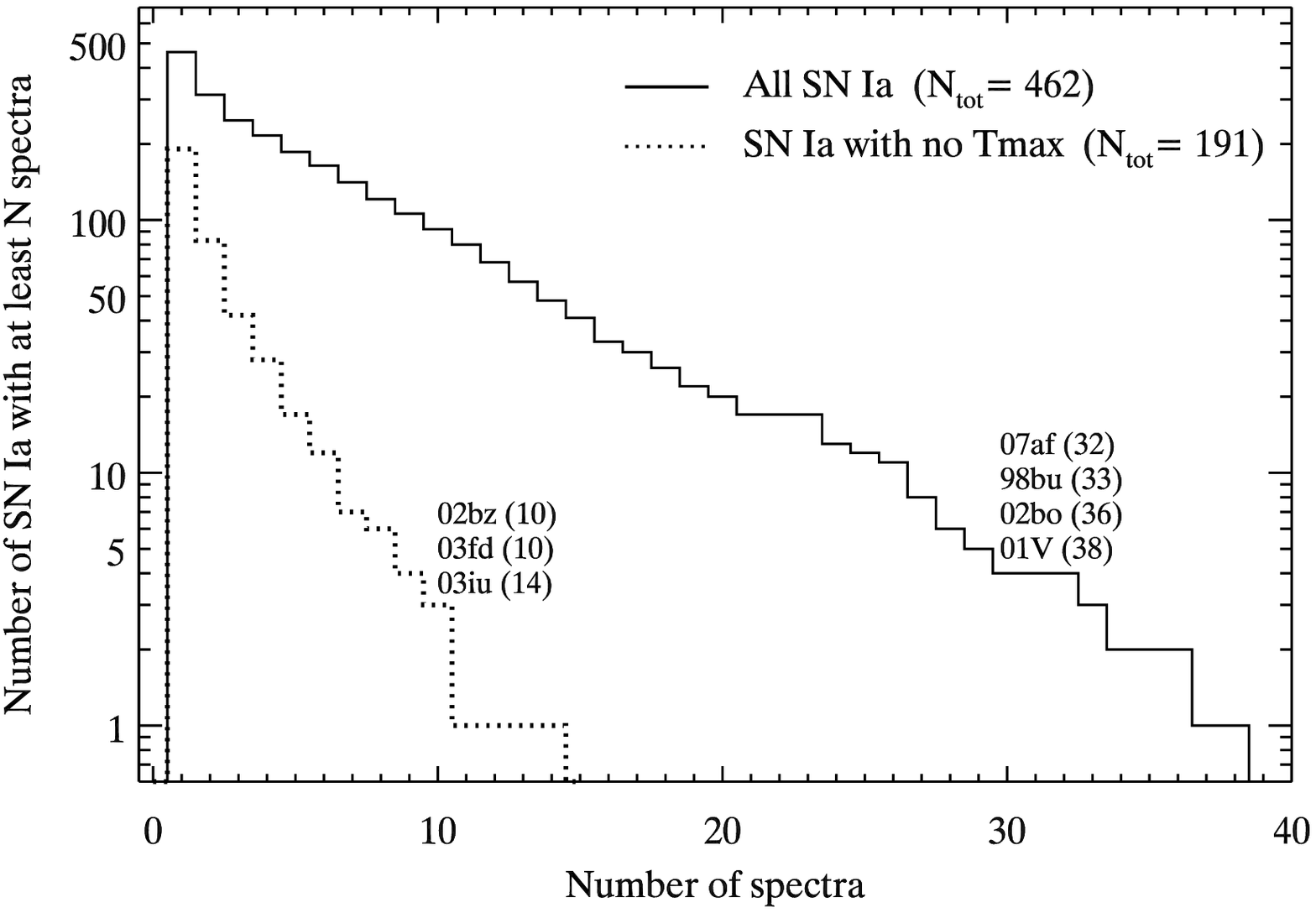}
\caption{\label{fig:nspeccumul}
Number of \sneia\ with a least $N$ spectra. The solid line is for
all \sneia, while the dotted line corresponds to \sneia\ with no
reliable estimate for the time of maximum.
}
\end{figure}

A clearer measure of the scientific potential of this sample is the
number of SN with at least $N$ spectra. Figure~\ref{fig:nspeccumul}
shows the corresponding distribution, for all \nsntot\ \sneia\ in our
sample ({\it solid line}) and for the \nsnnotmax\ with no time of
maximum ({\it dotted line}). There are \nsnwtenspec\ \sneia\ for which
we have at least 10 spectra, three of which have no reliable $t_{\rm
  max}(B)$ (SN~2002bz, SN~2003fd, and SN~2003iu). There are
\nsnwtwentyspec\ \sneia\ with at least 20 spectra, and 4 objects with over
30 spectra (SN~1998bu, SN~2001V, SN~2002bo, and SN~2007af). 
There are on average \meannspecpersnwtwospec\
spectra for each of the \nsnwtwospec\ \sneia\ with at least two
spectra.

\epsscale{1.15}
\epsscale{1.0}

\begin{figure}
\plotone{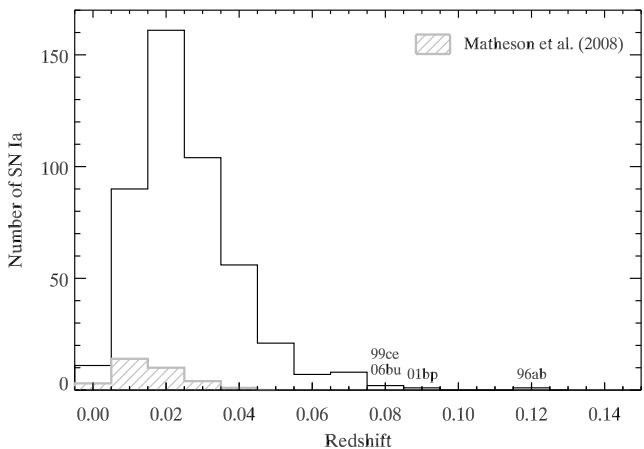}
\caption{\label{fig:zhisto}
Redshift distribution for the \nsntot\ \sneia\ in our sample. The
hatched histogram corresponds to the \snia\ sample published by
\cite{Matheson/etal:2008}. There are 4 \sneia\ at redshifts larger
than 0.075: SN~1999ce, SN~2006bu (part of the CfA3 light-curve
sample), SN~2001bp, and SN~1996ab (part of the CfA1 light-curve
sample). 
}
\end{figure}

The redshift distribution of our sample is shown in
Fig.~\ref{fig:zhisto}. The median redshift is $\sim0.023$, and \nsnzgt0p015\
\sneia\ are at $z>0.015$ (nearby Hubble flow). The highest-redshift
\snia\ in our sample, SN~1996ab at $z\approx0.123$, is part of the CfA1
light-curve sample of \cite{Riess/etal:1999a}.
Nearly all redshifts are from the corresponding
host galaxy as reported in the NASA/IPAC Extragalactic Database
(NED; $\sim40$\% are from the Updated Zwicky Catalog of
\citealt{UZC}). For a few objects there is no NED redshift for the
host galaxy, and we then determine our own redshift based on narrow
emission or absorption lines from the host galaxy (either in the SN
spectrum itself or on a separate host-galaxy spectrum; see
Table~\ref{tab:z}), using the IRAF RVSAO package of
  \cite{RVSAO}. When no such lines are visible in the SN spectrum,
we resort to cross-correlation with a library of supernova spectra
using SNID, and report the median redshift
based on the top-5 matches. For two \snia\ host galaxies the redshift
reported by NED is in error: MCG +07-36-33, the host of SN~2003fa, has
a NED redshift of 1800\,\kms\ \citep{RC3} where we measure $z=0.0404$
($cz=12125$\,\kms; see Table~\ref{tab:z}) based on five emission lines
(including [O\two]\,\l3727 and H$\alpha$). The host of SN~2002es,
UGC~2708, has a NED redshift of
8523\,\kms\ \citep{MonnierRagaigne/etal:2003} reported on the main
page, where the correct redshift of 5394\,\kms\ \citep{RC3} is
reported on the redshift subpage (this is confirmed by the Berkeley SN
Group using an optical spectrum of UGC~2708; Li 2011, priv. comm.).
All other NED redshifts were found to be consistent with our own
estimates.
Four of the five \sneia\ in our sample discovered by the SNfactory
occur in anonymous hosts. We derive their heliocentric redshifts based
on the CMB-frame redshifts given in Table~2 of \cite{Bailey/etal:2009}
and the coordinates as reported on the SNfactory Supernova Discoveries
webpage.

\begin{deluxetable}{lllc}
\tabletypesize{\scriptsize}
\tablewidth{0pt}
\tablecaption{\label{tab:z}SN redshifts not taken from NED.}
\tablehead{\colhead{SN} & \colhead{Host} & \colhead{$z$} & \colhead{Ref.\tablenotemark{a}}}
\startdata
1993ac & CGCG 307-023               & 0.0503\tablenotemark{b}  & abs   \\
1996ab & Anonymous                  & 0.123\tablenotemark{c}   & SNID  \\
1998ex & MCG +11-10-16              & 0.0365  & em    \\
2000dl & UGC 1191                   & 0.0611  & abs   \\
2001eo & UGC 3963                   & 0.066   & SNID  \\
2001es & 2X J2020834+1905246     & 0.0420  & em    \\
2002hu & MCG +06-6-12               & 0.0374  & em      \\
2003fa & MCG +07-36-33              & 0.0404\tablenotemark{d}  & em           \\
2003hw & 2X J03014982+3544343    & 0.0420  & abs   \\
2004gz & MCG +10-23-45              & 0.0137  & em    \\
2005ch & Anonymous                  & 0.027   & SNID  \\
2005lz & UGC 1666                   & 0.040   & SNID  \\
2006bu & 2X J13524703+0518496    & 0.081   & SNID  \\
2007H  & Anonymous                  & 0.0426  & em    \\
2007cq & 2X J22144070+0504435    & 0.0262  & em    \\
2007if & Anonymous                  & 0.0742\tablenotemark{e}  & 1  \\
2007kf & Anonymous                  & 0.044   & SNID                     \\
2007kg & 2MFGC 18005                & 0.006   & SNID                     \\
2007qe & Anonymous                  & 0.0239  & em    \\
2008E  & Anonymous                  & 0.034\tablenotemark{f}  & 2 \\
SNF20080522-000 & Anonymous         & 0.0450\tablenotemark{g} & 3 \\
SNF20080522-011 & Anonymous         & 0.0379\tablenotemark{g} & 3 \\
SNF20080623-001 & Anonymous         & 0.0430\tablenotemark{g} & 3 \\
SNF20080720-001 & Anonymous         & 0.0209\tablenotemark{g} & 3 \\[-.25cm]
\enddata
\tablecomments{2X~=~2MASX.}
\tablerefs{(1) \citealt{Scalzo/etal:2010}; (2) \citealt{Yuan_cbet1206}; (3) \citealt{Bailey/etal:2009}.}
\tablenotetext{a}{em=emission lines from the host galaxy in our own spectrum;
  abs=absorption lines from the host galaxy in our own spectrum; 
  SNID=cross-correlation with \snia\ spectral templates using the SNID
code of \cite{SNID} [typical error $<0.005$].}
\tablenotetext{b}{NED reports a redshift z=0.049 based on the
    host-galaxy redshift given in IAUC~5882
    \citep{Schmidt_iauc5882}. We have re-derived the redshift using
    the same spectrum of the host galaxy as \cite{Schmidt_iauc5882}.}
\tablenotetext{c}{NED reports an approximate redshift $z=0.13$ based on the
    blueshift of the Si\two\,\l5972 line (assumed to be 10000\,\kms\
    in IAUC~6405; \citealt{Garnavich_iauc6405}); 
    \cite{Riess/etal:1999a} give $\log(cz)=4.571$, or
    $cz=37239$\,\kms\ ($z=0.1242$), ``from [their] spectra of the host galaxies'';
    last, \cite{MLCS2k2} give $cz=37109$\,\kms\ ($z=0.1238$). There
    are no obvious galaxy lines in our spectrum of SN~1996ab, nor
    could we find a spectrum of its host galaxy.}
\tablenotetext{d}{NED reports an erroneous redshift of $cz=1800$\kms\
  ($z=0.0060$) for MCG +07-36-33.}
\tablenotetext{e}{\cite{Scalzo/etal:2010} report $z=0.07416\pm0.00082$ based on [O\two]\,\l3727 and H$\alpha$.}
\tablenotetext{f}{Redshift of a nearby galaxy cluster; the SN spectrum is consistent with this redshift.}
\tablenotetext{g}{\cite{Bailey/etal:2009} report the CMB-frame
  redshift for this SN. We derived a heliocentric redshift using its
  coordinates as reported on the SNfactory Supernova
  Discoveries webpage.}
\end{deluxetable}

\epsscale{1.15}
\epsscale{1.0}

\begin{figure}
\plotone{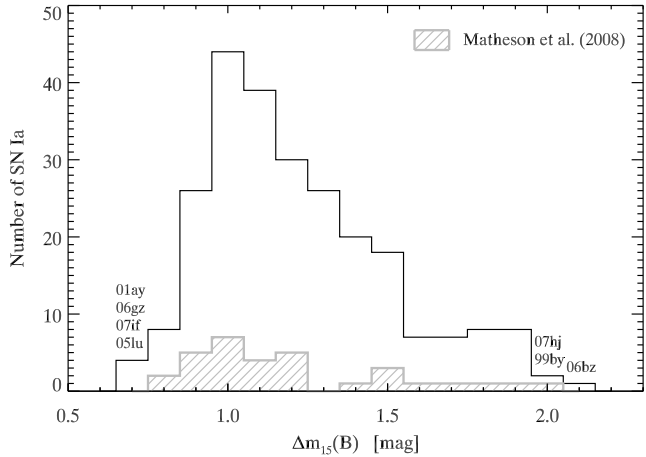}
\caption{\label{fig:dm15histo}
Histogram of the \nsndmft\ \sneia\ for which we have a reliable
\dmft\ measurement. The hatched histogram corresponds to the
\snia\ sample published by \cite{Matheson/etal:2008}. There are four
\sneia\ with $\dmft \approx 0.7$\,mag (SN~2001ay, SN~2005lu, SN~2006gz, and SN~2007if)
and three \sneia\ with $\dmft\approx 2$ (SN~1999by, SN~2006bz, and SN~2007hj).
}
\end{figure}

The distribution of the light-curve width parameter
\dmft\footnote{The difference in $B$-band magnitude between
maximum light and 15\,d after maximum \citep{Phillips:1993}.}
is shown in Fig.~\ref{fig:dm15histo} (see also
Table~\ref{tab:snparam}). Of the \nsntmax\ \sneia\ in our sample with
$t_{\rm max}(B)$ information, \nsndmft\ have a reliable \dmft\
estimate. For most SN we use the \dmft\ inferred from our own BayeSN
(\nsndmftbayesn) or SNooPy (\nsndmftsnpy) fits (we checked these were
consistent with previously-published values). For the remaining
\nsndmftother\ objects we take \dmft\ values published in 
the literature (mostly from \citealt{Jha/etal:2006},
\citealt{Ganeshalingam/etal:2010}, \citealt{Burns/etal:2011} and
\citealt{Stritzinger/etal:2011}). Our sample covers a large range in
\dmft, from luminous \sneia\ with $\dmft\approx0.7$\,mag to faint ones with
$\dmft\gtrsim2.0$\,mag. There are \nsnnot\ SN in our sample with
1991T/1999aa-like spectra, while \nsnnobg\ have spectra similar to
SN~1991bg or SN~1986G. Of the three \sneia\ with the smallest \dmft\
in our sample, two (SN~2006gz, \citealt{Hicken/etal:2007} and
SN~2007if, \citealt{Scalzo/etal:2010}) were speculated to originate from 
super-Chandrasekhar-mass progenitors.

\epsscale{1.15}
\epsscale{1.0}


\section{Spectroscopic subclasses}\label{sect:specsub}

We investigate how \sneia\ can be classified into various subclasses
based on their spectra, using the classification schemes of
\cite{Branch/etal:2006} and \cite{WangX/etal:2009b}. We also attempt
to quantify the spectroscopic and photometric variation within each
subclass. Last, we present updated correlations of several
spectroscopic indicators with \dmft. The results in this section are
largely based on spectra from the CfA SN Program (see
\S~\ref{sect:sample}), but we complement them with published data from
the literature, whose references are given in
  Table~\ref{tab:litspec}.


\subsection{Branch and Wang classification schemes}

\subsubsection{\cite{Branch/etal:2006} classification}\label{sect:branchclass}

\cite{Branch/etal:2006} presented a new classification scheme for
\sneia\ based on measurements of the {\it pseudo} equivalent widths
(pEW) of absorption features near 5750\,\AA\ and 6100\,\AA\ (generally
attributed to Si\two\,\l5972 and Si\two\,\l6355, respectively) in
their maximum-light spectra. The overall shape of the Si\two\,\l6355
absorption also enters this classification
scheme. They divided \sneia\ into four subgroups based on their
position in this two-dimensional parameter space, but noted the
absence of strict boundaries between them, suggesting that the
different subgroups were not physically distinct but instead
corresponded to a continuous distribution of properties
(with the possible exception of SN~2002cx-like events).

\begin{figure*}
\resizebox{\textwidth}{!}{
\includegraphics{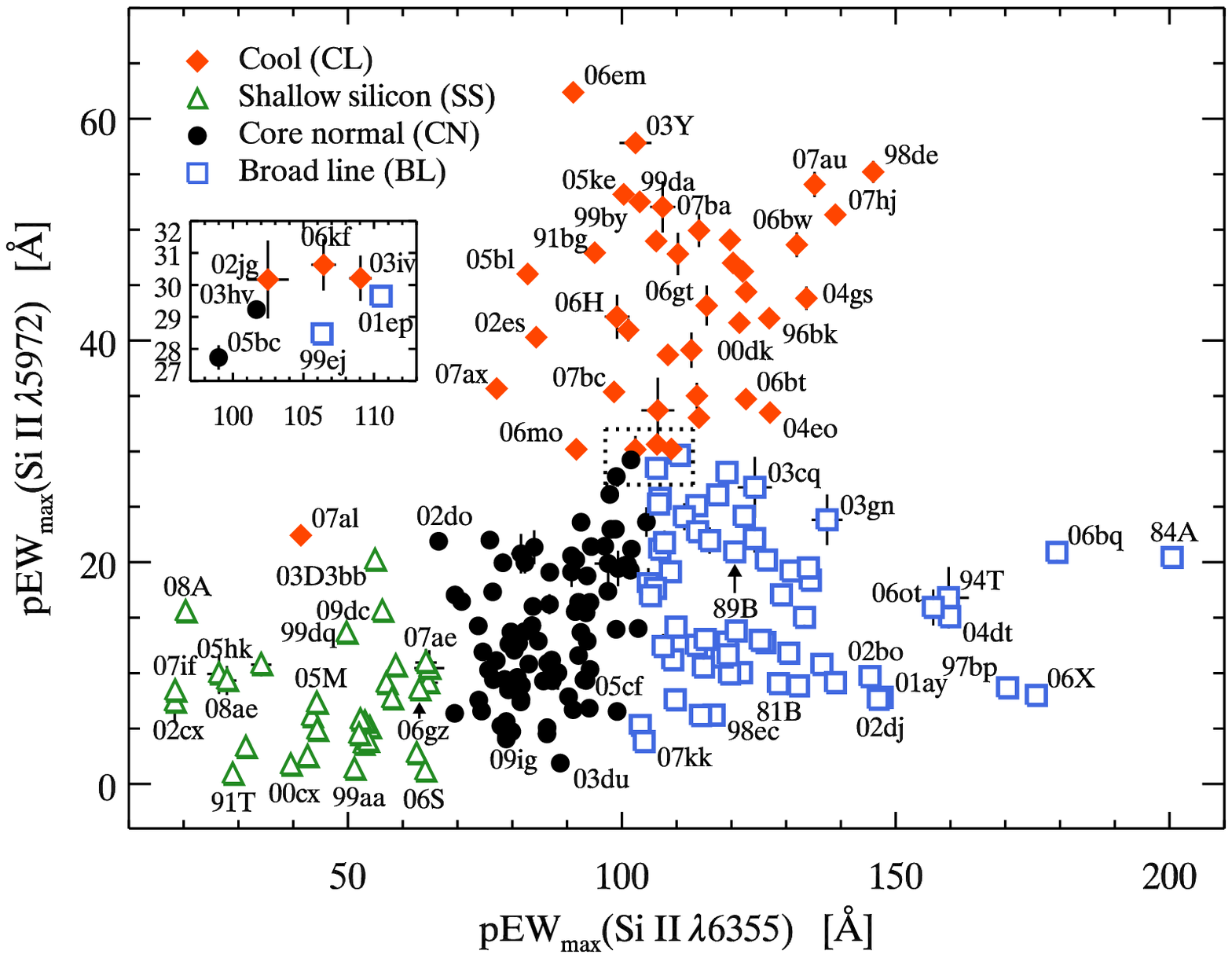}\hspace{1cm}
\includegraphics{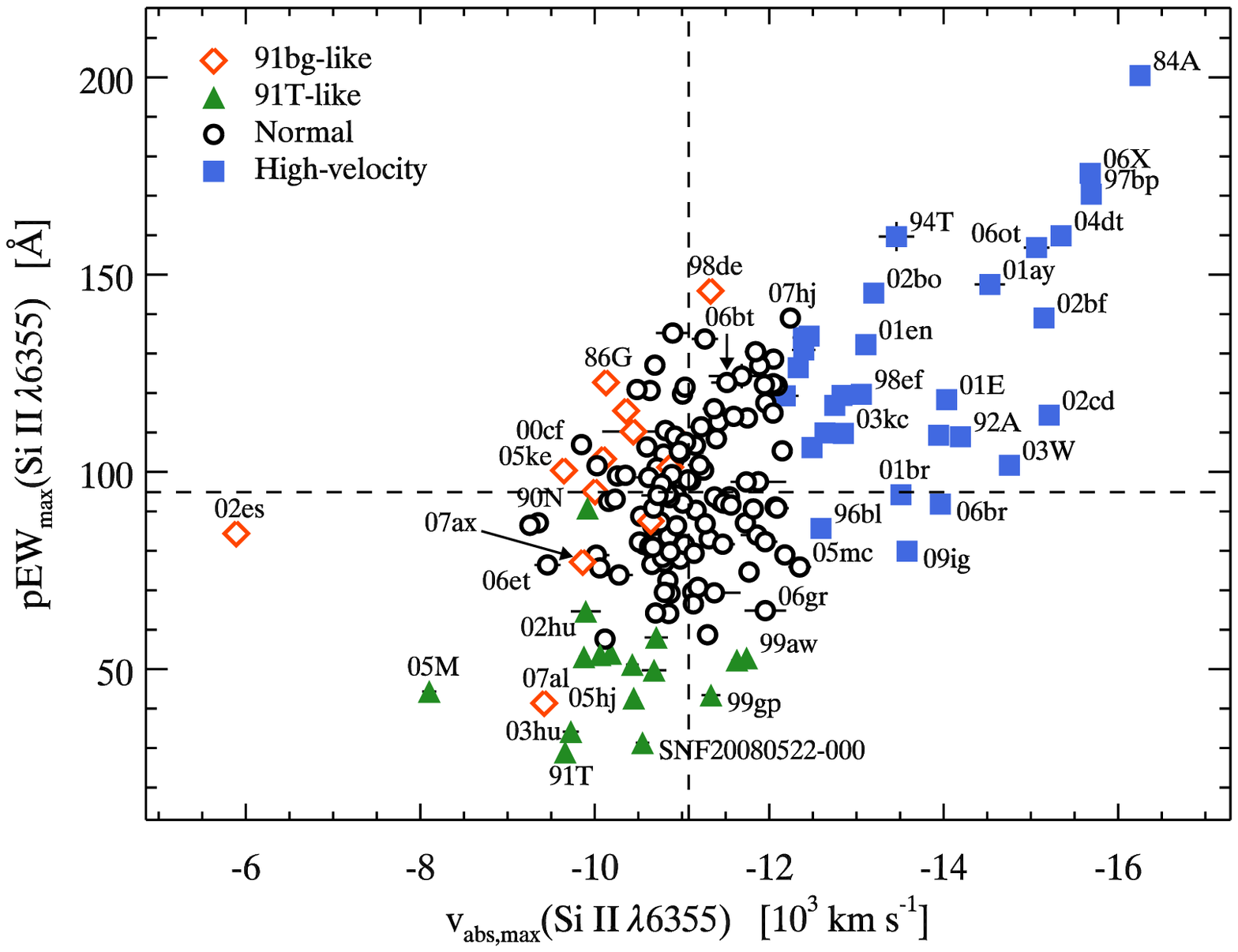}
}
\caption{\label{fig:branchwangclass}
{\it Left:}
Pseudo-EW of the Si\two\,\l5972 absorption {\it vs.} that of the
Si\two\,\l6355 absorption within five days from maximum light. When
several measurements are available for a given SN within this phase
range, we only show the measurement closest to maximum light. The
different symbols correspond to the various spectroscopic subclasses
defined by \cite{Branch/etal:2006}. The inset corresponds to the
region bounded by the dotted box in the main plot.
{\it Right:}
Pseudo-EW of the Si\two\,\l6355 absorption {\it vs.} the absorption
velocity of the same line within three days from maximum light. When
several measurements are available for a given SN within this phase
range, we only show the measurement closest to maximum light. The
different symbols correspond to the various spectroscopic subclasses
defined by \cite{WangX/etal:2009b}. Note that peculiar \sneia\ (e.g.,
SN~2002cx, SN~2005cc, SN~2005hk, SN~2008A) are not included in this
classification scheme. The dotted lines correspond to the mean
pseudo-EW and mean $v_{\rm abs}$ of the Si\two\,\l6355 line for the
Normal sample ({\it open circles}).
}
\end{figure*}

\begin{figure}
\centering
\plotone{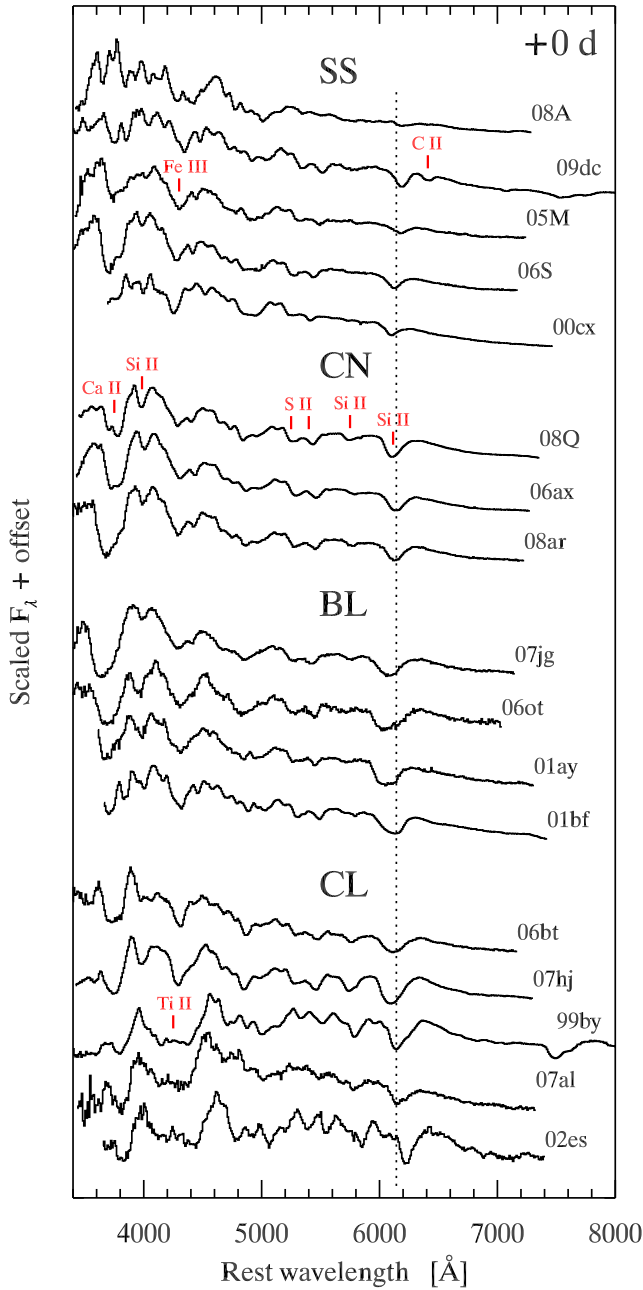}
\caption{\label{fig:montbranchmax}
Montage of representative \snia\ spectra at maximum light in each of
the spectroscopic subclasses defined by \cite{Branch/etal:2006}. We
indicate tentative line identifications corresponding to the 
dominant ion contribution to specific absorption features (the BL
subclass shares the same line identifications as the CN subclass). The
vertical dotted line corresponds to Si\two\,\l6355 blueshifted by
10000\,\kms. All the spectra are from the CfA SN Program, except for
the spectrum of SN~2009dc \citep{Taubenberger/etal:2011}.
}
\end{figure}

Figure~\ref{fig:branchwangclass} ({\it left}) shows the pEW of
the Si\two\,\l5972 line {\it vs.} that of the Si\two\,\l6355 line for
\nsnbranchclassfig\ \sneia\ (see Fig.~2 of
  \citealt{Branch/etal:2009}), based on spectra within 5 days from
maximum light (see Table~\ref{tab:branchwang}). The different
symbols correspond to the various spectroscopic subclasses defined by
\cite{Branch/etal:2006}. Example maximum-light spectra within each
subclass are shown in Fig.~\ref{fig:montbranchmax}.
The exact locations of the boundaries
between the different subclasses are ill-defined, but this does not
affect our analysis.

The cluster of black dots
forms the ``Core Normal'' (CN) subclass. \sneia\ with similar
pEW(Si\two\,\l5972) but with broader Si\two\,\l6355 absorptions
(characteristic of larger expansion velocities) are labeled ``Broad
Line'' (BL). Together, these subclasses constitute the 
bulk of \sneia\ loosely referred to as ``normal'' in the
literature. There is no strict boundary between the CN and BL
subclasses, although the \sneia\ we classify as BL have
pEW(Si\two\,\l6355)$\gtrsim105$\,\AA. As noted by
\cite{Branch/etal:2009}, there is not a one-parameter sequence from
the CN subclass to the more extreme BL \sneia\ (such as SN~1984A,
SN~2006X, and SN~2006bq).
Two \sneia\ classified as BL by \cite{Branch/etal:2009} are
re-classified by us as CN (SN~1999gd, SN~2002er; see
Table~\ref{tab:branchwang}) based on the magnitude of 
their pEW(Si\two\,\l6355).
 We also classify SN~2009ig in the CN subclass, as opposed to the BL
 subclass \citep[see][]{Parrent/etal:2011}.

\begin{deluxetable*}{lrrrrcc}
\tabletypesize{\scriptsize}
\tablewidth{0pt}
\tablecaption{\label{tab:branchwang}\snia\ classification using the schemes of \cite{Branch/etal:2006} and \cite{WangX/etal:2009b}}
\tablehead{\colhead{SN} & \colhead{pEW(\l5972)\tablenotemark{a}} & \colhead{pEW(\l6355)\tablenotemark{b}} & \colhead{$v_{\rm abs}$(\l6355)\tablenotemark{c}} & \colhead{Phase\tablenotemark{d}} & \colhead{Branch\tablenotemark{e}} & \colhead{Wang\tablenotemark{f}} \\
           \colhead{}   & \colhead{(\AA)}       & \colhead{(\AA)}       & \colhead{(\kms)}                & \colhead{(d)}   & \colhead{class}  & \colhead{class}}
\startdata
1981B                          &     9.1 &   128.6 & $-$12050 & $-$1.5  &      BL &       N \\
1984A                          &    20.4 &   200.5 & $-$16251 & $-$2.8  &      BL &      HV \\
1986G                          &    44.4 &   122.7 & $-$10128 & $-$0.2  &      CL &    91bg \\
1989B\tablenotemark{\ddag}     &    20.9 &   120.6 & $-$10634 & $-$1.3  &      BL &       N \\
1990N                          &    10.9 &    87.1 & $-$9352  & 1.6     &      CN &       N \\
1990O                          &     6.7 &    91.1 & $-$12069 & $-$0.9  &      CN &       N \\
1991M                          &    18.4 &   134.4 & $-$12458 & 2.7     &      BL &      HV \\
1991T                          &     0.9 &    29.0 & $-$9660  & $-$1.5  &      SS &     91T \\
1991bg                         &    47.9 &    95.0 & $-$10004 & 0.9     &      CL &    91bg \\
1992A                          &    19.1 &   108.9 & $-$14192 & $-$0.8  &      BL &      HV \\[-.25cm]
\enddata
\tablecomments{Table~\ref{tab:branchwang} is published in its entirety in the electronic edition of The Astronomical Journal. A portion is shown here for guidance regarding its form and content.}
\tablenotetext{a}{Pseudo-EW of the Si\two\,\l5972 absorption feature (formal uncertainty typically $<1$\,\AA).}
\tablenotetext{b}{Pseudo-EW of the Si\two\,\l6355 absorption feature (formal uncertainty typically $<1$\,\AA).}
\tablenotetext{c}{Absorption velocity the Si\two\,\l6355 absorption feature (formal uncertainty typically $<100$\,\kms).}
\tablenotetext{d}{Phase of measurements in rest-frame days from $B$-band maximum.}
\tablenotetext{e}{Branch class based on pEW(\l5972) and pEW(\l6355) within 5\,d from maximum light: CN=Core Normal, BL=Broad Line, SS=Shallow Silicon, CL=Cool. We also report a classification for \snia\ with no spectra within 5\,d from maximum (but within 7\,d) based on the overall shape of the Si\two\,\l6355 absorption feature.}
\tablenotetext{f}{Wang class based on $v_{\rm abs}$(\l6355) within 7\,d from maximum light (note that Fig.~\ref{fig:branchwangclass} only displays measurements within 3\,d from maximum): N=Normal, HV=High-velocity, 91T=1991T-like, 91bg=1991bg-like. Peculiar \sneia\ (SN~2000cx, SN~2002cx, SN~2003fg, SN~2005hk, SN~2006gz, SN~2007if, SN~2008A, SN~2008ae, SN~2009dc) are not part of this classification scheme.}
\tablenotetext{g}{SN~2003fg is also known as SNLS-03D3bb \citep{Howell/etal:2006}.}
\tablenotetext{\dag}{Reddening and contamination by the host galaxy renders the Branch classification uncertain.}
\tablenotetext{\ddag}{Branch classification differs from that reported by \cite{Branch/etal:2009} [indicated in parentheses]: SN~1989B=BL (CL), SN~1999ac=CN (SS), SN~1999ee=CN (SS), SN~1999gd=CN (BL), SN~2000E=CN (SS), SN~2002er=CN (BL), SN~2005cg=CN (SS).}
\end{deluxetable*}

\epsscale{1.15}
\epsscale{1.0}

The ``Cool'' (CL) subclass consists of \sneia\ with deeper
Si\two\,\l5972 absorptions, often associated with low-luminosity
events. They are referred to as ``Cool'' due to the notable absorption
band around $\sim4200$\,\AA\ caused by lines of Ti\two, a signature of
lower temperatures in the line-forming region \citep[see,
  e.g.,][]{Hatano/etal:1999}. The strength of this absorption band
varies greatly within the CL subclass: it is strong in SN~2005ke,
moderate in SN~1986G, and close to nonexistent in SN~2007au. The CL
subclass contains SN~2006bt, which has a slowly-declining light
curve characteristic of luminous \sneia\ but with spectra displaying
Ti\two\ absorption features characteristic of low-luminosity
\sneia\ \citep{SN2006bt}. It also includes SN~2004eo, labeled
``transitional'' by \cite{Pastorello/etal:2007b} due to its
intermediate properties between normal and low-luminosity \sneia.
Based on its resemblance to SN~2004eo, \cite{Branch/etal:2009}
classified SN~1989B as a CL \snia\ despite its shallower
Si\two\,\l5972 absorption. This SN also displays narrower features in
the 4600-5100\,\AA\ spectral region, indicative of lower ejecta
expansion velocities. However, the width of its Si\two\,\l6355
absorption is comparable to other BL \sneia, and we choose to include
SN~1989B in the BL subclass (the developer of this
  classification scheme concurs; Branch 2011, priv. comm.).

As seen from the inset in Fig.~\ref{fig:branchwangclass} ({\it left}),
there appears to be a continuous distribution between CL \sneia\ at
the low end of the pEW(Si\two\,\l5972) distribution and CN and BL
\sneia\ at the high end of their pEW(Si\two\,\l5972)
distributions. However, we identify one CL \snia\ (SN~2007al) with
a remarkably shallow silicon absorption. We include this SN in the CL
subclass based on the presence of a prominent Ti\two\ absorption
feature in its maximum-light spectrum (see
Fig.~\ref{fig:montbranchmax}).

The last subclass, ``Shallow Silicon'' (SS), consists of \sneia\ with
weak Si\two\,\l5972 and Si\,\two\,\l6355. It is a heterogeneous
category, including luminous 1991T/1999aa-like events characterized
by higher ionization lines (Fe\three), \sneia\
resulting from possible super-Chandrasekhar-mass progenitors
(SNLS-03D3bb or SN~2003fg, SN~2006gz, SN~2007if, SN~2009dc),
faint 2002cx-like SN (SN~2005hk, SN~2008A), and otherwise
spectroscopically ``normal'' events (e.g., SN~2006S). 
Several \sneia\ classified as SS by \cite{Branch/etal:2009} are
re-classified by us as CN (SN~1999ac, SN~1999ee, SN~2000E,
SN~2005cg; see Table~\ref{tab:branchwang}) based on the magnitude of
their pEW(Si\two\,\l6355). SN~1999ac was already noted as a
borderline SS/CN by \cite{Branch/etal:2007}, and its peculiar nature
discussed at length by \cite{Garavini/etal:2005} and
\cite{Phillips/etal:2006}.

Of the 246 \sneia\ to which we assign a \cite{Branch/etal:2006}
classification in Table~\ref{tab:branchwang}, 94 (38.2\%) are of the
CN subclass, 74 (30.1\%) are of the BL subclass, 43 (17.5\%) are of the CL
subclass, and 35 (14.2\%) are of the SS subclass. Note that
these numbers simply reflect the properties of the SN sample studied
in this paper and not of the \snia\ class as a whole.

\begin{figure}
\plotone{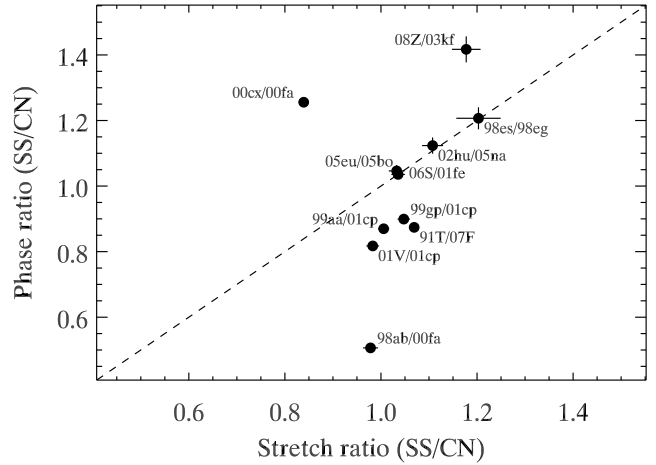}
\caption{\label{fig:ssagerate}
Ratio of phases of \snia\ spectra at three weeks past maximum in the SS
subclass to that of the best-match SNID template in the CN subclass
{\it vs.} the ratio of their $B$-band light-curve stretch parameters
from SALT2. The labels associated with each point indicate the name of
the SS supernova followed by the name of the best-match CN
template. The dashed line is the 1:1 correspondence.
}
\end{figure}

By $\sim3$ weeks past maximum light, spectra of SS \sneia\ are not
distinguishable from those of the CN subclass at similar phases. By
comparing the +19\,d spectrum of the SN~1999aa to spectra of Core
Normal \sneia, \cite{Branch/etal:2009} found the best match to be
SN~1994D at +14\,d. The corresponding phase ratio ($19/14\approx1.36$)
closely matched the ratio of their $B$-band light curve
stretch parameters ($1.143/0.838\approx1.36$), prompting
\cite{Branch/etal:2009} to suggest that \sneia\ from the CN and SS
subclasses ``age spectroscopically at the same rate that they decline
photometrically.'' We have tested this hypothesis by finding the
best-match CN \snia\ template to spectra of SS \sneia\ in the phase
range [+19,+23]\,d using SNID, and plot the ratio of their phases
against that of their stretch parameters (from SALT2) in
Fig.~\ref{fig:ssagerate}. While most SS \sneia\ indeed lie on or
close to the 1:1 relation, there are a few significant outliers
(including the peculiar SN~2000cx and the 1991T-like SN~1998ab), 
and the Pearson correlation coefficient between both ratios is only
$r=0.36$. We find similar results when considering SS \sneia\ at
two and four weeks past maximum light, respectively. The
hypothesis of \cite{Branch/etal:2009} therefore does not apply
universally to \sneia\ of the SS subclass.

\epsscale{1.15}
\epsscale{1.0}

\begin{figure}
\plotone{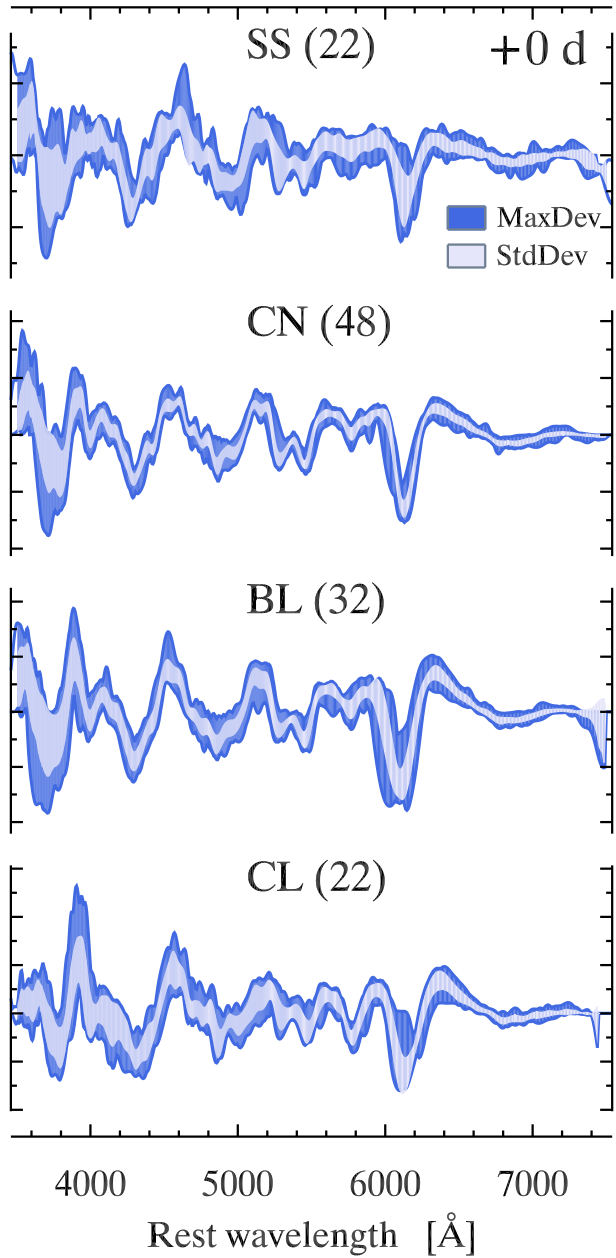}
\caption{\label{fig:branchsnid}
Composite maximum-light spectra for the various spectroscopic
subclasses defined by \cite{Branch/etal:2006}. Individual spectra have
been flattened through division by a {\it pseudo} continuum {\it (see
  text for details)}.
The shaded bands
correspond to the standard ({\it light blue}) and maximum ({\it dark
blue}) deviation about the mean maximum-light spectrum. The number in
between parentheses indicates the number of individual spectra that
were used to generate each composite spectrum.
}
\end{figure}

The spectroscopic properties of \sneia\ belonging to these different
subclasses have been discussed at length by
\cite{Branch/etal:2006,Branch/etal:2007,Branch/etal:2008} for
maximum-light, pre-maximum, and post-maximum epochs, respectively, and
by \cite{Branch/etal:2009} at all epochs based on a larger sample. Here we
focus on the spectroscopic variation at maximum-light. Within each
subclass, we generate composite spectra using the same
pre-processing as done by SNID: the individual spectra are
``flattened'' through division by a {\it pseudo} continuum. We then
compute the mean flux in each wavelength bin, as well as the standard
(and maximum) deviation from the mean. The result is a composite
spectrum with error bands for each subclass, which we show in
Fig.~\ref{fig:branchsnid}. The large variation within the SS subclass
is clearly visible throughout the optical range (3500-7500\,\AA). The
CN subclass displays the smallest variation at any given wavelength,
as expected, albeit with a noticeable variation in the strength of the
Ca\two\,H\&K absorption feature. The composite spectrum for the BL
subclass reveals the presence of high-velocity components to the
Si\two\,\l6355 line in the maximum-deviation spectrum ({\it dark blue}) in
addition to the large variation blueward of $\sim4000$\,\AA\ as for
the CN subclass (see, e.g, SN~2001bf and SN~2007jg in
Fig.~\ref{fig:montbranchmax}). The CL composite spectrum shows the
varying strength of the Ti\two\ absorption feature as well as the
large range in depth and position of the Si\two\,\l6355 absorption
(shallow in SN~2007al, less blueshifted in SN~2002es; see
Fig.~\ref{fig:montbranchmax}).

\epsscale{1.1}
\epsscale{1.0}

In a recent paper, \cite{Stritzinger/etal:2011} noted a similarity
between the photometric properties of SN~2006ot and SN~2006bt
\citep{SN2006bt}: a broad light curve characteristic of luminous
\sneia\ ($\dmft=0.84$\,mag), but a weak secondary $i$-band maximum
characteristic of low-luminosity events. They also noted major
differences between both SN in their maximum-light spectra. SN~2006bt
has a deep Si\two\,\l5972 absorption and evidence for absorption by
Ti\two, both of which are characteristic of low-luminosity
\sneia. SN~2006ot, on the other hand, is characterized by a shallower
Si\two\,\l5972 absorption and a broad Si\two\,\l6355 absorption, with
no evidence for features associated with Ti\two. In the classification
scheme of \cite{Branch/etal:2006}, SN~2006bt belongs to the CL
subclass while SN~2006ot is an extreme BL \snia, reflecting these
spectroscopic differences. Nonetheless, \cite{Stritzinger/etal:2011}
argue that at 3-4 weeks past maximum light, the spectra of SN~2006ot
are more similar to those of SN~2006bt than to those of the normal
SN~2006ax (part of the CN subclass), again stressing the similarity
between both SN. At such late phases, however, differences between the
various subclasses are less apparent than at maximum light, as noted
by \cite{SN2006bt}\footnote{In fact, \cite{SN2006bt} note the {\it
similarity} between the spectra of SN~2006bt and SN~2006ax at
  $\sim40$\,d past maximum light.}. We have run SNID on the +24\,d
spectrum of SN~2006ot 
and find good matches to spectra of \sneia\ from the CN, CL, and BL
subclasses at similar phases. The best-match spectrum of SN~2006bt (at
+33\,d) is ranked only 21$^{\rm st}$ overall. Similarly, the
best-match template spectrum for the +33\,d spectrum of the CL
SN~2006bt \citep{SN2006bt} is the CN SN~2004S
\citep{Krisciunas/etal:2007}, illustrating the difficulty in
distinguishing between different \snia\ subclasses at late times. If
anything, SN~2006bt and SN~2006ot demonstrate how two \sneia\ with
similar light-curve shapes can have very different maximum-light spectra.

\subsubsection{\cite{WangX/etal:2009b} classification}\label{sect:wangclass}

The classification scheme of \cite{WangX/etal:2009b} is based on the
velocity at maximum absorption (or absorption velocity, $v_{\rm abs}$)
of the Si\two\,\l6355 line around maximum light. It was designed
to study the spectroscopic diversity amongst ``normal'' \sneia.
We use our own measurements on the same sample of 10
\sneia\ used by \cite{WangX/etal:2009b} to define a mean $v_{\rm abs}$
trend within one week from $B$-band maximum. \sneia\ with
Si\two\,\l6355 absorption velocities 3$\sigma$ above this mean trend
(in absolute terms) are classified as ``High-velocity'' (HV), the
remaining objects forming the ``Normal'' (N) subclass (see
Table~\ref{tab:branchwang}). Note that this
classification scheme does not apply to luminous 1991T/1999aa-like
\sneia, nor to low-luminosity 1991bg-like events or peculiar objects
(2002cx-like, possibly super-Chandrasekhar etc.). 
Of the 246 \sneia\ to which we assign a \cite{WangX/etal:2009b}
classification in Table~\ref{tab:branchwang}, 143 (58.1\%) are of the
N subclass, while 51 (20.7\%) belong to the HV subclass. Again,
these numbers simply reflect the properties of the SN sample studied
in this paper and are not meant to reflect the \snia\ class as a
whole.

This classification scheme is illustrated in
Fig.~\ref{fig:branchwangclass} {\it (right)}, where we show the
relation between the pseudo-EW of the Si\two\,\l6355 line {\it vs.}
its absorption velocity for \sneia\ with spectra within 3\,d from
maximum light. This figure is largely similar to Fig.~2 of
\cite{WangX/etal:2009b}, the Normal \sneia\ forming a cluster in
the (pEW,$v_{\rm abs}$) plane, and a tail of High-velocity objects
following a trend of larger pEW for larger absorption
blueshifts. The mean pEW and mean $v_{\rm abs}$ for the Normal sample
are $\sim95$\,\AA\ and $\sim-11000$\,\kms, respectively, comparable to
the values found by \cite{WangX/etal:2009b}. At maximum light, the
boundary between the N and HV subclasses is located at
$\sim-12200$\,\kms. There is a large overlap between the Core Normal
subclass of \cite{Branch/etal:2006} and the Normal subclass of
\cite{WangX/etal:2009b}, as well as between the BL and HV
subclasses (with a few exceptions such as SN~2009ig, which is
  part of the CN and HV subclasses). The 1991T-like objects have
absorption velocities similar 
to the Normal subclass, but with smaller pEW. All 1991T-like \sneia\
belong to the Shallow Silicon subclass of
\cite{Branch/etal:2006}, except for the 1999aa-like SN~1999ac. \sneia\
with spectra similar to SN~1991bg 
also have similar velocities but on average larger pEW. Notable
exceptions include SN~2007al with similar velocities but much smaller
pEW (also an outlier in the CL subclass; see
Fig.\ref{fig:branchwangclass}, {\it left}), and SN~2002es with similar
pEW but almost a factor of two lower $|v_{\rm abs}|$. This latter SN
is discussed in more detail by \cite{Ganeshalingam/etal:2012}.
All 1991bg-like \sneia\ belong to the Cool subclass of
\cite{Branch/etal:2006}.

The largest measured Si\two\,\l6355 pEW and blueshift in
Fig.~\ref{fig:branchwangclass} is for SN~1984A (pEW$\approx 200$\,\AA,
$|v_{\rm abs}| \approx 16250$\,\kms). This SN is also the most extreme
BL \snia\ in the classification scheme of \cite{Branch/etal:2006}. We
note that \cite{WangX/etal:2009b} measure pEW$\approx170$\,\AA\ and
$|v_{\rm abs}|\approx 15200$\,\kms\ (see their Fig.~2), but their
measurement is based on a non-flux calibrated spectrum directly
scanned from Fig.~4 of \cite{Barbon/etal:1989}, biasing their pEW
measurement to lower values. Moreover, their use of the classical
Doppler formula (as opposed the relativistic version used in this
paper; see Eq.~6 in \citealt{Blondin/etal:2006}) leads to a
$\sim500$\,\kms\ lower blueshift when converting the 
wavelength at maximum absorption in Si\two\,\l6355 to a velocity. 


\subsection{Photometric properties as a function of spectroscopic class}

Based on a restricted sample of 9 \sneia\ from the CN subclass,
\cite{Branch/etal:2009} inferred a mean absolute peak $B$-band
magnitude of $-19.48$\,mag with a scatter of only 0.14\,mag,
suggesting that \sneia\ belonging to this subclass could be true
standard candles. They noted, however, that this should be tested on a
larger sample of \sneia\ in the Hubble flow. 

We use the BayeSN statistical models of
\cite{Mandel/etal:2009,Mandel/etal:2011} to infer the intrinsic peak
absolute $B$-band magnitude (and hence the total extinction along the
line of sight) of \sneia\ in a consistent manner (assuming a Hubble
constant of $H_0=72$\,\kms\,Mpc$^{-1}$; see Table~\ref{tab:snparam}).
These statistical models describe the apparent distribution of light
curves as a convolution of intrinsic \snia\ variations and a dust
distribution. \cite{Mandel/etal:2011} modeled the intrinsic covariance
structure of the full multi-band light curves, capturing population
correlations between the intrinsic absolute magnitudes, intrinsic
colors and light curve decline rates over multiple phases and
wavelengths, as well as the distribution of host galaxy dust and an
apparent correlation between the dust extinction $A_V$ and its
wavelength dependence, parameterized by $R_V$. The models fit
individual optical and NIR \snia\ light-curve data to estimate the
dust extinction, apparent and absolute light curves, and intrinsic
colors for each SN.  These models were trained on a nearby ($z <
0.07$) set of \sneia\ with optical (CfA3, \citealt{Hicken/etal:2009a};
Carnegie SN Program, \citealt{Contreras/etal:2010}) and 
NIR (PAIRITEL; \citealt{Wood-Vasey/etal:2008}) data, plus light curves
from the literature with joint optical and NIR observations. For the
fits used in this paper, we have employed the distance-redshift
constraint, except where alternative distance information was used
(e.g., Cepheids; see \S~\ref{sect:vnebfeni}).

\begin{figure}
\plotone{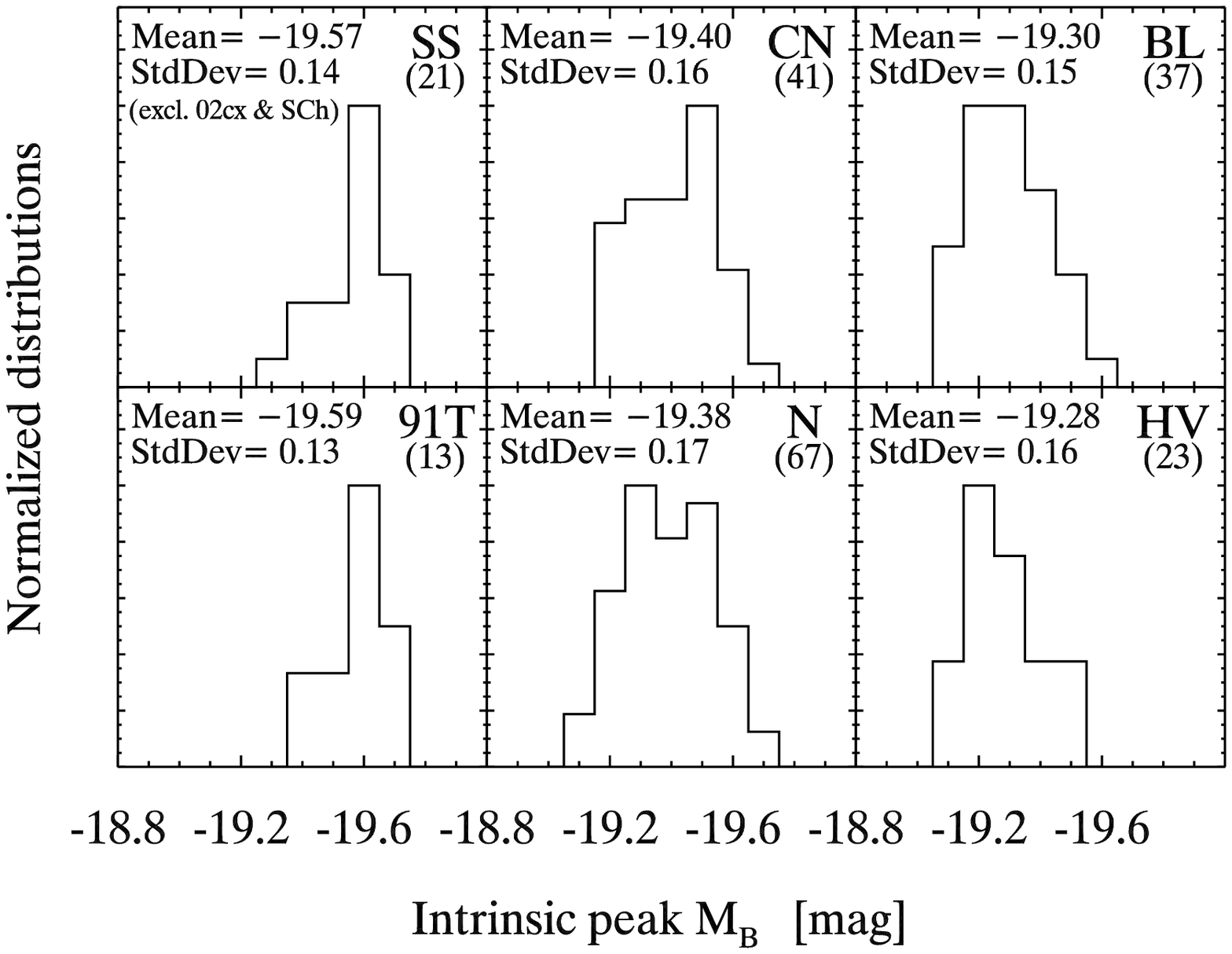}
\caption{\label{fig:mbclass}
Normalized distributions of intrinsic peak absolute $B$-band magnitude for the
various spectroscopic subclasses defined by
\cite{Branch/etal:2006} [{\it top}] and \cite{WangX/etal:2009b} [{\it
 bottom}]. The peak rest-frame magnitude and host-galaxy extinction
were determined using the BayeSN light-curve statistical model of
\cite{Mandel/etal:2011}. Only \sneia\ with a CMB-frame redshift
$z_{\rm CMB}>0.01$ and an inferred visual extinction $A_V<1$\,mag were
included. In each panel we indicate the sample mean and standard
deviation of the intrinsic peak $M_B$ (see also
Table~\ref{tab:mbclass}). The number of \sneia\ in each subclass is
indicated in between parentheses. We do not show the distributions for
the CL or 91bg subclasses since the BayeSN statistical model did not
include \sneia\ with $\dmft>1.6$\,mag in its training set. The same is
true of faint 2002cx-like \sneia\ and luminous (possibly
super-Chandrasekhar) events from the SS subclass. 
}
\end{figure}

Figure~\ref{fig:mbclass} shows the resulting intrinsic peak $M_B$
distribution for \sneia\ at CMB-frame redshifts $z_{\rm CMB}>0.01$
with an inferred visual extinction $A_V<1$\,mag, for the various
spectroscopic subclasses defined by 
\cite{Branch/etal:2006} [{\it top}] and \cite{WangX/etal:2009b} [{\it
 bottom}]. The 41 \sneia\ in the CN subclass that satisfy these criteria have
a mean intrinsic peak $M_B=-19.40$\,mag with $\sigma(M_B)=0.16$\,mag. This
is comparable to the values derived for the Normal subclass of
\cite{WangX/etal:2009b} [$M_B=-19.38$\,mag;
$\sigma(M_B)=0.17$\,mag]. The intrinsic peak $M_B$ scatter is small
for the CN (or Normal) subclass, but it is in fact larger than for the
SS ($\sigma(M_B)=0.13$\,mag) and BL subclasses ($\sigma(M_B)=0.15$\,mag) 
[the same is also true when comparing the Normal subclass to the 91T
and HV subclasses of \citealt{WangX/etal:2009b}]. We note, however,
that the SS distribution does not include faint 2002cx-like \sneia\
(which typically have $M_B\gtrsim-18$\,mag;
e.g., $M_B\approx-17.5$\,mag for SN~2002cx and $M_B\approx-18.0$\,mag
for SN~2005hk; see \citealt{Phillips/etal:2007}) nor luminous
(possibly super-Chandrasekhar) events (e.g., SN~2006gz with $M_B\approx-19.9$,
\citealt{SN2006gz}; or SN~2009dc with $M_B\approx-20.2$,
\citealt{Taubenberger/etal:2011}) since BayeSN did not include
such \sneia\ in its training set.

\epsscale{1.15}
\epsscale{1.0}

\begin{deluxetable}{lcccr}
\tablewidth{0pt}
\tablecaption{\label{tab:mbclass}Mean properties of \sneia\ in the classification schemes of \cite{Branch/etal:2006} and \cite{WangX/etal:2009b}}
\tablehead{\colhead{Class} & \colhead{$\langle M_B\rangle$ ($\sigma$)} & \colhead{$\langle \dmft\rangle$ ($\sigma$)} & \colhead{$\langle B-V\rangle$ ($\sigma$)} & \colhead{$N_{\rm SN}$}}
\startdata
SS\tablenotemark{a}      & $-19.57\ (0.14)$ & $0.94\ (0.13)$ & $-0.11\ (0.07)$ & 21 \\
CN      & $-19.40\ (0.16)$ & $1.12\ (0.18)$ & $-0.12\ (0.06)$ & 41 \\
BL      & $-19.30\ (0.15)$ & $1.22\ (0.15)$ & $-0.10\ (0.07)$ & 37 \\
\hline \\[-.2cm]
91T     & $-19.59\ (0.13)$ & $0.93\ (0.14)$ & $-0.12\ (0.07)$ & 13 \\
Normal  & $-19.38\ (0.17)$ & $1.15\ (0.21)$ & $-0.12\ (0.06)$ & 67 \\
HV      & $-19.28\ (0.16)$ & $1.21\ (0.14)$ & $-0.09\ (0.07)$ & 23 \\[-.25cm]
\enddata
\tablecomments{SS=Shallow Silicon; CN=Core Normal; BL=Broad Line; HV=High-velocity. Cool (CL) and 1991bg-like \sneia\ are not shown here {\it (see text for details)}.}
\tablenotetext{a}{excluding 2002cx-like \sneia\ and super-Chandrasekhar events.}
\end{deluxetable}

The mean intrinsic peak $M_B$ increases steadily along
the SS$\rightarrow$CN$\rightarrow$BL sequence, with a corresponding
increase in the mean \dmft\ [see Table~\ref{tab:mbclass}].
The same applies to the 91T$\rightarrow$N$\rightarrow$HV
sequence. There is a hint of redder mean intrinsic $B-V$ color for
the BL (HV) subclass compared to the CN (N) subclass, as expected from
the ``brighter-bluer'' relation of \cite{Tripp:1998}, but the
difference (0.02-0.03\,mag) is small compared to the scatter
(0.06-0.07\,mag) within each subclass and only measurable with
  sample averages.
Low-luminosity \sneia\ from the CL or 91bg subclasses are not included
in this comparison since the BayeSN statistical model did not include
\sneia\ with $\dmft>1.6$\,mag in its training set. Supernovae
similar to SN~1991bg have intrinsic peak 
$B$-band magnitudes $\sim2$\,mag fainter than normal \sneia, with
$\dmft\gtrsim1.9$\,mag, and much redder intrinsic $B-V$ colors at maximum
light. A prime example is SN~2005bl,
with $M_B\approx -17.2$\,mag, $\dmft\approx1.9$\,mag, and $(B-V)_{\rm
  max}\approx0.5$\,mag \citep{Taubenberger/etal:2008}.

\begin{figure*}
\resizebox{\textwidth}{!}{
\includegraphics{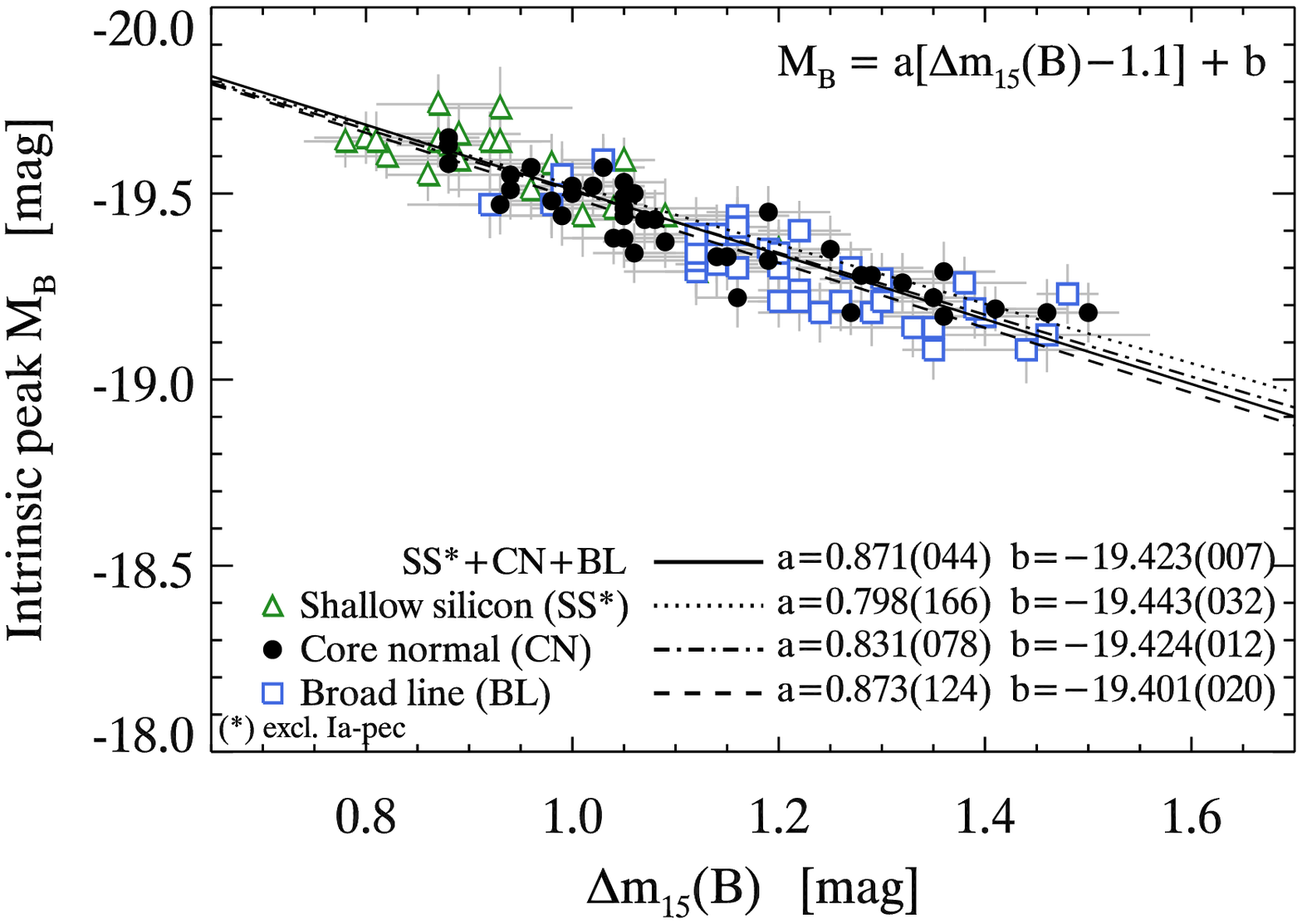}\hspace{1cm}
\includegraphics{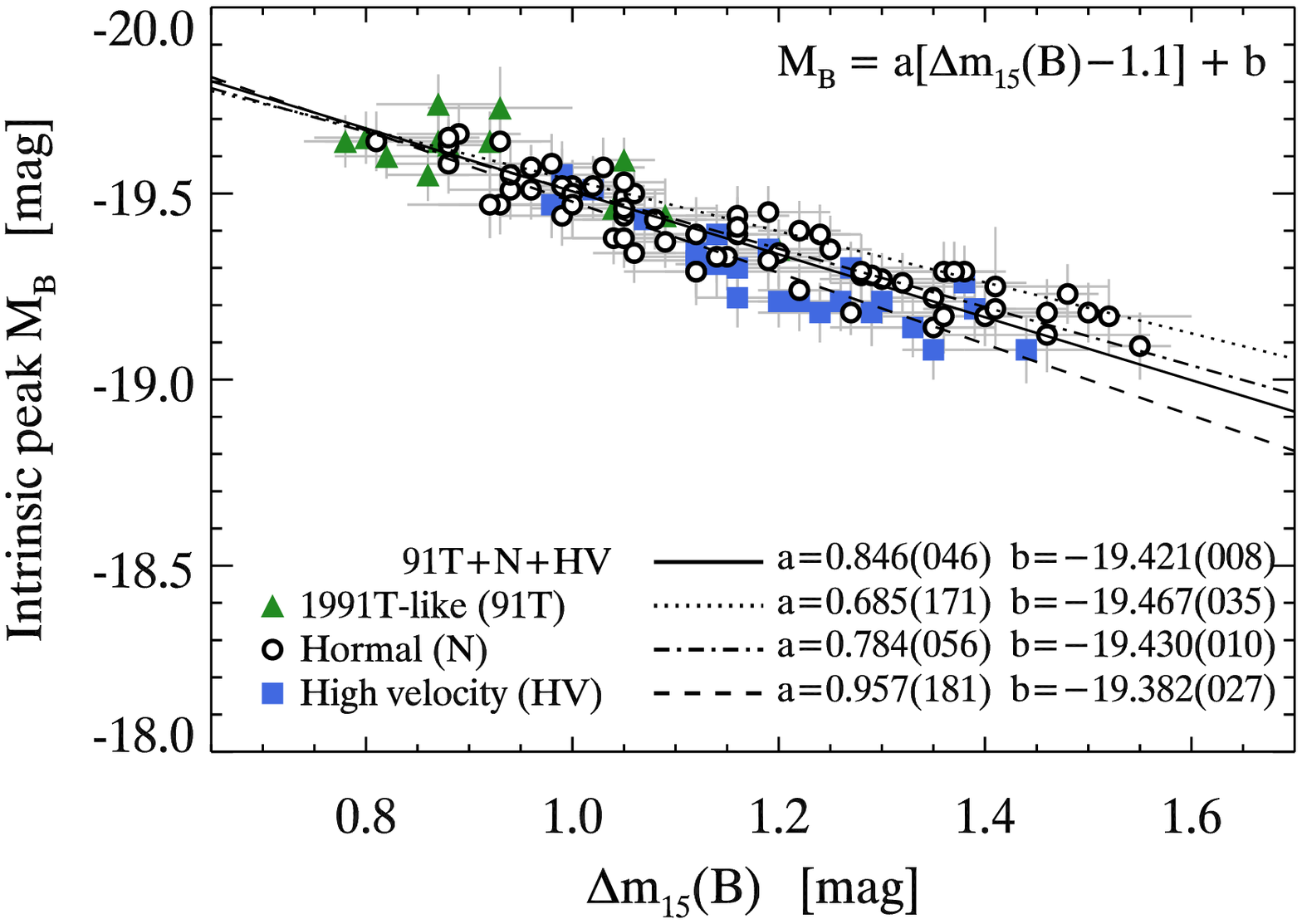}
}
\caption{\label{fig:wlrbranchwang}
Width-luminosity relation for the various spectroscopic subclasses
defined by \cite{Branch/etal:2006} ({\it left}) and
\cite{WangX/etal:2009b} ({\it right}). We report the linear
coefficients of the fit $M_B = a[\dmft-1.1]+b$ for the entire sample
({\it solid line}) and for the individual subclasses. Errors in
parentheses are given in units of 0.001 and 0.001\,mag for the
  slope and offset, respectively.
}
\end{figure*}

One might wonder whether \sneia\ in different subclasses obey
different width-luminosity relations
(WLR). Figure~\ref{fig:wlrbranchwang} ({\it left}) shows the intrinsic
peak $M_B$ inferred from BayeSN fits {\it vs.} \dmft\ for the SS, CN,
and BL subclasses of \cite{Branch/etal:2006}. Overplotted are best-fit
relations of the form $M_B = a[\dmft - 1.1] + b$
for the entire sample ({\it solid line}) and for the individual
subclasses ({\it dashed and dotted lines}). The coefficients of the
linear fits are given in each case. The slope of the WLR gets steeper
along the SS$\rightarrow$CN$\rightarrow$BL sequence, while the
intercept corresponds to a progressively fainter fiducial intrinsic
peak $M_B$ along this same sequence. The slopes and intercepts
for the individual subclasses are however consistent within
$\sim1\sigma$ of one another, such that a single
average WLR is adequate to describe all the subclasses (other
than the CL subclass). As seen from the right panel of
Fig.~\ref{fig:wlrbranchwang}, the same analysis holds when replacing
(SS,CN,BL) with (91T,N,HV) for the \cite{WangX/etal:2009b}
classification scheme, the only difference being the intercept
for the HV subclass which is $\sim2\sigma$ larger than those for the
Normal and 91T subclasses.

\subsection{Spectroscopic luminosity indicators}

\begin{figure*}
\resizebox{\textwidth}{!}{
\includegraphics{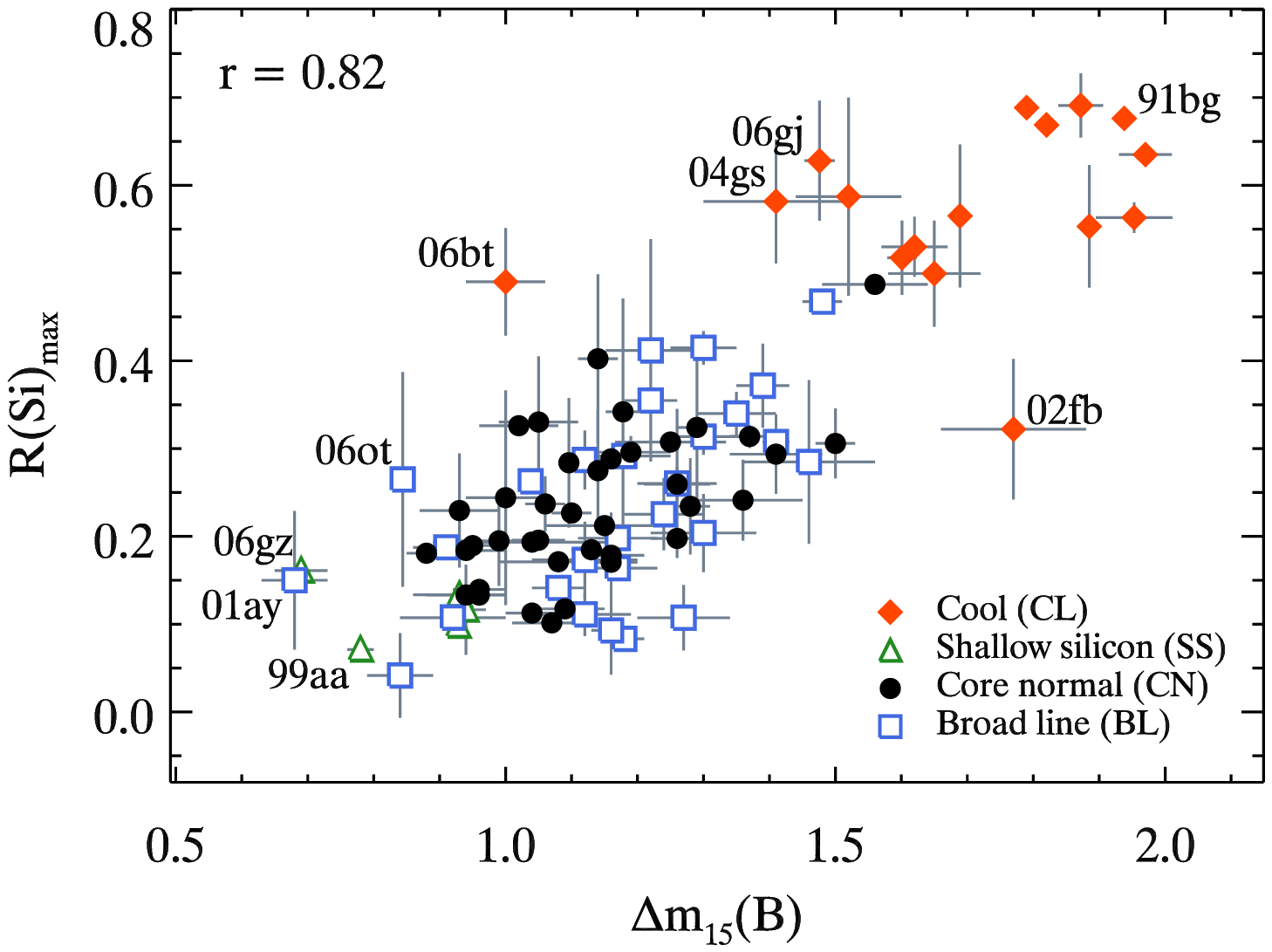}\hspace{1.5cm}
\includegraphics{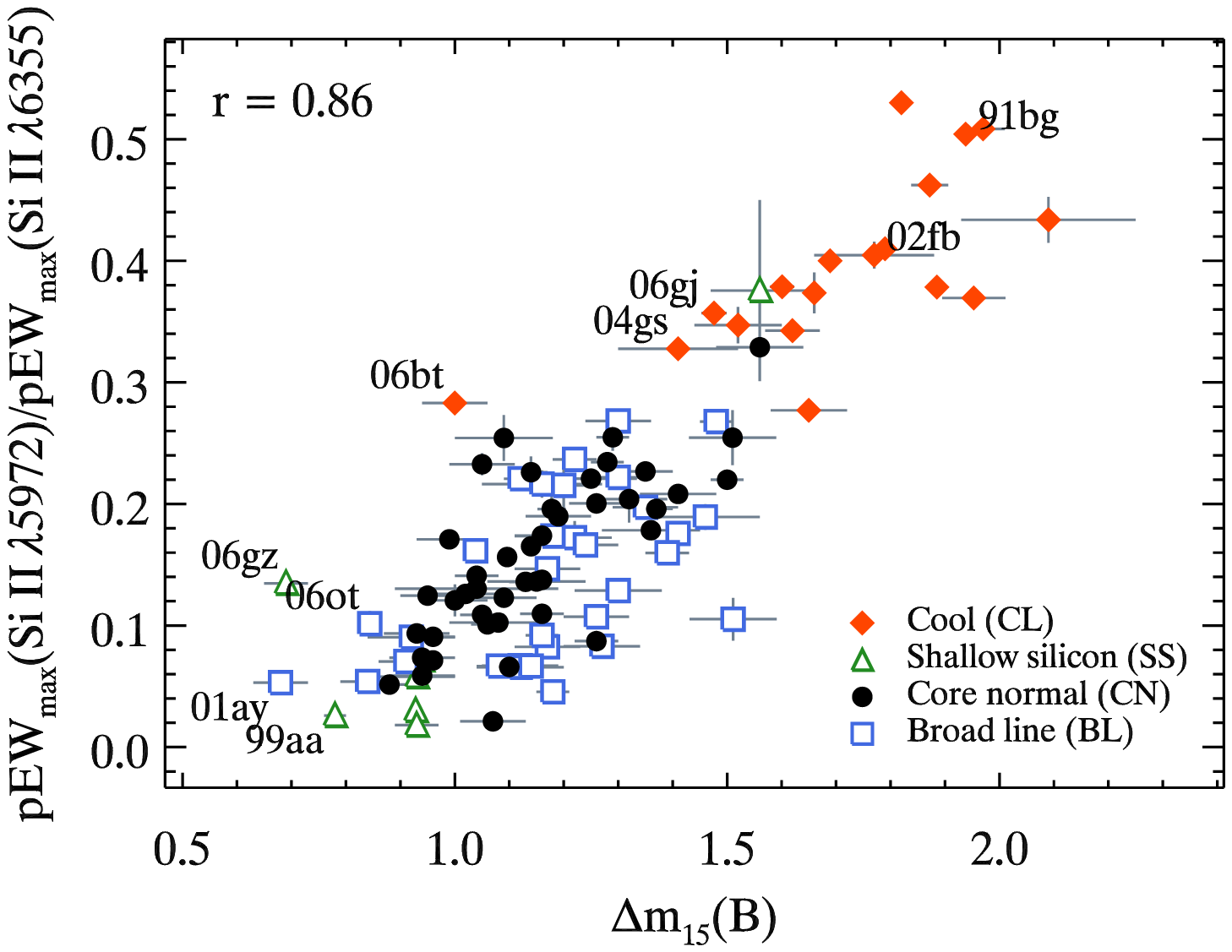}}

\vspace{.5cm}
\resizebox{\textwidth}{!}{
\includegraphics{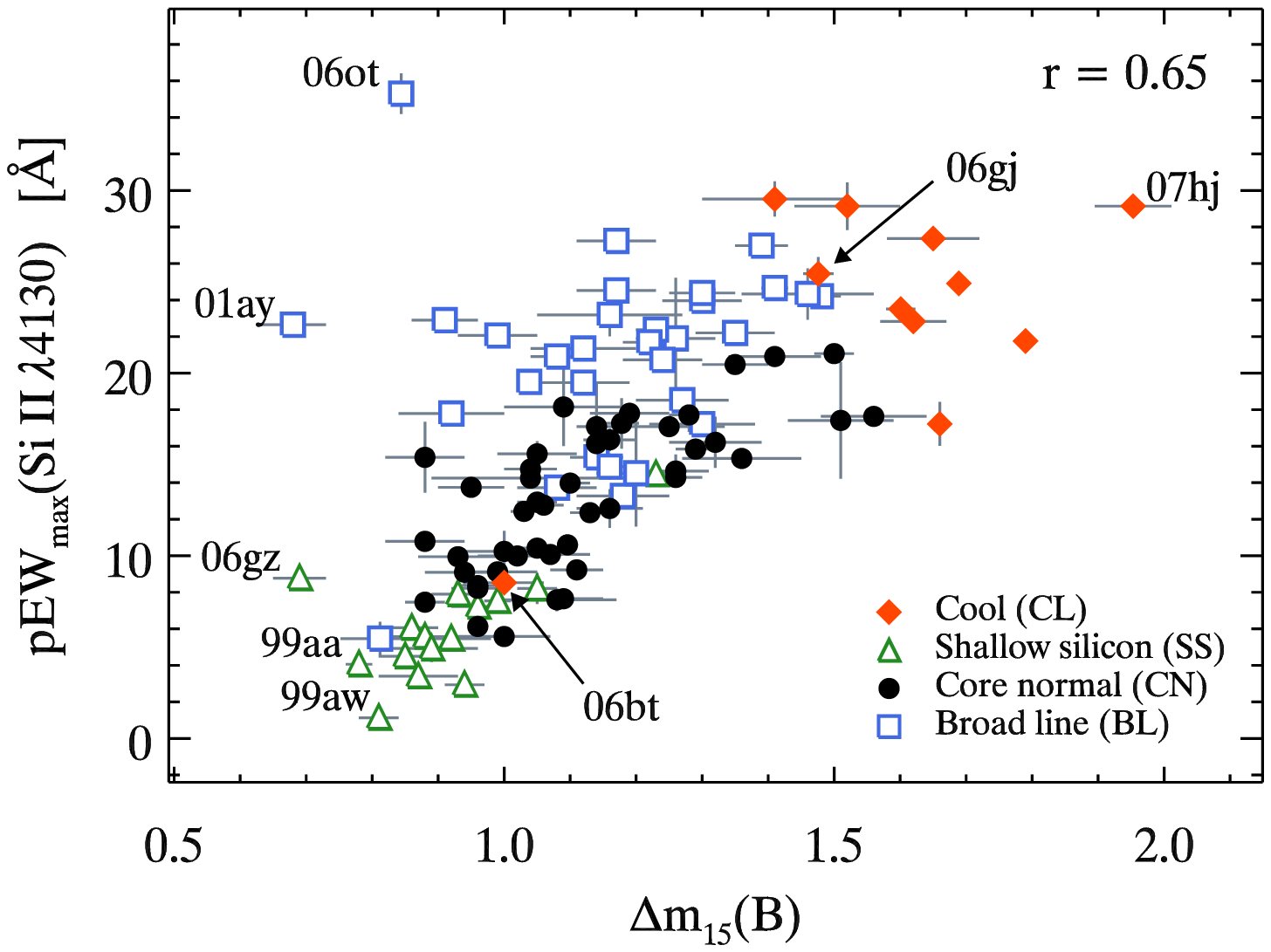}\hspace{1.5cm}
\includegraphics{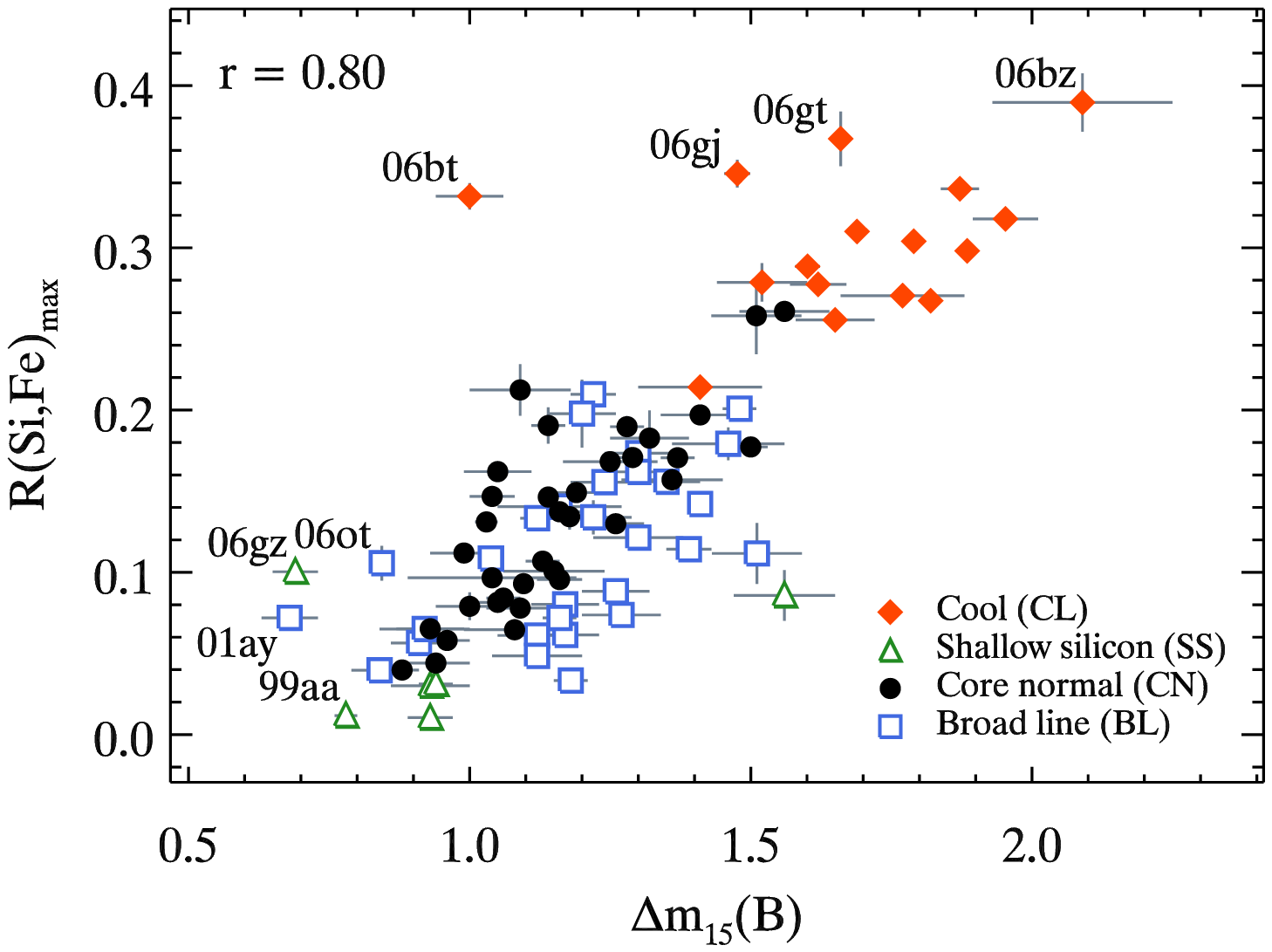}
}
\caption{\label{fig:specindic}
Various spectroscopic indicators {\it vs.} \dmft. The points are coded
according to the spectroscopic subclasses defined by
\cite{Branch/etal:2006}, as in the left panel of
Fig.~\ref{fig:branchwangclass}. The Pearson correlation coefficient is
indicated in the upper-left or upper-right corner of each plot.
{\it Top:} spectroscopic ratio \rsi\ of \cite{Nugent/etal:1995} [{\it
    left}] and its pseudo-EW equivalent ({\it right}). 
{\it Bottom:} pseudo-EW of the Si\two\,\l4130 line ({\it left}) and
spectroscopic ratio \rsife\ ({\it right};
\citealt{Hachinger/Mazzali/Benetti:2006}).
}
\end{figure*}

Several authors have studied the correlation between various
spectroscopic indicators and light-curve parameters, with the
aim to use such indicators to improve distance measurements to
\sneia. 
\cite{Blondin/Mandel/Kirshner:2011} showed that spectroscopic
indicators alone could compete with the standard light-curve
parameters for distance measurements, but that combining
spectra with photometry yielded no statistically significant
improvement \citep[see also][]{Silverman/etal:2012c}.
Their analysis was based on a small number of supernovae,
however, so that larger samples may reveal spectroscopic
measurements that provide independent information on the luminosity
of \sneia. Such correlations are also
potentially useful to constrain \snia\ models \citep[see,
  e.g.,][]{Blondin/etal:2011}, and we investigate the relation of three
spectroscopic indicators with \dmft\ in Fig.~\ref{fig:specindic}.

\epsscale{1.15}
\epsscale{1.0}

The upper-left panel shows the \rsi\ ratio of \cite{Nugent/etal:1995}
(defined as the ratio of the relative absorption depth of the
Si\two\,\l5972 line to that of Si\two\,\l6355), which is thought to
define a temperature sequence from low-luminosity 1991bg-like \sneia\
to luminous 1991T-like events \citep[see
    also][]{Hachinger/etal:2008}. The different symbols correspond to
the spectroscopic subclasses defined by \cite{Branch/etal:2006}, as in
the left panel of Fig.~\ref{fig:branchwangclass}. There is a clear
correlation of \rsi\ with \dmft\ (the Pearson correlation coefficient
is $r=0.82$), with some notable outliers including two \sneia\ from the
CL subclass: SN~2002fb and SN~2006bt (whose peculiar nature has been
discussed at length by \citealt{SN2006bt}). At low \dmft, the BL
SN~2001ay (the most slowly-declining \snia\ to date;
\citealt{SN2001ay}) and the peculiar SS SN~2006gz (possibly resulting
from a double WD merger; \citealt{SN2006gz}) are modest outliers. 

The \rsi\ ratio has a relatively large associated measurement
error, since the relative absorption depth is measured at a specific
wavelength. Using the ratio of pseudo-EW of both lines instead not only
significantly reduces the error on individual measurements (since we
are integrating over many wavelength bins), it also leads to a
stronger and tighter relation with \dmft\ ($r=0.86$; {\it
  upper-right}). SN~2006bt remains an outlier using the pseudo-EW
ratio, but SN~2002fb now follows the
linear relation defined by the bulk of the sample.
As noted empirically by \cite{Hachinger/Mazzali/Benetti:2006}, and
later confirmed theoretically by \citealt{Hachinger/etal:2008}, it is
the Si\two\,\l5972 line that drives the correlation of \rsi\ and its
pseudo-EW equivalent with \dmft. Using the relative absorption depth
or the pseudo-EW of the Si\two\,\l5972 line alone leads to equally
strong correlations with \dmft\ ($r=0.79$ and $r=0.85$, respectively;
{\it not shown}).

The lower-left panel shows the pseudo-EW of the Si\two\,\l4130
line {\it vs.} \dmft\
\citep[see][]{Bronder/etal:2008,Arsenijevic/etal:2008,Walker/etal:2011,Blondin/Mandel/Kirshner:2011,Chotard/etal:2011}.
The correlation is very clear, although a non-linear relation
may need to be invoked to accommodate the CL \sneia\ at
$\dmft\gtrsim1.5$\,mag. The peculiar SN~2006gz is again a modest
outlier, but SN~2001ay is now a significant outlier in this
relation. The largest outlier however is the BL SN~2006ot (which was 
not a significant outlier in the \rsi\ {\it vs.} \dmft\ relation),
whose similarity in terms of photometric properties with SN~2006bt was
noted by \cite{Stritzinger/etal:2011}. We argued in
\S~\ref{sect:branchclass} that both objects are spectroscopically
distinct, and this is clearly visible in this plot, where SN~2006bt
follows the same relation as the bulk of the \snia\ sample. 

Finally, the lower-right panel shows the \rsife\ ratio of
\cite{Hachinger/Mazzali/Benetti:2006}, defined as the ratio of the
pseudo-EW of the Si\two\,\l5972 line to that of the Fe\two\,\l4800
feature (see \citealt{Blondin/Mandel/Kirshner:2011},
their Fig.~14). The correlation with \dmft\ is again very strong
($r=0.80$). In contrast to pEW(Si\two\,\l4130),
SN~2006bt is this time the largest outlier, while SN~2006ot lies close
to the mean relation. As was the case for the \rsi\ ratio, both
SN~2001ay and SN~2006gz are modest outliers. Since \rsife\ and
the pseudo-EW equivalent of \rsi\ are both positively correlated
with \dmft, one expects the pseudo-EWs of Fe\two\,\l4800 and
Si\two\,\l6355 (at maximum light) to also be correlated. There is
indeed a modest correlation ($r=0.60$) between pEW(Si\two\,\l6355)
and pEW(Fe\two\,\l4800), but the scatter is large and the relation
flattens off for pEW(Fe\two\,\l4800)$\gtrsim150$\,\AA.


\section{Line velocities}\label{sect:linevel}

We now focus on the defining Si\two\,\l6355 line of Type Ia
supernovae, studying the variation in the time evolution of its
absorption velocity (\S~\ref{sect:si2evol}). We propose an
alternative (and in our view more robust) method to
\cite{Benetti/etal:2005} for measuring velocity gradients
(\S~\ref{sect:vabsgrad}). Last, we investigate the correlation of
Si\two\ velocity with intrinsic color at maximum light suggested
by \cite{Foley/Kasen:2011} in \S~\ref{sect:vsibmv}. 


\subsection{Evolution of the Si\two\,\l6355 absorption velocity}\label{sect:si2evol}

\begin{figure*}
\resizebox{\textwidth}{!}{
\includegraphics{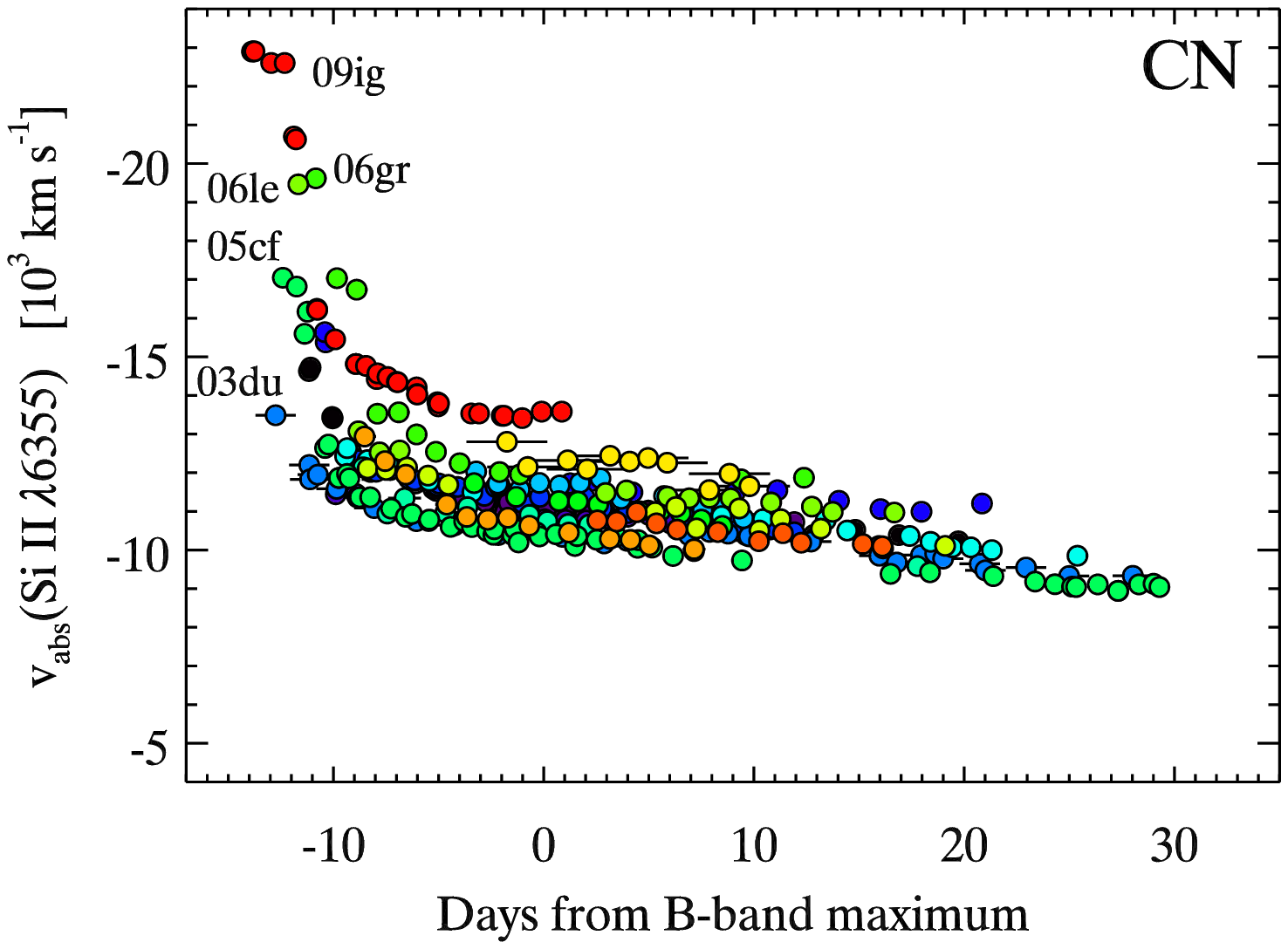}\hspace{.5cm}
\includegraphics{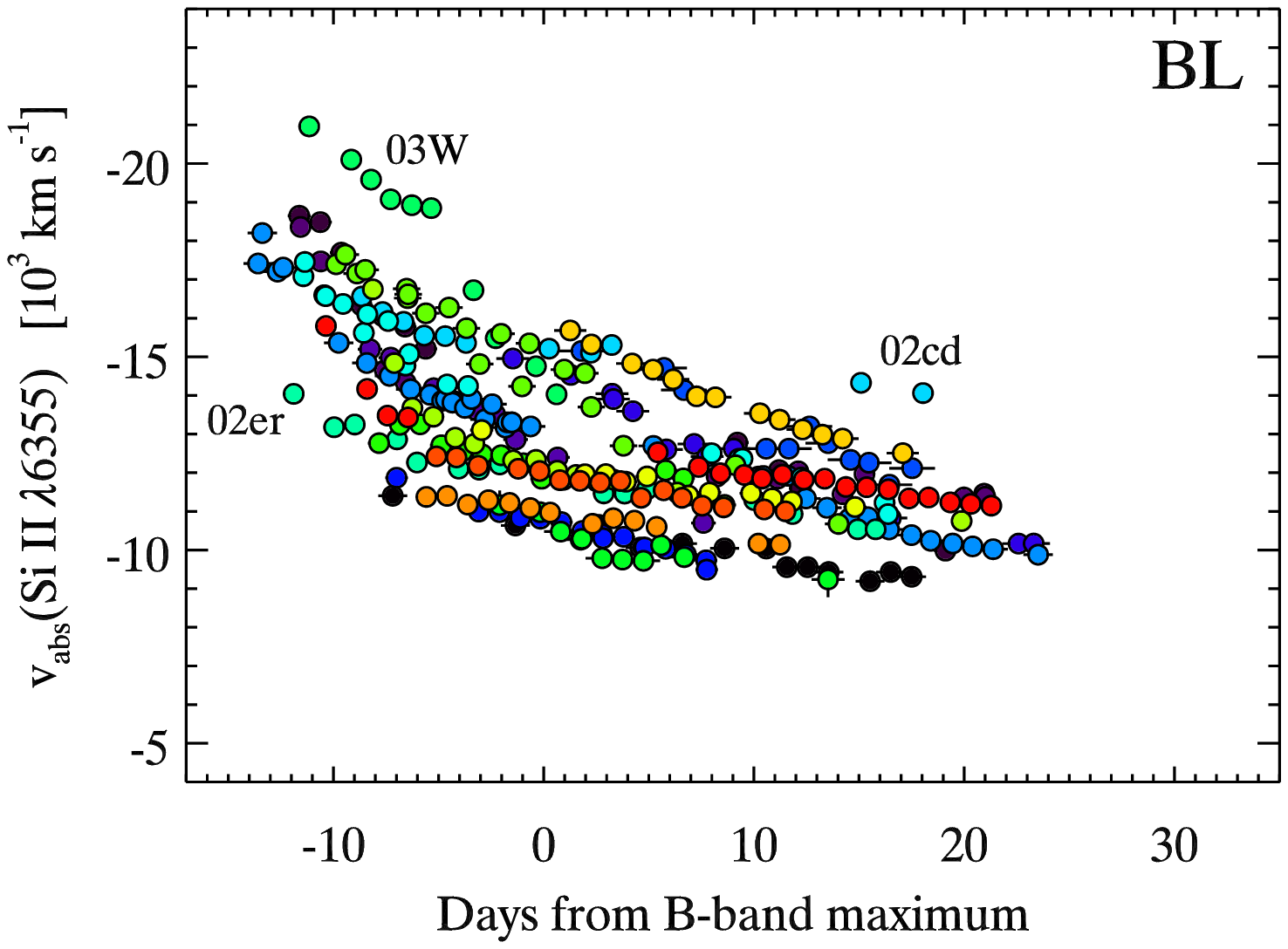}}

\vspace{.5cm}
\resizebox{\textwidth}{!}{
\includegraphics{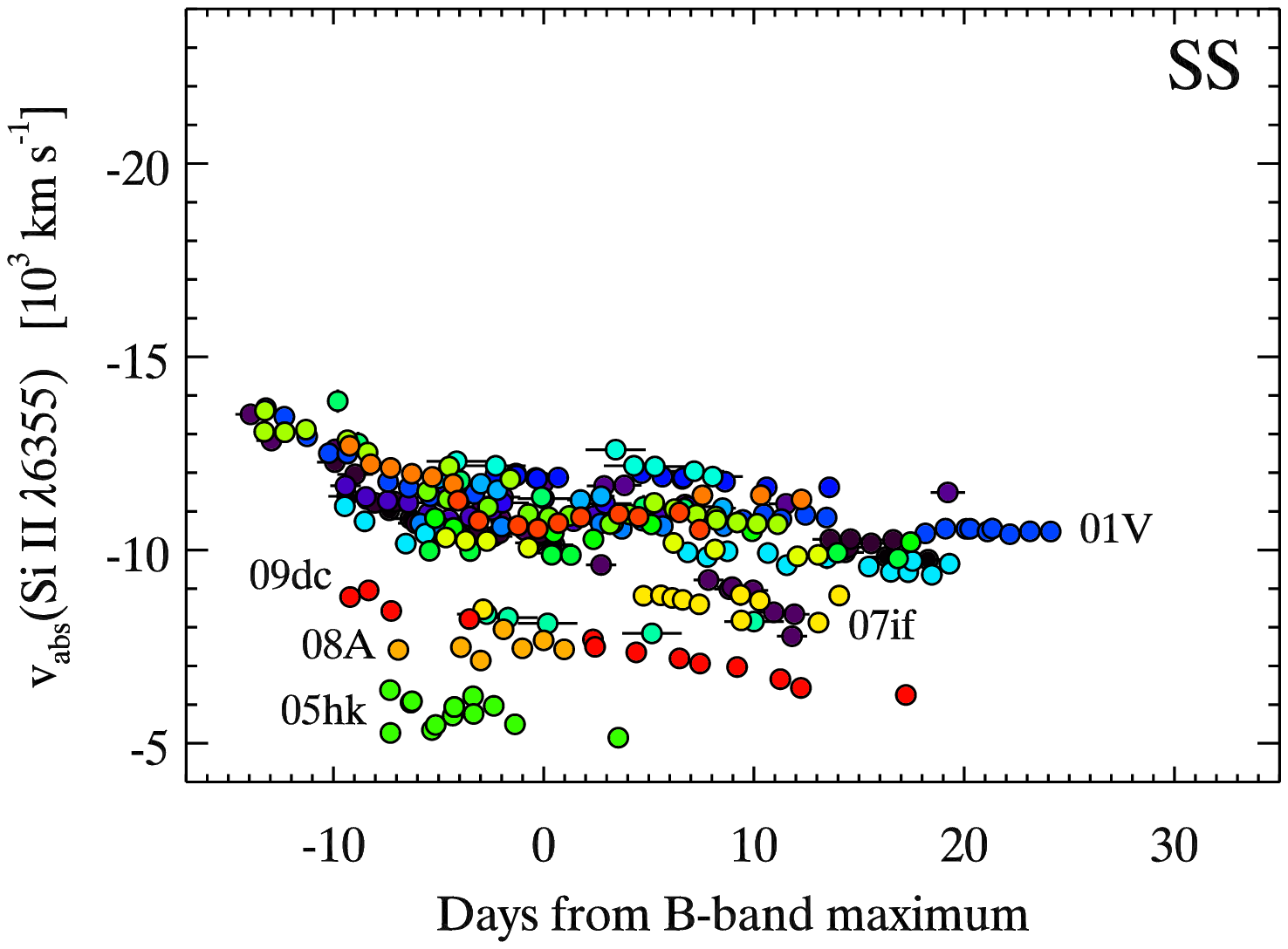}\hspace{.5cm}
\includegraphics{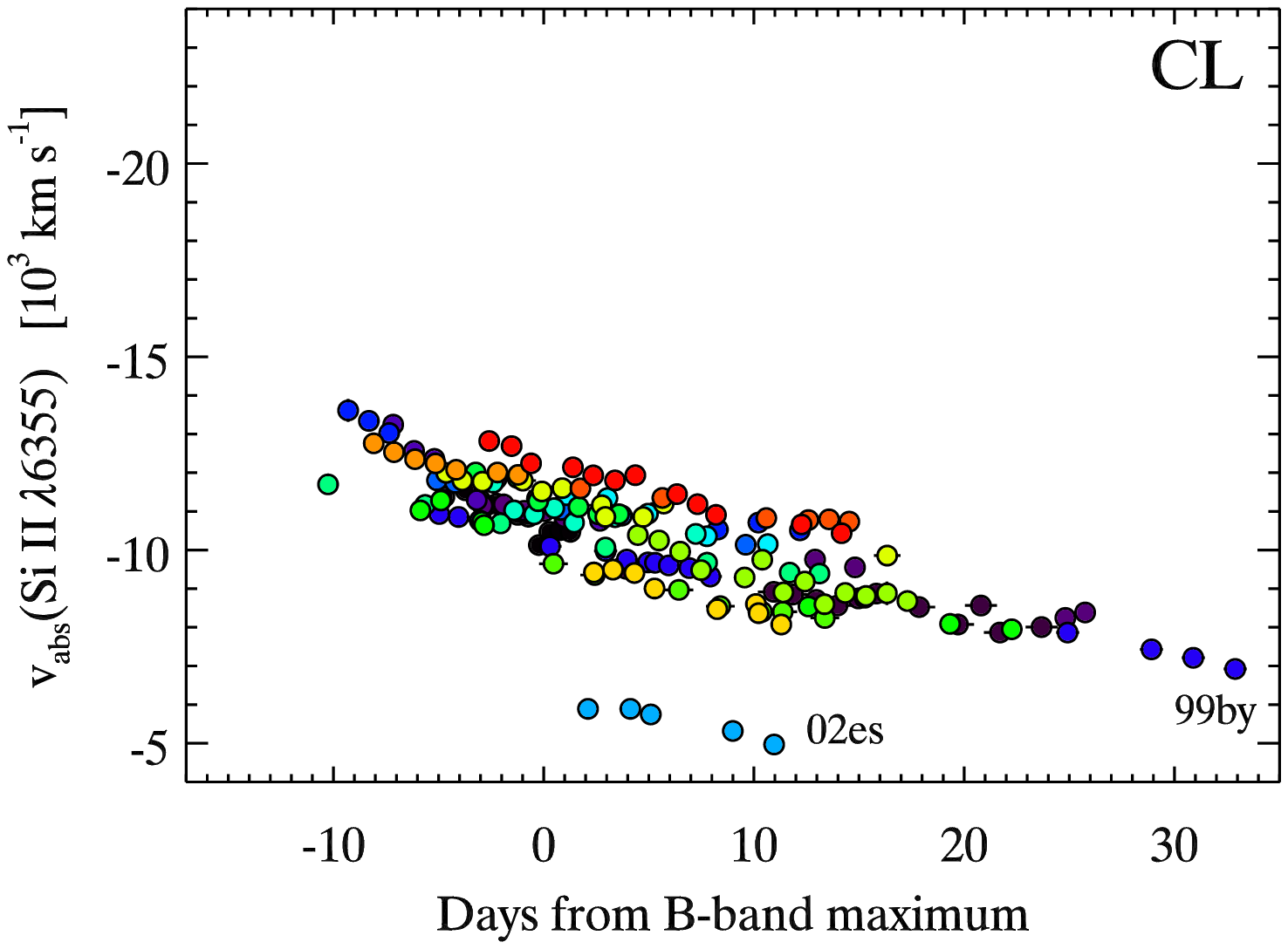}
}
\caption{\label{fig:vSiII_branch}
Evolution of the Si\two\,\l6355 absorption velocity with time for the
four spectroscopic subclasses defined by \cite{Branch/etal:2006}. The
color-coding denotes individual \sneia.
 For the CN and BL subclasses we
have only included \sneia\ for which we have at least 10 $v_{\rm abs}$
measurements prior to +30\,d from maximum light, while for the SS and
CL subclasses only \sneia\ with at least 5 measurements are shown.
}
\end{figure*}

Figure~\ref{fig:vSiII_branch} shows the time evolution of the
Si\two\,\l6355 absorption velocity for the different spectroscopic
subclasses of \cite{Branch/etal:2006}. \cite{Blondin/etal:2006}
presented similar measurements (in different \dmft\ bins) along with a
physical interpretation in terms of microscopic atomic-transition
properties and macroscopic properties of the ejecta (density and
velocity distributions). We refer the reader to that paper for more
detail on the impact of these properties on the morphology of line
profiles in \snia\ ejecta.

For the Core Normal subclass ({\it top left}), there is a large
variation in $v_{\rm abs}$(Si\two\,\l6355) at phases $\lesssim-10$\,d
(e.g., SN~2003du and SN~2006gr have $v_{\rm abs}\approx-12000$\,\kms\
and $-20000$\,\kms\ at $-10$\,d, respectively). By maximum light, the
spread in $v_{\rm abs}$ has decreased to $\lesssim3000$\,\kms, 
SN~2009ig being a clear outlier (it is part of the High-velocity
subclass of \cite{WangX/etal:2009b}, but the pseudo-EW of its
Si\two\,\l6355 line at maximum light places it unambiguously in the
CN subclass of \cite{Branch/etal:2006}; see
Fig.~\ref{fig:branchwangclass}). The 
evolution is fast at early times and more gradual at post-maximum
epochs, reflecting the recession of the line-forming region in the
co-moving frame of the ejecta. There is also a large spread in $v_{\rm
 abs}$ at early times for \sneia\ in the Broad Line subclass ({\it top
right}), but unlike the CN subclass it remains large at later phases
($>5000$\,\kms\ variation at maximum light). At phases later than
$-10$\,d, the absorption velocity is on average larger at any given
time when compared to the CN subclass, illustrating the large overlap
between the BL subclass and the High-velocity subclass defined by
\cite{WangX/etal:2009b}. 

The bulk of the Shallow Silicon subclass
({\it bottom left}) has velocities comparable to the CN subclass,
except at early times where there are no measurements $|v_{\rm
 abs}|>15000$\,\kms. A handful of \sneia\ display $-5000\gtrsim v_{\rm
abs}\gtrsim-10000$\,\kms\ at maximum light. These include 2002cx-like
\sneia\ (SN~2005hk, $v_{\rm abs,max}\approx-5500$\,\kms; SN~2008A,
$v_{\rm abs,max}\approx-7500$\,\kms), \sneia\ resulting from possible
super-Chandrasekhar progenitors (SN~2009dc, $v_{\rm
abs,max}\approx-8000$\,\kms; SN~2007if, $v_{\rm
abs,max}\approx-7500$\,\kms), and the 1991T-like SN~2005M ($v_{\rm
abs,max}\approx-8000$\,\kms). This last SN stands out amongst other
1991T-like \sneia\ which all display $\gtrsim10000$\,\kms\ absorption
blueshifts at maximum light.

The Cool subclass ({\it bottom right}) displays velocities similar to
the CN and SS subclasses around maximum light, but with a more rapid
evolution at post-maximum epochs. One significant outlier is SN~2002es
\citep{Ganeshalingam/etal:2012}, with $v_{\rm
  abs}\approx-6000$\,\kms\ at +2\,d (see also
Fig.~\ref{fig:branchwangclass}, {\it right}), comparable to the SS
SN~2005hk at a similar phase.

\cite{Benetti/etal:2005} noted the large variation in post-maximum
evolution of the Si\two\,\l6355 absorption velocity for different
\sneia. This is clearly visible in Fig.~\ref{fig:vSiII_branch}, not
only across the different subclasses but also within each subclass
(the largest variation being apparent within the BL subclass).
\cite{Benetti/etal:2005} introduced the velocity gradient,
noted $\dot{v}$, representing the average $|v_{\rm abs}|$ decline rate
past maximum light (units of \kmsd), and divided their \snia\ sample
into three groups (FAINT for 1991bg-like \sneia, HVG and LVG for
\sneia\ with high and low velocity gradients, respectively). They
suggested that the LVG and HVG groups could consist of \sneia\ with
different properties of their outer ejecta, resulting from different
degrees of mixing or from interaction with circumstellar material. A
recent study by \cite{Maeda/etal:2010c} associates the observed
diversity in the velocity gradients with viewing-angle effects in
off-center delayed-detonation models. \cite{Blondin/etal:2011}
showed that varying the criterion for deflagration-to-detonation
transition in explosions resulting from an isotropic distribution of
ignition points (i.e., {\it not} off-center explosions) resulted in a range of
velocity gradients comparable to that observed by
\cite{Benetti/etal:2005}.

The notion of velocity gradients has received much attention since the
paper by \cite{Benetti/etal:2005}, and there are different physical
interpretations as to the origin of their variation. This however can
only be properly addressed if the actual parameter is defined in a
robust way, which we set out to do in the following section.


\subsection{Redefining velocity gradients}\label{sect:vabsgrad}

According to \cite{Benetti/etal:2005}, the velocity gradient is
derived ``from a least-squares fit of the measurements taken between
maximum and either the time the Si\two\ feature disappears or the last
available spectrum, whichever comes first.'' This definition makes two
implicit (and related) assumptions: (1) the evolution of $v_{\rm abs}$
is a linear function of time at post-maximum epochs, and (2) the phase
of the last $v_{\rm abs}$ measurement (or the range of available
measurements) has little impact on the derived velocity gradient. 

\begin{figure*}
\resizebox{\textwidth}{!}{
\includegraphics{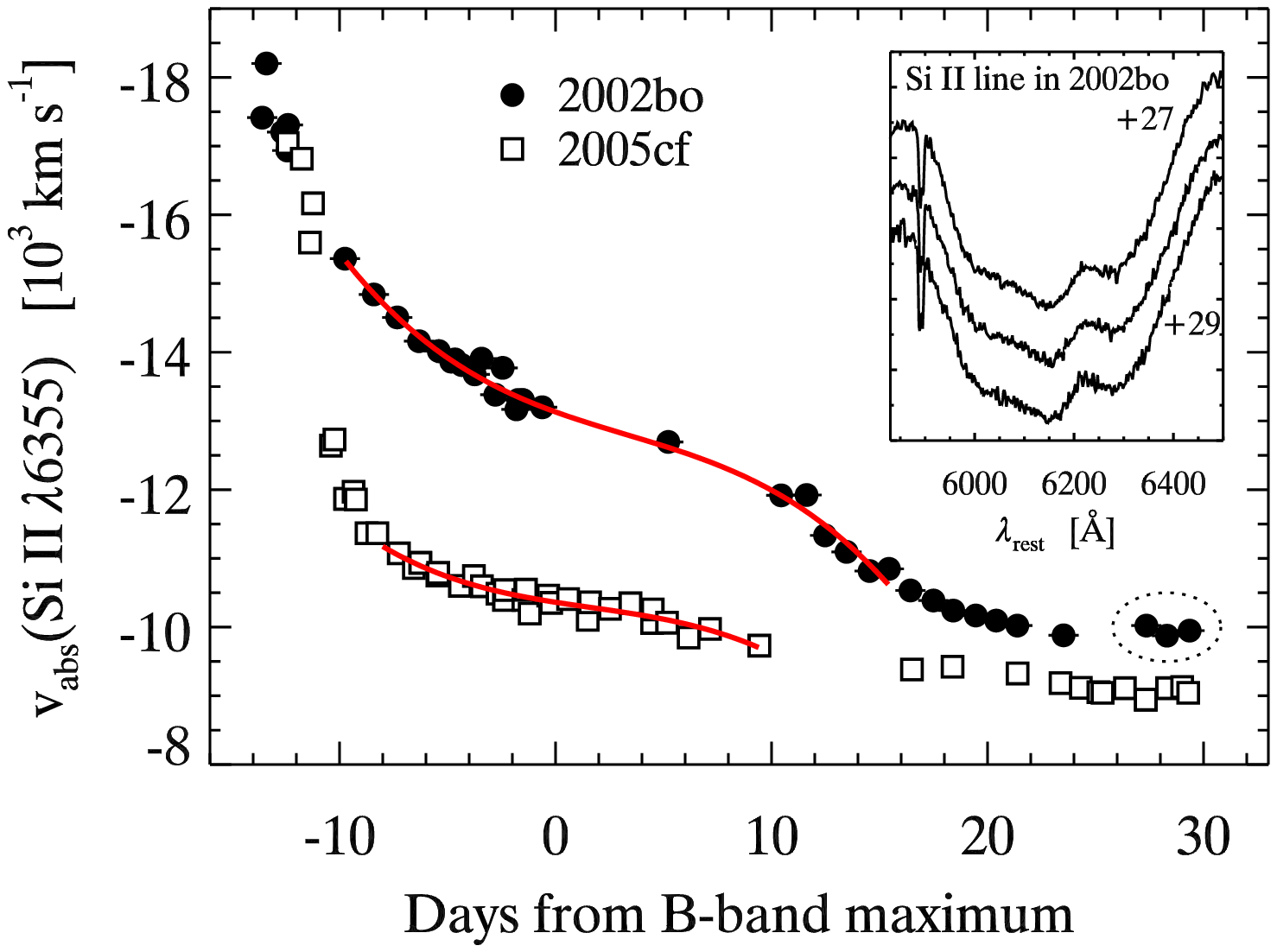}\hspace{1cm}
\includegraphics{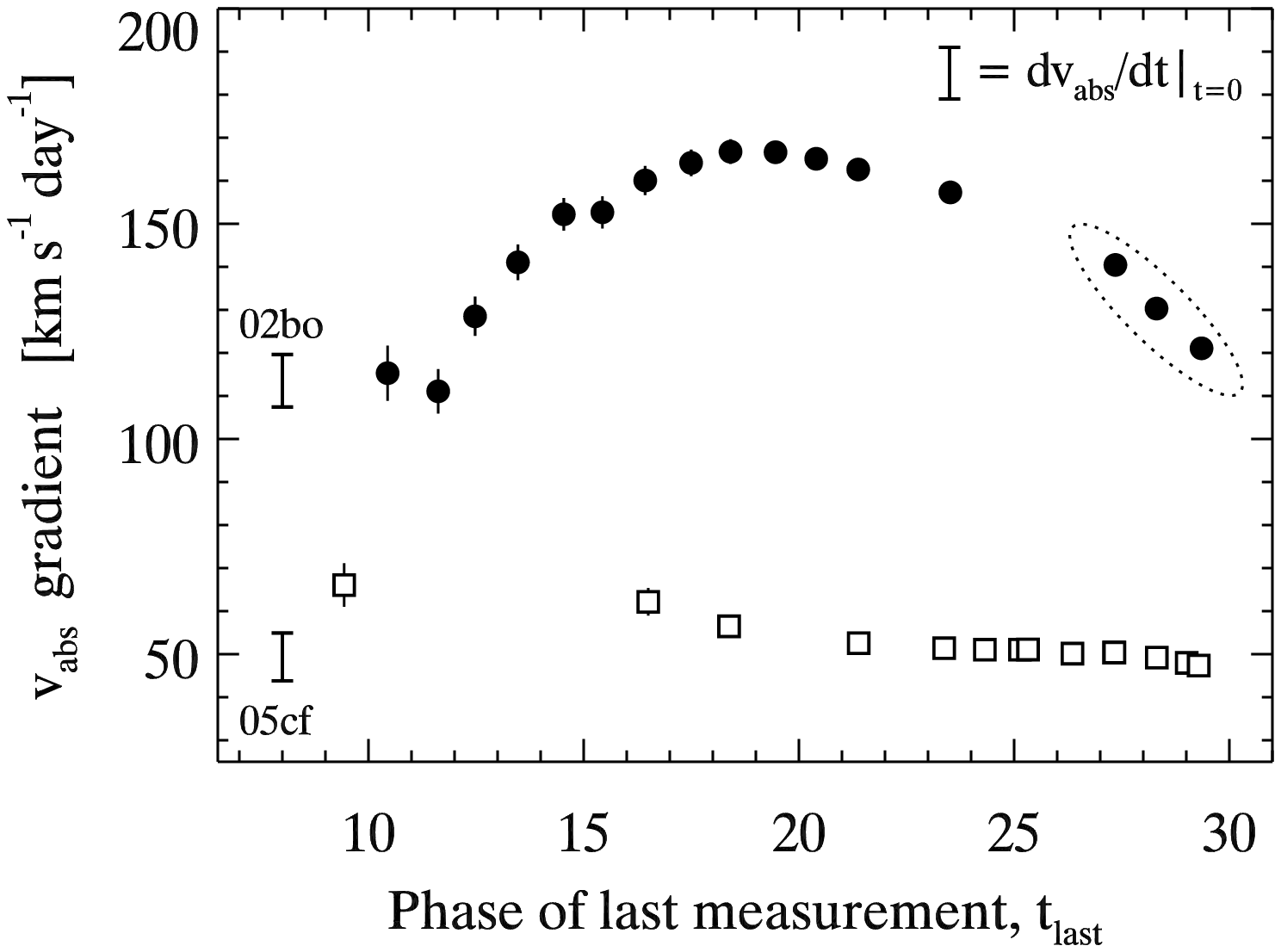}
}
\caption{\label{fig:vabsgradbias}
{\it Left:}
Time evolution of the Si\two\,\l6355 absorption velocity for SN~2002bo
({\it filled circles}) and SN~2005cf ({\it open squares}). The solid
lines are 3$^{\rm rd}$-order polynomial fits to the data. Past +25\,d
the Si\two\ absorption profile is increasingly contaminated in
SN~2002bo ({\it see inset}) and the velocity measurements at +27, +28,
and +29\,d are biased ({\it dotted ellipse}).
{\it Right:}
Variation of the Si\two\ velocity gradient as defined by
\cite{Benetti/etal:2005} with the phase of the last $v_{\rm abs}$
measurement for the same supernovae. The dotted ellipse again
highlights the biased measurements for SN~2002bo ({\it right
  panel}). The vertical error bars indicate the instantaneous velocity
decline rate at maximum light, i.e., the slope of the tangents to the
red curves in the right panel at $t=+0$\,d.
}
\end{figure*}

Figure~\ref{fig:vabsgradbias} shows that both assumptions can have a
significant impact on the derived gradient. In the left panel we show
the time evolution of $v_{\rm abs}$(Si\two\,\l6355) for the BL
SN~2002bo ({\it filled circles}) and the CN SN~2005cf ({\it open
  squares}). While the $v_{\rm abs}$ evolution for SN~2005cf appears
to be almost linear at post-maximum epochs, that of SN~2002bo is
clearly not, with a more rapid change in the interval
[+10,+15]\,d. Moreover, at phases $\gtrsim+25$\,d the Si\two\,\l6355
absorption profile is increasingly contaminated by the emergence of
lines from different ions ({\it see inset}), such that the $v_{\rm
  abs}$ measurements at these times are not reliable. The right panel
of Fig.~\ref{fig:vabsgradbias} shows the variation in the measured
velocity gradient in the interval [+0,$t_{\rm last}$]\,d as a function
of the phase of the last available measurement, $t_{\rm last}$. The
velocity gradient for SN~2002bo is $\sim115$\,\kmsd\ when measured
over the interval [+0,+10]\,d, and $\sim165$\,\kmsd\ when measured
over the interval [+0,+20]\,d. Including the biased $v_{\rm abs}$
measurements at $>+25$\,d yields a velocity gradient $\sim120$\,\kmsd,
comparable to that obtained for [+0,+10]\,d, but with a $\chi^2$ for
the linear fit that is two orders of magnitude larger.
\cite{Benetti/etal:2005}
report a velocity gradient of $110\pm7$\,\kmsd\ for this SN, due to
the inclusion of measurements out to +30\,d (Benetti 2008,
priv. comm.). For SN~2005cf, the quasi-linear variation of $v_{\rm
  abs}$ with time results in a minor variation of the velocity
gradient with $t_{\rm last}$, from $\sim65$\,\kmsd\ for $t_{\rm
last}\approx+10$\,d to $\sim50$\,\kmsd\ for $t_{\rm
last}\approx+30$\,d. \cite{WangX/etal:2009a} report a velocity
gradient of $38\pm5$\,\kmsd\ for this SN using measurements out to
+30\,d.

The classification of SN~2002bo and SN~2005cf respectively in the HVG
and LVG groups still holds, regardless of $t_{\rm last}$ and whether
or not the biased $v_{\rm abs}$ measurements of SN~2002bo are
included. This is not always the case. For SN~2002er,
\cite{Benetti/etal:2005} report a velocity gradient of
$92\pm5$\,\kmsd, including $v_{\rm abs}$ measurements past +20\,d
when the Si\two\,\l6355 absorption profile is clearly contaminated by
other lines. Using measurements in the phase interval [+0,+16]\,d we
find $81\pm3$\,\kmsd, consistent with their value, and confirming the
classification of SN~2002er as a HVG \snia. However, restricting the
phase interval to [+0,+10]\,d, we infer a velocity gradient of
$47\pm7$\,\kmsd, consistent with the LVG group. One can think of a
variety of reasons why no spectra would have been obtained at later
times (poor weather, telescope problems, conflict with other
observations etc.), none of which should affect the inferred intrinsic
properties of \sneia.

We also note that the velocity gradient for SN~2006bt reported by
\cite{SN2006bt} is in error. We measure a velocity gradient of
$93\pm16$\,\kmsd\ over the time interval [+0,+16]\,d, while
\cite{SN2006bt} report 51\,\kmsd\ over the same interval. Foley (2011,
priv. comm.) has revised this measurement to 96\,\kmsd. According to
Fig.~7 of \cite{SN2006bt}, this new measurement places SN~2006bt well
into the HVG group, where \cite{SN2006bt} originally included this SN
in the LVG group (albeit close to the border with HVG \sneia).

A more robust definition of the velocity gradient should consider
$v_{\rm abs}$ measurements over a {\it fixed} phase range, as done for
instance by \cite{Foley/Sanders/Kirshner:2011} over the interval
$-6\le t \le +10$\,d using linear functions to describe the $v_{\rm
 abs}(t)$ curves. This still does not account for non-linear
variations in absorption velocity. As an alternative, we suggest the
measurement of the mean velocity decline rate $\Delta v_{\rm
  abs}/\Delta t$ over some fixed time interval $[t_0,t_1]$:

\begin{equation}
\label{eqn:vabsdec}
\left.\frac{\Delta v_{\rm abs}}{\Delta t}\right\vert_{[t_0,t_1]} = \frac{v_{\rm
    abs}(t_1)-v_{\rm abs}(t_0)}{t_1-t_0}.
\end{equation}

\begin{figure*}
\resizebox{\textwidth}{!}{
\includegraphics{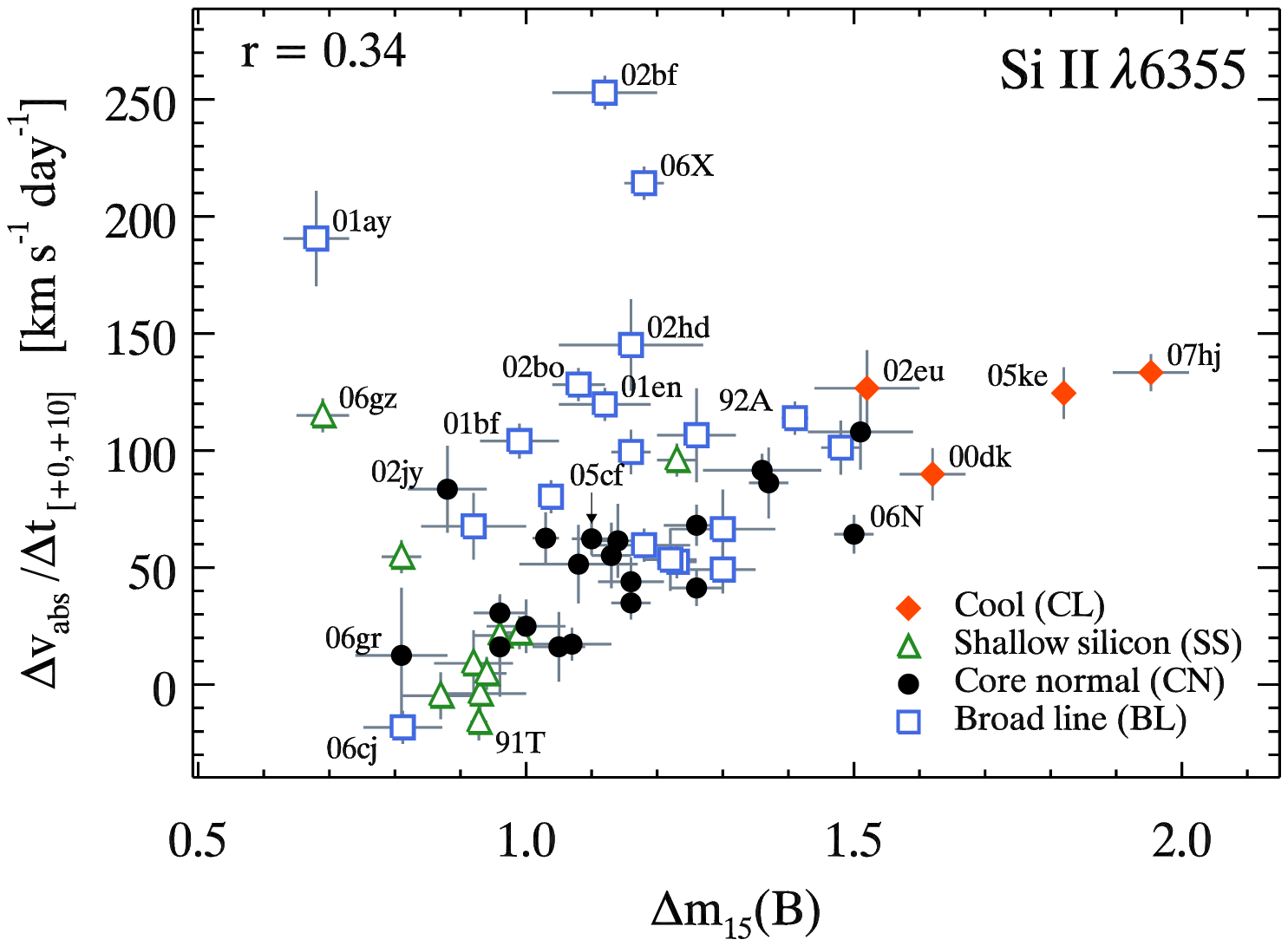}\hspace{1cm}
\includegraphics{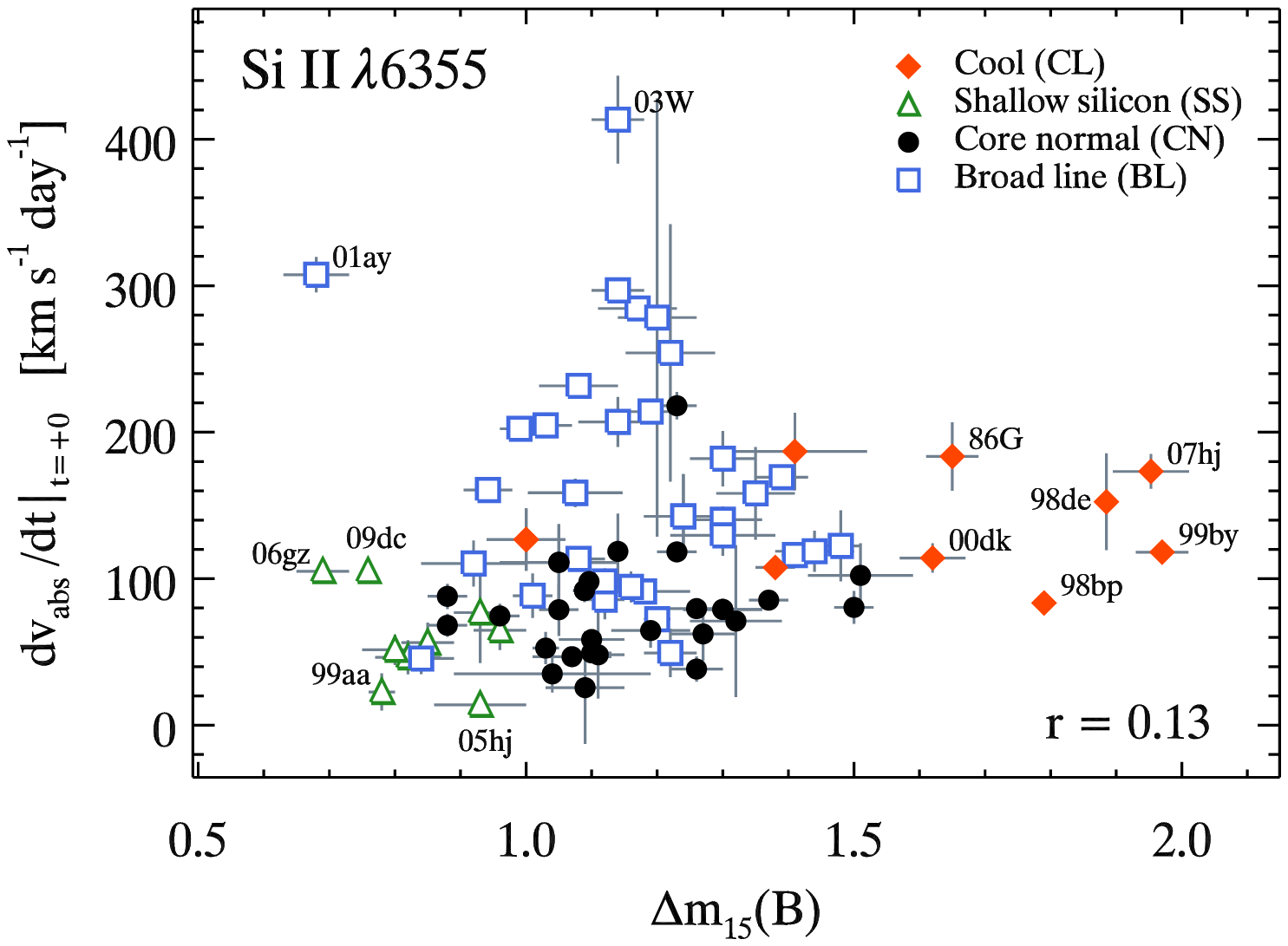}
}
\caption{\label{fig:vabsdecdm15}
{\it Left:} 
Mean velocity decline rate $\Delta v_{\rm abs}/\Delta t$
between maximum light and 10\,d past maximum for the Si\two\,\l6355
line {\it vs.} \dmft.
The different symbols correspond to the various
spectroscopic subclasses defined by \cite{Branch/etal:2006}, as in
Fig.~\ref{fig:branchwangclass}.
{\it Right:} Instantaneous velocity decline rate $dv_{\rm abs}/dt$ at
maximum light for the Si\two\,\l6355 line {\it vs.} \dmft.
}
\end{figure*}

\noindent
In practice there are not always measurements at precisely $t_0$ and
$t_1$, so one may choose to either select the $v_{\rm abs}$
measurement closest in time to $t_0$ and $t_1$ (within a small time 
interval, e.g., $\pm2$\,d), or to interpolate the $v_{\rm abs}$
measurements at the desired phases, again making sure the $v_{\rm
 abs}(t)$ data is well sampled around $t_0$ and
$t_1$. Figure~\ref{fig:vabsdecdm15} ({\it left}) shows the mean
velocity decline rate for the Si\two\,\l6355 line for the time
interval [+0,+10]\,d {\it vs.} \dmft, where we choose the $v_{\rm
abs}$ measurement closest in time and within 2\,d from these
time boundaries. The clustering into the three groups (FAINT, LVG, HVG)
is less obvious than in Fig.~3 of \cite{Benetti/etal:2005}, but the
points occupy similar regions of parameter space, with a smaller
scatter in $\Delta v_{\rm abs}/\Delta t_{[+0,+10]}$ for
$\dmft\gtrsim1.5$\,mag. The Broad Line subclass displays the largest
variation in mean velocity decline rate (confirming what was visually
apparent in Fig.~\ref{fig:vSiII_branch}), ranging from $\sim
-20$\,\kmsd\ for SN~2006cj to +215\,\kmsd\ for SN~2006X. Two other SN
have $\Delta v_{\rm abs}/\Delta t_{[+0,+10]} < 0$, SN~1991T and
SN~2005M (both part of the SS subclass). A combination of optical-depth
effects (e.g., a Si\three$\rightarrow$Si\two\ recombination
wave) and measurement error (the Si\two\ absorption is
relatively weak and occurs on a steep pseudo-continuum) could
explain why the absorption 
blueshift apparently increases over this phase range. Several objects
have $\Delta v_{\rm abs}/\Delta t_{[+0,+10]}$ consistent with zero,
including the CN SN~2006gr. This was also the case for SN~2005hj, for
which \cite{Quimby/etal:2007a} invoke a shell-like density structure
(formed within the context of a pulsating delayed detonation or WD-WD
merger scenario) to explain the $\sim10$\,d-long plateau in absorption
velocity of the Si\two\,\l6355 line.

The CN and CL subclasses appear to form a sequence of linearly
increasing $\Delta v_{\rm abs}/\Delta t_{[+0,+10]}$ with \dmft. Both
quantities are indeed strongly correlated ($r=0.80$) for these two
subclasses. Within the BL subclass, there is a strong correlation
($r=-0.80$) between the Si\two\,\l6355 absorption velocity at maximum
light and its subsequent mean decline over the interval
[+0,+10]\,d. In other words, \sneia\ with the largest absorption
blueshifts at maximum light also display steeper velocity
gradients \citep[see
  also][]{Benetti/etal:2005,WangX/etal:2009b,Foley/Sanders/Kirshner:2011}.

In the limit $(t_1-t_0)\rightarrow0$, Eq.~\ref{eqn:vabsdec} yields an
instantaneous velocity decline rate at a given time, $dv_{\rm
  abs}/dt|_{t=t_0}$. We determine this using second or
  third-order polynomial fits
to the $v_{\rm abs}(t)$ data, as shown in Fig.~\ref{fig:vabsgradbias}
({\it left; solid curves}). The instantaneous velocity decline rates at
$t=+0$\,d for Si\two\,\l6355 in SN~2002bo and SN~2005cf are shown as
the error bars in the right panel of
Fig.~\ref{fig:vabsgradbias}. Figure~\ref{fig:vabsdecdm15} ({\it right})
shows the same quantity for a larger sample {\it vs.} \dmft. The
different subclasses occupy similar regions of parameter space as when
considering the mean velocity decline rate (not surprisingly, since
the mean and instantaneous decline rates are strongly correlated), but
reveals an even larger variation within the BL subclass, with
instantaneous velocity decline rates at maximum light as large as
$\sim300$\,\kmsd\ for SN~2001ay and $\sim400$\,\kmsd\ for
SN~2003W. SN~2005hj has the lowest instantaneous velocity decline rate
at maximum light ($14\pm5$\,\kmsd), in line with the results of
\cite{Quimby/etal:2007a}.


\subsection{Intrinsic color and Si\two\ velocity}\label{sect:vsibmv}

Based on the results by \cite{WangX/etal:2009b},
\cite{Foley/Kasen:2011} noted that \sneia\ in the High-velocity
subclass were intrinsically redder (in $B^{\rm max}-V^{\rm max}$
pseudo-color\footnote{difference between the $B$ magnitude at
$B$-band maximum and the $V$ magnitude at $V$-band maximum.}) than
\sneia\ in the Normal subclass. They provided a simple physical
explanation using radiative
transfer calculations by \cite{Kasen/Plewa:2007} based on the
detonating failed deflagration (DFD) model of \cite{Plewa:2007},
although \cite{Blondin/etal:2011} showed that this relation did not
hold in the 2D delayed-detonation models of \cite{KRW09}. Using
the Si\two\,\l6355 absorption velocity measurements presented in this
paper, \cite{Foley/Sanders/Kirshner:2011} quantified the relation
between Si\two\ velocity and intrinsic $B^{\rm max}-V^{\rm max}$
pseudo-color at maximum light for Normal and High-velocity \sneia\
with $1.0\le\dmft\le 1.5$\,mag. They find that the slope of the
relation between both quantities is non-zero at the $\sim9\sigma$
level using a linear weighted least-squares fit (but only at the
$\sim3.5\sigma$ level when using a Bayesian approach to linear
regression; see \citealt{Kelly:2007}), with a Pearson correlation
coefficient $r=-0.39$.

\begin{figure*}
\plotone{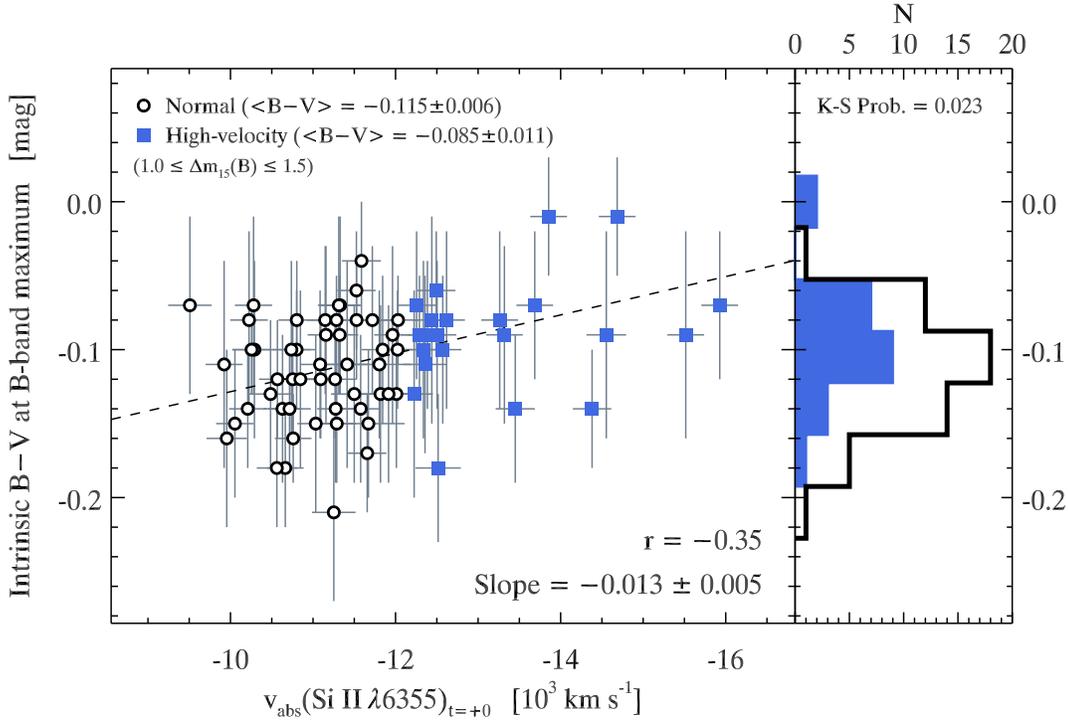}
\caption{\label{fig:vsi_intcol}
{\it Left:}
Intrinsic $B-V$ color at $B$-band maximum light, as
derived from BayeSN fits to \snia\ light curves, {\it vs.} the 
absorption velocity of the Si\two\,\l6355 line, for \sneia\ in the
Normal and High-velocity subclasses with $1.0\le \dmft \le 1.5$\,mag.
The Pearson correlation coefficient is $r=-0.35$ ($r\approx-0.20$ when
taking into account the uncertainties in intrinsic color).
The dashed line is the result of a least-squares linear fit, with a
slope of $0.013\pm0.005$ ($\sim2\sigma$ different from zero).
We also indicate the weighted mean intrinsic color ($\langle
B-V\rangle$) for both samples.
{\it Right:}
Distribution of intrinsic $B-V$ color at maximum light for the Normal
and High-velocity subsamples. A K-S test 
shows that the two samples are discrepant at the $p=0.023
< 0.05$ significance level.
}
\end{figure*}

We show the relation between the intrinsic $B-V$ color at $B$-band
maximum light, as derived from BayeSN fits to \snia\ light curves (see
Table~\ref{tab:snparam}), and the absorption velocity of the
Si\two\,\l6355 line in Fig.~\ref{fig:vsi_intcol} ({\it left}) for the
Normal and High-velocity subclasses. We restrict the \dmft\ range to
$1.0\le \dmft \le 1.5$\,mag, as done by
\cite{Foley/Sanders/Kirshner:2011}. Moreover we use their interpolated
velocity measurements at maximum light (their Table~1). The right panel
shows histograms of intrinsic $B-V$ color at maximum light for the
Normal ($N_{\rm SN}=51$; {\it open}) and High-velocity ($N_{\rm
 SN}=22$; {\it filled}) subsamples. Of the 65 \sneia\ 
used by \cite{Foley/Sanders/Kirshner:2011} in their analysis, 42 (27
Normal and 15 High-velocity) are part of our sample.

\epsscale{.9}
\epsscale{1.0}

\begin{deluxetable*}{cccrrcc}
\tablewidth{0pt}
\tabletypesize{\scriptsize}
\tablecaption{\label{tab:vsi_intcol}Intrinsic color {\it vs.} Si\two\,\l6355 absorption velocity.}
\tablehead{\colhead{Intrinsic color} & \colhead{Slope} & \colhead{$r$} & \colhead{$\langle$color$\rangle_N$} & \colhead{$\langle$color$\rangle_{HV}$} & \colhead{$\Delta\langle$color$\rangle_{HV,N}$} & \colhead{K-S Prob.}}
\startdata
\multicolumn{7}{c}{\bf This paper $(N_{\rm N}=51$, $N_{\rm HV}=22)$}\\
$B-V$                          & $-$0.013$\pm$0.005 (2.9$\sigma$) & $-$0.35 & $-$0.115$\pm$0.006 & $-$0.085$\pm$0.011 &  0.030$\pm$0.013 (2.3$\sigma$) & 0.023 \\
$B^{\rm max}-V^{\rm max}$      & $-$0.011$\pm$0.005 (2.3$\sigma$) & $-$0.26 & $-$0.101$\pm$0.007 & $-$0.070$\pm$0.012 &  0.030$\pm$0.013 (2.3$\sigma$) & 0.140 \\
\hline
\multicolumn{7}{c}{} \\[-.2cm]
\multicolumn{7}{c}{\bf \citealt{Foley/Sanders/Kirshner:2011} $(N_{\rm N}=46$, $N_{\rm HV}=19)$\tablenotemark{a}} \\
$B^{\rm max}-V^{\rm max}$      & $-$0.033$\pm$0.004 (8.8$\sigma$) & $-$0.39 & $-$0.022$\pm$0.005 &  0.047$\pm$0.008 &  0.069$\pm$0.010 (7.0$\sigma$) & 0.003 \\[-.25cm]
\enddata
\tablenotetext{a}{We have used our own derived boundary between the N and HV subclasses at $\sim-12200$\,\kms\ (see \S~\ref{sect:wangclass}).}
\end{deluxetable*}

A linear weighted least-squares fit to the data in Fig.~\ref{fig:vsi_intcol}
yields the following relation between intrinsic $B-V$ color and
Si\two\ velocity at maximum light:

\begin{equation}
B-V = -0.26(0.05) - 0.013(0.005) \times [v_{\rm Si}/10^3\,\kms]\mathrm{\ mag},
\end{equation}

\noindent
with a Pearson correlation coefficient $r=-0.35$ (see
Table~\ref{tab:vsi_intcol}), consistent with that found by
\cite{Foley/Sanders/Kirshner:2011}. Taking into account the
uncertainty in the intrinsic color estimates via a Monte Carlo
simulation results in almost no correlation ($r\approx-0.20$). 
The slope of the best-fit linear relation is shallower than that found
in their paper, and with a larger associated error that reflects the
larger uncertainty we derive on intrinsic color estimates (typically
$0.05$\,mag using BayeSN cf. 0.03\,mag for the estimates reported
by \citealt{Foley/Sanders/Kirshner:2011}). The statistical
significance of the linear relation we find is also significantly
lower than reported by \cite{Foley/Sanders/Kirshner:2011}. Our fit
rejects a non-zero slope at the $\sim3\sigma$ level\footnote{Unlike
  Foley et al. (2011), we obtain the {\it same} statistical
  significance when using the Bayesian approach to linear regression
  of Kelly (2007).}, compared to $\sim9\sigma$ in their paper.
The scatter about the best-fit linear relation ({\it dashed line}) is
the same to within $\sim10$\%\ for both samples, in conflict with
the results of \cite{Foley/Sanders/Kirshner:2011}, who infer a
$\sim50$\%\ larger scatter for the HV sample.

The robustness of the relation between Si\two\ velocity and intrinsic
color remains questionable given the large uncertainty associated with
the determination of the latter quantity. 
\cite{Foley/Sanders/Kirshner:2011} fit for a relation between
light-curve shape corrected peak absolute $M_V$ (not corrected for
host-galaxy reddening) and observed $B^{\rm max}-V^{\rm max}$ pseudo-color
(their Fig.~15), based on Eq.~1 of \cite{WangX/etal:2009b}. They
assume that the horizontal scatter about this relation is entirely due
to intrinsic variations in $B^{\rm max}-V^{\rm max}$ pseudo-color,
where it in fact also includes some intrinsic scatter about the
luminosity-\dmft\ relation. It is thus unlikely that their ``intrinsic
color estimates'' are simply offset from the true intrinsic color, as
postulated in their paper. Moreover,
\cite{Foley/Sanders/Kirshner:2011} do not propagate the errors
associated with their nuisance parameters ($\alpha$, $M_{\rm
ZP}$, $R_V$, and $C$ in their Eq.~15) in the intrinsic color error
bars, which artificially enhances the statistical significance of
their correlation. The same limitations apply to the recent study
of this relation for high-redshift \sneia\ by \cite{Foley:2012},
with the additional caveat that their analysis makes use of an
indirect measure of the observed $B^{\rm max}-V^{\rm max}$
pseudo-color.

BayeSN uses a more elaborate statistical treatment of the population
correlations between the intrinsic absolute magnitudes ($BVRI$,
including $JH$ when available), intrinsic colors and light-curve
shape, as well as the distribution and properties of host galaxy dust.  
BayeSN therefore provides an estimate of the true intrinsic
color, with realistic error bars reflecting the uncertainty in the
various population correlations. It is then not surprising that our
estimates for the intrinsic $B^{\rm max}-V^{\rm max}$ pseudo-color
are only weakly correlated ($r=0.21$) with those of
\cite{Foley/Sanders/Kirshner:2011}. Likewise, the mean intrinsic
colors we derive for both Normal and High-velocity samples are
significantly bluer (by $\sim0.08$\,mag and $\sim0.12$\,mag,
respectively, when using the intrinsic $B^{\rm max}-V^{\rm max}$
pseudo-color from BayeSN) than those using the color estimates of
\cite{Foley/Sanders/Kirshner:2011}.
The larger uncertainties derived from BayeSN fits reflect the
difficulty in obtaining accurate estimates of intrinsic color,
properly accounting for the population distributions (or variance) of
intrinsic color and dust extinction.

The difference in the weighted mean intrinsic $B-V$ color between
the High-velocity and Normal samples is $0.030\pm0.013$\,mag, corresponding
to a $\sim2\sigma$ difference (cf. $\sim7\sigma$ in
\citealt{Foley/Sanders/Kirshner:2011}). A Kolmogorov-Smirnov (K-S)
test shows that the two samples are discrepant at the
$p=0.023$  significance level, which is less than the
canonical cutoff of 0.05 for statistical significance\footnote{Note that a
smaller $p$-value indicates a higher level of statistical significance
for the K-S test.}, with the caveat that this statistic does not
consider the (large) uncertainty in intrinsic color. This is still
significantly larger (i.e., less significant) than the K-S
probability inferred for the \cite{Foley/Sanders/Kirshner:2011}
estimates of intrinsic $B^{\rm max}-V^{\rm max}$ pseudo-color
($p=0.003$).

Using the intrinsic $B-V$ color at $B$-band maximum as opposed to
the $B^{\rm max}-V^{\rm max}$ pseudo-color does not affect our results nor
their interpretation, as both quantities are strongly correlated
($r=0.72$; see Table~\ref{tab:snparam}). The weighted mean difference
between intrinsic $B^{\rm max}-V^{\rm max}$ pseudo-color and intrinsic
$B-V$ color at $B$-band maximum is only $0.02$\,mag with an rms of
0.04\,mag. 
Pseudo-colors are difficult to interpret since, for an individual
supernova, the $V$-band may reach its maximum brightness between
$\sim0$\,d and $\sim4$\,d later than the $B$-band \citep[see,
  e.g.,][their Fig.~9]{Blondin/etal:2011}. We prefer to 
use the true color at a given time.
The resulting correlation between Si\two\ absorption 
velocity and intrinsic $B^{\rm max}-V^{\rm max}$ pseudo-color is
slightly weaker than when using the $B-V$ color at $B$-band
maximum light ($r=-0.26$). While the High-velocity sample is
found to have a redder average intrinsic $B^{\rm max}-V^{\rm
  max}$ pseudo-color than the 
Normal sample at the $\sim2\sigma$ level (i.e. the same significance
than when using the $B-V$ color), a K-S test shows that the two
samples are consistent at the $p=0.14 > 0.05$ significance level.

A larger sample of \sneia\ with combined optical and NIR photometry
should enable a more precise determination of the intrinsic colors of
\sneia\ \citep[see][]{Mandel/etal:2011} and a more reliable study of
the relation with Si\two\ velocity.


\section{Early-time spectra}\label{sect:early}

Spectra taken shortly (within $\sim10$\,d) past explosion probe the
outer layers of the SN ejecta and hence the completeness of C/O
burning in the progenitor WD. The velocity interval over which
intermediate-mass elements absorb (in particular Si and Ca,
see \S~\ref{sect:SiII} and \S~\ref{sect:HVF}) and the detection of lines
from (unburnt) carbon (see \S~\ref{sect:CII}) both provide constraints on
possible explosion mechanisms.

\subsection{Maximum velocity of Si\two\ absorption}\label{sect:SiII}

In \S~\ref{sect:si2evol} we investigated the time evolution of the
Si\two\,\l6355 absorption velocity as a function of spectroscopic
subclass (see Fig.~\ref{fig:vSiII_branch}). This velocity corresponds
to the location of {\it maximum absorption}. Here we are interested in
the {\it maximum velocity} at which Si\two\ absorbs, since this value
gives an estimate of the extent of incomplete C/O nuclear burning in
the progenitor WD star.

\begin{figure}
\plotone{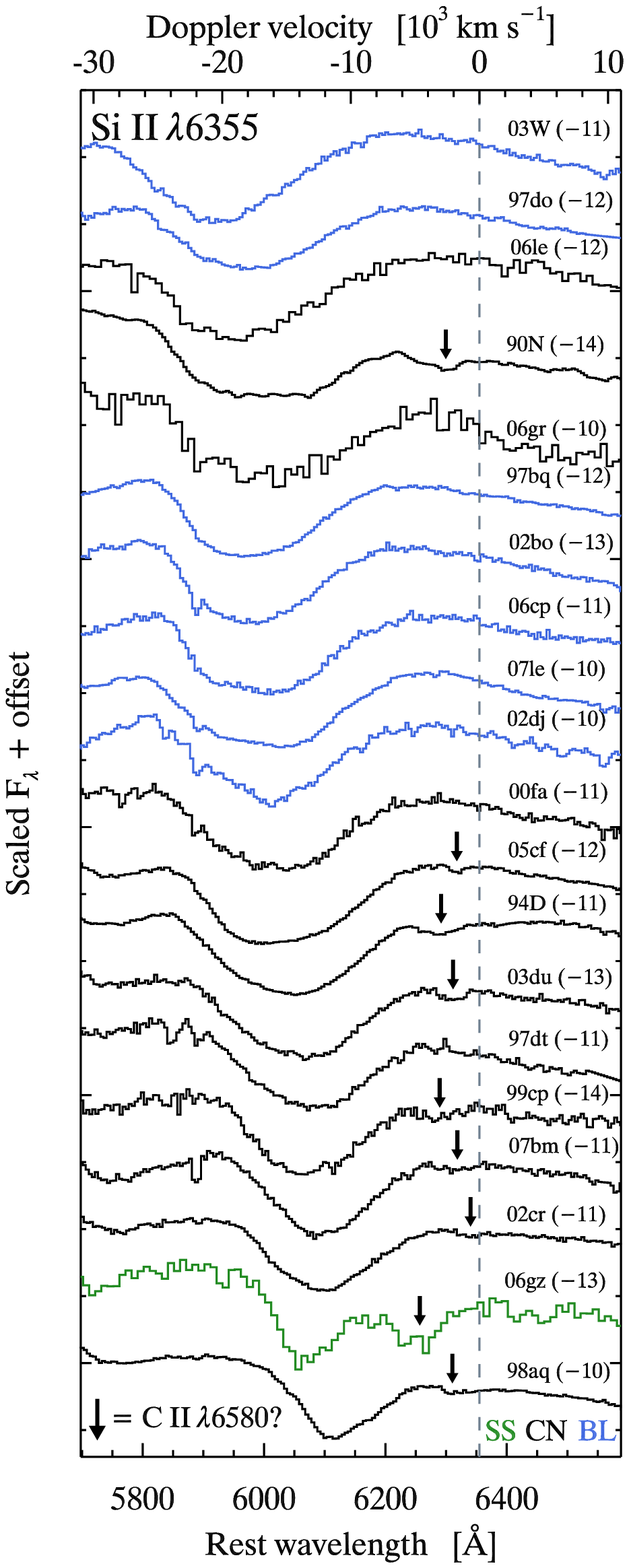}
\caption{\label{fig:earlyspec_SiII}
\scriptsize
Zoom in on the Si\two\,\l6355 line for \sneia\ for which we have
spectra $-10$\,d from $B$-band maximum or earlier.
The spectra are roughly ordered
according to the maximum velocity at which the Si\two\,\l6355 line
absorbs (from $\sim -28000$\,\kms\ for SN~2003W to $\sim
-18000$\,\kms\ for SN~1998aq), and are color-coded according to their
spectroscopic subclass as defined by \cite{Branch/etal:2006}. 
We do not show
spectra of 1991T/1999aa-like SN which generally have weak to
non-existent Si\two\ absorptions at these epochs, nor the low-S/N
highly-reddened spectrum of SN~2006cc. 
Tentative identification of a C\two\,\l6580 absorption is indicated
for some \sneia\ ({\it downward-pointing arrows}).
The spectrum of SN~1990N \citep{Leibundgut/etal:1991} is not part of
the CfA sample presented in this paper, but it was obtained through
the CfA SN Program (see Table~\ref{tab:litspec}).
}
\end{figure}

In Fig.~\ref{fig:earlyspec_SiII} we show the Si\two\,\l6355 line
profile for \sneia\ in the CfA sample for which we have spectra prior
to $-10$\,d from $B$-band maximum light (excluding 1991T/1999aa-like
objects and SN~2006cc; {\it see caption}) . There is a
large variation in the maximum Si velocity (roughly corresponding to
the blue edge of the Si\two\,\l6355 absorption profile), from $\sim
-28000$\,\kms\ for SN~2003W to $\sim -18000$\,\kms\ for
SN~1998aq. These values are consistent with the lower limit of
$\sim12000$\,\kms\ found by \cite{Mazzali/etal:2007} for the outer
extent of the Si-rich layer in \snia\ ejecta, although as noted by
\cite{Branch/etal:2007} this layer generally extends to much higher
velocities, in conflict with pure deflagration models
\citep[e.g.,][]{Maeda/etal:2010b}.

\epsscale{1.1}
\epsscale{1.0}

\cite{Mazzali/etal:2007} assume that pre-maximum Si\two\,\l6355
profiles are affected by high-velocity features (see next
section) that are not representative of the bulk of intermediate-mass
elements. In Fig.~\ref{fig:earlyspec_SiII} we clearly see variations
in the line-profile morphology of the Si\two\ line, from single
absorptions (both broad as in SN~2003W and narrow as in SN~1998aq) to
evidence for multiple components (e.g., SN~2006cp, SN2007le,
SN~2005cf). In the latter case the absorption profile is interpreted
as consisting of a photospheric and a ``detached'' high-velocity
component, as in the analysis of SN~2005cf by
\cite{WangX/etal:2009a}. In this interpretation both components would
then be blended together in \sneia\ displaying a single broad
absorption profile \citep[see, e.g.,][]{Tanaka/etal:2008}.

The emission component of the P-cygni profiles in
Fig.~\ref{fig:earlyspec_SiII} appears blueshifted with respect to the
rest-frame wavelength of the Si\two\ transition (6355\,\AA). In
SN~2003W, the peak emission occurs at a Doppler velocity of
$\sim -6000$\,\kms. Such peak-emission blueshifts are routinely observed
in Type II supernovae, and result from the steep density gradient in
the ejecta, confining the absorbing and emitting regions (as seen by
an external observer) to a narrow range of impact parameters
\citep{DH05a}. Sizeable peak-emission blueshifts are also expected
theoretically in \sneia\ \citep{Blondin/etal:2006}, although line
overlap renders their identification problematic
\citep[see discussion by][]{Branch/etal:2007}. In several cases we note a small
absorption notch in the Si\two\,\l6355 emission profile ({\it
downward-pointing arrows}), usually attributed to C\two\,\l6580 (see
\S~\ref{sect:CII}), which obviously biases the measurement of the
velocity at peak emission to higher apparent blueshifts.

\subsection{High-velocity features}\label{sect:HVF}

The Ca\two\ near-infrared triplet (IRT; 8498\,\AA, \l8542\,\AA, and
\l8662\,\AA) is less subject to overlap with lines from other ions
than Ca\two\,\l3945 and is hence more appropriate to search for
high-velocity absorption features (HVF). \cite{Mazzali/etal:2005b}
found HVFs to be present in all \snia\ spectra with early-time data
covering the Ca\two\ IRT, with evidence of similar features in the 
Si\two\,\l6355 line for some. They suggested that an abundance enhancement
in Ca and Si was unlikely to be the sole cause, as extreme
enhancements were required in the outer ejecta (e.g., factor of 20
increase in the Ca abundance above 20000\,\kms\ for SN~1999ee;
\citealt{Mazzali/etal:2005a}). A density enhancement was found to be a
more likely explanation, either intrinsic to the explosion or through
an interaction with circumstellar material. In the former case one
expects a large diversity in HVFs due to line-of-sight effects
dependent on the geometry of the high-velocity material (e.g., thick
torus or clumpy structure; see \citealt{Tanaka/etal:2006}). There is
direct evidence for the multi-dimensional nature of HVFs in one \snia,
SN~2001el, where spectro-polarimetric observations revealed the
high-velocity component of the Ca\two\ IRT to be polarized at
the $\sim1$\%\ level \citep{WangL/etal:2003,Kasen/etal:2003}.

\begin{figure*}
\resizebox{\textwidth}{!}{
\includegraphics{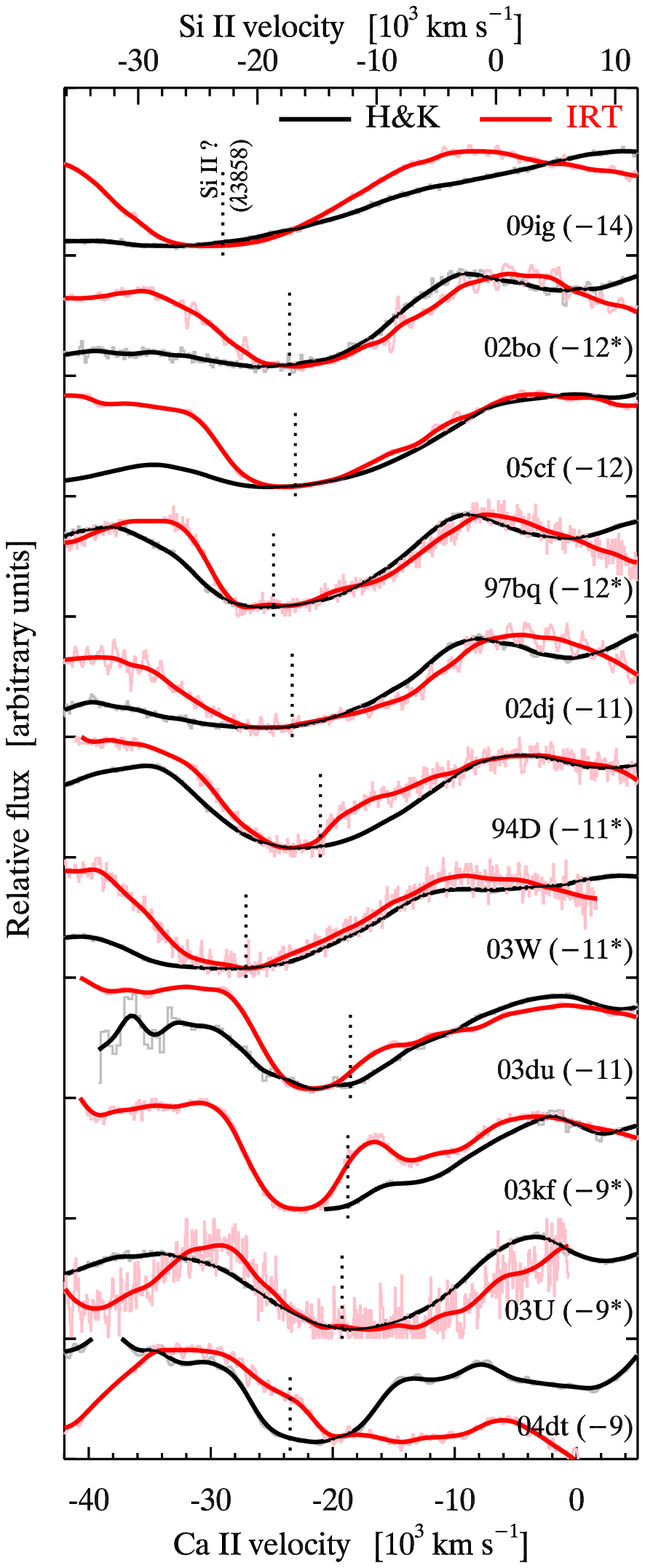}\hspace{1cm}
\includegraphics{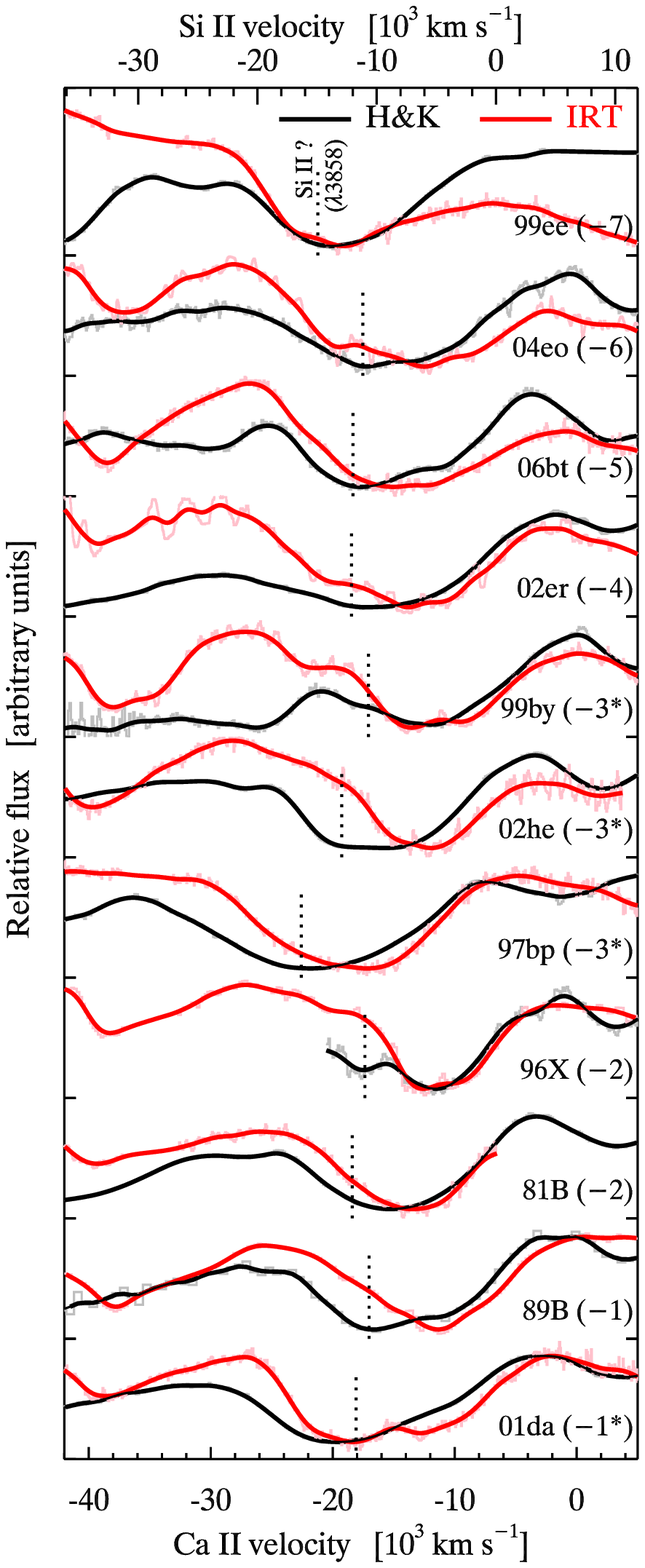}
}
\caption{\label{fig:comp_ca2}
\scriptsize
Ca\two\,\l3945 (H\&K; {\it black}) and \l8579 (IR triplet, or IRT;
{\it red}) line profiles in pre-maximum \snia\ spectra, ordered by
phase, and plotted in velocity space ({\it bottom x-axis}). The spectra
have been smoothed using an inverse-variance Gaussian weighted filter
with $d\l/\l=0.003$. The vertical
dotted lines indicate the absorption velocity of the Si\two\,\l3858 line
({\it top x-axis}) assuming it is equal to that of the Si\two\,\l6355
line. By comparing the H\&K and IRT profiles and taking into account
the likely location of maximum absorption by Si\two\,\l3858 one can
interpret absorptions in the blue wing of the Ca\two\,\l3945 line as
being due to high-velocity (``detached'') Ca\two\ or normal-velocity
(``photospheric'') Si\two, although a definitive answer can only come
from detailed spectroscopic modeling. Spectra marked with an asterisk
are from the CfA SN Program.
}
\end{figure*}

Figure~\ref{fig:comp_ca2} shows Ca\two\,\l3945 ({\it black}) and IRT
({\it red}) line profiles in pre-maximum \snia\ spectra, ordered by
phase. The diversity in line-profile morphology noted by other authors
(e.g., \citealt{Tanaka/etal:2008}) is clearly seen, ranging from
single to multiple detached absorption profiles. The correspondence
between the Ca\two\,\l3945 and IRT absorption profiles is obvious in
some cases (e.g., SN~1997bq), more questionable in others (e.g.,
SN~2001da) and nonexistent for a few cases (e.g., SN~2002he,
SN~1996X). One possible source of confusion comes from the overlap of
the Si\two\,\l3858 line with Ca\two\,\l3945, which can lead to an
apparent HVF in the H\&K line where none is visible in the IRT (e.g.,
SN~1996X, SN~1989B). The dotted lines in Fig.~\ref{fig:comp_ca2} show
the location of maximum absorption by Si\two\,\l3858, assuming it has
the same absorption velocity as Si\two\,\l6355. In several cases it
coincides with a blue absorption in the Ca\two\,\l3945 absorption
profile, sometimes resulting in a flat-bottomed absorption (e.g.,
SN~2002he). Another difficulty when interpreting HVFs associated with
Ca\two\ is that a mass fraction as low as $\sim10^{-5}$
is sufficient to yield a strong Ca\two\ absorption at high velocities
\citep[see, e.g.,][]{Blondin/etal:2011}.

\cite{Tanaka/etal:2006} noted that if HVFs result from a density
enhancement (regardless of its origin), other lines
(e.g., Si\two\,\l6355) should be affected, and the relative strengths
and velocities of these HVFs should be correlated in some
way. \cite{Stanishev/etal:2007} suggest that the relative strengths of
HVFs in Ca\two\,\l3945, Si\two\,\l6355, and Ca\two\ IRT are indeed
correlated in SN~2003du, but \cite{Tanaka/etal:2008} find that the
velocities of the Si\two\,\l6355 and Ca\two\ IRT high-velocity
features in several \sneia\ are at odds with this hypothesis. We note
based on Figs.~\ref{fig:earlyspec_SiII} and \ref{fig:comp_ca2} that
five of the \sneia\ with the largest extent of Ca\two\ IRT absorption
(SN~2002bo, SN~2005cf, SN~1997bq, SN~2002dj, SN~2003W) also have some
of the largest Si\two\,\l6355 absorption velocities at early times
(the same is also true for SN~2009ig, shown in
  Fig.~\ref{fig:comp_ca2} but not in Fig.~\ref{fig:earlyspec_SiII}).
Interestingly, \cite{Tanaka/etal:2006} also suggest that a combination
of density and abundance enhancements might provide a better
explanation for the origin of HVFs, on the account that dense blobs
covering the entire photosphere would result in Si\two\,\l6355
absorption velocities in excess of 20000\,\kms, ``never observed in
any SN''. In fact, such high absorption blueshifts are 
seen in one supernova from the CfA sample, SN~2003W (see
Fig.~\ref{fig:vSiII_branch}), and more recently in SN~2009ig
\citep{SN2009ig}.

\subsection{Evidence for unburnt carbon}\label{sect:CII}

Inferring the presence of unburnt carbon in \snia\ spectra can yield
important clues on the properties of the explosion. Pure deflagration
models predict that large amounts of unburnt C/O fuel are left over,
sometimes mixed to the innermost ejecta regions (see, e.g.,
\citealt{Roepke:2005}), although one then predicts strong forbidden
lines of C\one\ and O\one\ in late-time spectra that are not
observed \citep{Kozma/etal:2005}. Delayed-detonation
models, on the other hand, consume all but the outermost regions of
the SN ejecta, leaving varying amounts of unburnt C/O at high
expansion velocities \citep[e.g.,][]{Maeda/etal:2010b} which can
affect early-time \snia\ spectra.

As noted by several authors \citep[e.g.,][]{Tanaka/etal:2008}, carbon
is expected to be once-ionized in the ejecta of \sneia\ at early
times, such that the ``best bet'' at constraining the abundance of
carbon, let alone inferring its presence, is through its strongest
optical line C\two\,\l6580 (although see the analysis by
\citealt{Marion/etal:2006} using C\one\ lines in NIR spectra). Until
recently, firm detections of this line had been reported in only a
handful of \sneia, the most convincing cases being those of SN~2006D
\citep{SN2006D} and SN~2006gz \citep{SN2006gz}, both of which show
gradually fading C\two\,\l6580 lines with respect to the earliest
detection. The scarcity of C\two\ detections led to the suggestion of
a clumpy carbon distribution in the outer ejecta of \sneia\ combined
with line-of-sight effects \citep[e.g.,][]{Branch/etal:2007}.

In a recent survey of C\two\ features in \snia\ spectra, 
\cite{Parrent/etal:2011} find evidence for C\two\,\l6580 absorption in
$\sim30$\% of \sneia\ in their sample, a significantly larger fraction
than previously suspected \citep[e.g.,][]{SN2006D}. Moreover, they show that
these features are most common in SN in the low-velocity gradient
group, in line with the idea that LVG \sneia\ correspond to
off-center delayed-detonations seen from the deflagration side
(\citealt{Maeda/etal:2010c}; although see \citealt{Blondin/etal:2011}
for an interpretation of velocity gradients in the context of
symmetric explosions), or at least to WDs that underwent less intense
burning \citep{Tanaka/etal:2008}.

\cite{Thomas/etal:2011} investigated the presence of C\two\,\l6580
in early-time \snia\ spectra from the SNfactory collaboration
and found definite detections in 5 \sneia. All five SN
are part of the CN/LVG subclass and are
otherwise unremarkable, apart from SN~2005cf which displays clear
high-velocity features in the Si\two\,\l6355 and Ca\two\ IRT
lines. Of the remaining four SN, three (SN~2005el, SN~2005ki, and
SNF20080514-002) are characterized by blue colors and narrow
light-curve widths, suggesting that carbon-positive \sneia\ may be
preferentially associated with a specific photometric behavior, and
one (SN~2005di) is highly reddened by host-galaxy
dust. \cite{Thomas/etal:2011} estimate that $\sim20$\% of ``normal''
\sneia\ display a C\two\,\l6580 absorption as late as $-5$\,d from
$B$-band maximum light, in line with the results of
\cite{Parrent/etal:2011} [see also the paper recently submitted 
  by \citealt{Silverman/Filippenko:2012}]. They also illustrate using a
toy model that 
a spherically symmetric distribution of carbon can account for the
C\two\,\l6580 line-profile morphology, although they cannot rule out
the large-scale asymmetries inferred by
\cite{Maeda/etal:2010c} to explain the spectroscopic diversity of
\sneia.

\begin{figure*}
\resizebox{\textwidth}{!}{
\includegraphics{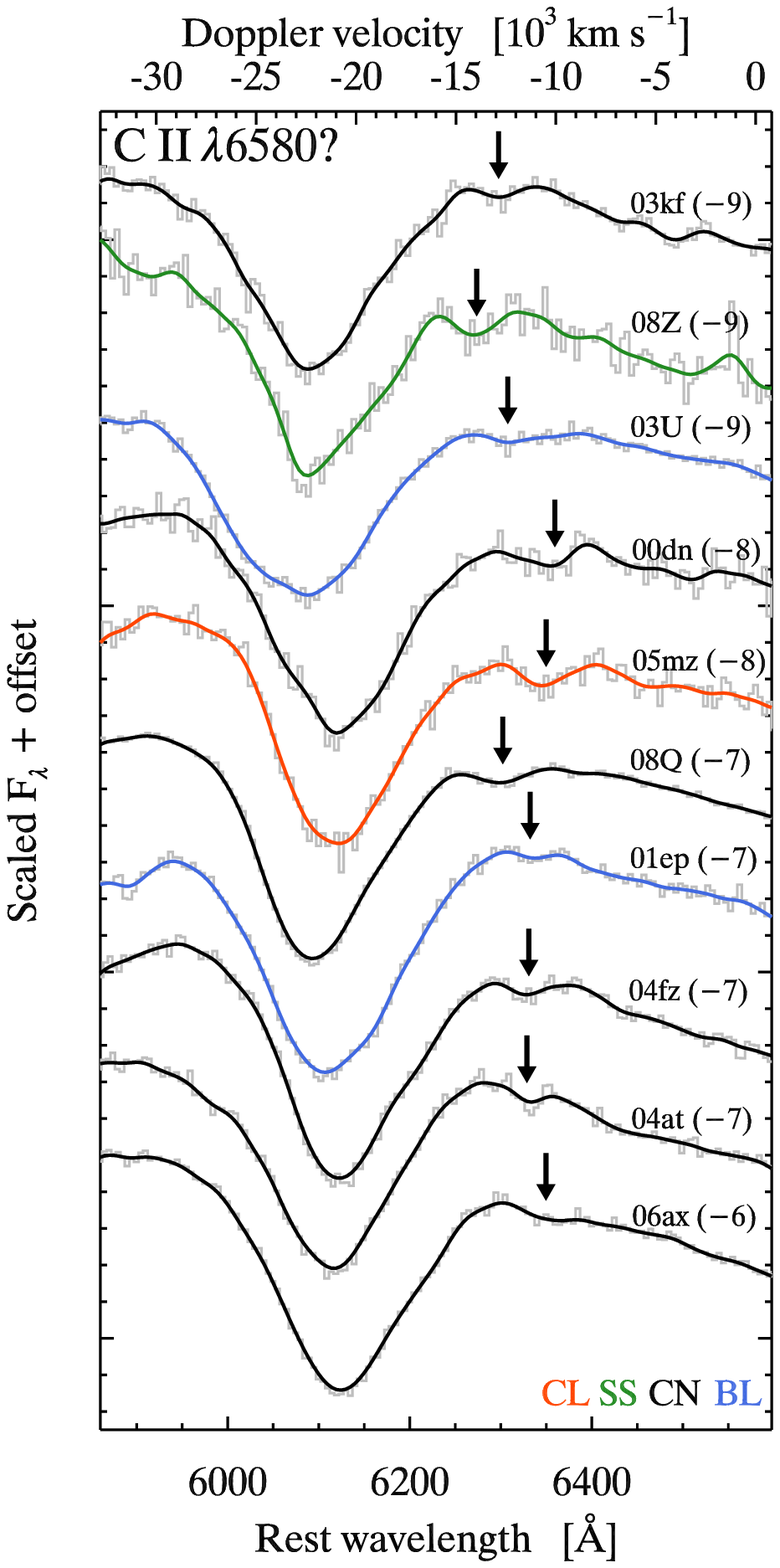}\hspace{1cm}
\includegraphics{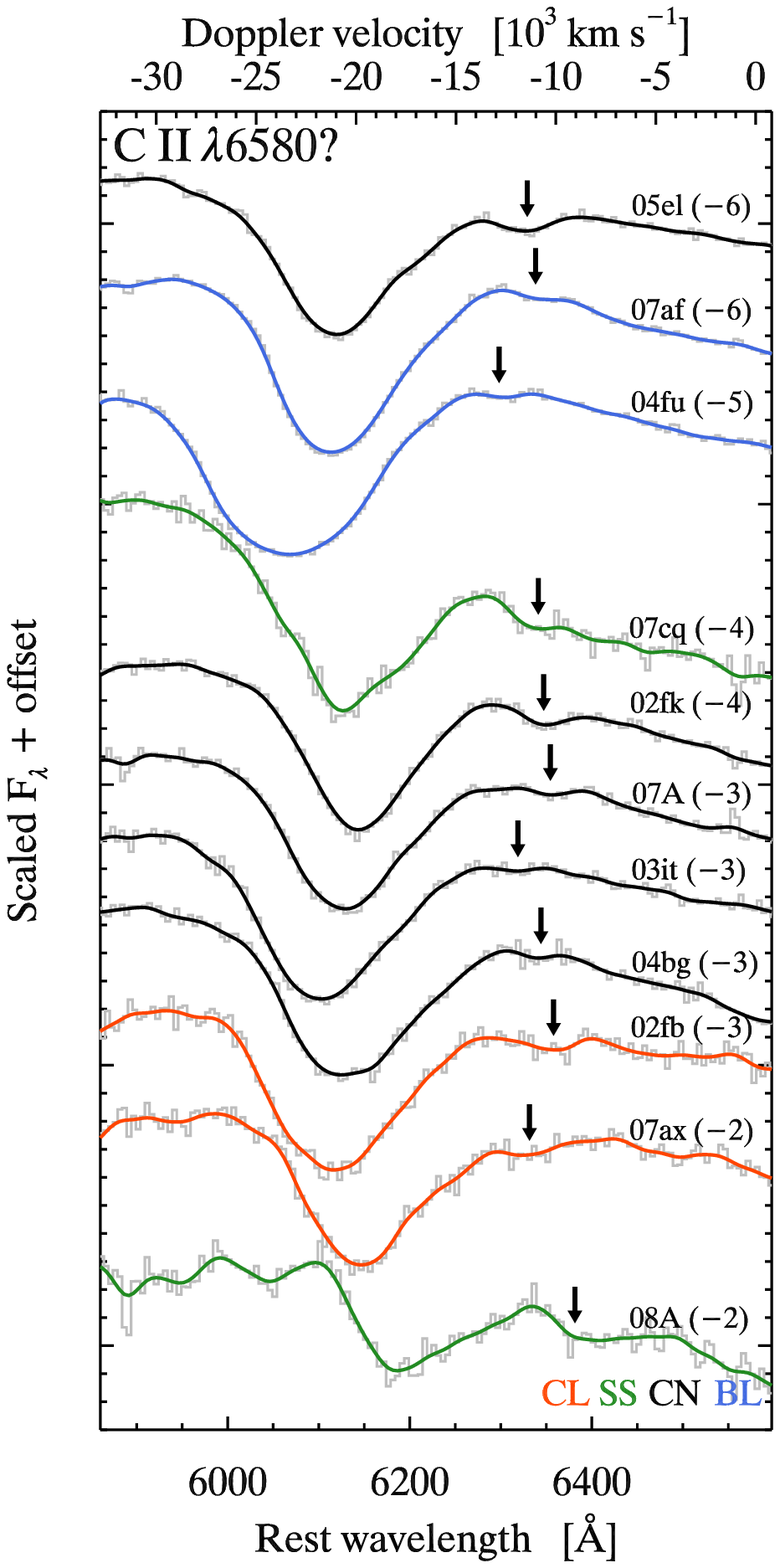}
}
\caption{\label{fig:earlyspec_CII}
Tentative identification of a C\two\,\l6580 absorption feature ({\it
downward-pointing arrows}) in previously unpublished pre-maximum
\snia\ spectra from the CfA SN Program, ordered by phase. The spectra
have been smoothed using an inverse-variance Gaussian weighted filter
with $d\l/\l=0.002$, and are color-coded according to their
spectroscopic subclass as defined by \cite{Branch/etal:2006}. 
We do not include the spectra with possible C\two\,\l6580 absorption
features already shown in Fig.~\ref{fig:earlyspec_SiII}. 
}
\end{figure*}

In Fig.~\ref{fig:earlyspec_CII} we show the Si\two\,\l6355 region in
previously unpublished spectra from the CfA SN Program where we
tentatively identify an absorption due to C\two\,\l6580. For each of
the 21 \sneia\ in this figure we only show the earliest spectrum where
the C\two\,\l6580 absorption is seen, except for a few cases where
the earliest spectra had too low S/N (SN~2004at, SN~2004fu, SN~2006ax,
SN~2007cq, and SN~2008A). Furthermore, we do not show spectra of \sneia\
already included in Fig.~\ref{fig:earlyspec_SiII}. In all this
represents a total of 24 \sneia\ not included in the study by
\cite{Parrent/etal:2011}\footnote{An even more recent paper by
\cite{Folatelli/etal:2012} confirms C\two\ detections in 4
\sneia\ from  Fig.~\ref{fig:earlyspec_CII} (SN~2005el, SN~2006ax,
SN~2007A, and SN~2007af). They argue, however, that SN~2007ax shows no
such absorption. They also report a C\two\ detection in pre-maximum
spectra of SN~2008bf. We have spectra of this SN, but only
past maximum light, by which time the C\two\ absorption feature is no
longer detected.}. A C\two\,\l6580 line in
SN~2005el was already noted by \cite{Thomas/etal:2011} in their
earliest spectrum\footnote{\cite{Thomas/etal:2011} also report a
C\two\ detection in SN~2005ki, for which we have one FLWO
1.5\,m+FAST spectrum taken at +1\,d past $B$-band maximum light, and
SNF20080514-002, for which we have spectra only starting at +2\,d (see
Table~\ref{tab:obs}), cf. $-8$\,d for the latest detection by 
\cite{Thomas/etal:2011}.}. We do not attempt to confirm these
detections with parameterized spectrum synthesis codes as done by
\cite{Parrent/etal:2011} using SYNOW
\citep[e.g.,][]{Branch/etal:2005} or by \cite{Thomas/etal:2011} using
SYNAPPS \citep{SYNAPPS}, but most of the 
tentative C\two\ identifications (17 out of 24) correspond to firm
detections in spectra from different nights. Of the remaining 7
\sneia, 3 only have a single pre-maximum spectrum (SN~2004bg,
SN~2004fz, and SN~2005mz), and 2 have possible detections in multiple
spectra (SN~2000dn and SN~2003U). For SN~2001ep, we have a definite
detection in the $-7$\,d spectrum but none in the $-3$\,d
spectrum. The small notch seen in the $-2$\,d spectrum of SN~2007ax is
no longer visible one day later.

The majority of \sneia\ with C\two\,\l6580 detections in
Figs.~\ref{fig:earlyspec_SiII} and \ref{fig:earlyspec_CII} are part of
the Core Normal subclass (19 out of 30). Moreover, 7/9 CN \sneia\ with
spectra prior to $-10$\,d from $B$-band maximum (see
Fig.~\ref{fig:earlyspec_SiII}) show C\two\ absorptions, suggesting
that this feature is a generic property of this subclass, and
highlighting the need for very early-time spectroscopy.
The other 11 \sneia\ are almost
equally divided amongst the SS (4 objects), CL (3), and BL (4)
subclasses. The carbon-positive SS \sneia\ include the 1991T-like
SN~2007cq, the 2002cx-like SN~2008A, and the possibly
super-Chandrasekhar SN~2006gz. One of the carbon-positive CL
(SN~2007ax) has an obvious 1991bg-like spectrum, while the remaining
two (SN~2002fb and SN~2005mz; both similar to SN~2006D) show evidence
for more moderate Ti\two\ absorption. Of the 4 carbon-positive BL
\sneia, only one (SN~2004fu) is part of the high-velocity subclass of
\cite{WangX/etal:2009b}, with a high velocity gradient in the
Si\two\,\l6355 line ($\dot{v}\approx120$\,\kmsd\ measured between
+2\,d and +22\,d). This also confirms the results of
\cite{Parrent/etal:2011} who find only one carbon-positive \snia\ of
the HVG group in their sample (SN~2009ig; see also
\citealt{SN2009ig}). \cite{Parrent/etal:2011} suggest line blending as
a possible explanation for the apparent paucity of high-velocity
carbon-positive \sneia.

The C\two\,\l6580 blueshifts at maximum absorption in this sample of
30 \sneia\ are typically in the range $-10000\gtrsim v_{\rm
 abs}\gtrsim-14000$\,\kms, with the notable exceptions of the
2002cx-like SN~2008A ($v_{\rm abs}\approx-8000$\,\kms\ at $-6$\,d,
similar to the Si\two\,\l6355 absorption velocity at the same phase; see
Fig.~\ref{fig:vSiII_branch}) and the peculiar SN~2006gz ($v_{\rm
 abs}\approx-15500$\,\kms\ between $-13$\,d and $-11$\,d), both part
of the SS subclass. The absorption velocities of C\two\,\l6580 and
Si\two\,\l6355 are similar for most objects at any given pre-maximum
epoch, their ratio being within 10\%\ of unity (see also Fig.~8 of
\citealt{Parrent/etal:2011}). Two \sneia\ (SN~1990N at $-14$\,d and
SN~2005cf between $-13$\,d and $-11$\,d) have $v_{\rm
  abs}($C\two$)/v_{\rm abs}($Si\two$) < 0.8$, but both display 
high-velocity features in the Si\two\,\l6355 line (see
Fig.~\ref{fig:earlyspec_SiII}) which biases this ratio to lower
values. Conversely, SN~2006gz, SN~2008Z (both SS \sneia), and the
1991bg-like SN~2007ax display $v_{\rm abs}$ ratios closer to 1.2.

The maximum velocity at which C\two\,\l6580 absorbs ($v_{\rm max}$) is
in the range $-12000\gtrsim v_{\rm max}\gtrsim-16000$\,\kms\ for
most \sneia. Since $v_{\rm max}$ and $v_{\rm abs}$ are strongly
correlated, we find the same two outliers as above (SN~2008A, $v_{\rm
 max}\approx-10000$\,\kms; SN~2006gz, $|v_{\rm
 max}|\gtrsim18000$\,\kms). Within the same sample, the maximum
velocity at which Si\two\,\l6355 absorbs is typically significantly
higher ($-20000\gtrsim v_{\rm max}\gtrsim-28000$\,\kms). This
suggests that some silicon is ejected at larger velocities than the
bulk of unburnt carbon, incompatible with current delayed-detonation
models \citep[see, e.g.,][]{Maeda/etal:2010b}. As noted by
\cite{Branch/etal:2007}, the pulsating reverse detonation models of
\cite{Bravo/Garcia-Senz:2006} could provide a viable explanation for
the low carbon velocities compared to the bulk of IME.

\cite{Thomas/etal:2011} suggest that the presence of C\two\ lines in
normal \sneia\ could be preferentially associated with blue colors and
narrow light-curve widths, based on the SALT2 parameters
$(x_1,c)$. They note, however, that this trend does not appear to hold
for the sample analyzed by \cite{Parrent/etal:2011}. We have reliable
SALT2 fits for 24 out of 30 \sneia\ in our carbon-positive sample, and
do not find evidence for bluer colors with respect to the
carbon-negative sample (171 \sneia\ with reliable SALT2 fits). The
median SALT2 $x_1$ is approximately $-0.6$ for both samples. Since the
SALT2 color parameter does not distinguish between intrinsic color and
extinction by dust, we re-ran this analysis using the intrinsic $B-V$
color at $B$-band maximum and \dmft\ values from BayeSN. Within the
Core Normal sample, the carbon-positive \sneia\ appear to occupy a
narrower \dmft\ range ($1\lesssim\dmft\lesssim1.5$\,mag) than the
carbon-negative sample ($0.8\lesssim\dmft\lesssim1.6$\,mag), 
but a K-S test shows that the two distributions are consistent at
the $p=0.37 > 0.05$ significance level. The distribution of 
intrinsic $B-V$ color is the same for both subsamples ($p=0.34$). These
conclusions hold when considering only \sneia\ with persistent C\two\
features (i.e., present later than $-5$\,d), although the numbers
become then too small for a robust analysis.



\section{Nebular spectra}\label{sect:neb}

\subsection{Nebular line widths and luminosity}

We have \nspecgt150\ spectra of \nsnwspecgt150\ \sneia\ with spectra
taken later than 150\,d past maximum in the CfA sample presented in
this paper (see Table~\ref{tab:nebspec}). At these
epochs, the spectra are dominated by forbidden lines of Fe\two\ and
Fe\three, reflecting the NSE composition of the inner ejecta
regions. Based on the observed correlation of the FWHM of the iron
emission feature centered at $\sim4700$\,\AA\ with \dmft\ in nebular \snia\
spectra from the Asiago SN archive, \cite{Mazzali/etal:1998} found a
relation between the kinetic energy of the explosion and the mass of
synthesized \nifs. In particular, the low-luminosity SN~1991bg was
associated with a low mass of \nifs\ but also a low explosion energy,
suggesting that the fast-evolving light curve was the result of a
small (sub-Chandrasekhar) exploding mass.

\begin{figure}
\plotone{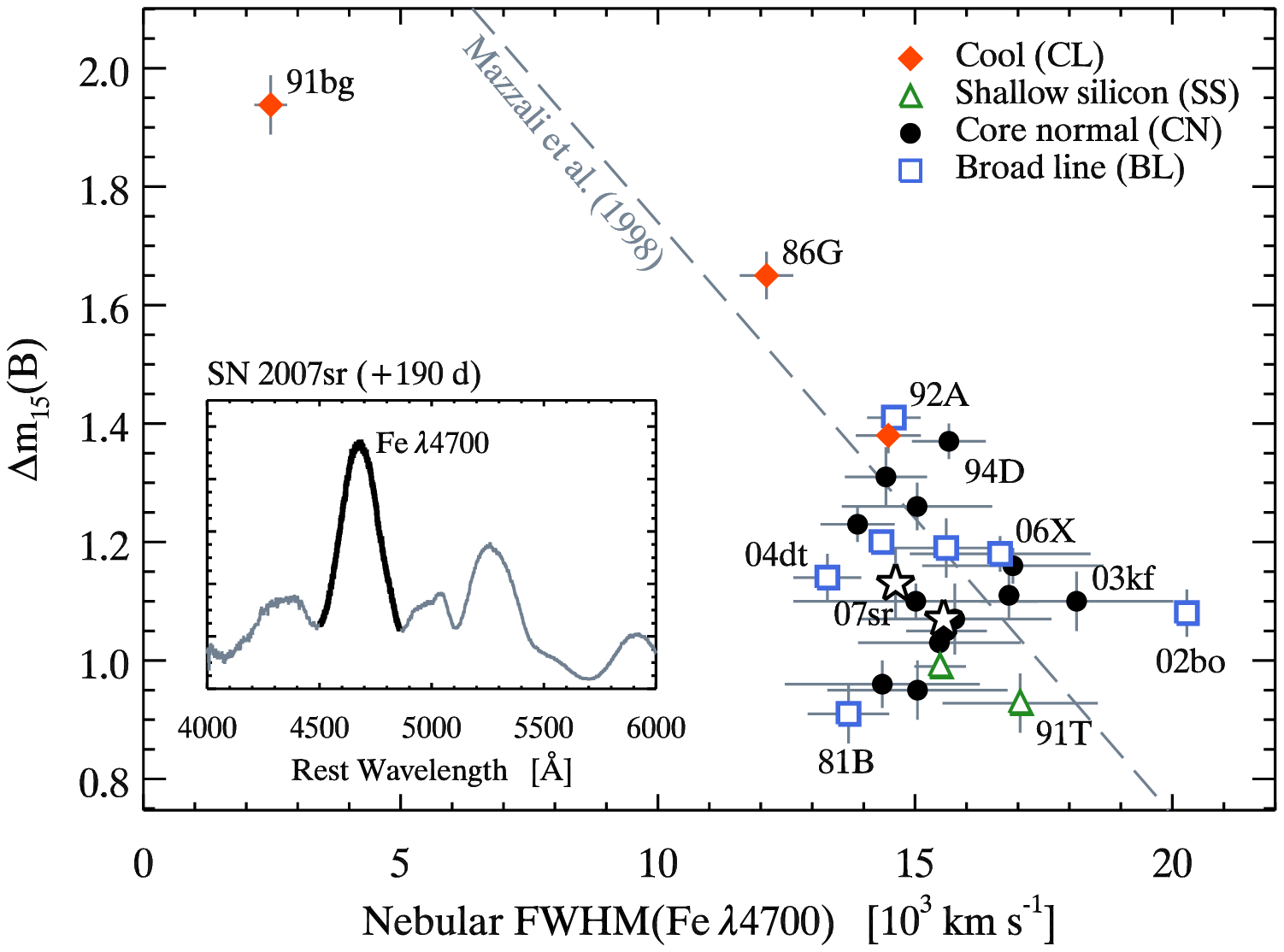}
\caption{\label{fig:nebvel}
Relation between decline-rate parameter \dmft\ and FWHM of the Fe
emission feature at $\sim4700$\,\AA\ (which includes emission from
[Fe\two] and [Fe\three]) in nebular \snia\ spectra. 
The points are color-coded according to the spectroscopic subclasses
defined by \cite{Branch/etal:2006}.
Two SN with no Branch class (SN~2006dd and SN~2007sr) are shown as
open stars.
We have used previously unpublished data from the CfA SN Program
supplemented with our own measurements on publicly-available data from
the literature, where we have excluded SN~1989B and SN~2003cg
due to the large associated error ($\sigma_{\rm
  FWHM}\approx6000$\,\kms\ and 3500\,\kms, respectively), and
SN~2002cx due to the low S/N of the spectrum.
The dashed line corresponds to the original 
relation published by \cite{Mazzali/etal:1998}.
The inset shows a Magellan Clay+LDSS3 spectrum of SN~2007sr at 190\,d
past maximum with the Fe\,\l4700 feature highlighted.
}
\end{figure}

We revisit the relation between the FWHM of the nebular iron feature
at $\sim4700$\,\AA\ and \dmft\ in Fig.~\ref{fig:nebvel} (see Fig.~2 of
\citealt{Mazzali/etal:1998}). We have used our own FWHM measurements
(see Table~\ref{tab:nebvel_data})
on the publicly-available data from \cite{Mazzali/etal:1998}
(i.e., with the exception of SN~1989M, SN~1991M, and SN~1993L),
supplemented with 10 of the 12 \sneia\ from the CfA SN Program with
nebular spectra (we exclude SN~2002cx due to low S/N and SN~2003cg due
to the large associated error on FWHM), and additional data from the
literature (including a nebular spectrum of SN~1991T from
\citealt{Schmidt/etal:1994}, publicly available in the CfA SN Archive;
see Table~\ref{tab:litspec}).
When several measurements are available for a given
SN, we plot the error-weighted mean FWHM. We also show the original
relation published by \cite{Mazzali/etal:1998} ({\it long-dashed
  line}).

\epsscale{1.15}
\epsscale{1.0}

\begin{deluxetable*}{lccr@{ $\pm$ }ll}
\tabletypesize{\scriptsize}
\tablewidth{0pt}
\tablecaption{\label{tab:nebvel_data}FWHM measurements of the Fe\,\l4700 feature in nebular \snia\ spectra.}
\tablehead{\colhead{SN} & \colhead{Branch} & \colhead{\dmft} & \multicolumn{2}{c}{FWHM}   & \colhead{Phase(s)} \\
           \colhead{}   & \colhead{class}  & \colhead{(mag)} & \multicolumn{2}{c}{(\kms)} & \colhead{(d)}    }
\startdata
1981B  & BL                        & 0.91 (0.05) & 13705 &  790 & +266                                    \\
1986G  & CL                        & 1.65 (0.04) & 12112 &  521 & +256,+323                               \\
1990N  & CN                        & 0.95 (0.05) & 15047 & 1750 & +212,+227,+244,+245,+254,+279,+308,+331 \\
1991T  & SS                        & 0.93 (0.05) & 17044 & 1507 & +259                                    \\
1991bg & CL                        & 1.94 (0.05) &  2474 &  317 & +201                                    \\
1992A  & BL                        & 1.41 (0.02) & 14588 &  519 & +225                                    \\
1994D  & CN                        & 1.37 (0.03) & 15656 &  717 & +611                                    \\
1994ae & CN                        & 0.96 (0.04) & 14363 & 1895 & +152,+366                               \\
1995D  & CN                        & 1.05 (0.03) & 15612 &  782 & +283                                    \\
1996X  & CN                        & 1.26 (0.04) & 15039 & 1459 & +296                                    \\
1998aq & CN                        & 1.11 (0.04) & 16824 &  204 & +229,+239                               \\
1998bu & CN                        & 1.03 (0.02) & 15473 & 1577 & +178,+189,+207,+216,+242,+248,+328      \\
2000cx & SS                        & 0.99 (0.02) & 15486 &  499 & +154,+181                               \\
2001el & CN                        & 1.16 (0.03) & 16901 & 1759 & +397                                    \\
2002bo & BL                        & 1.08 (0.04) & 20281 &   35 & +310                                    \\
2002dj & BL                        & 1.19 (0.05) & 15606 & 1081 & +219,+271                               \\
2002er & CN                        & 1.23 (0.03) & 13882 &  720 & +212                                    \\
2003du & CN                        & 1.07 (0.06) & 15772 & 1876 & +219,+374                               \\
2003hv & CN                        & 1.31 (0.05) & 14433 &  797 & +322                                    \\
2003kf & CN                        & 1.10 (0.05) & 18139 & 1877 & +398                                    \\
2004dt & BL                        & 1.14 (0.04) & 13294 &  657 & +347                                    \\
2004eo & CL                        & 1.38 (0.03) & 14480 &  630 & +225                                    \\
2005cf & CN                        & 1.10 (0.03) & 15013 & 2377 & +317                                    \\
2006X  & BL                        & 1.18 (0.03) & 16653 & 1754 & +150,+275                               \\
2006dd & \nodata\tablenotemark{a}  & 1.07 (0.03) & 15554 & 1142 & +186,+193                               \\
2007af & BL                        & 1.20 (0.02) & 14347 &   36 & +301                                    \\
2007sr & \nodata\tablenotemark{a}  & 1.13 (0.06) & 14621 &   32 & +190                                    
\enddata
\tablecomments{When several FWHM measurements are available (see the Phase(s) column), we report the error-weighted mean FWHM.}
\tablenotetext{a}{No Branch class for SN~2006dd or SN~2007sr since there are no available spectra around maximum light.}
\end{deluxetable*}

We have performed multiple least-squares fits of the relation
$\dmft = a + b ({\rm FWHM}-15)$ [FWHM in units of 10$^3$\,\kms] of
\cite{Mazzali/etal:1998} to the data, the results of which are reported in
Table~\ref{tab:nebvel}, including the reduced $\chi^2_\nu$ and Pearson
correlation coefficient ($r$). We also include the results of fits to
the original data of \cite{Mazzali/etal:1998} (see their
Table~1). The entire CfA+Literature sample yields a slope ($b=-0.069$)
that is somewhat shallower than the relation published by
\cite{Mazzali/etal:1998} ($b=-0.103$), although the correlation is
strong in both cases ($r=-0.71$ and $-0.80$, respectively). The
large $\chi^2_\nu$ indicates that the scatter about the best-fit linear
relation is real and not the result of measurement
uncertainties. \cite{Mazzali/etal:1998} suggested that this intrinsic
scatter could be taken as evidence for a second parameter affecting
the light-curve widths (in addition to the \nifs\ mass) or the
velocities of the iron-dominated nebula (in addition to the kinetic
energy of the explosion).

\begin{deluxetable}{lcccc}
\tablewidth{0pt}
\tablecaption{\label{tab:nebvel}Coefficients of least-squares fits of $\dmft=a+b({\rm FWHM\ [10^3\kms]}-15)$ to nebular \snia\ spectra.}
\tablehead{\colhead{Sample} & \colhead{$a$} & \colhead{$b$} & \colhead{$\chi^2_\nu$} & \colhead{$r$}}
\startdata
\multicolumn{5}{c}{\bf CfA + Literature sample} \\
All                  & 1.20 (0.01) & $-$0.069 (0.003) &  7.5 & $-$0.71 \\
Excl. 91bg           & 1.21 (0.01) & $-$0.079 (0.004) &  7.4 & $-$0.37 \\
Excl. 91bg+86G       & 1.20 (0.01) & $-$0.072 (0.005) &  6.9 & $-$0.17 \\
\hline
\multicolumn{5}{c}{} \\[-.2cm]
\multicolumn{5}{c}{{\bf Mazzali et al. (1998) sample}\tablenotemark{a}} \\
All                  & 1.24 (0.03) & $-$0.103 (0.012) &  4.4 & $-$0.80 \\
Excl. 91bg           & 1.26 (0.04) & $-$0.169 (0.024) &  2.7 & $-$0.69 \\
Excl. 91bg+86G       & 1.25 (0.04) & $-$0.168 (0.034) &  3.0 & $-$0.58 \\[-.25cm]
\enddata
\tablecomments{1$\sigma$ errors are given in between parentheses.}
\tablenotetext{a}{Measurements taken from Table~1 of \cite{Mazzali/etal:1998}.}
\end{deluxetable}

As noted by \cite{Mazzali/etal:1998}, the low-luminosity SN~1991bg
appears to be a clear outlier, being the only \snia\
with ${\rm FWHM}<10000$\,\kms\ (${\rm
  FWHM}\approx2500$\,\kms). Excluding this single SN from our sample
results in a similar slope ($b=-0.079$) but a significantly degraded
correlation ($r=-0.37$). This is somewhat different to the result
obtained using the original measurements of \cite{Mazzali/etal:1998},
where the exclusion of SN~1991bg leads to a steeper
slope ($b=-0.169$) and a minor impact on the correlation strength
($r=-0.69$). The new CfA+Literature sample indicates that the scatter
about the best-fit linear relation (excluding SN~1991bg) is $\sim30$\%
larger than for the original \cite{Mazzali/etal:1998} sample
($\sim50$\% larger when including SN~1991bg). Further excluding the
low-luminosity SN~1986G (i.e., only considering \sneia\ with
$\dmft<1.6$\,mag), there is essentially no correlation between \dmft\ and
the FWHM of the 4700\,\AA\ Fe line ($r=-0.17$). If the latter quantity
is indeed a proxy for the kinetic energy of the explosion, as suggested
by \cite{Mazzali/etal:1998}, our measurements imply that the kinetic
energy of the explosion does not strongly depend on \dmft, and hence
on the \snia\ luminosity, for all but the least luminous
events. This is expected in delayed-detonation 
models where the WD star is almost completely incinerated, as burning
of C/O to Ni or Si-group elements releases similar amounts of energy
\citep[see, e.g.,][]{Hoeflich/Khokhlov/Wheeler:1995}.
One will need to see whether additional data
(in particular for low-luminosity \sneia) continue to disfavor a
strong relation between both quantities.

\subsection{Nebular line shifts and intrinsic color}\label{sect:vnebfeni}

A recent study of nebular-phase \snia\ spectra by
\cite{Maeda/etal:2010a} revealed that some forbidden lines associated
with Ni\two\ and Fe\two\ displayed up to $\pm3000$\,\kms\ velocity
shifts about the central wavelength, while others did not (this
was first noted in mid-IR spectra of SN~2003hv by
\citealt{Gerardy/etal:2007}). This result 
was interpreted by \cite{Maeda/etal:2010a} in the context of a simple
kinematic model crafted to reproduce basic characteristics of an
off-center delayed-detonation model. These include an inner
high-density region of stable iron-group elements and \nifs, and an
outer low-density \nifs-rich region. The former results from the
deflagration phase, and hence presents a global velocity offset in the
case of an off-center ignition, while the latter results from the
detonation phase, resulting in a spherically symmetric distribution of
\nifs\ in the outer region. Different spectral lines in nebular \snia\
spectra thus trace different zones of the inner ejecta, and the
diversity in velocity shifts for lines tracing the deflagration ash
(in particular [Fe\two]\,\l7155 and [Ni\two]\,\l7378) then stems from
viewing-angle effects: the lines are blueshifted when the SN is
viewed from the ignition side and redshifted when it is viewed from
the opposite direction.

Subsequent investigations by Maeda and co-workers
showed that these nebular line shifts were related to the velocity
evolution of the Si\two\,\l6355 line around maximum light
\citep{Maeda/etal:2010c} and to the intrinsic color at maximum light
\citep{Maeda/etal:2011}, suggesting that both latter quantities were
also related to viewing angle effects in off-center delayed-detonation 
models.  

\epsscale{1.15}
\begin{figure}
\plotone{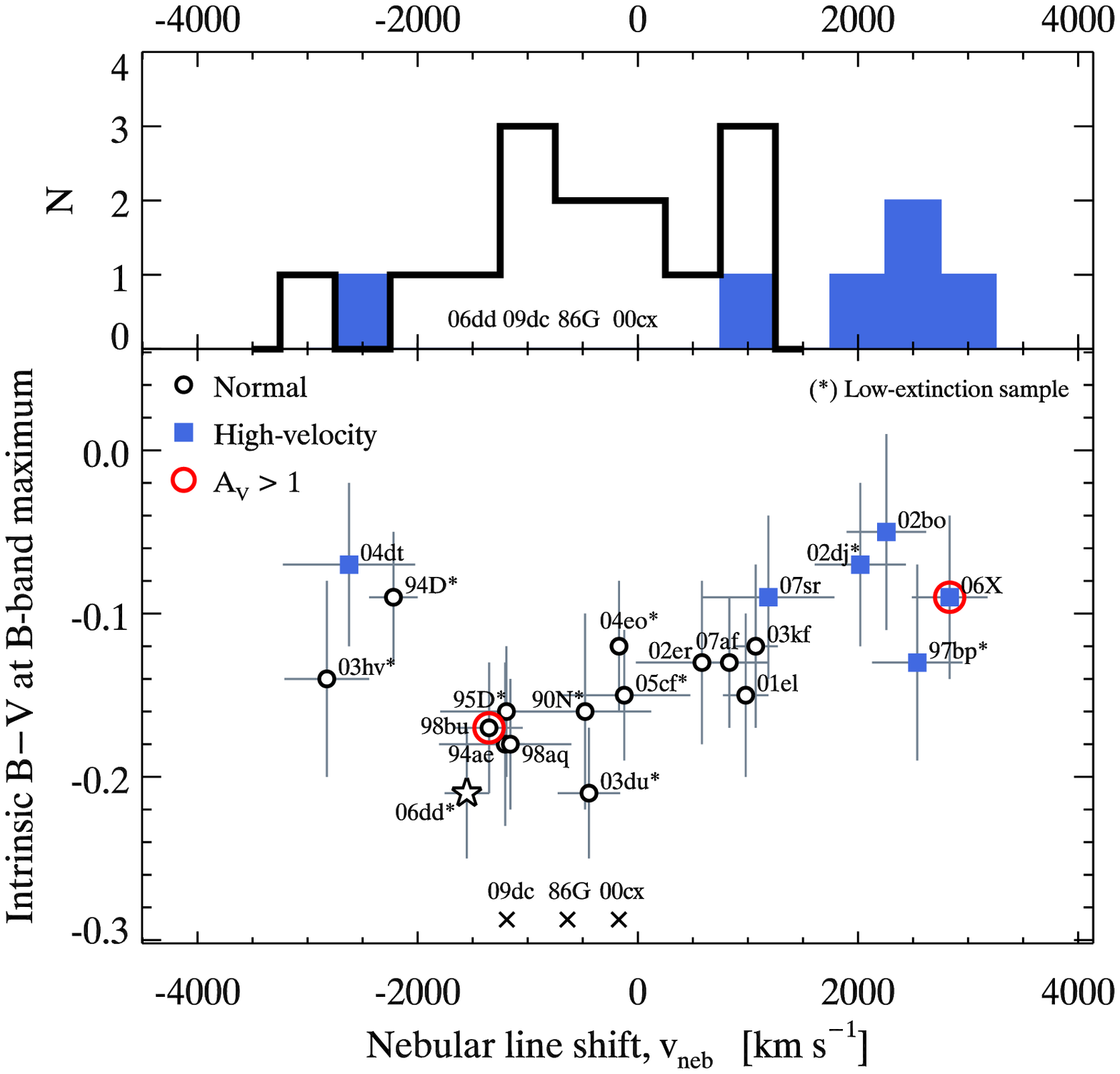}
\caption{\label{fig:vnebfeni}
{\it Bottom:}
Intrinsic $B-V$ color at $B$-band maximum light, as
derived from BayeSN fits to \snia\ light curves, {\it vs.} the 
nebular line shift ($v_{\rm neb}$) derived from [Fe\two]\,\l7155 and
[Ni\two]\,\l7378 (see Table~\ref{tab:vnebfeni}), for \sneia\ in the
Normal and High-velocity subclasses of \cite{WangX/etal:2009b}.
Highly-reddened \sneia\ ($A_V>1$\,mag) are circled in red. Objects
part of the low-extinction sample are marked with an asterisk ({\it
  see text for details}).
We could not determine the spectroscopic subclass of SN~2006dd ({\it
  open star}) since no spectra are available around maximum light
\citep{Stritzinger/etal:2010}.
We have no intrinsic color estimates for the low-luminosity
SN~1986G, the peculiar SN~2000cx, or the possibly
super-Chandrasekhar-mass SN~2009dc, but we nonetheless indicate their
nebular line shifts ({\it crosses}). 
{\it Top:}
Distribution of $v_{\rm neb}$ for the Normal ({\it open}) and
High-velocity ({\it filled}) subsamples. SN~2006dd and the three
\sneia\ with no intrinsic color estimates mentioned above are not
included in either distribution but their names appear in the
corresponding velocity bin.
}
\end{figure}
\epsscale{1.0}

We investigate the relation between nebular line shifts and
intrinsic color in Fig.~\ref{fig:vnebfeni}. We use the
intrinsic $B-V$ color at $B$-band maximum light derived from BayeSN
fits as opposed to the intrinsic $B^{\rm max}-V^{\rm max}$
pseudo-color used by
\cite{Maeda/etal:2010a,Maeda/etal:2010c,Maeda/etal:2011} [see also
discussion in \S~\ref{sect:vsibmv}]. We were not able to determine
reliable intrinsic colors for SN~1986G, SN~2000cx, or SN~2009dc, since
BayeSN does not include such objects in its training sample. We also
highlight two objects with an inferred visual extinction $A_V>1$\,mag,
and note that objects at redshifts $<0.01$ are subject to a larger
uncertainty given the impact of peculiar velocities on the
redshift-based distance. We have checked, however, that using
external distance priors based on Cepheids or Surface Brightness
Fluctuations results in inferred intrinsic colors that are consistent
within $<1\sigma$ with those reported here.
We computed nebular line shifts based on a two-Gaussian fit to the
emission profiles of [Fe\two]\,\l7155 and [Ni\two]\,\l7378. The
nebular velocity, $v_{\rm neb}$, is simply computed as the mean
velocity of both lines, while the error is set to the standard
deviation of both measurements (see Table~\ref{tab:vnebfeni}). For
cases where only one of two lines has a well-defined emission profile
(usually [Fe\two]\,\l7155), we set the error in $v_{\rm neb}$ to
600\,\kms, as done by \cite{Maeda/etal:2011}. 
Our measurements are consistent within the errors with those reported by
\cite{Maeda/etal:2010c,Maeda/etal:2011}, with a scatter about zero
difference of $\sim150$\,\kms\ when the same spectrum was used
(marked with a ``$\dag$'' in Table~\ref{tab:vnebfeni}), and of
$\sim700$\,\kms\ when using a different spectrum of the same
object. One notable exception is SN~2006X for which
we infer a $\sim1500$\,\kms\ larger $v_{\rm neb}$ ($2832\pm343$\,\kms\
cf. $1331\pm164$\,\kms\ reported by \citealt{Maeda/etal:2011}). This
difference is suspiciously close to the host-galaxy redshift
($cz=1550$\,\kms\ from \citealt{UZC}) and results from
Maeda et al.'s accidental use of spectra of
SN~2006X available in the
SUSPECT\footnote{http://suspect.nhn.ou.edu/$\sim$suspect/} archive
that were already de-redshifted by 1587\,\kms\ (Maeda 2011,
  priv. comm.; this error does not 
affect their conclusions, however). We include four \sneia\
(SN~1994ae, SN~1995D, SN~2003kf, SN~2007af) that are not part of the
sample studied by
\cite{Maeda/etal:2010a,Maeda/etal:2010c,Maeda/etal:2011} and for which
we have spectra past +150\,d from $B$-band maximum light (see
Table~\ref{tab:nebspec}).

The data points in Fig.~\ref{fig:vnebfeni} are color-coded according
to their spectroscopic subclass (``Normal'' and ``High-velocity'' in
the classification scheme of \citealt{WangX/etal:2009b}). Our
spectroscopic coverage for SN~2007sr only starts $\sim2$ weeks past
$B$-band maximum (see Table~\ref{tab:obs}), so we infer its
spectroscopic subclass (HV) based on the Si\two\,\l6355 absorption
velocity of $-12300$\,\kms\ at +5\,d reported in CBET 1173 by
\cite{Naito_cbet1173}. We could not determine a spectroscopic subclass
for SN~2006dd, since there are no available spectra around maximum
light \citep{Stritzinger/etal:2010}. 

The distribution of nebular line shifts (Fig.~\ref{fig:vnebfeni}, {\it
 top}) shows a clear separation at $v_{\rm neb}\approx 1000$\,\kms\
between the N/HV spectroscopic subclasses. One notable exception is
the HV SN~2004dt, whose nebular spectrum was shown by
\cite{Maeda/etal:2010c} to resemble most that of the faint SN~1991bg
than other HV \sneia. We thus reproduce the main result of
\cite{Maeda/etal:2010c}, although they base their spectroscopic
classification on the velocity gradient of the Si\two\,\l6355 line
defined by \cite{Benetti/etal:2005} as opposed to its absorption
velocity at maximum (all the HV \sneia\ in Fig.~\ref{fig:vnebfeni} are
classified as HVG by \citealt{Maeda/etal:2010c}). In the context of
the simple kinematic model of \cite{Maeda/etal:2010a}, the HV subclass
consists of \sneia\ viewed from the direction opposite to the initial
off-center ignition (hence the pronounced redshift of nebular lines),
where intermediate-mass elements synthesized during the detonation
phase are ejected at higher velocities.

The overall correlation between intrinsic $B-V$ color at $B$-band
maximum and nebular line shift is rather modest ($r=0.36$; $r=0.58$
when excluding the peculiar HV SN~2004dt). Considering only the 10
\sneia\ part of our low-extinction sample (by which we mean \sneia\
with E/S0 hosts or in 
the outskirts of spiral galaxies, following \citealt{Maeda/etal:2011};
see \sneia\ marked with an asterisk in Fig.~\ref{fig:vnebfeni} and
Table~\ref{tab:vnebfeni}), the correlation is even lower than for the
entire sample ($r=0.30$). The bulk of \sneia\ in our sample with
$v_{\rm neb}>-2000$\,\kms\ (18 out of 21 with intrinsic color
estimates) do nonetheless appear to follow a clear trend of redder
intrinsic color for larger $v_{\rm neb}$ ($r=0.82$). It is unclear
whether the two Normal SN~1994D and SN~2003hv at $v_{\rm
  neb}<-2000$\,\kms\ simply reflect a larger scatter at large nebular
line blueshifts, or reveal a non-monotonic relation between
nebular line shifts and intrinsic color (which would then
also accommodate SN~2004dt).

Our results appear to differ slightly from those of
\cite{Maeda/etal:2011}, who find a
strong correlation and an obvious linear relation between intrinsic
color and nebular line shift for their low-extinction sample (which
includes SN~1994D and SN~2003hv, both modest outliers in the relation
defined by the bulk of our sample), and which also applies to all
other \sneia\ in their sample (including SN~2004dt) after excluding
three highly-reddened objects (SN~1998bu, SN~2002bo, and SN~2006X;
their Fig.~4). In fact, \cite{Maeda/etal:2011} use the relation
between ``intrinsic'' $B^{\rm max}-V^{\rm max}$
pseudo-color\footnote{really the {\it apparent} $B^{\rm max}-V^{\rm
 max}$ pseudo-color for a low-extinction sample.} and \dmft\
found by \cite{Folatelli/etal:2010} to correct their own pseudo-color
measurement. Using this same approach we too find a strong correlation
between this color residual and $v_{\rm neb}$ for our low-extinction
sample ($r=0.72$). We note, however, that the \dmft-color relation of
\cite{Folatelli/etal:2010} has a low statistical significance
($\sim2\sigma$), and that there is no correlation between the $B^{\rm
  max}-V^{\rm max}$ pseudo-color and \dmft\ for the low-extinction
sample of \cite{Maeda/etal:2011} ($r=0.05$; their Fig.~3). Whether or
not we apply this correction makes little difference to the strength of
the correlation between $B^{\rm max}-V^{\rm max}$ pseudo-color and
$v_{\rm neb}$ for our low-extinction sample ($r=0.68$ with no
correction). 

In any case these nebular line shifts do seem to provide
information on the intrinsic properties of \sneia\ independently from
\dmft. We confirm the lack of correlation between $v_{\rm neb}$
and \dmft\ ($|r|<0.1$ whether or not we exclude peculiar \sneia) noted
by \cite{Maeda/etal:2010c}, although \cite{Maeda/etal:2011} argue that
such a correlation is expected theoretically given the correlation
they find between \dmft\ and viewing angle (for their model with
0.3\,\msun\ of \nifs; their Fig.~12). The recent study of 2D
delayed-detonation models of \cite{KRW09} by \cite{Blondin/etal:2011}
shows, however, that the relation between \dmft\ and viewing angle is a
complex one, both non-monotonic and non-linear in character regardless
of the \nifs\ mass. While we agree with \cite{Maeda/etal:2011} that
more data are needed to study the relation between nebular line shifts
and \dmft, we also note the limiting predictive power of simple
models crafted to reproduce the observed trends.

\begin{deluxetable*}{@{\extracolsep{-.1cm}}lcccrlrrr@{ $\pm$ }lcc}
\tablewidth{0pt}
\tablecaption{\label{tab:vnebfeni}Intrinsic color {\it vs.} nebular line shift.}
\tablehead{\colhead{SN\tablenotemark{a}} & \colhead{$z_{\rm CMB}$} & \colhead{Wang\tablenotemark{b}} & \colhead{\dmft} & \colhead{$(B-V)_0$\tablenotemark{c}} & \colhead{Phase\tablenotemark{d}} & \colhead{$v_{\rm Fe}$} & \colhead{$v_{\rm Ni}$} & \multicolumn{2}{c}{$v_{\rm neb}$\tablenotemark{e}} & \colhead{Ref.} & \colhead{Notes} \\
 \colhead{} & \colhead{} & \colhead{class} & \colhead{(mag)} & \colhead{(mag)} & \colhead{(d)} & \colhead{(\kms)} & \colhead{(\kms)} & \multicolumn{2}{c}{(\kms)} & \colhead{} & \colhead{} }
\startdata
1986G    &  0.003 & 91bg                           & 1.65 (0.04)                    & \multicolumn{1}{c}{$\cdots$}   & +256$^\dag$                    & \multicolumn{1}{c}{$\cdots$}   & $-$639                           &  $-$639 &  600 & This paper      &                               \\
1990N*   &  0.004 & N                              & 0.95 (0.05)                    & $-$0.16 (0.06)                   & +309                           & $-$478                           & \multicolumn{1}{c}{$\cdots$}   &  $-$478 &  600 & This paper      &                               \\
1994D*   &  0.004 & N                              & 1.37 (0.03)                    & $-$0.09 (0.04)                   & +306$^\dag$                    & \multicolumn{1}{c}{$\cdots$}   & \multicolumn{1}{c}{$\cdots$}   & $-$2220 &  220 & Maeda11         &                               \\
1994ae   &  0.005 & N                              & 0.96 (0.04)                    & $-$0.18 (0.05)                   & +152                           & $-$1204                          & \multicolumn{1}{c}{$\cdots$}   & $-$1204 &  600 & This paper      &                               \\
1995D*   &  0.008 & N                              & 1.05 (0.03)                    & $-$0.16 (0.04)                   & +283                           & $-$1193                          & \multicolumn{1}{c}{$\cdots$}   & $-$1193 &  600 & This paper      &                               \\
1997bp*  &  0.009 & HV                             & 1.08 (0.06)                    & $-$0.13 (0.06)                   & +300                           & \multicolumn{1}{c}{$\cdots$}   & \multicolumn{1}{c}{$\cdots$}   &  2539 &  410 & Maeda11         &                               \\
1998aq   &  0.005 & N                              & 1.11 (0.04)                    & $-$0.18 (0.04)                   & +239$^\dag$                    & $-$1249                          & $-$1065                          & $-$1157 &  129 & This paper      &                               \\
1998bu   &  0.004 & N                              & 1.03 (0.02)                    & $-$0.17 (0.04)                   & +328$^\dag$                    & $-$1567                          & $-$1135                          & $-$1351 &  305 & This paper      & $A_V>1$\,mag                  \\
2000cx   &  0.007 & \multicolumn{1}{c}{$\cdots$}   & 0.99 (0.02)                    & 0.06 (0.05)                    & +146$^\dag$                    & \multicolumn{1}{c}{$\cdots$}   & $-$172                           &  $-$172 &  600 & This paper      & Peculiar \snia                \\
2001el   &  0.004 & N                              & 1.16 (0.03)                    & $-$0.15 (0.05)                   & +397$^\dag$                    & 834                            & 1126                           &   980 &  206 & This paper      &                               \\
2002bo   &  0.005 & HV                             & 1.08 (0.04)                    & $-$0.05 (0.06)                   & +310                           & 2002                           & 2513                           &  2258 &  361 & This paper      &                               \\
2002dj*  &  0.010 & HV                             & 1.19 (0.05)                    & $-$0.07 (0.05)                   & +271$^\dag$                    & 1728                           & 2313                           &  2021 &  414 & This paper      &                               \\
2002er   &  0.009 & N                              & 1.23 (0.03)                    & $-$0.13 (0.05)                   & +212$^\dag$                    & \multicolumn{1}{c}{$\cdots$}   & 582                            &   582 &  600 & This paper      &                               \\
2003du*  &  0.007 & N                              & 1.07 (0.06)                    & $-$0.21 (0.04)                   & +374$^\dag$                    & $-$644                           & $-$243                           &  $-$443 &  283 & This paper      &                               \\
2003hv*  &  0.005 & N                              & 1.31 (0.05)                    & $-$0.14 (0.06)                   & +321$^\dag$                    & $-$2553                          & $-$3097                          & $-$2825 &  384 & This paper      &                               \\
2003kf   &  0.008 & N                              & 1.10 (0.05)                    & $-$0.12 (0.05)                   & +398                           & 928                            & 1213                           &  1070 &  201 & This paper      &                               \\
2004dt   &  0.019 & HV                             & 1.14 (0.04)                    & $-$0.07 (0.05)                   & +347                           & $-$2623                          & \multicolumn{1}{c}{$\cdots$}   & $-$2623 &  600 & This paper      &                               \\
2004eo*  &  0.015 & N                              & 1.38 (0.03)                    & $-$0.12 (0.04)                   & +225$^\dag$                    & $-$188                           & $-$152                           &  $-$170 &   25 & This paper      &                               \\
2005cf*  &  0.007 & N                              & 1.10 (0.03)                    & $-$0.15 (0.04)                   & +317                           & $-$123                           & \multicolumn{1}{c}{$\cdots$}   &  $-$123 &  600 & This paper      &                               \\
2006X    &  0.006 & HV                             & 1.18 (0.03)                    & $-$0.09 (0.05)                   & +275$^\dag$                    & 2589                           & 3075                           &  2832 &  343 & This paper      & $A_V>1$\,mag                  \\
2006dd*  &  0.005 & \multicolumn{1}{c}{$\cdots$}   & 1.07 (0.03)                    & $-$0.21 (0.04)                   & +193$^\dag$                    & $-$1696                          & $-$1410                          & $-$1553 &  201 & This paper      & No spectra around max         \\
2007af   &  0.006 & N                              & 1.20 (0.02)                    & $-$0.13 (0.04)                   & +301                           & 583                            & 1079                           &   831 &  350 & This paper      &                               \\
2007sr   &  0.007 & HV                             & 1.13 (0.06)                    & $-$0.09 (0.05)                   & +190                           & 1185                           & \multicolumn{1}{c}{$\cdots$}   &  1185 &  600 & This paper      &                               \\
2009dc   &  0.022 & \multicolumn{1}{c}{$\cdots$}   & 0.76 (0.01)                    & \multicolumn{1}{c}{$\cdots$}   & +281                           & $-$1190                          & \multicolumn{1}{c}{$\cdots$}   & $-$1190 &  600 & This paper      & Peculiar \snia                
\enddata
\tablenotetext{a}{SN names marked with an asterisk are part of the low-extinction sample ({\it see text for details}).}
\tablenotetext{b}{Spectroscopic class of \cite{WangX/etal:2009b}. SN~2000cx and SN~2009dc are both peculiar \sneia, while SN~2006dd lacks spectra around maximum light.}
\tablenotetext{c}{Intrinsic $B-V$ color at $B$-band maximum light from BayeSN.}
\tablenotetext{d}{Rest-frame phase of the spectrum used to determine the nebular line shift, in days relative to $B$-band maximum. For SN~1994D and SN~1997bp we give the phase as reported by \cite{Maeda/etal:2010c}. A ``$\dag$'' indicates the same spectrum was used as \cite{Maeda/etal:2010c,Maeda/etal:2011}.}
\tablenotetext{e}{Mean velocity derived from [Fe\two]\,\l7155 ($v_{\rm Fe}$) and [Ni\two]\,\l7378 ($v_{\rm Ni}$).
The associated error is simply the standard deviation of both measurements. When only a single measurement is available, we set the associated error to 600\,\kms, as done by \cite{Maeda/etal:2011}.
We use our own measurements when available, and those reported by \cite{Maeda/etal:2011} [noted Maeda11] otherwise.}
\end{deluxetable*}


\section{Conclusions}\label{sect:ccl}

We have presented \nspectot\ spectra of \nsntot\ nearby \sneia\ out to
$z\approx0.12$ obtained through the CfA Supernova Program between
October 1993 and August 2008 (\nspecunpub\ spectra of \nsnunpub\
\sneia\ are published for the first time in this paper). 
For comparison, the entire SUSPECT archive contains 1741
spectra of 185 supernovae of all types, with 867 spectra of 81
\sneia\ (including the 432 spectra of 32 \sneia\ published by
\citealt{Matheson/etal:2008} and other spectra from the CfA SN
Archive).
The SN group at Berkeley is also preparing a public release of
their sample of 1298 spectra of 582 \sneia\
\citep{Silverman/etal:2012a}, complementary in many ways to the data
set presented here.
The sample of \nspecfast\
\snia\ spectra taken with the FAST spectrograph at the FLWO 1.5\,m
telescope constitutes the largest homogeneous supernova data set to
date. All spectra were reduced in a consistent manner (except for
\nspecmmt9397\ MMT spectra from 1993-1997), making this a unique
sample for detailed spectroscopic studies of \sneia. The sample spans
a large range in \dmft\ (between $\sim0.7$\,mag and $\sim2$\,mag)
and includes spectra as early as $-15$\,d and as late as +611\,d from
$B$-band maximum. There are \nsnwspecltmax\ \sneia\ with spectra
before maximum light, and \nsnwtenspec\ \sneia\ with at least 10
epochs of spectroscopy.

Using this unique data set, complemented with additional data from the
literature, we have studied the diversity of \snia\ properties as a
function of spectroscopic class, using the classification schemes of
\cite{Branch/etal:2006} and \cite{WangX/etal:2009b}. We find that
there is a continuum of properties (i.e., no strict boundaries) between
the different subclasses, but that a single parameter is unlikely to
explain the spectroscopic variation between extreme members of each
subclass. We have studied two hypotheses of \cite{Branch/etal:2009},
namely that (1) \sneia\ from the Core Normal (CN) subclass have a
significantly smaller scatter in intrinsic peak $M_B$, and (2) \sneia\
from the Core Normal and Shallow Silicon (SS) subclasses age at the same
rate that they decline photometrically. We reject the first and
show that the second is not universally true. Last, we examine the
width-luminosity relation [intrinsic peak $M_B$ {\it vs.} \dmft] for
individual spectroscopic subclasses, and find that
SS$\rightarrow$CN$\rightarrow$Broad Line (BL) form a sequence 
of increasingly steeper relations, although a single relation is
found to adequately describe all subclasses (other than the faint CL
subclass). The same is true for the
91T$\rightarrow$Normal$\rightarrow$High-velocity sequence using the
\cite{WangX/etal:2009b} classification scheme.

We provide updated versions of the relations between several
spectroscopic indicators (the \rsi\ ratio of
\citealt{Nugent/etal:1995} and its pseudo-EW equivalent, the
pseudo-EW of the Si\two\,\l4130 line, 
and the \rsife\ ratio of \citealt{Hachinger/Mazzali/Benetti:2006}) and
\dmft. We confirm the strong correlation reported in previous
publications with a significantly larger sample, but note a
significant scatter and possible non-linearities, as well as
notable outliers, e.g., SN~2006ot in the pEW(Si\two\,\l4130) {\it vs.}
\dmft\ relation (which we argue is spectroscopically distinct from
SN~2006bt, contrary to claims by \citealt{Stritzinger/etal:2011}).

We then focus on the evolution of the velocity at maximum absorption
($v_{\rm abs}$) of the defining Si\two\,\l6355 line with spectral phase,
for the different spectroscopic subclasses of
\cite{Branch/etal:2006}. The CN and BL subclasses display a large
variation in $v_{\rm abs}$ at early times (prior to $-10$\,d from
$B$-band maximum), which is largely suppressed at maximum light in the
former but persists in the latter. The BL subclass has the largest
variation in the post-maximum evolution of $v_{\rm abs}$ (the velocity
gradient). The SS subclass has similar $v_{\rm
  abs}$ as the CN subclass around maximum light, except for
2002cx-like \sneia\ and possibly super-Chandrasekhar events. The same
applies to the Cool (CL) subclass, with the exception of the
1991bg-like SN~2002es which displays a factor of $\sim2$ lower
Si\two\ blueshifts \citep{Ganeshalingam/etal:2012}.

We propose alternative measurements of velocity gradients that do not
depend on the phase of the last available measurement and can
accommodate a non-linear evolution of the absorption velocity with
time. These are the average decline rate over a fixed phase interval
$[t_0,t_1]$ and the instantaneous decline rate at a given phase (i.e., in
the limit $t_0\rightarrow t_1$). The clustering into the three groups
defined by \cite{Benetti/etal:2005} [FAINT, LVG, HVG] is less obvious
than in their Fig.~3, and the use of the instantaneous velocity
decline rate at maximum reveals a significantly larger range in
velocity gradients (from $\sim0$\,\kmsd\ for SN~2005hj to
$\sim400$\,\kmsd\ for SN~2003W).

We revisit the relation of \cite{Foley/Kasen:2011} between Si\two\
velocity and intrinsic color, re-analyzed by
\cite{Foley/Sanders/Kirshner:2011} using our velocity measurements. A
key difference in our approach is the use of intrinsic color
estimates inferred from the BayeSN statistical model for
\snia\ light curves \citep{Mandel/etal:2011}, which allows for
variations in the host-galaxy dust properties on a SN-by-SN basis, as
opposed to assuming a single reddening law for all \sneia. The
intrinsic colors estimated using BayeSN have significantly larger
(and probably more realistic) uncertainties, resulting in a very weak
correlation between Si\two\ velocity and intrinsic color
($r\approx-0.20$) when these are taken into account. 
\sneia\ from the High-velocity subclass have a redder mean
intrinsic color that those from the Normal subclass, but the
difference is only at the $\sim2\sigma$ level of statistical
significance. 
Moreover, we find that the scatter in intrinsic color is
comparable for both N and HV samples, contrary to claims made by
\cite{Foley/Sanders/Kirshner:2011}.

Through a study of early-time spectra, which probe the outermost
ejecta, we attempt to place constraints on the extent and completeness
of nuclear burning in \sneia. Silicon is found to absorb at
significantly larger velocities ($-28000\lesssim v_{\rm max}\lesssim
-18000$\,\kms) than the lower limit of $\sim12000$\,\kms\ found by
\cite{Mazzali/etal:2007} for the outer extent of the Si-rich layer in
\snia\ ejecta, in conflict with pure deflagration models. We study the
impact of high-velocity features (HVF) on the Ca\two\ near-infrared
triplet (IRT), and find that \sneia\ with the largest extent of Ca
absorption correspond to those with the largest Si\two\ absorption
velocity. By comparing the line-profile morphology of the Ca\two\ IRT
and Ca\two\,H\&K absorptions, we illustrate the contamination of the
latter profile by Si\two\,\l3858, which can be confused with an HVF
in the absence of spectra covering the IRT absorption feature.

We searched for signs of unburnt carbon in pre-maximum spectra,
a subject that has received much attention lately with the studies by 
\cite{Parrent/etal:2011} and \cite{Thomas/etal:2011}, and find
evidence for a C\two\,\l6580 absorption feature in 23 new \sneia\ 
(4 of which have confirmed detections by \citealt{Folatelli/etal:2012}), the
majority of which are from the CN subclass. Moreover, 7/9 CN \sneia\
with spectra prior to $-10$\,d show C\two\ absorptions, suggesting
that this feature is a generic property of this subclass, and
highlighting the need for very early-time spectroscopy. The absorption
velocity of C\two\,\l6580 is in general within $\sim10$\% of that of
Si\two\,\l6355, but silicon is present at significantly larger
velocities than carbon, in conflict with current delayed-detonation
models. Contrary to \cite{Thomas/etal:2011}, we find no preference for
carbon detections in \sneia\ with blue intrinsic colors and narrow
light curves, although we are currently lacking data to study this
association for \sneia\ with persistent C\two\ lines.

Using nebular spectra of 27 \sneia, we investigate the relation
between \dmft\ and the FWHM of the iron emission feature at
$\sim4700$\,\AA\ found by \cite{Mazzali/etal:1998}, interpreted as a
correlation between luminosity and explosion kinetic energy, more
luminous \sneia\ corresponding to more energetic explosions. We find
the correlation is largely driven by the low-luminosity SN~1991bg,
which is clearly separated from the rest of the sample in the (\dmft,
FWHM) plane. Further excluding SN~1986G from the sample (i.e., only
considering \sneia\ with $\dmft<1.6$\,mag), we find no correlation
between both quantities ($r=-0.17$), suggesting that the peak
luminosity is only weakly dependent on explosion kinetic energy for
most \sneia, as in delayed-detonation models in which the WD star is
almost completely incinerated.

Last, we studied the relation between velocity shifts of certain
nebular emission lines and intrinsic \snia\ properties at maximum
light, following the recent work of
\cite{Maeda/etal:2010a,Maeda/etal:2010c,Maeda/etal:2011}, and
interpreted in the context of an off-center delayed-detonation
model. Excluding the spectroscopically peculiar SN~2004dt, the five
\sneia\ in our sample that are part of the High-velocity subclass of 
\cite{WangX/etal:2009b} display pronounced redshifts in the nebular
lines of [Fe\two]\,\l7155 and [Ni\two]\,\l7378 ($v_{\rm
  neb}>+1000$\,\kms) and are thus consistent with \sneia\ viewed from
the direction opposite to an initial off-center ignition. We find a
strong correlation ($r=0.82$) of nebular line shifts with intrinsic
$B-V$ color at $B$-band maximum light for \sneia\ with $v_{\rm
  neb}>-2000$\,\kms\ (18 out of 21 in our sample), in line with the
results presented by \cite{Maeda/etal:2011}, but note the
possibility of a non-monotonic behavior for the largest
blueshifts. We caution that
their toy models do not capture the complexity of the variation of
\dmft\ with viewing angle in off-center delayed detonation models, as
inferred by \cite{Blondin/etal:2011} based on the 2D models of
\cite{KRW09}.

The study presented here is far from exhaustive, and we have still a
lot to learn from the study of \snia\ spectra, from constraining the
physics of the explosion and the nature of their progenitors to
determining more precise distances on cosmological scales. All
\nspectot\ spectra will be made publicly available through the CfA
Supernova Archive to enable further studies by the rest of the SN
community.


\begin{acknowledgments}
We would like to thank the staffs of the F.~L.~Whipple, MMT, and Las
Campanas observatories for their extensive assistance and support
during this project, as well as everyone who assisted with
observations (see Table~\ref{tab:obs}).
We acknowledge useful discussions with David Branch, Luc Dessart,
Gast\'on Folatelli, Ryan Foley, Keiichi Maeda, and Xiaofeng
Wang. Additional thanks to  
Mohan Ganeshalingam, Weidong Li, \& Jeff Silverman from
the Berkeley SN Group for sending us $t_{\rm max}(B)$
values for 6 \sneia, and to Stephen Bailey from the SNfactory
collaboration for sending us the results of SALT2 fits to 5 \sneia;
to Giuseppe Altavilla, Stefano Benetti, Seppo Mattila, Jesper
Sollerman, and Max Stritzinger, for sending us spectroscopic data.
Some of the non-CfA spectra used in this
paper were downloaded from the SUSPECT archive.
Support for supernova research at Harvard University, including the
CfA Supernova Archive, is provided in part by NSF grant AST
09-07903.
SWJ is supported at Rutgers University in part by NSF CAREER
award AST-0847157.
This research has made use of the NASA/IPAC Extragalactic Database
(NED) which is operated by the Jet Propulsion Laboratory, California
Institute of Technology, under contract with the National Aeronautics
and Space Administration.
\end{acknowledgments}

{\it Facilities:} FLWO:1.5m, MMT, Magellan:Baade, Magellan:Clay


\bibliographystyle{apj}
\bibliography{ms}

\begin{thebibliography}{179}
\expandafter\ifx\csname natexlab\endcsname\relax\def\natexlab#1{#1}\fi

\bibitem[{{Aldering} {et~al.}(2002){Aldering}, {Adam}, {Antilogus}, {Astier},
  {Bacon}, {Bongard}, {Bonnaud}, {Copin}, {Hardin}, {Henault}, {Howell},
  {Lemonnier}, {Levy}, {Loken}, {Nugent}, {Pain}, {Pecontal}, {Pecontal},
  {Perlmutter}, {Quimby}, {Schahmaneche}, {Smadja}, \&
  {Wood-Vasey}}]{Aldering/etal:2002}
{Aldering}, G., et~al.\ 2002, in Society of Photo-Optical Instrumentation Engineers (SPIE) Conference  Series, Vol. 4836, Society of Photo-Optical Instrumentation Engineers (SPIE)  Conference Series, ed. {J.~A.~Tyson \& S.~Wolff}, 61--72

\bibitem[{{Altavilla} {et~al.}(2007){Altavilla}, {Stehle}, {Ruiz-Lapuente},
  {Mazzali}, {Pignata}, {Balastegui}, {Benetti}, {Blanc}, {Canal},
  {Elias-Rosa}, {Goobar}, {Harutyunyan}, {Pastorello}, {Patat}, {Rich},
  {Salvo}, {Schmidt}, {Stanishev}, {Taubenberger}, {Turatto}, \&
  {Hillebrandt}}]{Altavilla/etal:2007}
{Altavilla}, G., et~al.\ 2007, \aap, 475, 585

\bibitem[{{Anupama} {et~al.}(2005){Anupama}, {Sahu}, \&
  {Jose}}]{Anupama/Sahu/Jose:2005}
{Anupama}, G.~C., {Sahu}, D.~K., \& {Jose}, J. 2005, \aap, 429, 667

\bibitem[{{Arsenijevic} {et~al.}(2008){Arsenijevic}, {Fabbro}, {Mour{\~a}o}, \&
  {Rica da Silva}}]{Arsenijevic/etal:2008}
{Arsenijevic}, V., {Fabbro}, S., {Mour{\~a}o}, A.~M., \& {Rica da Silva}, A.~J.
  2008, \aap, 492, 535

\bibitem[{{Bailey} {et~al.}(2009){Bailey}, {Aldering}, {Antilogus}, {Aragon},
  {Baltay}, {Bongard}, {Buton}, {Childress}, {Chotard}, {Copin}, {Gangler},
  {Loken}, {Nugent}, {Pain}, {Pecontal}, {Pereira}, {Perlmutter}, {Rabinowitz},
  {Rigaudier}, {Runge}, {Scalzo}, {Smadja}, {Swift}, {Tao}, {Thomas}, {Wu}, \&
  {The Nearby Supernova Factory}}]{Bailey/etal:2009}
{Bailey}, S., et~al.\ 2009, \aap, 500, L17

\bibitem[{{Barbon} {et~al.}(1989){Barbon}, {Rosino}, \&
  {Iijima}}]{Barbon/etal:1989}
{Barbon}, R., {Rosino}, L., \& {Iijima}, T. 1989, \aap, 220, 83

\bibitem[{{Benetti} {et~al.}(2005){Benetti}, {Cappellaro}, {Mazzali},
  {Turatto}, {Altavilla}, {Bufano}, {Elias-Rosa}, {Kotak}, {Pignata}, {Salvo},
  \& {Stanishev}}]{Benetti/etal:2005}
{Benetti}, S., et~al.\ 2005, \apj, 623, 1011

\bibitem[{{Benetti} {et~al.}(2004){Benetti}, {Meikle}, {Stehle}, {Altavilla},
  {Desidera}, {Folatelli}, {Goobar}, {Mattila}, {Mendez}, {Navasardyan},
  {Pastorello}, {Patat}, {Riello}, {Ruiz-Lapuente}, {Tsvetkov}, {Turatto},
  {Mazzali}, \& {Hillebrandt}}]{Benetti/etal:2004}
{Benetti}, S., et~al.\ 2004,  \mnras, 348, 261

\bibitem[{{Blondin} {et~al.}(2006){Blondin}, {Dessart}, {Leibundgut}, {Branch},
  {H{\"o}flich}, {Tonry}, {Matheson}, {Foley}, {Chornock}, {Filippenko},
  {Sollerman}, {Spyromilio}, {Kirshner}, {Wood-Vasey}, {Clocchiatti},
  {Aguilera}, {Barris}, {Becker}, {Challis}, {Covarrubias}, {Davis},
  {Garnavich}, {Hicken}, {Jha}, {Krisciunas}, {Li}, {Miceli}, {Miknaitis},
  {Pignata}, {Prieto}, {Rest}, {Riess}, {Salvo}, {Schmidt}, {Smith}, {Stubbs},
  \& {Suntzeff}}]{Blondin/etal:2006}
{Blondin}, S., et~al.\ 2006, \aj, 131, 1648

\bibitem[{{Blondin} {et~al.}(2011{\natexlab{a}}){Blondin}, {Kasen},
  {R{\"o}pke}, {Kirshner}, \& {Mandel}}]{Blondin/etal:2011}
{Blondin}, S., {Kasen}, D., {R{\"o}pke}, F.~K., {Kirshner}, R.~P., \& {Mandel},
  K.~S. 2011{\natexlab{a}}, \mnras, 417, 1280

\bibitem[{{Blondin} {et~al.}(2011{\natexlab{b}}){Blondin}, {Mandel}, \&
  {Kirshner}}]{Blondin/Mandel/Kirshner:2011}
{Blondin}, S., {Mandel}, K.~S., \& {Kirshner}, R.~P. 2011{\natexlab{b}}, \aap,
  526, A81+

\bibitem[{{Blondin} \& {Tonry}(2007)}]{SNID}
{Blondin}, S. \& {Tonry}, J.~L. 2007, \apj, 666, 1024

\bibitem[{{Bloom} {et~al.}(2012){Bloom}, {Kasen}, {Shen}, {Nugent}, {Butler},
  {Graham}, {Howell}, {Kolb}, {Holmes}, {Haswell}, {Burwitz}, {Rodriguez}, \&
  {Sullivan}}]{Bloom/etal:2012}
{Bloom}, J.~S., et~al.\ 2012, \apjl, 744, L17

\bibitem[{{Branch} {et~al.}(2005){Branch}, {Baron}, {Hall}, {Melakayil}, \&
  {Parrent}}]{Branch/etal:2005}
{Branch}, D., {Baron}, E., {Hall}, N., {Melakayil}, M., \& {Parrent}, J. 2005,
  \pasp, 117, 545

\bibitem[{{Branch} {et~al.}(1982){Branch}, {Buta}, {Falk}, {McCall}, {Uomoto},
  {Wheeler}, {Wills}, \& {Sutherland}}]{Branch/etal:1982}
{Branch}, D., {Buta}, R., {Falk}, S.~W., {McCall}, M.~L., {Uomoto}, A.,
  {Wheeler}, J.~C., {Wills}, B.~J., \& {Sutherland}, P.~G. 1982, \apjl, 252,
  L61

\bibitem[{{Branch} {et~al.}(2009){Branch}, {Dang}, \&
  {Baron}}]{Branch/etal:2009}
{Branch}, D., {Dang}, L.~C., \& {Baron}, E. 2009, \pasp, 121, 238

\bibitem[{{Branch} {et~al.}(2006){Branch}, {Dang}, {Hall}, {Ketchum},
  {Melakayil}, {Parrent}, {Troxel}, {Casebeer}, {Jeffery}, \&
  {Baron}}]{Branch/etal:2006}
{Branch}, D., et~al.\ 2006, \pasp, 118, 560

\bibitem[{{Branch} {et~al.}(2003){Branch}, {Garnavich}, {Matheson}, {Baron},
  {Thomas}, {Hatano}, {Challis}, {Jha}, \& {Kirshner}}]{Branch/etal:2003}
{Branch}, D., et~al.\ 2003, \aj, 126,  1489

\bibitem[{{Branch} {et~al.}(2008){Branch}, {Jeffery}, {Parrent}, {Baron},
  {Troxel}, {Stanishev}, {Keithley}, {Harrison}, \&
  {Bruner}}]{Branch/etal:2008}
{Branch}, D., et~al.\ 2008, \pasp,  120, 135

\bibitem[{{Branch} {et~al.}(1983){Branch}, {Lacy}, {McCall}, {Sutherland},
  {Uomoto}, {Wheeler}, \& {Wills}}]{Branch/etal:1983}
{Branch}, D., {Lacy}, C.~H., {McCall}, M.~L., {Sutherland}, P.~G., {Uomoto},
  A., {Wheeler}, J.~C., \& {Wills}, B.~J. 1983, \apj, 270, 123

\bibitem[{{Branch} {et~al.}(2007){Branch}, {Troxel}, {Jeffery}, {Hatano},
  {Musco}, {Parrent}, {Baron}, {Dang}, {Casebeer}, {Hall}, \&
  {Ketchum}}]{Branch/etal:2007}
{Branch}, D., et~al.\ 2007, \pasp, 119, 709

\bibitem[{{Bravo} \& {Garc{\'{\i}}a-Senz}(2006)}]{Bravo/Garcia-Senz:2006}
{Bravo}, E. \& {Garc{\'{\i}}a-Senz}, D. 2006, \apjl, 642, L157

\bibitem[{{Bronder} {et~al.}(2008){Bronder}, {Hook}, {Astier}, {Balam},
  {Balland}, {Basa}, {Carlberg}, {Conley}, {Fouchez}, {Guy}, {Howell}, {Neill},
  {Pain}, {Perrett}, {Pritchet}, {Regnault}, {Sullivan}, {Baumont}, {Fabbro},
  {Filliol}, {Perlmutter}, \& {Ripoche}}]{Bronder/etal:2008}
{Bronder}, T.~J., et~al.\ 2008, \aap, 477, 717

\bibitem[{{Bufano} {et~al.}(2005){Bufano}, {Turatto}, {Benetti}, {Harutyunyan},
  {Elias de La Rosa}, \& {Cappellaro}}]{Bufano/etal:2005}
{Bufano}, F., {Turatto}, M., {Benetti}, S., {Harutyunyan}, A., {Elias de La
  Rosa}, N., \& {Cappellaro}, E. 2005, in Astronomical Society of the Pacific
  Conference Series, Vol. 342, 1604-2004: Supernovae as Cosmological
  Lighthouses, ed. {M.~Turatto, S.~Benetti, L.~Zampieri, \& W.~Shea}, 255--+

\bibitem[{{Burns} {et~al.}(2011{\natexlab{a}}){Burns}, {Stritzinger},
  {Phillips}, {Kattner}, {Persson}, {Madore}, {Freedman}, {Boldt}, {Campillay},
  {Contreras}, {Folatelli}, {Gonzalez}, {Krzeminski}, {Morrell}, {Salgado}, \&
  {Suntzeff}}]{SNooPy}
{Burns}, C.~R., et~al.\ 2011{\natexlab{a}}, \aj,  141, 19

\bibitem[{{Burns} {et~al.}(2011{\natexlab{b}}){Burns}, {Stritzinger},
  {Phillips}, {Kattner}, {Persson}, {Madore}, {Freedman}, {Boldt}, {Campillay},
  {Contreras}, {Folatelli}, {Gonzalez}, {Krzeminski}, {Morrell}, {Salgado}, \&
  {Suntzeff}}]{Burns/etal:2011}
---. 2011{\natexlab{b}}, \aj, 141, 19

\bibitem[{{Cappellaro} {et~al.}(2001){Cappellaro}, {Patat}, {Mazzali},
  {Benetti}, {Danziger}, {Pastorello}, {Rizzi}, {Salvo}, \&
  {Turatto}}]{Cappellaro/etal:2001}
{Cappellaro}, E., et~al.\ 2001,  \apjl, 549, L215

\bibitem[{{Chotard} {et~al.}(2011){Chotard}, {Gangler}, {Aldering},
  {Antilogus}, {Aragon}, {Bailey}, {Baltay}, {Bongard}, \&
  {Buton}}]{Chotard/etal:2011}
{Chotard}, N., et~al.\ 2011, \aap, 529,  L4+

\bibitem[{{Contreras} {et~al.}(2010){Contreras}, {Hamuy}, {Phillips},
  {Folatelli}, {Suntzeff}, {Persson}, {Stritzinger}, {Boldt}, {Gonz{\'a}lez},
  {Krzeminski}, {Morrell}, {Roth}, {Salgado}, {Jos{\'e} Maureira}, {Burns},
  {Freedman}, {Madore}, {Murphy}, {Wyatt}, {Li}, \&
  {Filippenko}}]{Contreras/etal:2010}
{Contreras}, C., {Hamuy}, M., {Phillips}, M.~M., {Folatelli}, G., {Suntzeff},
  N.~B., {Persson}, S.~E., {Stritzinger}, M., {Boldt}, L., {Gonz{\'a}lez}, S.,
  {Krzeminski}, W., {Morrell}, N., {Roth}, M., {Salgado}, F., {Jos{\'e}
  Maureira}, M., {Burns}, C.~R., {Freedman}, W.~L., {Madore}, B.~F., {Murphy},
  D., {Wyatt}, P., {Li}, W., \& {Filippenko}, A.~V. 2010, \aj, 139, 519

\bibitem[{{Cristiani} {et~al.}(1992){Cristiani}, {Cappellaro}, {Turatto},
  {Bergeron}, {Bues}, {Buson}, {Danziger}, {di Serego-Alighieri}, {Duerbeck},
  {Heydari-Malayeri}, {Krautter}, {Schmutz}, \&
  {Schulte-Ladbeck}}]{Cristiani/etal:1992}
{Cristiani}, S., et~al.\ 1992, \aap, 259, 63

\bibitem[{{de Vaucouleurs} {et~al.}(1991){de Vaucouleurs}, {de Vaucouleurs},
  {Corwin}, {Buta}, {Paturel}, \& {Fouqu{\'e}}}]{RC3}
{de Vaucouleurs}, G., {de Vaucouleurs}, A., {Corwin}, Jr., H.~G., {Buta},
  R.~J., {Paturel}, G., \& {Fouqu{\'e}}, P. 1991, {Third Reference Catalogue of
  Bright Galaxies. Volume I: Explanations and references. Volume II: Data for
  galaxies between $0^{h}$ and $12^{h}$. Volume III: Data for galaxies between
  $12^{h}$ and $24^{h}$.}, ed. {de Vaucouleurs, G., de Vaucouleurs, A., Corwin,
  H.~G., Jr., Buta, R.~J., Paturel, G., \& Fouqu{\'e}, P.}

\bibitem[{{Dessart} {et~al.}(2008){Dessart}, {Blondin}, {Brown}, {Hicken},
  {Hillier}, {Holland}, {Immler}, {Kirshner}, {Milne}, {Modjaz}, \&
  {Roming}}]{Dessart/etal:2008}
{Dessart}, L., et~al.\ 2008, \apj, 675, 644

\bibitem[{{Dessart} \& {Hillier}(2005)}]{DH05a}
{Dessart}, L. \& {Hillier}, D.~J. 2005, \aap, 437, 667

\bibitem[{{Elias-Rosa} {et~al.}(2006){Elias-Rosa}, {Benetti}, {Cappellaro},
  {Turatto}, {Mazzali}, {Patat}, {Meikle}, {Stehle}, {Pastorello}, {Pignata},
  {Kotak}, {Harutyunyan}, {Altavilla}, {Navasardyan}, {Qiu}, {Salvo}, \&
  {Hillebrandt}}]{SN2003cg}
{Elias-Rosa}, N., et~al.\ 2006, \mnras,  369, 1880

\bibitem[{{Fabricant} {et~al.}(1998){Fabricant}, {Cheimets}, {Caldwell}, \&
  {Geary}}]{Fabricant/etal:1998}
{Fabricant}, D., {Cheimets}, P., {Caldwell}, N., \& {Geary}, J. 1998, \pasp,
  110, 79

\bibitem[{{Falco} {et~al.}(1999){Falco}, {Kurtz}, {Geller}, {Huchra}, {Peters},
  {Berlind}, {Mink}, {Tokarz}, \& {Elwell}}]{UZC}
{Falco}, E.~E., et~al.\ 1999, \pasp,  111, 438

\bibitem[{{Filippenko}(1982)}]{Filippenko:1982}
{Filippenko}, A.~V. 1982, \pasp, 94, 715

\bibitem[{{Filippenko} {et~al.}(2001){Filippenko}, {Li}, {Treffers}, \&
  {Modjaz}}]{Filippenko/etal:2001}
{Filippenko}, A.~V., {Li}, W.~D., {Treffers}, R.~R., \& {Modjaz}, M. 2001, in
  Astronomical Society of the Pacific Conference Series, Vol. 246, IAU Colloq.
  183: Small Telescope Astronomy on Global Scales, ed. {B.~Paczynski,
  W.-P.~Chen, \& C.~Lemme}, 121--+

\bibitem[{{Filippenko} {et~al.}(1992{\natexlab{a}}){Filippenko}, {Richmond},
  {Branch}, {Gaskell}, {Herbst}, {Ford}, {Treffers}, {Matheson}, {Ho}, {Dey},
  {Sargent}, {Small}, \& {van Breugel}}]{Filippenko/etal:1992b}
{Filippenko}, A.~V., et~al.\ 1992{\natexlab{a}}, \aj, 104, 1543

\bibitem[{{Filippenko} {et~al.}(1992{\natexlab{b}}){Filippenko}, {Richmond},
  {Matheson}, {Shields}, {Burbidge}, {Cohen}, {Dickinson}, {Malkan}, {Nelson},
  {Pietz}, {Schlegel}, {Schmeer}, {Spinrad}, {Steidel}, {Tran}, \&
  {Wren}}]{Filippenko/etal:1992a}
{Filippenko}, A.~V., et~al.\ 1992{\natexlab{b}}, \apjl,  384, L15

\bibitem[{{Folatelli} {et~al.}(2010){Folatelli}, {Phillips}, {Burns},
  {Contreras}, {Hamuy}, {Freedman}, {Persson}, {Stritzinger}, {Suntzeff},
  {Krisciunas}, {Boldt}, {Gonz{\'a}lez}, {Krzeminski}, {Morrell}, {Roth},
  {Salgado}, {Madore}, {Murphy}, {Wyatt}, {Li}, {Filippenko}, \&
  {Miller}}]{Folatelli/etal:2010}
{Folatelli}, G., et~al.\ 2010, \aj, 139,  120

\bibitem[{{Folatelli} {et~al.}(2012){Folatelli}, {Phillips}, {Morrell},
  {Tanaka}, {Maeda}, {Nomoto}, {Stritzinger}, {Burns}, {Hamuy}, {Mazzali},
  {Boldt}, {Campillay}, {Contreras}, {Gonz{\'a}lez}, {Roth}, {Salgado},
  {Freedman}, {Madore}, {Persson}, \& {Suntzeff}}]{Folatelli/etal:2012}
{Folatelli}, G., et~al.\ 2012, \apj, 745, 74

\bibitem[{{Foley}(2012)}]{Foley:2012}
{Foley}, R.~J. 2012, arXiv:1202.0003 [astro-ph.CO]

\bibitem[{{Foley} {et~al.}(2012){Foley}, {Challis}, {Filippenko},
  {Ganeshalingam}, {Landsman}, {Li}, {Marion}, {Silverman}, {Beaton},
  {Bennert}, {Cenko}, {Childress}, {Guhathakurta}, {Jiang}, {Kalirai},
  {Kirshner}, {Stockton}, {Tollerud}, {Vink{\'o}}, {Wheeler}, \&
  {Woo}}]{SN2009ig}
{Foley}, R.~J., et~al.\ 2012, \apj, 744, 38

\bibitem[{{Foley} \& {Kasen}(2011)}]{Foley/Kasen:2011}
{Foley}, R.~J. \& {Kasen}, D. 2011, \apj, 729, 55

\bibitem[{{Foley} {et~al.}(2010){Foley}, {Narayan}, {Challis}, {Filippenko},
  {Kirshner}, {Silverman}, \& {Steele}}]{SN2006bt}
{Foley}, R.~J., {Narayan}, G., {Challis}, P.~J., {Filippenko}, A.~V.,
  {Kirshner}, R.~P., {Silverman}, J.~M., \& {Steele}, T.~N. 2010, \apj, 708,
  1748

\bibitem[{{Foley} {et~al.}(2011){Foley}, {Sanders}, \&
  {Kirshner}}]{Foley/Sanders/Kirshner:2011}
{Foley}, R.~J., {Sanders}, N.~E., \& {Kirshner}, R.~P. 2011, \apj, 742, 89

\bibitem[{{Gamezo} {et~al.}(2005){Gamezo}, {Khokhlov}, \&
  {Oran}}]{Gamezo/Khokhlov/Oran:2005}
{Gamezo}, V.~N., {Khokhlov}, A.~M., \& {Oran}, E.~S. 2005, \apj, 623, 337

\bibitem[{{Ganeshalingam} {et~al.}(2010){Ganeshalingam}, {Li}, {Filippenko},
  {Anderson}, {Foster}, {Gates}, {Griffith}, {Grigsby}, {Joubert}, {Leja},
  {Lowe}, {Macomber}, {Pritchard}, {Thrasher}, \&
  {Winslow}}]{Ganeshalingam/etal:2010}
{Ganeshalingam}, M., et~al.\ 2010, \apjs, 190, 418

\bibitem[{{Ganeshalingam} {et~al.}(2012){Ganeshalingam}, {Li}, {Filippenko},
  {Silverman}, {Chornock}, {Foley}, {Matheson}, {Kirshner}, {Milne}, {Calkins},
  \& {Shen}}]{Ganeshalingam/etal:2012}
{Ganeshalingam}, M., et~al.\ 2012, ArXiv:1202.3140 [astro-ph.CO]

\bibitem[{{Garavini} {et~al.}(2005){Garavini}, {Aldering}, {Amadon},
  {Amanullah}, {Astier}, {Balland}, {Blanc}, {Conley}, {Dahl{\'e}n}, {Deustua},
  {Ellis}, {Fabbro}, {Fadeyev}, {Fan}, {Folatelli}, {Frye}, {Gates}, {Gibbons},
  {Goldhaber}, {Goldman}, {Goobar}, {Groom}, {Haissinski}, {Hardin}, {Hook},
  {Howell}, {Kent}, {Kim}, {Knop}, {Kowalski}, {Kuznetsova}, {Lee}, {Lidman},
  {Mendez}, {Miller}, {Moniez}, {Mouchet}, {Mour{\~a}o}, {Newberg}, {Nobili},
  {Nugent}, {Pain}, {Perdereau}, {Perlmutter}, {Quimby}, {Regnault}, {Rich},
  {Richards}, {Ruiz-Lapuente}, {Schaefer}, {Schahmaneche}, {Smith},
  {Spadafora}, {Stanishev}, {Thomas}, {Walton}, {Wang}, \&
  {Wood-Vasey}}]{Garavini/etal:2005}
{Garavini}, G., et~al.\ 2005, \aj, 130, 2278

\bibitem[{{Garavini} {et~al.}(2004){Garavini}, {Folatelli}, {Goobar}, {Nobili},
  {Aldering}, {Amadon}, {Amanullah}, {Astier}, {Balland}, {Blanc}, {Burns},
  {Conley}, {Dahl{\'e}n}, {Deustua}, {Ellis}, {Fabbro}, {Fan}, {Frye}, {Gates},
  {Gibbons}, {Goldhaber}, {Goldman}, {Groom}, {Haissinski}, {Hardin}, {Hook},
  {Howell}, {Kasen}, {Kent}, {Kim}, {Knop}, {Lee}, {Lidman}, {Mendez},
  {Miller}, {Moniez}, {Mour{\~a}o}, {Newberg}, {Nugent}, {Pain}, {Perdereau},
  {Perlmutter}, {Prasad}, {Quimby}, {Raux}, {Regnault}, {Rich}, {Richards},
  {Ruiz-Lapuente}, {Sainton}, {Schaefer}, {Schahmaneche}, {Smith}, {Spadafora},
  {Stanishev}, {Walton}, {Wang}, \& {Wood-Vasey}}]{Garavini/etal:2004}
{Garavini}, G., et~al.\ 2004, \aj, 128, 387

\bibitem[{{Garavini} {et~al.}(2007){Garavini}, {Nobili}, {Taubenberger},
  {Pastorello}, {Elias-Rosa}, {Stanishev}, {Blanc}, {Benetti}, {Goobar},
  {Mazzali}, {Sanchez}, {Salvo}, {Schmidt}, \&
  {Hillebrandt}}]{Garavini/etal:2007b}
{Garavini}, G., et~al.\ 2007, \aap, 471, 527

\bibitem[{{Garnavich} {et~al.}(1996){Garnavich}, {Riess}, {Challis},
  {Kirshner}, \& {Berlind}}]{Garnavich_iauc6405}
{Garnavich}, P., {Riess}, A., {Challis}, P., {Kirshner}, R., \& {Berlind}, P.
  1996, \iaucirc, 6405, 1

\bibitem[{{Garnavich} {et~al.}(2004){Garnavich}, {Bonanos}, {Krisciunas},
  {Jha}, {Kirshner}, {Schlegel}, {Challis}, {Macri}, {Hatano}, {Branch},
  {Bothun}, \& {Freedman}}]{Garnavich/etal:2004}
{Garnavich}, P.~M., et~al.\ 2004, \apj, 613, 1120

\bibitem[{{Gerardy} {et~al.}(2007){Gerardy}, {Meikle}, {Kotak}, {H{\"o}flich},
  {Farrah}, {Filippenko}, {Foley}, {Lundqvist}, {Mattila}, {Pozzo},
  {Sollerman}, {Van Dyk}, \& {Wheeler}}]{Gerardy/etal:2007}
{Gerardy}, C.~L., et~al.\ 2007,  \apj, 661, 995

\bibitem[{{G{\'o}mez} \& {L{\'o}pez}(1998)}]{Gomez/Lopez:1998}
{G{\'o}mez}, G. \& {L{\'o}pez}, R. 1998, \aj, 115, 1096

\bibitem[{{Guy} {et~al.}(2007){Guy}, {Astier}, {Baumont}, {Hardin}, {Pain},
  {Regnault}, {Basa}, {Carlberg}, {Conley}, {Fabbro}, {Fouchez}, {Hook},
  {Howell}, {Perrett}, {Pritchet}, {Rich}, {Sullivan}, {Antilogus}, {Aubourg},
  {Bazin}, {Bronder}, {Filiol}, {Palanque-Delabrouille}, {Ripoche}, \&
  {Ruhlmann-Kleider}}]{SALT2}
{Guy}, J., et~al.\ 2007, \aap, 466, 11

\bibitem[{{Hachinger} {et~al.}(2006){Hachinger}, {Mazzali}, \&
  {Benetti}}]{Hachinger/Mazzali/Benetti:2006}
{Hachinger}, S., {Mazzali}, P.~A., \& {Benetti}, S. 2006, \mnras, 370, 299

\bibitem[{{Hachinger} {et~al.}(2008){Hachinger}, {Mazzali}, {Tanaka},
  {Hillebrandt}, \& {Benetti}}]{Hachinger/etal:2008}
{Hachinger}, S., {Mazzali}, P.~A., {Tanaka}, M., {Hillebrandt}, W., \&
  {Benetti}, S. 2008, \mnras, 389, 1087

\bibitem[{{Hamuy} {et~al.}(2002){Hamuy}, {Maza}, {Pinto}, {Phillips},
  {Suntzeff}, {Blum}, {Olsen}, {Pinfield}, {Ivanov}, {Augusteijn}, {Brillant},
  {Chadid}, {Cuby}, {Doublier}, {Hainaut}, {Le Floc'h}, {Lidman},
  {Petr-Gotzens}, {Pompei}, \& {Vanzi}}]{Hamuy/etal:2002}
{Hamuy}, M., et~al.\ 2002, \aj, 124, 417

\bibitem[{{Hamuy} {et~al.}(1996){Hamuy}, {Phillips}, {Suntzeff}, {Schommer},
  {Maza}, {Antezan}, {Wischnjewsky}, {Valladares}, {Muena}, {Gonzales},
  {Aviles}, {Wells}, {Smith}, {Navarrete}, {Covarrubias}, {Williger}, {Walker},
  {Layden}, {Elias}, {Baldwin}, {Hernandez}, {Tirado}, {Ugarte}, {Elston},
  {Saavedra}, {Barrientos}, {Costa}, {Lira}, {Ruiz}, {Anguita}, {Gomez},
  {Ortiz}, {della Valle}, {Danziger}, {Storm}, {Kim}, {Bailyn}, {Rubenstein},
  {Tucker}, {Cersosimo}, {Mendez}, {Siciliano}, {Sherry}, {Chaboyer},
  {Koopmann}, {Geisler}, {Sarajedini}, {Dey}, {Tyson}, {Rich}, {Gal},
  {Lamontagne}, {Caldwell}, {Guhathakurta}, {Phillips}, {Szkody}, {Prosser},
  {Ho}, {McMahan}, {Baggley}, {Cheng}, {Havlen}, {Wakamatsu}, {Janes},
  {Malkan}, {Baganoff}, {Seitzer}, {Shara}, {Sturch}, {Hesser}, {Hartig},
  {Hughes}, {Welch}, {Williams}, {Ferguson}, {Francis}, {French}, {Bolte},
  {Roth}, {Odewahn}, {Howell}, \& {Krzeminski}}]{Hamuy/etal:1996}
{Hamuy}, M., et~al.\ 1996, \aj, 112, 2408

\bibitem[{{Hatano} {et~al.}(1999){Hatano}, {Branch}, {Fisher}, {Millard}, \&
  {Baron}}]{Hatano/etal:1999}
{Hatano}, K., {Branch}, D., {Fisher}, A., {Millard}, J., \& {Baron}, E. 1999,
  \apjs, 121, 233

\bibitem[{{Hicken} {et~al.}(2009){Hicken}, {Challis}, {Jha}, {Kirshner},
  {Matheson}, {Modjaz}, {Rest}, {Michael Wood-Vasey}, {Bakos}, {Barton},
  {Berlind}, {Bragg}, {Brice{\~n}o}, {Brown}, {Caldwell}, {Calkins}, {Cho},
  {Ciupik}, {Contreras}, {Dendy}, {Dosaj}, {Durham}, {Eriksen}, {Esquerdo},
  {Everett}, {Falco}, {Fernandez}, {Gaba}, {Garnavich}, {Graves}, {Green},
  {Groner}, {Hergenrother}, {Holman}, {Hradecky}, {Huchra}, {Hutchison},
  {Jerius}, {Jordan}, {Kilgard}, {Krauss}, {Luhman}, {Macri}, {Marrone},
  {McDowell}, {McIntosh}, {McNamara}, {Megeath}, {Mochejska}, {Munoz},
  {Muzerolle}, {Naranjo}, {Narayan}, {Pahre}, {Peters}, {Peterson}, {Rines},
  {Ripman}, {Roussanova}, {Schild}, {Sicilia-Aguilar}, {Sokoloski}, {Smalley},
  {Smith}, {Spahr}, {Stanek}, {Barmby}, {Blondin}, {Stubbs}, {Szentgyorgyi},
  {Torres}, {Vaz}, {Vikhlinin}, {Wang}, {Westover}, {Woods}, \&
  {Zhao}}]{Hicken/etal:2009a}
{Hicken}, M., et~al.\ 2009, \apj, 700, 331

\bibitem[{{Hicken} {et~al.}(2007{\natexlab{a}}){Hicken}, {Garnavich}, {Prieto},
  {Blondin}, {DePoy}, {Kirshner}, \& {Parrent}}]{SN2006gz}
{Hicken}, M., {Garnavich}, P.~M., {Prieto}, J.~L., {Blondin}, S., {DePoy},
  D.~L., {Kirshner}, R.~P., \& {Parrent}, J. 2007{\natexlab{a}}, \apjl, 669,
  L17

\bibitem[{{Hicken} {et~al.}(2007{\natexlab{b}}){Hicken}, {Garnavich}, {Prieto},
  {Blondin}, {DePoy}, {Kirshner}, \& {Parrent}}]{Hicken/etal:2007}
---. 2007{\natexlab{b}}, \apjl, 669, L17

\bibitem[{{Hoeflich} {et~al.}(1995){Hoeflich}, {Khokhlov}, \&
  {Wheeler}}]{Hoeflich/Khokhlov/Wheeler:1995}
{Hoeflich}, P., {Khokhlov}, A.~M., \& {Wheeler}, J.~C. 1995, \apj, 444, 831

\bibitem[{{Horne}(1986)}]{Horne:1986}
{Horne}, K. 1986, \pasp, 98, 609

\bibitem[{{Howell}(2011)}]{Howell:2011}
{Howell}, D.~A. 2011, Nature Communications, 2

\bibitem[{{Howell} {et~al.}(2006){Howell}, {Sullivan}, {Nugent}, {Ellis},
  {Conley}, {Le Borgne}, {Carlberg}, {Guy}, {Balam}, {Basa}, {Fouchez}, {Hook},
  {Hsiao}, {Neill}, {Pain}, {Perrett}, \& {Pritchet}}]{Howell/etal:2006}
{Howell}, D.~A., et~al.\ 2006, \nat, 443, 308

\bibitem[{{Hoyle} \& {Fowler}(1960)}]{Hoyle/Fowler:1960}
{Hoyle}, F. \& {Fowler}, W.~A. 1960, \apj, 132, 565

\bibitem[{{Hsiao} {et~al.}(2007){Hsiao}, {Conley}, {Howell}, {Sullivan},
  {Pritchet}, {Carlberg}, {Nugent}, \& {Phillips}}]{Hsiao/etal:2007}
{Hsiao}, E.~Y., {Conley}, A., {Howell}, D.~A., {Sullivan}, M., {Pritchet},
  C.~J., {Carlberg}, R.~G., {Nugent}, P.~E., \& {Phillips}, M.~M. 2007, \apj,
  663, 1187

\bibitem[{{Iben} \& {Tutukov}(1984)}]{Iben/Tutukov:1984}
{Iben}, Jr., I. \& {Tutukov}, A.~V. 1984, \apjs, 54, 335

\bibitem[{{Jeffery} {et~al.}(1992){Jeffery}, {Leibundgut}, {Kirshner},
  {Benetti}, {Branch}, \& {Sonneborn}}]{Jeffery/etal:1992}
{Jeffery}, D.~J., {Leibundgut}, B., {Kirshner}, R.~P., {Benetti}, S., {Branch},
  D., \& {Sonneborn}, G. 1992, \apj, 397, 304

\bibitem[{{Jha} {et~al.}(1999{\natexlab{a}}){Jha}, {Garnavich}, {Challis},
  {Kirshner}, {Calkins}, {Brown}, \& {Filippenko}}]{Jha_iauc7150}
{Jha}, S., {Garnavich}, P., {Challis}, P., {Kirshner}, R., {Calkins}, M.,
  {Brown}, W., \& {Filippenko}, A.~V. 1999{\natexlab{a}}, \iaucirc, 7150, 3

\bibitem[{{Jha} {et~al.}(1999{\natexlab{b}}){Jha}, {Garnavich}, {Kirshner},
  {Challis}, {Soderberg}, {Macri}, {Huchra}, {Barmby}, {Barton}, {Berlind},
  {Brown}, {Caldwell}, {Calkins}, {Kannappan}, {Koranyi}, {Pahre}, {Rines},
  {Stanek}, {Stefanik}, {Szentgyorgyi}, {V{\"a}is{\"a}nen}, {Wang}, {Zajac},
  {Riess}, {Filippenko}, {Li}, {Modjaz}, {Treffers}, {Hergenrother}, {Grebel},
  {Seitzer}, {Jacoby}, {Benson}, {Rizvi}, {Marschall}, {Goldader}, {Beasley},
  {Vacca}, {Leibundgut}, {Spyromilio}, {Schmidt}, \& {Wood}}]{Jha/etal:1999}
{Jha}, S., et~al.\ 1999{\natexlab{b}},  \apjs, 125, 73

\bibitem[{{Jha} {et~al.}(2006){Jha}, {Kirshner}, {Challis}, {Garnavich},
  {Matheson}, {Soderberg}, {Graves}, {Hicken}, {Alves}, {Arce}, {Balog},
  {Barmby}, {Barton}, {Berlind}, {Bragg}, {Brice{\~n}o}, {Brown}, {Buckley},
  {Caldwell}, {Calkins}, {Carter}, {Concannon}, {Donnelly}, {Eriksen},
  {Fabricant}, {Falco}, {Fiore}, {Garcia}, {G{\'o}mez}, {Grogin}, {Groner},
  {Groot}, {Haisch}, {Hartmann}, {Hergenrother}, {Holman}, {Huchra},
  {Jayawardhana}, {Jerius}, {Kannappan}, {Kim}, {Kleyna}, {Kochanek},
  {Koranyi}, {Krockenberger}, {Lada}, {Luhman}, {Luu}, {Macri}, {Mader},
  {Mahdavi}, {Marengo}, {Marsden}, {McLeod}, {McNamara}, {Megeath}, {Moraru},
  {Mossman}, {Muench}, {Mu{\~n}oz}, {Muzerolle}, {Naranjo}, {Nelson-Patel},
  {Pahre}, {Patten}, {Peters}, {Peters}, {Raymond}, {Rines}, {Schild},
  {Sobczak}, {Spahr}, {Stauffer}, {Stefanik}, {Szentgyorgyi}, {Tollestrup},
  {V{\"a}is{\"a}nen}, {Vikhlinin}, {Wang}, {Willner}, {Wolk}, {Zajac}, {Zhao},
  \& {Stanek}}]{Jha/etal:2006}
{Jha}, S., et~al.\ 2006, \aj, 131, 527

\bibitem[{{Jha} {et~al.}(2007){Jha}, {Riess}, \& {Kirshner}}]{MLCS2k2}
{Jha}, S., {Riess}, A.~G., \& {Kirshner}, R.~P. 2007, \apj, 659, 122

\bibitem[{{Kasen} {et~al.}(2003){Kasen}, {Nugent}, {Wang}, {Howell}, {Wheeler},
  {H{\"o}flich}, {Baade}, {Baron}, \& {Hauschildt}}]{Kasen/etal:2003}
{Kasen}, D., et~al.\ 2003,  \apj, 593, 788

\bibitem[{{Kasen} \& {Plewa}(2007)}]{Kasen/Plewa:2007}
{Kasen}, D. \& {Plewa}, T. 2007, \apj, 662, 459

\bibitem[{{Kasen} {et~al.}(2009){Kasen}, {R{\"o}pke}, \& {Woosley}}]{KRW09}
{Kasen}, D., {R{\"o}pke}, F.~K., \& {Woosley}, S.~E. 2009, \nat, 460, 869

\bibitem[{{Kasliwal} {et~al.}(2008){Kasliwal}, {Ofek}, {Gal-Yam}, {Rau},
  {Brown}, {Cenko}, {Cameron}, {Quimby}, {Kulkarni}, {Bildsten}, {Milne}, \&
  {Bryngelson}}]{Kasliwal/etal:2008}
{Kasliwal}, M.~M., et~al.\ 2008, \apjl, 683, L29

\bibitem[{{Kelly}(2007)}]{Kelly:2007}
{Kelly}, B.~C. 2007, \apj, 665, 1489

\bibitem[{{Khokhlov}(1991)}]{Khokhlov:1991}
{Khokhlov}, A.~M. 1991, \aap, 245, 114

\bibitem[{{Kirshner} {et~al.}(1993){Kirshner}, {Jeffery}, {Leibundgut},
  {Challis}, {Sonneborn}, {Phillips}, {Suntzeff}, {Smith}, {Winkler}, {Winge},
  {Hamuy}, {Hunter}, {Roth}, {Blades}, {Branch}, {Chevalier}, {Fransson},
  {Panagia}, {Wagoner}, {Wheeler}, \& {Harkness}}]{Kirshner/etal:1993}
{Kirshner}, R.~P., et~al.\ 1993, \apj, 415,  589

\bibitem[{{Kirshner} \& {Kwan}(1974)}]{Kirshner/Kwan:1974}
{Kirshner}, R.~P. \& {Kwan}, J. 1974, \apj, 193, 27

\bibitem[{{Kirshner} {et~al.}(1973){Kirshner}, {Willner}, {Becklin},
  {Neugebauer}, \& {Oke}}]{Kirshner/etal:1973}
{Kirshner}, R.~P., {Willner}, S.~P., {Becklin}, E.~E., {Neugebauer}, G., \&
  {Oke}, J.~B. 1973, \apjl, 180, L97+

\bibitem[{{Kotak} {et~al.}(2005){Kotak}, {Meikle}, {Pignata}, {Stehle},
  {Smartt}, {Benetti}, {Hillebrandt}, {Lennon}, {Mazzali}, {Patat}, \&
  {Turatto}}]{SN2002er}
{Kotak}, R., et~al.\ 2005, \aap, 436, 1021

\bibitem[{{Kozma} {et~al.}(2005){Kozma}, {Fransson}, {Hillebrandt},
  {Travaglio}, {Sollerman}, {Reinecke}, {R{\"o}pke}, \&
  {Spyromilio}}]{Kozma/etal:2005}
{Kozma}, C., {Fransson}, C., {Hillebrandt}, W., {Travaglio}, C., {Sollerman},
  J., {Reinecke}, M., {R{\"o}pke}, F.~K., \& {Spyromilio}, J. 2005, \aap, 437,
  983

\bibitem[{{Krisciunas} {et~al.}(2007){Krisciunas}, {Garnavich}, {Stanishev},
  {Suntzeff}, {Prieto}, {Espinoza}, {Gonzalez}, {Salvo}, {Elias de la Rosa},
  {Smartt}, {Maund}, \& {Kudritzki}}]{Krisciunas/etal:2007}
{Krisciunas}, K., et~al.\ 2007, \aj,  133, 58

\bibitem[{{Krisciunas} {et~al.}(2011{\natexlab{a}}){Krisciunas}, {Li},
  {Matheson}, {Howell}, {Stritzinger}, {Aldering}, {Berlind}, {Calkins},
  {Challis}, {Chornock}, {Conley}, {Filippenko}, {Ganeshalingam}, {Germany},
  {Gonz{\'a}lez}, {Gooding}, {Hsiao}, {Kasen}, {Kirshner}, {Howie Marion},
  {Muena}, {Nugent}, {Phelps}, {Phillips}, {Qiu}, {Quimby}, {Rines},
  {Silverman}, {Suntzeff}, {Thomas}, \& {Wang}}]{Krisciunas/etal:2011}
{Krisciunas}, K., et~al.\ 2011{\natexlab{a}}, \aj,  142, 74

\bibitem[{{Krisciunas} {et~al.}(2011{\natexlab{b}}){Krisciunas}, {Li},
  {Matheson}, {Howell}, {Stritzinger}, {Aldering}, {Berlind}, {Calkins},
  {Challis}, {Chornock}, {Conley}, {Filippenko}, {Ganeshalingam}, {Germany},
  {Gonz{\'a}lez}, {Gooding}, {Hsiao}, {Kasen}, {Kirshner}, {Howie Marion},
  {Muena}, {Nugent}, {Phelps}, {Phillips}, {Qiu}, {Quimby}, {Rines},
  {Silverman}, {Suntzeff}, {Thomas}, \& {Wang}}]{SN2001ay}
---. 2011{\natexlab{b}}, \aj, 142, 74

\bibitem[{{Kurtz} \& {Mink}(1998)}]{RVSAO}
{Kurtz}, M.~J. \& {Mink}, D.~J. 1998, \pasp, 110, 934

\bibitem[{{Leibundgut} {et~al.}(1991){Leibundgut}, {Kirshner}, {Filippenko},
  {Shields}, {Foltz}, {Phillips}, \& {Sonneborn}}]{Leibundgut/etal:1991}
{Leibundgut}, B., {Kirshner}, R.~P., {Filippenko}, A.~V., {Shields}, J.~C.,
  {Foltz}, C.~B., {Phillips}, M.~M., \& {Sonneborn}, G. 1991, \apjl, 371, L23

\bibitem[{{Leibundgut} {et~al.}(1993){Leibundgut}, {Kirshner}, {Phillips},
  {Wells}, {Suntzeff}, {Hamuy}, {Schommer}, {Walker}, {Gonzalez}, {Ugarte},
  {Williams}, {Williger}, {Gomez}, {Marzke}, {Schmidt}, {Whitney}, {Coldwell},
  {Peters}, {Chaffee}, {Foltz}, {Rehner}, {Siciliano}, {Barnes}, {Cheng},
  {Hintzen}, {Kim}, {Maza}, {Parker}, {Porter}, {Schmidtke}, \&
  {Sonneborn}}]{Leibundgut/etal:1993}
{Leibundgut}, B., et~al.\ 1993, \aj, 105, 301

\bibitem[{{Leloudas} {et~al.}(2009){Leloudas}, {Stritzinger}, {Sollerman},
  {Burns}, {Kozma}, {Krisciunas}, {Maund}, {Milne}, {Filippenko}, {Fransson},
  {Ganeshalingam}, {Hamuy}, {Li}, {Phillips}, {Schmidt}, {Skottfelt},
  {Taubenberger}, {Boldt}, {Fynbo}, {Gonzalez}, {Salvo}, \&
  {Thomas-Osip}}]{SN2003hv}
{Leloudas}, G., et~al.\ 2009,  \aap, 505, 265

\bibitem[{{Li} {et~al.}(2003){Li}, {Filippenko}, {Chornock}, {Berger},
  {Berlind}, {Calkins}, {Challis}, {Fassnacht}, {Jha}, {Kirshner}, {Matheson},
  {Sargent}, {Simcoe}, {Smith}, \& {Squires}}]{Li/etal:2003}
{Li}, W., et~al.\ 2003, \pasp, 115, 453

\bibitem[{{Li} {et~al.}(2001){Li}, {Filippenko}, {Gates}, {Chornock},
  {Gal-Yam}, {Ofek}, {Leonard}, {Modjaz}, {Rich}, {Riess}, \&
  {Treffers}}]{Li/etal:2001b}
{Li}, W., et~al.\ 2001, \pasp, 113, 1178

\bibitem[{{Maeda} {et~al.}(2010{\natexlab{a}}){Maeda}, {Benetti},
  {Stritzinger}, {R{\"o}pke}, {Folatelli}, {Sollerman}, {Taubenberger},
  {Nomoto}, {Leloudas}, {Hamuy}, {Tanaka}, {Mazzali}, \&
  {Elias-Rosa}}]{Maeda/etal:2010c}
{Maeda}, K., et~al.\ 2010{\natexlab{a}}, \nat, 466, 82

\bibitem[{{Maeda} {et~al.}(2011){Maeda}, {Leloudas}, {Taubenberger},
  {Stritzinger}, {Sollerman}, {Elias-Rosa}, {Benetti}, {Hamuy}, {Folatelli}, \&
  {Mazzali}}]{Maeda/etal:2011}
{Maeda}, K., et~al.\ 2011, \mnras, 413, 3075

\bibitem[{{Maeda} {et~al.}(2010{\natexlab{b}}){Maeda}, {R{\"o}pke}, {Fink},
  {Hillebrandt}, {Travaglio}, \& {Thielemann}}]{Maeda/etal:2010b}
{Maeda}, K., {R{\"o}pke}, F.~K., {Fink}, M., {Hillebrandt}, W., {Travaglio},
  C., \& {Thielemann}, F. 2010{\natexlab{b}}, \apj, 712, 624

\bibitem[{{Maeda} {et~al.}(2010{\natexlab{c}}){Maeda}, {Taubenberger},
  {Sollerman}, {Mazzali}, {Leloudas}, {Nomoto}, \&
  {Motohara}}]{Maeda/etal:2010a}
{Maeda}, K., {Taubenberger}, S., {Sollerman}, J., {Mazzali}, P.~A., {Leloudas},
  G., {Nomoto}, K., \& {Motohara}, K. 2010{\natexlab{c}}, \apj, 708, 1703

\bibitem[{{Mandel} {et~al.}(2011){Mandel}, {Narayan}, \&
  {Kirshner}}]{Mandel/etal:2011}
{Mandel}, K.~S., {Narayan}, G., \& {Kirshner}, R.~P. 2011, \apj, 731, 120

\bibitem[{{Mandel} {et~al.}(2009){Mandel}, {Wood-Vasey}, {Friedman}, \&
  {Kirshner}}]{Mandel/etal:2009}
{Mandel}, K.~S., {Wood-Vasey}, W.~M., {Friedman}, A.~S., \& {Kirshner}, R.~P.
  2009, \apj, 704, 629

\bibitem[{{Marion} {et~al.}(2006){Marion}, {H{\"o}flich}, {Wheeler},
  {Robinson}, {Gerardy}, \& {Vacca}}]{Marion/etal:2006}
{Marion}, G.~H., {H{\"o}flich}, P., {Wheeler}, J.~C., {Robinson}, E.~L.,
  {Gerardy}, C.~L., \& {Vacca}, W.~D. 2006, \apj, 645, 1392

\bibitem[{{Matheson} {et~al.}(2000){Matheson}, {Filippenko}, {Ho}, {Barth}, \&
  {Leonard}}]{Matheson/etal:2000}
{Matheson}, T., {Filippenko}, A.~V., {Ho}, L.~C., {Barth}, A.~J., \& {Leonard},
  D.~C. 2000, \aj, 120, 1499

\bibitem[{{Matheson} {et~al.}(2001){Matheson}, {Filippenko}, {Li}, {Leonard},
  \& {Shields}}]{Matheson/etal:2001}
{Matheson}, T., {Filippenko}, A.~V., {Li}, W., {Leonard}, D.~C., \& {Shields},
  J.~C. 2001, \aj, 121, 1648

\bibitem[{{Matheson} {et~al.}(2008){Matheson}, {Kirshner}, {Challis}, {Jha},
  {Garnavich}, {Berlind}, {Calkins}, {Blondin}, {Balog}, {Bragg}, {Caldwell},
  {Dendy Concannon}, {Falco}, {Graves}, {Huchra}, {Kuraszkiewicz}, {Mader},
  {Mahdavi}, {Phelps}, {Rines}, {Song}, \& {Wilkes}}]{Matheson/etal:2008}
{Matheson}, T., et~al.\ 2008, \aj, 135,  1598

\bibitem[{{Mattila} {et~al.}(2005){Mattila}, {Lundqvist}, {Sollerman}, {Kozma},
  {Baron}, {Fransson}, {Leibundgut}, \& {Nomoto}}]{Mattila/etal:2005}
{Mattila}, S., {Lundqvist}, P., {Sollerman}, J., {Kozma}, C., {Baron}, E.,
  {Fransson}, C., {Leibundgut}, B., \& {Nomoto}, K. 2005, \aap, 443, 649

\bibitem[{{Mazzali} {et~al.}(2005{\natexlab{a}}){Mazzali}, {Benetti},
  {Altavilla}, {Blanc}, {Cappellaro}, {Elias-Rosa}, {Garavini}, {Goobar},
  {Harutyunyan}, {Kotak}, {Leibundgut}, {Lundqvist}, {Mattila}, {Mendez},
  {Nobili}, {Pain}, {Pastorello}, {Patat}, {Pignata}, {Podsiadlowski},
  {Ruiz-Lapuente}, {Salvo}, {Schmidt}, {Sollerman}, {Stanishev}, {Stehle},
  {Tout}, {Turatto}, \& {Hillebrandt}}]{Mazzali/etal:2005b}
{Mazzali}, P.~A., et~al.\ 2005{\natexlab{a}}, \apjl, 623, L37

\bibitem[{{Mazzali} {et~al.}(2005{\natexlab{b}}){Mazzali}, {Benetti}, {Stehle},
  {Branch}, {Deng}, {Maeda}, {Nomoto}, \& {Hamuy}}]{Mazzali/etal:2005a}
{Mazzali}, P.~A., et~al.\ 2005{\natexlab{b}}, \mnras, 357,  200

\bibitem[{{Mazzali} {et~al.}(1998){Mazzali}, {Cappellaro}, {Danziger},
  {Turatto}, \& {Benetti}}]{Mazzali/etal:1998}
{Mazzali}, P.~A., {Cappellaro}, E., {Danziger}, I.~J., {Turatto}, M., \&
  {Benetti}, S. 1998, \apjl, 499, L49+

\bibitem[{{Mazzali} {et~al.}(1993){Mazzali}, {Lucy}, {Danziger}, {Gouiffes},
  {Cappellaro}, \& {Turatto}}]{Mazzali/etal:1993}
{Mazzali}, P.~A., {Lucy}, L.~B., {Danziger}, I.~J., {Gouiffes}, C.,
  {Cappellaro}, E., \& {Turatto}, M. 1993, \aap, 269, 423

\bibitem[{{Mazzali} {et~al.}(2007){Mazzali}, {R{\"o}pke}, {Benetti}, \&
  {Hillebrandt}}]{Mazzali/etal:2007}
{Mazzali}, P.~A., {R{\"o}pke}, F.~K., {Benetti}, S., \& {Hillebrandt}, W. 2007,
  Science, 315, 825

\bibitem[{{Milne} {et~al.}(2010){Milne}, {Brown}, {Roming}, {Holland},
  {Immler}, {Filippenko}, {Ganeshalingam}, {Li}, {Stritzinger}, {Phillips},
  {Hicken}, {Kirshner}, {Challis}, {Mazzali}, {Schmidt}, {Bufano}, {Gehrels},
  \& {Vanden Berk}}]{Milne/etal:2010}
{Milne}, P.~A., et~al.\ 2010, \apj, 721, 1627

\bibitem[{{Modjaz} {et~al.}(2008){Modjaz}, {Kewley}, {Kirshner}, {Stanek},
  {Challis}, {Garnavich}, {Greene}, {Kelly}, \& {Prieto}}]{Modjaz/etal:2008}
{Modjaz}, M., et~al.\ 2008,  \aj, 135, 1136

\bibitem[{{Modjaz} {et~al.}(2009){Modjaz}, {Li}, {Butler}, {Chornock},
  {Perley}, {Blondin}, {Bloom}, {Filippenko}, {Kirshner}, {Kocevski},
  {Poznanski}, {Hicken}, {Foley}, {Stringfellow}, {Berlind}, {Barrado y
  Navascues}, {Blake}, {Bouy}, {Brown}, {Challis}, {Chen}, {de Vries},
  {Dufour}, {Falco}, {Friedman}, {Ganeshalingam}, {Garnavich}, {Holden},
  {Illingworth}, {Lee}, {Liebert}, {Marion}, {Olivier}, {Prochaska},
  {Silverman}, {Smith}, {Starr}, {Steele}, {Stockton}, {Williams}, \&
  {Wood-Vasey}}]{Modjaz/etal:2009}
{Modjaz}, M., et~al.\ 2009, \apj, 702, 226

\bibitem[{{Modjaz} {et~al.}(2006){Modjaz}, {Stanek}, {Garnavich}, {Berlind},
  {Blondin}, {Brown}, {Calkins}, {Challis}, {Diamond-Stanic}, {Hao}, {Hicken},
  {Kirshner}, \& {Prieto}}]{Modjaz/etal:2006}
{Modjaz}, M., et~al.\ 2006, \apjl,  645, L21

\bibitem[{{Monnier Ragaigne} {et~al.}(2003){Monnier Ragaigne}, {van Driel},
  {Schneider}, {Balkowski}, \& {Jarrett}}]{MonnierRagaigne/etal:2003}
{Monnier Ragaigne}, D., {van Driel}, W., {Schneider}, S.~E., {Balkowski}, C.,
  \& {Jarrett}, T.~H. 2003, \aap, 408, 465

\bibitem[{{Naito} {et~al.}(2007){Naito}, {Sakane}, {Anan}, {Kouzuma}, \&
  {Yamaoka}}]{Naito_cbet1173}
{Naito}, H., {Sakane}, Y., {Anan}, T., {Kouzuma}, S., \& {Yamaoka}, H. 2007,
  Central Bureau Electronic Telegrams, 1173, 1

\bibitem[{{Nomoto} {et~al.}(1984){Nomoto}, {Thielemann}, \& {Yokoi}}]{W7}
{Nomoto}, K., {Thielemann}, F.-K., \& {Yokoi}, K. 1984, \apj, 286, 644

\bibitem[{{Nugent} {et~al.}(1995){Nugent}, {Phillips}, {Baron}, {Branch}, \&
  {Hauschildt}}]{Nugent/etal:1995}
{Nugent}, P., {Phillips}, M., {Baron}, E., {Branch}, D., \& {Hauschildt}, P.
  1995, \apjl, 455, L147+

\bibitem[{{Nugent} {et~al.}(2011){Nugent}, {Sullivan}, {Cenko}, {Thomas},
  {Kasen}, {Howell}, {Bersier}, {Bloom}, {Kulkarni}, {Kandrashoff},
  {Filippenko}, {Silverman}, {Marcy}, {Howard}, {Isaacson}, {Maguire},
  {Suzuki}, {Tarlton}, {Pan}, {Bildsten}, {Fulton}, {Parrent}, {Sand},
  {Podsiadlowski}, {Bianco}, {Dilday}, {Graham}, {Lyman}, {James}, {Kasliwal},
  {Law}, {Quimby}, {Hook}, {Walker}, {Mazzali}, {Pian}, {Ofek}, {Gal-Yam}, \&
  {Poznanski}}]{Nugent/etal:2011}
{Nugent}, P.~E., et~al.\ 2011, \nat, 480, 344

\bibitem[{{Parrent} {et~al.}(2011){Parrent}, {Thomas}, {Fesen}, {Marion},
  {Challis}, {Garnavich}, {Milisavljevic}, {Vink{\`o}}, \&
  {Wheeler}}]{Parrent/etal:2011}
{Parrent}, J.~T., et~al.\ 2011, \apj, 732, 30

\bibitem[{{Pastorello} {et~al.}(2007){Pastorello}, {Mazzali}, {Pignata},
  {Benetti}, {Cappellaro}, {Filippenko}, {Li}, {Meikle}, {Arkharov}, {Blanc},
  {Bufano}, {Derekas}, {Dolci}, {Elias-Rosa}, {Foley}, {Ganeshalingam},
  {Harutyunyan}, {Kiss}, {Kotak}, {Larionov}, {Lucey}, {Napoleone},
  {Navasardyan}, {Patat}, {Rich}, {Ryder}, {Salvo}, {Schmidt}, {Stanishev},
  {Sz{\'e}kely}, {Taubenberger}, {Temporin}, {Turatto}, \&
  {Hillebrandt}}]{Pastorello/etal:2007b}
{Pastorello}, A., et~al.\ 2007, \mnras, 377, 1531

\bibitem[{{Patat} {et~al.}(1996){Patat}, {Benetti}, {Cappellaro}, {Danziger},
  {della Valle}, {Mazzali}, \& {Turatto}}]{Patat/etal:1996}
{Patat}, F., {Benetti}, S., {Cappellaro}, E., {Danziger}, I.~J., {della Valle},
  M., {Mazzali}, P.~A., \& {Turatto}, M. 1996, \mnras, 278, 111

\bibitem[{{Perlmutter} {et~al.}(1999){Perlmutter}, {Aldering}, {Goldhaber},
  {Knop}, {Nugent}, {Castro}, {Deustua}, {Fabbro}, {Goobar}, {Groom}, {Hook},
  {Kim}, {Kim}, {Lee}, {Nunes}, {Pain}, {Pennypacker}, {Quimby}, {Lidman},
  {Ellis}, {Irwin}, {McMahon}, {Ruiz-Lapuente}, {Walton}, {Schaefer}, {Boyle},
  {Filippenko}, {Matheson}, {Fruchter}, {Panagia}, {Newberg}, {Couch}, \& {The
  Supernova Cosmology Project}}]{P99}
{Perlmutter}, S., et~al.\ 1999, \apj, 517, 565

\bibitem[{{Phillips}(1993)}]{Phillips:1993}
{Phillips}, M.~M. 1993, \apjl, 413, L105

\bibitem[{{Phillips} {et~al.}(2006){Phillips}, {Krisciunas}, {Suntzeff},
  {Abraham}, {Beckett}, {Bonati}, {Candia}, {Corwin}, {Depoy}, {Espinoza},
  {Firth}, {Freedman}, {Galaz}, {Germany}, {Gonzalez}, {Hamuy}, {Hastings},
  {Hungerford}, {Ivanov}, {Labb{\'e}}, {Marzke}, {McCarthy}, {McMahon},
  {McMillan}, {Muena}, {Persson}, {Roth}, {Ruiz}, {Smith}, {Smith}, {Strolger},
  \& {Stubbs}}]{Phillips/etal:2006}
{Phillips}, M.~M., et~al.\ 2006, \aj, 131, 2615

\bibitem[{{Phillips} {et~al.}(2007){Phillips}, {Li}, {Frieman}, {Blinnikov},
  {DePoy}, {Prieto}, {Milne}, {Contreras}, {Folatelli}, {Morrell}, {Hamuy},
  {Suntzeff}, {Roth}, {Gonz{\'a}lez}, {Krzeminski}, {Filippenko}, {Freedman},
  {Chornock}, {Jha}, {Madore}, {Persson}, {Burns}, {Wyatt}, {Murphy}, {Foley},
  {Ganeshalingam}, {Serduke}, {Krisciunas}, {Bassett}, {Becker}, {Dilday},
  {Eastman}, {Garnavich}, {Holtzman}, {Kessler}, {Lampeitl}, {Marriner},
  {Frank}, {Marshall}, {Miknaitis}, {Sako}, {Schneider}, {van der Heyden}, \&
  {Yasuda}}]{Phillips/etal:2007}
{Phillips}, M.~M., et~al.\ 2007, \pasp, 119,  360

\bibitem[{{Phillips} {et~al.}(1999){Phillips}, {Lira}, {Suntzeff}, {Schommer},
  {Hamuy}, \& {Maza}}]{Phillips/etal:1999}
{Phillips}, M.~M., {Lira}, P., {Suntzeff}, N.~B., {Schommer}, R.~A., {Hamuy},
  M., \& {Maza}, J. 1999, \aj, 118, 1766

\bibitem[{{Phillips} {et~al.}(1987){Phillips}, {Phillips}, {Heathcote},
  {Blanco}, {Geisler}, {Hamilton}, {Suntzeff}, {Jablonski}, {Steiner},
  {Cowley}, {Schmidtke}, {Wyckoff}, {Hutchings}, {Tonry}, {Strauss},
  {Thorstensen}, {Honey}, {Maza}, {Ruiz}, {Landolt}, {Uomoto}, {Rich},
  {Grindlay}, {Cohn}, {Smith}, {Lutz}, {Lavery}, \&
  {Saha}}]{Phillips/etal:1987}
{Phillips}, M.~M., et~al.\ 1987, \pasp, 99, 592

\bibitem[{{Pignata} {et~al.}(2008){Pignata}, {Benetti}, {Mazzali}, {Kotak},
  {Patat}, {Meikle}, {Stehle}, {Leibundgut}, {Suntzeff}, {Buson}, {Cappellaro},
  {Clocchiatti}, {Hamuy}, {Maza}, {Mendez}, {Ruiz-Lapuente}, {Salvo},
  {Schmidt}, {Turatto}, \& {Hillebrandt}}]{SN2002dj}
{Pignata}, G., et~al.\ 2008, \mnras, 388, 971

\bibitem[{{Plewa}(2007)}]{Plewa:2007}
{Plewa}, T. 2007, \apj, 657, 942

\bibitem[{{Poludnenko} {et~al.}(2011){Poludnenko}, {Gardiner}, \&
  {Oran}}]{Poludnenko/etal:2011}
{Poludnenko}, A.~Y., {Gardiner}, T.~A., \& {Oran}, E.~S. 2011, Physical Review
  Letters, 107, 054501

\bibitem[{{Prieto} {et~al.}(2006){Prieto}, {Rest}, \&
  {Suntzeff}}]{Prieto/Rest/Suntzeff:2006}
{Prieto}, J.~L., {Rest}, A., \& {Suntzeff}, N.~B. 2006, \apj, 647, 501

\bibitem[{{Pskovskii}(1977)}]{Pskovskii:1977}
{Pskovskii}, I.~P. 1977, Soviet Astronomy, 21, 675

\bibitem[{{Quimby} {et~al.}(2006){Quimby}, {H{\"o}flich}, {Kannappan},
  {Rykoff}, {Rujopakarn}, {Akerlof}, {Gerardy}, \&
  {Wheeler}}]{Quimby/etal:2006}
{Quimby}, R., {H{\"o}flich}, P., {Kannappan}, S.~J., {Rykoff}, E.,
  {Rujopakarn}, W., {Akerlof}, C.~W., {Gerardy}, C.~L., \& {Wheeler}, J.~C.
  2006, \apj, 636, 400

\bibitem[{{Quimby} {et~al.}(2007){Quimby}, {H{\"o}flich}, \&
  {Wheeler}}]{Quimby/etal:2007a}
{Quimby}, R., {H{\"o}flich}, P., \& {Wheeler}, J.~C. 2007, \apj, 666, 1083

\bibitem[{{Riess} {et~al.}(1998){Riess}, {Filippenko}, {Challis},
  {Clocchiatti}, {Diercks}, {Garnavich}, {Gilliland}, {Hogan}, {Jha},
  {Kirshner}, {Leibundgut}, {Phillips}, {Reiss}, {Schmidt}, {Schommer},
  {Smith}, {Spyromilio}, {Stubbs}, {Suntzeff}, \& {Tonry}}]{R98}
{Riess}, A.~G., et~al.\ 1998, \aj, 116, 1009

\bibitem[{{Riess} {et~al.}(1999){Riess}, {Kirshner}, {Schmidt}, {Jha},
  {Challis}, {Garnavich}, {Esin}, {Carpenter}, {Grashius}, {Schild}, {Berlind},
  {Huchra}, {Prosser}, {Falco}, {Benson}, {Brice{\~n}o}, {Brown}, {Caldwell},
  {dell'Antonio}, {Filippenko}, {Goodman}, {Grogin}, {Groner}, {Hughes},
  {Green}, {Jansen}, {Kleyna}, {Luu}, {Macri}, {McLeod}, {McLeod}, {McNamara},
  {McLean}, {Milone}, {Mohr}, {Moraru}, {Peng}, {Peters}, {Prestwich},
  {Stanek}, {Szentgyorgyi}, \& {Zhao}}]{Riess/etal:1999a}
{Riess}, A.~G., et~al.\ 1999, \aj, 117, 707

\bibitem[{{R{\"o}pke}(2005)}]{Roepke:2005}
{R{\"o}pke}, F.~K. 2005, \aap, 432, 969

\bibitem[{{R{\"o}pke} \& {Niemeyer}(2007)}]{Roepke/Niemeyer:2007}
{R{\"o}pke}, F.~K. \& {Niemeyer}, J.~C. 2007, \aap, 464, 683

\bibitem[{{Salvo} {et~al.}(2001){Salvo}, {Cappellaro}, {Mazzali}, {Benetti},
  {Danziger}, {Patat}, \& {Turatto}}]{Salvo/etal:2001}
{Salvo}, M.~E., {Cappellaro}, E., {Mazzali}, P.~A., {Benetti}, S., {Danziger},
  I.~J., {Patat}, F., \& {Turatto}, M. 2001, \mnras, 321, 254

\bibitem[{{Sauer} {et~al.}(2008){Sauer}, {Mazzali}, {Blondin}, {Stehle},
  {Benetti}, {Challis}, {Filippenko}, {Kirshner}, {Li}, \&
  {Matheson}}]{Sauer/etal:2008}
{Sauer}, D.~N., et~al.\ 2008, \mnras, 391, 1605

\bibitem[{{Scalzo} {et~al.}(2010){Scalzo}, {Aldering}, {Antilogus}, {Aragon},
  {Bailey}, {Baltay}, {Bongard}, {Buton}, {Childress}, {Chotard}, {Copin},
  {Fakhouri}, {Gal-Yam}, {Gangler}, {Hoyer}, {Kasliwal}, {Loken}, {Nugent},
  {Pain}, {P{\'e}contal}, {Pereira}, {Perlmutter}, {Rabinowitz}, {Rau},
  {Rigaudier}, {Runge}, {Smadja}, {Tao}, {Thomas}, {Weaver}, \&
  {Wu}}]{Scalzo/etal:2010}
{Scalzo}, R.~A., et~al.\ 2010, \apj, 713, 1073

\bibitem[{{Schmidt} {et~al.}(1993){Schmidt}, {Kirshner}, {Huchra}, \&
  {Brodie}}]{Schmidt_iauc5882}
{Schmidt}, B., {Kirshner}, R., {Huchra}, J., \& {Brodie}, J. 1993, \iaucirc,
  5882, 3

\bibitem[{{Schmidt} {et~al.}(1994){Schmidt}, {Kirshner}, {Leibundgut}, {Wells},
  {Porter}, {Ruiz-Lapuente}, {Challis}, \& {Filippenko}}]{Schmidt/etal:1994}
{Schmidt}, B.~P., {Kirshner}, R.~P., {Leibundgut}, B., {Wells}, L.~A.,
  {Porter}, A.~C., {Ruiz-Lapuente}, P., {Challis}, P., \& {Filippenko}, A.~V.
  1994, \apjl, 434, L19

\bibitem[{{Schmidt} {et~al.}(1989){Schmidt}, {Weymann}, \&
  {Foltz}}]{Schmidt/etal:1989}
{Schmidt}, G.~D., {Weymann}, R.~J., \& {Foltz}, C.~B. 1989, \pasp, 101, 713

\bibitem[{{Silverman} \& {Filippenko}(2012)}]{Silverman/Filippenko:2012}
{Silverman}, J.~M. \& {Filippenko}, A.~V. 2012, arXiv:1202.3788 [astro-ph.CO]

\bibitem[{{Silverman} {et~al.}(2012{\natexlab{a}}){Silverman}, {Foley},
  {Filippenko}, {Ganeshalingam}, {Barth}, {Chornock}, {Griffith}, {Kong},
  {Lee}, {Leonard}, {Matheson}, {Miller}, {Steele}, {Barris}, {Bloom}, {Cobb},
  {Coil}, {Desroches}, {Gates}, {Ho}, {Jha}, {Kandrashoff}, {Li}, {Mandel},
  {Modjaz}, {Moore}, {Mostardi}, {Papenkova}, {Park}, {Perley}, {Poznanski},
  {Reuter}, {Scala}, {Serduke}, {Shields}, {Swift}, {Tonry}, {Van Dyk}, {Wang},
  \& {Wong}}]{Silverman/etal:2012a}
{Silverman}, J.~M., et~al.\ 2012{\natexlab{a}}, arXiv:1202.2128 [astro-ph.CO]

\bibitem[{{Silverman} {et~al.}(2012{\natexlab{b}}){Silverman}, {Ganeshalingam},
  {Li}, \& {Filippenko}}]{Silverman/etal:2012c}
{Silverman}, J.~M., {Ganeshalingam}, M., {Li}, W., \& {Filippenko}, A.~V.
  2012{\natexlab{b}}, arXiv:1202.2130 [astro-ph.CO]

\bibitem[{{Stanishev} {et~al.}(2007){Stanishev}, {Goobar}, {Benetti}, {Kotak},
  {Pignata}, {Navasardyan}, {Mazzali}, {Amanullah}, {Garavini}, {Nobili},
  {Qiu}, {Elias-Rosa}, {Ruiz-Lapuente}, {Mendez}, {Meikle}, {Patat},
  {Pastorello}, {Altavilla}, {Gustafsson}, {Harutyunyan}, {Iijima},
  {Jakobsson}, {Kichizhieva}, {Lundqvist}, {Mattila}, {Melinder}, {Pavlenko},
  {Pavlyuk}, {Sollerman}, {Tsvetkov}, {Turatto}, \&
  {Hillebrandt}}]{Stanishev/etal:2007}
{Stanishev}, V., et~al.\ 2007, \aap, 469, 645

\bibitem[{{Stritzinger} {et~al.}(2010){Stritzinger}, {Burns}, {Phillips},
  {Folatelli}, {Krisciunas}, {Kattner}, {Persson}, {Boldt}, {Campillay},
  {Contreras}, {Krzeminski}, {Morrell}, {Salgado}, {Freedman}, {Hamuy},
  {Madore}, {Roth}, \& {Suntzeff}}]{Stritzinger/etal:2010}
{Stritzinger}, M., et~al.\ 2010, \aj, 140, 2036

\bibitem[{{Stritzinger} {et~al.}(2011){Stritzinger}, {Phillips}, {Boldt},
  {Burns}, {Campillay}, {Contreras}, {Gonzalez}, {Folatelli}, {Morrell},
  {Krzeminski}, {Roth}, {Salgado}, {DePoy}, {Hamuy}, {Freedman}, {Madore},
  {Marshall}, {Persson}, {Rheault}, {Suntzeff}, {Villanueva}, {Li}, \&
  {Filippenko}}]{Stritzinger/etal:2011}
{Stritzinger}, M.~D., et~al.\ 2011, \aj, 142, 156

\bibitem[{{Strolger} {et~al.}(2002){Strolger}, {Smith}, {Suntzeff}, {Phillips},
  {Aldering}, {Nugent}, {Knop}, {Perlmutter}, {Schommer}, {Ho}, {Hamuy},
  {Krisciunas}, {Germany}, {Covarrubias}, {Candia}, {Athey}, {Blanc},
  {Bonacic}, {Bowers}, {Conley}, {Dahl{\'e}n}, {Freedman}, {Galaz}, {Gates},
  {Goldhaber}, {Goobar}, {Groom}, {Hook}, {Marzke}, {Mateo}, {McCarthy},
  {M{\'e}ndez}, {Muena}, {Persson}, {Quimby}, {Roth}, {Ruiz-Lapuente},
  {Seguel}, {Szentgyorgyi}, {von Braun}, {Wood-Vasey}, \&
  {York}}]{Strolger/etal:2002}
{Strolger}, L.-G., et~al.\ 2002, \aj, 124, 2905

\bibitem[{{Tanaka} {et~al.}(2008){Tanaka}, {Mazzali}, {Benetti}, {Nomoto},
  {Elias-Rosa}, {Kotak}, {Pignata}, {Stanishev}, \&
  {Hachinger}}]{Tanaka/etal:2008}
{Tanaka}, M., et~al.\ 2008, \apj,  677, 448

\bibitem[{{Tanaka} {et~al.}(2006){Tanaka}, {Mazzali}, {Maeda}, \&
  {Nomoto}}]{Tanaka/etal:2006}
{Tanaka}, M., {Mazzali}, P.~A., {Maeda}, K., \& {Nomoto}, K. 2006, \apj, 645,
  470

\bibitem[{{Taubenberger} {et~al.}(2011){Taubenberger}, {Benetti}, {Childress},
  {Pakmor}, {Hachinger}, {Mazzali}, {Stanishev}, {Elias-Rosa}, {Agnoletto},
  {Bufano}, {Ergon}, {Harutyunyan}, {Inserra}, {Kankare}, {Kromer},
  {Navasardyan}, {Nicolas}, {Pastorello}, {Prosperi}, {Salgado}, {Sollerman},
  {Stritzinger}, {Turatto}, {Valenti}, \&
  {Hillebrandt}}]{Taubenberger/etal:2011}
{Taubenberger}, S., et~al.\ 2011, \mnras, 412, 2735

\bibitem[{{Taubenberger} {et~al.}(2008){Taubenberger}, {Hachinger}, {Pignata},
  {Mazzali}, {Contreras}, {Valenti}, {Pastorello}, {Elias-Rosa},
  {B{\"a}rnbantner}, {Barwig}, {Benetti}, {Dolci}, {Fliri}, {Folatelli},
  {Freedman}, {Gonzalez}, {Hamuy}, {Krzeminski}, {Morrell}, {Navasardyan},
  {Persson}, {Phillips}, {Ries}, {Roth}, {Suntzeff}, {Turatto}, \&
  {Hillebrandt}}]{Taubenberger/etal:2008}
{Taubenberger}, S., et~al.\ 2008, \mnras, 385, 75

\bibitem[{{Thomas} {et~al.}(2007){Thomas}, {Aldering}, {Antilogus}, {Aragon},
  {Bailey}, {Baltay}, {Baron}, {Bauer}, {Buton}, {Bongard}, {Copin}, {Gangler},
  {Gilles}, {Kessler}, {Loken}, {Nugent}, {Pain}, {Parrent}, {P{\'e}contal},
  {Pereira}, {Perlmutter}, {Rabinowitz}, {Rigaudier}, {Runge}, {Scalzo},
  {Smadja}, {Wang}, \& {Weaver}}]{SN2006D}
{Thomas}, R.~C., et~al.\ 2007, \apjl, 654, L53

\bibitem[{{Thomas} {et~al.}(2011{\natexlab{a}}){Thomas}, {Aldering},
  {Antilogus}, {Aragon}, {Bailey}, {Baltay}, {Bongard}, {Buton}, {Canto},
  {Childress}, {Chotard}, {Copin}, {Fakhouri}, {Gangler}, {Hsiao},
  {Kerschhaggl}, {Kowalski}, {Loken}, {Nugent}, {Paech}, {Pain}, {Pecontal},
  {Pereira}, {Perlmutter}, {Rabinowitz}, {Rigault}, {Rubin}, {Runge}, {Scalzo},
  {Smadja}, {Tao}, {Weaver}, {Wu}, {(The Nearby Supernova Factory)}, {Brown},
  \& {Milne}}]{Thomas/etal:2011}
{Thomas}, R.~C., et~al.\ 2011{\natexlab{a}}, \apj, 743, 27

\bibitem[{{Thomas} {et~al.}(2011{\natexlab{b}}){Thomas}, {Nugent}, \&
  {Meza}}]{SYNAPPS}
{Thomas}, R.~C., {Nugent}, P.~E., \& {Meza}, J.~C. 2011{\natexlab{b}}, \pasp,
  123, 237

\bibitem[{{Tripp}(1998)}]{Tripp:1998}
{Tripp}, R. 1998, \aap, 331, 815

\bibitem[{{Turatto} {et~al.}(1996){Turatto}, {Benetti}, {Cappellaro},
  {Danziger}, {Della Valle}, {Gouiffes}, {Mazzali}, \&
  {Patat}}]{Turatto/etal:1996}
{Turatto}, M., {Benetti}, S., {Cappellaro}, E., {Danziger}, I.~J., {Della
  Valle}, M., {Gouiffes}, C., {Mazzali}, P.~A., \& {Patat}, F. 1996, \mnras,
  283, 1

\bibitem[{{Turatto} {et~al.}(1998){Turatto}, {Piemonte}, {Benetti},
  {Cappellaro}, {Mazzali}, {Danziger}, \& {Patat}}]{Turatto/etal:1998}
{Turatto}, M., {Piemonte}, A., {Benetti}, S., {Cappellaro}, E., {Mazzali},
  P.~A., {Danziger}, I.~J., \& {Patat}, F. 1998, \aj, 116, 2431

\bibitem[{{Valentini} {et~al.}(2003){Valentini}, {Di Carlo}, {Massi}, {Dolci},
  {Arkharov}, {Larionov}, {Pastorello}, {Di Paola}, {Benetti}, {Cappellaro},
  {Turatto}, {Pedichini}, {D'Alessio}, {Caratti o Garatti}, {Li Causi},
  {Speziali}, {Danziger}, \& {Tornamb{\'e}}}]{Valentini/etal:2003}
{Valentini}, G., et~al.\ 2003, \apj, 595, 779

\bibitem[{{Wade} \& {Horne}(1988)}]{Wade/Horne:1988}
{Wade}, R.~A. \& {Horne}, K. 1988, \apj, 324, 411

\bibitem[{{Walker} {et~al.}(2011){Walker}, {Hook}, {Sullivan}, {Howell},
  {Astier}, {Balland}, {Basa}, {Bronder}, {Carlberg}, {Conley}, {Fouchez},
  {Guy}, {Hardin}, {Pain}, {Perrett}, {Pritchet}, {Regnault}, {Rich},
  {Aldering}, {Fakhouri}, {Kronborg}, {Palanque-Delabrouille}, {Perlmutter},
  {Ruhlmann-Kleider}, \& {Zhang}}]{Walker/etal:2011}
{Walker}, E.~S., et~al.\ 2011, \mnras, 410, 1262

\bibitem[{{Wang} {et~al.}(2003){Wang}, {Baade}, {H{\"o}flich}, {Khokhlov},
  {Wheeler}, {Kasen}, {Nugent}, {Perlmutter}, {Fransson}, \&
  {Lundqvist}}]{WangL/etal:2003}
{Wang}, L., et~al.\ 2003, \apj, 591, 1110

\bibitem[{{Wang} {et~al.}(2009{\natexlab{a}}){Wang}, {Filippenko},
  {Ganeshalingam}, {Li}, {Silverman}, {Wang}, {Chornock}, {Foley}, {Gates},
  {Macomber}, {Serduke}, {Steele}, \& {Wong}}]{WangX/etal:2009b}
{Wang}, X., et~al.\ 2009{\natexlab{a}}, \apjl, 699, L139

\bibitem[{{Wang} {et~al.}(2009{\natexlab{b}}){Wang}, {Li}, {Filippenko},
  {Foley}, {Kirshner}, {Modjaz}, {Bloom}, {Brown}, {Carter}, {Friedman},
  {Gal-Yam}, {Ganeshalingam}, {Hicken}, {Krisciunas}, {Milne}, {Silverman},
  {Suntzeff}, {Wood-Vasey}, {Cenko}, {Challis}, {Fox}, {Kirkman}, {Li}, {Li},
  {Malkan}, {Moore}, {Reitzel}, {Rich}, {Serduke}, {Shang}, {Steele}, {Swift},
  {Tao}, {Wong}, \& {Zhang}}]{WangX/etal:2009a}
{Wang}, X., et~al.\ 2009{\natexlab{b}}, \apj, 697, 380

\bibitem[{{Wang} {et~al.}(2008){Wang}, {Li}, {Filippenko}, {Krisciunas},
  {Suntzeff}, {Li}, {Zhang}, {Deng}, {Foley}, {Ganeshalingam}, {Li}, {Lou},
  {Qiu}, {Shang}, {Silverman}, {Zhang}, \& {Zhang}}]{WangX/etal:2008a}
{Wang}, X., et~al.\ 2008, \apj, 675, 626

\bibitem[{{Webbink}(1984)}]{Webbink:1984}
{Webbink}, R.~F. 1984, \apj, 277, 355

\bibitem[{{Wells} {et~al.}(1994){Wells}, {Phillips}, {Suntzeff}, {Heathcote},
  {Hamuy}, {Navarrete}, {Fernandez}, {Weller}, {Schommer}, {Kirshner},
  {Leibundgut}, {Willner}, {Peletier}, {Schlegel}, {Wheeler}, {Harkness},
  {Bell}, {Matthews}, {Filippenko}, {Shields}, {Richmond}, {Jewitt}, {Luu},
  {Tran}, {Appleton}, {Robson}, {Tyson}, {Guhathakurta}, {Eder}, {Bond},
  {Potter}, {Veilleux}, {Porter}, {Humphreys}, {Janes}, {Williams}, {Costa},
  {Ruiz}, {Lee}, {Lutz}, {Rich}, {Winkler}, \& {Tyson}}]{Wells/etal:1994}
{Wells}, L.~A., et~al.\ 1994, \aj, 108, 2233

\bibitem[{{Wood-Vasey} {et~al.}(2008){Wood-Vasey}, {Friedman}, {Bloom},
  {Hicken}, {Modjaz}, {Kirshner}, {Starr}, {Blake}, {Falco}, {Szentgyorgyi},
  {Challis}, {Blondin}, {Mandel}, \& {Rest}}]{Wood-Vasey/etal:2008}
{Wood-Vasey}, W.~M., et~al.\ 2008, \apj, 689, 377

\bibitem[{{Woosley} {et~al.}(2009){Woosley}, {Kerstein}, {Sankaran}, {Aspden},
  \& {R{\"o}pke}}]{Woosley/etal:2009}
{Woosley}, S.~E., {Kerstein}, A.~R., {Sankaran}, V., {Aspden}, A.~J., \&
  {R{\"o}pke}, F.~K. 2009, \apj, 704, 255

\bibitem[{{Yuan} {et~al.}(2008){Yuan}, {Quimby}, {Akerlof}, {Wheeler},
  {Odewahn}, \& {Terrazas}}]{Yuan_cbet1206}
{Yuan}, F., {Quimby}, R., {Akerlof}, C., {Wheeler}, J.~C., {Odewahn}, S., \&
  {Terrazas}, E. 2008, Central Bureau Electronic Telegrams, 1206, 1

\bibitem[{{Yuan} {et~al.}(2010){Yuan}, {Quimby}, {Wheeler}, {Vink{\'o}},
  {Chatzopoulos}, {Akerlof}, {Kulkarni}, {Miller}, {McKay}, \&
  {Aharonian}}]{Yuan/etal:2010}
{Yuan}, F., et~al.\ 2010, \apj, 715, 1338


\end{thebibliography}


\appendix

\section{ADDITIONAL TABLES}\label{sect:app}

\clearpage

\LongTables
\begin{landscape}
\begin{deluxetable}{l@{\hspace{-3.25cm}}crccccrrcccccc}
\tabletypesize{\scriptsize}
\tablewidth{0pt}
\tablecaption{Journal of Observations\label{tab:obs}}
\tablehead{
\colhead{UT Date\tablenotemark{a}} &
\colhead{HJD\tablenotemark{b}} &
\colhead{Phase\tablenotemark{c}} &
\colhead{Tel./Instr.\tablenotemark{d}} &
\colhead{Range\tablenotemark{e}} &
\colhead{Disp.\tablenotemark{f}} &
\colhead{Res.\tablenotemark{g}} &
\colhead{P.A.\tablenotemark{h}} &
\colhead{$|\Delta\Phi|$\tablenotemark{i}} &
\colhead{Air.\tablenotemark{j}} &
\colhead{Flux Std.\tablenotemark{k}} &
\colhead{See.\tablenotemark{l}} &
\colhead{Slit\tablenotemark{m}} &
\colhead{Exp.\tablenotemark{n}} &
\colhead{Observer(s)\tablenotemark{o}} \\
\colhead{} &
\colhead{} &
\colhead{(d)} &
\colhead{} &
\colhead{(\AA)} &
\colhead{(\AA/pix)} &
\colhead{(\AA)} &
\colhead{(\degr)} &
\colhead{(\degr)} &
\colhead{} &
\colhead{} &
\colhead{(\arcsec)} &
\colhead{(\arcsec)} &
\colhead{(s)} &
\colhead{}
}
\startdata
\multicolumn{15}{c}{\bf SN 1993ac} \\
1993-10-16.49 & 2449276.99 &    +6.9 &              MMTblue &  3150-8172 & 1.95 &   7-8 &    \nodata &       83.5 & 1.18 &      \nodata & \nodata &     1.5 &                  900 &      CF, ND, RW \\
1993-10-20.52 & 2449281.02 &   +10.8 &              MMTblue &  3567-7145 & 3.18 &   7-8 &    \nodata &       81.7 & 1.21 &      \nodata & \nodata &     5.0 &                  420 &         \nodata \\
\hline
\multicolumn{15}{c}{} \\[-.2cm]
\multicolumn{15}{c}{\bf SN 1993ae} \\
1993-12-13.11 & 2449334.61 &    @0.0 &              MMTblue & 3562-10536 & 6.29 &   7-8 &    \nodata &        0.0 & 1.24 &      \nodata & \nodata & \nodata &                  900 &              BS \\
\hline
\multicolumn{15}{c}{} \\[-.2cm]
\multicolumn{15}{c}{\bf SN 1994D} \\
1994-03-10.36 & 2449421.86 & $-$11.1 &              MMTblue &  3161-9061 & 1.92 &   7-8 &    \nodata &       79.7 & 1.09 &      \nodata & \nodata & \nodata &                  120 &      RK, PC, AR \\
1994-03-11.41 & 2449422.91 & $-$10.0 &                 FAST &  3703-7633 & 1.47 &   6-7 &       90.0 &       60.7 & 1.14 &      F67/F56 &     2-3 &     3.0 &                  300 &              ST \\
1994-03-13.31 & 2449424.81 &  $-$8.1 &                 FAST &  3701-7631 & 1.47 &   6-7 &       90.0 &       52.0 & 1.15 &      F67/F56 &       2 &     3.0 &                  600 &              ST \\
1994-03-15.36 & 2449426.86 &  $-$6.1 &              MMTblue &  3560-7964 & 1.90 &   7-8 &    \nodata &       79.7 & 1.09 &      \nodata & \nodata & \nodata &                  120 &    RC, JHuc, ST \\
1994-03-16.37 & 2449427.87 &  $-$5.1 &                 FAST &  3702-7635 & 1.47 &   6-7 &       90.0 &       81.3 & 1.23 &      F67/F56 &     1-2 &     3.0 &     600,2$\times$420 &              ST \\
1994-03-17.35 & 2449428.85 &  $-$4.1 &                 FAST &  3705-7635 & 1.47 &   6-7 &       90.0 &       80.4 & 1.09 &      F67/F56 &     1-2 &     3.0 &         2$\times$600 &              ST \\
1994-03-18.36 & 2449429.86 &  $-$3.1 &                 FAST &  3705-7635 & 1.47 &   6-7 &       90.0 &       87.5 & 1.09 &      F67/F56 &     2-3 &     3.0 &              420,480 &              ST \\
1994-03-21.35 & 2449432.85 &  $-$0.1 &                 FAST &  3704-7635 & 1.47 &   6-7 &       90.0 &       87.0 & 1.09 &      F67/F56 &     2-3 &     3.0 &     420,2$\times$480 &             JPe \\
1994-04-01.28 & 2449443.78 &   +10.8 &                 FAST &  3706-7636 & 1.47 &   6-7 &       90.0 &       65.0 & 1.11 &      F67/F56 &     2-3 &     3.0 &                  600 &        JPe, SMu \\
1994-04-03.36 & 2449445.86 &   +12.8 &                 FAST &  3706-7636 & 1.47 &   6-7 &       90.0 &       61.0 & 1.15 &      F67/F56 &       2 &     3.0 &                  660 &             JPe \\
1994-04-05.34 & 2449447.84 &   +14.8 &                 FAST &  3707-7636 & 1.47 &   6-7 &       90.0 &       71.1 & 1.12 &      F67/F56 &     2-3 &     3.0 &                  900 &             PBe \\
1994-04-07.42 & 2449449.92 &   +16.9 &                 FAST &  3708-7638 & 1.47 &   6-7 &       90.0 &       37.8 & 1.48 &      F67/F56 &     1-2 &     3.0 &              600,900 &             PBe \\
1994-04-10.26 & 2449452.76 &   +19.7 &                 FAST &  3706-7636 & 1.47 &   6-7 &       90.0 &       62.2 & 1.11 &      F67/F56 &     1-2 &     3.0 &                  900 &             JPe \\
1994-04-11.23 & 2449453.73 &   +20.7 &                 FAST &  3706-7636 & 1.47 &   6-7 &       90.0 &       54.2 & 1.14 &      F67/F56 &     1-2 &     3.0 &                  600 &             PBe \\
1994-04-30.32 & 2449472.82 &   +39.7 &                 FAST &  3708-7632 & 1.47 &   6-7 &      110.0 &       64.4 & 1.30 &      F67/F56 &     1-2 &     3.0 &                  900 &             PBe \\
1994-05-03.25 & 2449475.75 &   +42.6 &                 FAST &  3707-7631 & 1.47 &   6-7 &      110.0 &       77.8 & 1.11 &      F67/F56 &     1-2 &     3.0 &                 1800 &             JPe \\
1994-05-06.28 & 2449478.78 &   +45.7 &                 FAST &  3708-7632 & 1.47 &   6-7 &      110.0 &       75.3 & 1.20 &      F67/F56 &     1-2 &     3.0 &                 1200 &             PBe \\
1994-05-10.16 & 2449482.66 &   +49.5 &                 FAST &  3707-7631 & 1.47 &   6-7 &      110.0 &       30.2 & 1.13 &      F67/F56 &     1-2 &     3.0 &                 1800 &             JPe \\
1994-05-15.20 & 2449487.70 &   +54.6 &                 FAST &  3721-7645 & 1.47 &   6-7 &      110.0 &       73.1 & 1.10 &      F67/F56 &     2-3 &     3.0 &                  600 &             SMu \\
1994-05-16.18 & 2449488.68 &   +55.5 &                 FAST &  3710-7634 & 1.47 &   6-7 &      110.0 &       58.8 & 1.09 &      F67/F56 &     2-3 &     3.0 &         2$\times$600 &             SMu \\
1994-05-17.17 & 2449489.67 &   +56.5 &                 FAST &  3739-7666 & 1.47 &   6-7 &      110.0 &       58.5 & 1.09 &      F67/F56 &     2-3 &     3.0 &                  600 &             SMu \\
1994-05-19.18 & 2449491.68 &   +58.5 &                 FAST &  3748-7671 & 1.47 &   6-7 &      110.0 &       58.6 & 1.09 &      F67/F56 &     2-3 &     3.0 &                 1200 &             JPe \\
1994-06-02.18 & 2449505.68 &   +72.5 &                 FAST &  3797-7721 & 1.47 &   6-7 &      110.0 &       87.4 & 1.12 &      F67/F56 &     2-3 &     3.0 &                 1200 &             JPe \\
1994-06-04.22 & 2449507.72 &   +74.5 &                 FAST &  3801-7724 & 1.47 &   6-7 &      110.0 &       65.0 & 1.29 &      F67/F56 &     2-3 &     3.0 &                  900 &             SMu \\
1994-06-12.23\tablenotemark{p} & 2449515.73 &   +82.5 &              MMTblue &  3217-8572 & 1.91 &   7-8 &    \nodata &       77.1 & 1.50 &      \nodata & \nodata &     2.0 &         2$\times$900 &          RK, PC \\
1995-11-24.46 & 2450045.96 &  +611.2 &              MMTblue &  3167-8170 & 1.95 &   7-8 &    \nodata &       76.4 & 1.23 &      \nodata & \nodata & \nodata &                 1200 &         \nodata \\[-.25cm]
\enddata
\tablecomments{Table~\ref{tab:obs} is published in its entirety in the electronic edition of The Astronomical Journal. A portion is shown here for guidance regarding its form and content.}
\tablenotetext{a}{UT at midpoint of observation(s).}
\tablenotetext{b}{Heliocentric Julian date at midpoint of observation(s).}
\tablenotetext{c}{Rest-frame phase of spectrum in days relative to $B$-band maximum. For \sneia\ with no reliable estimate for the time of maximum, we indicate the rest-frame days relative to the first spectrum preceded by an ``@'' symbol}
\tablenotetext{d}{Telescope and instrument used for this spectrum:
FAST~=~FLWO 1.5\,m+FAST,
IMACS~=~Magellan Baade+IMACS,
LDSS2~=~Magellan Clay+LDSS2,
LDSS3~=~Magellan Clay+LDSS3,
MMTblue~=~MMT+Blue Channel,
MMTred~=~MMT+Red Channel.}
\tablenotetext{e}{Observed wavelength range of spectrum.}
\tablenotetext{f}{Spectral dispersion in \AA\ per pixel.}
\tablenotetext{g}{Approximate FWHM spectral resolution in \AA.}
\tablenotetext{h}{Observed position angle during the observation(s).}
\tablenotetext{i}{Absolute difference between the observed position angle and the average parallactic angle over the course of the observation(s).}
\tablenotetext{j}{Airmass of the observation.}
\tablenotetext{k}{Standard stars: 
BD17~=~BD+17$^{\circ}$4708,
BD26~=~BD+26$^{\circ}$2606,
BD28~=~BD+28$^{\circ}$4211,
BD33~=~BD+33$^{\circ}$2642,
CD32~=~CD-32~9927,
EG131~=~EG~131,
EG274~=~EG~274,
F15~=~Feige~15,
F25~=~Feige~25,
F34~=~Feige~34,
F56~=~Feige~56,
F66~=~Feige~66,
F67~=~Feige~67,
F110 ~=~Feige~110,
G191~=~G191B2B,
H102~=~Hiltner~102,
H600~=~Hiltner~600,
HD19~=~HD~192281,
HD21~=~HD~217086,
HD84~=~HD~84937,
HZ44~=~HZ~44,
HZ14~=~HZ~14,
L3218~=~LTT~3218,
L3864~=~LTT~3864,
L4816~=~LTT~4816,
vMa2~=~van Maanen 2.
}
\tablenotetext{l}{Seeing is based upon estimates by the observers.}
\tablenotetext{m}{Spectroscopic slit width.}
\tablenotetext{n}{Exposure time. Separate exposures are indicated.}
\tablenotetext{o}{Observers:
EA~=~E.~Adams,
VA~=~V.~Antoniou,
HA~=~H.~Arce,
JA~=~J.~P.~Anderson,
ZB~=~Z.~Balog,
PBa~=~P.~Barmby,
EB~=~E.~Barton,
JB~=~J.~Battat,
PBe~=~P.~Berlind,
GB~=~G.~Bernstein,
WBl~=~W.~P.~Blair,
SB~=~S.~Blondin,
AB~=~A.~E.~Bragg,
CB~=~C.~Brice\~no,
WBr~=~W.~Brown,
NC~=~N.~Caldwell,
MC~=~M.~L.~Calkins,
BC~=~B.~J.~Carter,
PC~=~P.~Challis,
JC~=~J.~R.~Cho,
CC~=~C.~Clemens,
ACo~=~A.~Cody,
ACr~=~A.~Crook,
TC~=~T.~Currie,
RC~=~R.~M.~Cutri,
KD~=~K.~Dendy,
AD~=~A.~Diamond-Stanic,
ND~=~N.~Dinshaw,
JDon~=~J.~L.~Donley,
JDow~=~J.~J.~Downes,
KE~=~K.~Eriksen,
GE~=~G.~Esquerdo,
EF~=~E.~E.~Falco,
RF~=~R.~Fesen,
CF~=~C.~B.~Foltz,
JF~=~J.~Foster,
JGa~=~J.~Gallagher,
AG~=~A.~Garg,
PG~=~P.~M.~Garnavich,
IG~=~I.~Ginsburg,
JGr~=~J.~Graves,
NG~=~N.~Grogin,
TG~=~T.~Groner,
VH~=~V.~Hradecky,
HH~=~H.~Hao,
CHei~=~C.~Heinke,
CHel~=~C.~Heller,
JHe~=~J.~Hernandez,
MH~=~M.~Hicken,
CHi~=~C.~Hill,
JHuc~=~J.~P.~Huchra,
JHug~=~J.~P.~Hughes,
CHu~=~C.~Hutcheson,
RH~=~R.~Hutchins,
RJ~=~R.~Jansen,
SJ~=~S.~Jha,
SKa~=~S.~J.~Kannapan,
SKe~=~S.~Kenyon,
RK~=~R.~P.~Kirshner,
DK~=~D.~M.~Koranyi,
JK~=~J.~Kuraszkiewicz,
HL~=~H.~Landt,
TL~=~T.~Lappin,
NL~=~N.~Lepore,
LM~=~L.~Macri,
JM~=~J.~A.~Mader,
AM~=~ A.~Mahdavi,
EM~=~E.~Mamajek,
SMa~=~S.~A.~Mao,
NM~=~N.~Martimbeau,
TM~=~T.~Matheson,
MM~=~M.~Modjaz,
FM~=~F.~Munshi,
SMu~=~S.~Muscarella,
GN~=~G.~Narayan,
PN~=~P.~Nutzman,
CP~=~C.~A.~Pantoja,
BP~=~B.~M.~Patten,
KP~=~K.~Penev,
JPe~=~J.~Peters,
WP~=~W.~Peters,
MP~=~M.~Phelps,
JPi~=~J.~Pi\~neda,
AR~=~A.~G.~Riess,
KR~=~K.~Rines,
BS~=~B.~P.~Schmidt,
MS~=~M.~Schr\"odter,
JS~=~J.~D.~Silverman,
IS~=~I.~Song,
ST~=~S.~Tokarz,
MT~=~M.~Torres,
CT~=~C.~Tremonti,
AV~=~A.~Vaz,
LW~=~L.~Wells,
MW~=~M.~Westover,
RW~=~R.~J.~Weymann.
}
\tablenotetext{p}{Spectra with $\sim400$\,\AA-wide gaps ($\sim$6150-6550\,\AA) between blue and red halves.}
\tablenotetext{q}{The first two standard stars were used to calibrate a spectrum taken with our standard grating tilt, while the third was used to calibrate a spectrum taken with a red tilt.}
\tablenotetext{r}{Spectra accidently ommitted from \cite{Matheson/etal:2008}.}
\tablenotetext{s}{Spectra strongly affected by dark-current problems following UV flashing, for which we have trimmed off a portion of the spectrum.}
\end{deluxetable}
\clearpage
\end{landscape}

\begin{deluxetable}{ll}
\tablewidth{0pt}
\tablecaption{\label{tab:litspec}References for spectroscopic data from the literature used in this paper.}
\tablehead{\colhead{SN} & \colhead{References}}
\startdata
1981B         & \citealt{Branch/etal:1983}        \\
1984A         & \citealt{Barbon/etal:1989}        \\
1986G         & \citealt{Phillips/etal:1987}, \citealt{Cristiani/etal:1992}\tablenotemark{a} \\
1989B         & \citealt{Wells/etal:1994}         \\
1990N         & \citealt{Leibundgut/etal:1991}\tablenotemark{b}, \citealt{Mazzali/etal:1993}\tablenotemark{a}, \citealt{Gomez/Lopez:1998}\tablenotemark{a} \\
1991M         & \citealt{Gomez/Lopez:1998}\tablenotemark{a}        \\
1991T         & \citealt{Jeffery/etal:1992}\tablenotemark{b}, \citealt{Schmidt/etal:1994}\tablenotemark{b}    \\
1991bg        & \citealt{Turatto/etal:1996}\tablenotemark{a}       \\
1992A         & \citealt{Kirshner/etal:1993}\tablenotemark{b}, \citealt{Mazzali/etal:1998}   \\
1994D         & \citealt{Patat/etal:1996}\tablenotemark{a}         \\
1996X         & \citealt{Salvo/etal:2001}\tablenotemark{a}         \\
1997cn        & \citealt{Turatto/etal:1998}\tablenotemark{a}       \\
1998bu        & \citealt{Cappellaro/etal:2001}\tablenotemark{a}    \\
1999aa        & \citealt{Garavini/etal:2004}\tablenotemark{a}      \\
1999ac        & \citealt{Garavini/etal:2005}\tablenotemark{a}, \citealt{Phillips/etal:2006}   \\
1999aw        & \citealt{Strolger/etal:2002}      \\
1999ee        & \citealt{Hamuy/etal:2002}\tablenotemark{a}         \\
2000E         & \citealt{Valentini/etal:2003}\tablenotemark{a}     \\
2000cx        & \citealt{Li/etal:2001b}\tablenotemark{a}           \\
2001ay        & \citealt{SN2001ay}                \\
2001el        & \citealt{WangL/etal:2003}\tablenotemark{a}, \citealt{Mattila/etal:2005}         \\
2002bo        & \citealt{Benetti/etal:2004}\tablenotemark{a}       \\
2002dj        & \citealt{SN2002dj}\tablenotemark{a}                \\
2002er        & \citealt{SN2002er}                \\
2003cg        & \citealt{SN2003cg}\tablenotemark{a}                \\
2003du        & \citealt{Anupama/Sahu/Jose:2005}\tablenotemark{a}, \citealt{Stanishev/etal:2007} \\
2003fg        & \citealt{Howell/etal:2006}        \\
2003hv        & \citealt{SN2003hv}                \\
2004S         & \citealt{Krisciunas/etal:2007}\tablenotemark{a}    \\
2004dt        & \citealt{Altavilla/etal:2007}     \\
2004eo        & \citealt{Pastorello/etal:2007b}\tablenotemark{a}   \\
2005bl        & \citealt{Taubenberger/etal:2008}\tablenotemark{a}  \\
2005cf        & \citealt{Garavini/etal:2007b}\tablenotemark{a}, \citealt{WangX/etal:2009a}       \\
2005cg        & \citealt{Quimby/etal:2006}\tablenotemark{a}        \\
2005hj        & \citealt{Quimby/etal:2007a}       \\
2005hk        & \citealt{Phillips/etal:2007}      \\
2006D         & \citealt{SN2006D}                 \\
2006X         & \citealt{WangX/etal:2008a}        \\
2006bt        & \citealt{SN2006bt}                \\
2006dd        & \citealt{Stritzinger/etal:2010}   \\
2006gz        & \citealt{SN2006gz}\tablenotemark{b}                \\
2006ot        & \citealt{Stritzinger/etal:2011}   \\
2007if        & \citealt{Scalzo/etal:2010}, \citealt{Yuan/etal:2010}            \\
2009dc        & \citealt{Taubenberger/etal:2011}        
\enddata
\tablenotetext{a}{Spectra dowloaded from the SUSPECT supernova archive.}
\tablenotetext{b}{Spectra available at the CfA Supernova Archive.}
\tablecomments{We do not include references to \snia\ spectra from our
  sample that have previously been published (see references in
  \S~\ref{sect:intro}).}
\end{deluxetable}

\end{document}